\documentclass[12pt,letterpaper,openright,oneside]{thesis}
\usepackage{natbib}
\usepackage{deluxetable} 
\usepackage[it,normalsize]{subfigure}
\usepackage{xspace}
\usepackage{amsmath}
\usepackage{setspace}
\usepackage{url}
\usepackage{color}
\usepackage{wrapfig2}

\bibpunct{(}{)}{;}{a}{}{,}%
\footnotesep \baselineskip
\newcommand{\sigT}{\mbox{$\sigma_{\mbox{\tiny T}}$}}
\newcommand{\Tcmb}{\mbox{$T_{\mbox{\tiny CMB}}$}}

\newcommand{\Lamee}{\mbox{$\Lambda_{ee}$}}
\newcommand{\Mgas}{\mbox{$M_{\mbox{\scriptsize gas}}$}}
\newcommand{\Mtot}{\mbox{$M_{\mbox{\scriptsize tot}}$}}
\newcommand{\Yint}{\mbox{$Y_{\mbox{\scriptsize int}}$}}
\newcommand{\Ylos}{\mbox{$Y_{\mbox{\scriptsize los}}$}}
\newcommand{\Yvol}{\mbox{$Y_{\mbox{\scriptsize vol}}$}}
\newcommand{\fgas}{\mbox{$f_{\mbox{\scriptsize gas}}$}}
\newcommand{\LCDM}{\mbox{$\Lambda$CDM}}

\begin{document}


\pagestyle{empty}      
\singlespace
\mbox{}   
\vspace{1.3in}
\begin{center}
{\Large The Sunyaev-Zel'dovich Array: \\ \medskip
\large Constraining a New Pressure Profile for Fitting SZE Observations of Galaxy Clusters\\ \medskip
}
\end{center}
\bigskip
\bigskip
\begin{center}
Tony Mroczkowski\\
~\\
Professor Amber Miller
\end{center}
\vspace{2.8in}
\begin{center}
Submitted in partial fulfillment of the\\
requirements for the degree\\
of Doctor of Philosophy\\
in the Graduate School of Arts and Sciences.
\end{center}
\bigskip
\bigskip
\bigskip
\begin{center}
COLUMBIA UNIVERSITY
\end{center}
\bigskip
\begin{center}
2009
\end{center}
 
\newpage
\mbox{}  
\vspace{7.in}
\begin{center}
\copyright 2009
\end{center}
\begin{center}
Tony Mroczkowski\\
All Rights Reserved
\end{center}
 
\doublespace  
\newpage
\begin{center}
{\bf Abstract}
\end{center}
\smallskip
\begin{center}
{\Large The Sunyaev-Zel'dovich Array: \\ \medskip
\large Constraining a New Pressure Profile for Fitting SZE Observations of Galaxy Clusters \\
}
Tony Mroczkowski
\end{center}

The Sunyaev-Zel'dovich Array (SZA), an eight element interferometer designed
to probe the Sunyaev-Zel'dovich effect (SZE) from galaxy clusters, which I 
helped construct and operate, is described here (Part I).  
I then use SZA observations to investigate the utility of 
a new, self-similar pressure profile for fitting SZE observations of galaxy
clusters (Part II). 

The SZA 30-GHz receiver system probes angular scales $\sim$ 1--5\arcmin.  
A model that can accurately describe a cluster's pressure profile 
over a correspondingly broad range of radii
is therefore required.  In the analysis presented here, I fit a
2-parameter, radial pressure profile, derived from simulations and
detailed X-ray analysis of relaxed clusters, to SZA observations of
three clusters with exceptionally high quality X-ray data.  From the
joint analysis of the SZE and X-ray data, I derive physical
properties of the cluster, such as gas and total mass, gas fraction 
and the integrated Compton $y$-parameter. 

The parameters derived from the joint fit to SZE+X-ray data
agree well with a detailed, independent, X-ray-only analysis of these
same clusters.  When combined with X-ray imaging data,
this new pressure profile yields an independent estimate of the electron temperature
profile that is in good agreement with spectroscopic X-ray determinations.
In addition to yielding relationships between cluster observables
and physical cluster properties, this model could prove to be a useful tool in 
helping to constrain the temperatures of high redshift clusters, for which X-ray spectroscopic 
data are difficult to obtain. 


\clearpage


\singlespace
\pagenumbering{roman} 
\pagestyle{plain} 
\tableofcontents
\listoftables
\listoffigures

\addtocontents{toc}{\protect\thispagestyle{plain}}
\addtocontents{lot}{\protect\thispagestyle{plain}}
\addtocontents{lof}{\protect\thispagestyle{plain}}
\clearpage
\doublespace

\begin{center}{\large Acknowledgements}\end{center}

I want to start off by saying this thesis has been an ``it takes a village'' 
effort, much more so than simply an individual's pursuit.
I think this is particularly true about those doing instrumentation and 
experimental astrophysics, unlike those in more traditional student--advisor 
hierarchies; our experience is necessarily broad in range, and the 
cultivation of a decent experimentalist requires a great effort by many.
Throughout graduate school, I have sought and received advice and guidance 
from so many people, all of whom are experts in their fields.

First, of course, I would like to thank my advisor Amber Miller, 
for her guidance and diligence in helping bring this work to fruition.  
Without her persistence and insistence upon understanding everything on 
a fundamental level, this work could not possibly have become the 
complete tome it is today.

I also thank Frits Paerels and Caleb Scharf, who first got me interested in 
cluster astrophysics, and who were always there when I had questions, doubts, 
results to share, or just needed someone to put science in perspective.  

I thank everyone at the Owens Valley Radio Observatory, 
all of whom also advised me and helped me keep my sanity.  
I especially thank David Woody, James Lamb, Ben Reddall, and 
Dave Hawkins for all their useful direction and for bringing me into their lives
at an otherwise remote and alienating place; I know I often was stubborn and 
unreceptive to their advice, but their persistence has helped to form me into 
the scientist I am today.
I also thank Cecil Patrick, who helped make OVRO fun 
and gave me plenty of advice on life and cooking; along with Cecil, I also
thank his dog Clementine just for being who she is.
And of course I thank Dianne Shirley and the girls --  Xena, Katie, Aukee, 
and Minnie -- for all the wild times in Bishop, California.

I thank everyone in the Sunyaev-Zel'dovich Array (SZA) Collaboration.
I especially thank Marshall Joy and Max Bonamente for helping me understand the 
many levels of cluster analysis required for this work.
I thank Daisuke Nagai for all his patient conversations that guided me in 
testing these analysis routines on simulated clusters and helped produce 
relevant models for fitting cluster data.
I thank Erik Leitch for all his useful discussions that helped me understand
radio interferometry, and for all his help with computers.
I thank John Carlstrom, who always raised good questions and was fun to work
with in the field.
And finally, I thank my fellow students in the SZA -- particularly Matthew Sharp
and Stephen Muchovej -- who helped make this more fun than it often should have 
been.

And while those in science have been so crucial to the completion of this work, 
I especially thank those who were there for me just because they like me.  
Foremost, I thank my wife, Natalia Holstein, who has put up with a lot
(I'm sorry I have to be out of town so much to do this stuff).  This 
would not have been possible without her love, support, and welcome 
distractions.  I thank her mother too, for being curious and brave enough to ask
questions about my work.  I thank my friends -- particularly Brian Cleary, 
Nathaniel Stern, Dave Spiegel, and Paul Kim -- who kept things interesting 
and convinced me I was smart enough to get this done.
And I thank my family, who helped make me what I am today.

Last, but not least,\footnote{Well, as a dog, she is the smallest creature 
being thanked.} I would like to thank Nina for being the only one who never 
questioned the value of this work, and for always waiting patiently by my 
desk throughout.

\clearpage


\doublespace
\pagenumbering{arabic}
\pagestyle{myheadings}


\chapter{Introduction \label{sec-intro}}

\indent ``What's so amazing that keeps us stargazing? What do we think we might see?'' 
-- Kermit the Frog ({\it The Rainbow Connection}, written by Paul Williams and Kenneth Ascher)

\section{Clusters of Galaxies}\label{clustercosmology}

The detailed expansion history of the Universe and the growth of large scale 
structure are two of the most important topics in cosmology.  
Clusters of galaxies are the largest gravitationally-bound systems in the Universe, 
and thus provide a unique handle on cosmic expansion and structure formation.
Measurements of the growth of structure, traceable using large cluster surveys, 
provide critical clues to the nature and abundance of dark matter and dark energy.\footnote{
For recent reviews of how cluster studies can be used to constrain cosmology, see 
the Dark Energy Task Force (DETF) report \citep{albrecht2006} and \citet{rapetti2007}.}

At the time of this writing, a low density, cold dark matter (CDM) cosmology, dominated by
a dark energy or cosmological constant ($\Lambda$) component, is heavily favored.\footnote{It is unknown what this
dark energy component is, whether it truly acts as a ``cosmological constant,'' or whether
it evolves over time.}  This is called \LCDM\ cosmology, where `$\Lambda$' refers to dark energy's
contribution, expressed as $\Omega_\Lambda$, to the total energy density of the Universe, $\Omega$.  Cold dark matter is a form of 
non-baryonic matter travelling much slower than the speed of light, which -- along with the baryonic
component $\Omega_{\rm b}$ -- provides matter's energy density contribution $\Omega_{\rm M}$.
In general, any quantity $\Omega_x$ is the ratio of $x$'s energy density to the critical energy 
density it would take to close the Universe.

\LCDM\ cosmology started to become the favored cosmology in the 1990's, when it displaced
the then-favored -- and now inappropriately-named -- Standard Cold Dark Matter (SCDM) 
picture of the Universe.  SCDM holds that $\Omega_{\rm M} = \Omega_\text{tot} = 1$ ($\Omega_\Lambda = 0$).
\LCDM\ cosmology reconciles two results that cannot be resolved in the SCDM paradigm:
the flatness of the Universe combined with the low value for the universal matter density $\Omega_{\rm M}$.

In the mid-1990's, X-ray gas mass measurements began to demonstrate that $\Omega_{\rm M} < 1$.\footnote{
Optical measurements of the stellar and total masses of galaxies and galaxy clusters 
also showed this, though I focus here on the dominant baryonic component of the largest
collapsed structures in the Universe -- the X-ray emitting gas in galaxy clusters.}
Because galaxy clusters collapse out of representatively large, comoving volumes of the Universe ($\sim 10$~Mpc), 
we assume their baryonic/dark matter ratio approaches the universal value.  
Using X-ray measurements of galaxy clusters, one can obtain both the hot gas mass (\Mgas) and an
estimate of the cluster total mass \Mtot, and use these to compute the gas mass fraction
$\fgas \equiv \Mgas/\Mtot$.\footnote{How \fgas\ is obtained from X-ray and X-ray+SZE observations 
is discussed in \S \ref{fgas}.}
Provided that the bulk of a cluster's baryons are in the hot gas probed by 
X-ray observations, we can approximately equate the gas mass fraction 
$\fgas \approx \Omega_\text{gas}/\Omega_{\rm M} \sim \Omega_\text{b}/\Omega_{\rm M}$ 
(within a factor of $\sim 2$, this is well-supported by the simulations).

The combination of \fgas\ measurements with constraints from Cosmic Microwave Background (CMB) measurements and
Big Bang Nucleosynthesis (BBN) predictions yields an estimate for $\Omega_{\rm M}$.  
Since CMB measurements, coupled with BBN theory, independently constrain $\Omega_\text{b} h^2$ (where $h$ is the local 
Hubble constant divided by $100~\rm km ~ s^{-1} ~ Mpc^{-1}$), one can use cluster gas fraction measurements
to approximately constrain $\Omega_{\rm M}$.
By accounting for the composition of the Coma cluster -- adding up the observable mass in the hot gas and
optically luminous stars -- \citet{white1993a} argued that the measurements implied a low value for 
$\Omega_{\rm M}$ in this way. Later, \citet{david1995} applied this method to many more
clusters observed by \emph{ROSAT}, and used this to estimate that $\Omega_{\rm M} \sim$ 0.1--0.2.

Furthermore, X-ray cluster surveys -- such as those provided by the \emph{ROSAT} All-Sky Survey (\emph{RASS})
and the Wide Angle \emph{ROSAT} Pointed Survey (\emph{WARPS}) -- provided independent indications
that we live in a low density universe with $\Omega_{\rm M} \sim$ 0.2--0.3. 
This low density was inferred from the lack of strong cluster number evolution, which
would have been seen if $\Omega_{\rm M} = 1$ \citep[see, e.g.][]{mushotzky1997},
as the matter density strongly affects when clusters form 
\citep[see, e.g.][and references therein]{holder2000, haiman2001}.

Whilst cluster measurements consistently implied we live in a low density universe, direct evidence for dark energy 
was provided by those who set out to measure the deceleration of the Universe's expansion, using type Ia supernovae (SNIa) as 
``standard candles'' (objects with a known luminosity).  
Surprisingly, their measurements indicated that the expansion of Universe is accelerating \citep{riess1998}, implying
it is dominated by some form of dark energy. 
Soon after, measurements of the CMB made by the \emph{BOOMERanG}, \emph{TOCO}, 
and \emph{MAXIMA} 
strongly constrained the Universe to be spatially flat, a result that has been confirmed by WMAP and other CMB 
measurements since. With no curvature component ($\Omega_{\rm k} = 0$), ``flatness'' means that the sum of the 
angles within a triangle is $180^\circ$, and that $\Omega_{\rm M} + \Omega_\Lambda = 1$.  

With strong evidence now in place for \LCDM, and many more recent results confirming this, 
a ``concordance cosmology'' with  $\Omega_\Lambda=0.7$, $\Omega_{\rm M}=0.3$, and $\Omega_{\rm k} = 0$ can be defined.
This cosmology is assumed for most of the results presented in this thesis, and is in agreement with the parameters published 
in the first year results of the Wilkinson Microwave Anisotropy Probe \citep[][\emph{WMAP}]{spergel2003}, which 
jointly fit data from a number of independent tests (e.g. their own and finer scale probes of the CMB, weak lensing, cluster counts, 
galactic velocity field, and HST's Key Project) described in \citet{spergel2003}. 
Defining a ``concordance cosmology'' provides a convenient framework for both observers and theorists.  As 
cosmological constraints are refined, any results assuming 
this cosmology can be scaled to reflect the updated parameters.\footnote{This assumes, 
of course, that new results remain consistent with \LCDM.} 

Recently, a number of studies have used clusters alone to constrain the dark energy component of the Universe
\citep[see][and references therein]{allen2004, laroque2006, allen2007, vikhlinin2008}.  This is done primarily
using X-ray measurements of the intracluster medium (ICM), the hot ($\gtrsim 10^7~\rm K$) gas that comprises 
the majority of the baryons in a galaxy cluster.  Assuming a theoretically-motivated 
functional form for the gas fraction, one can solve for the cosmological parameters that 
force the data to fit the expected $\fgas(z)$ evolution for redshift $z$.  

In this thesis, I explore the joint constraints provided by two independent probes of the ICM, using these 
to measure \fgas, \Mgas, \Mtot, and other cluster astrophysical properties.  In the next section, I describe
the primary tool used here to constrain the properties of the ICM -- the Sunyaev-Zel'dovich effect.

\section{The Sunyaev-Zel'dovich Effect}\label{sze}

\begin{figure}
\centerline{\includegraphics[width=3in]{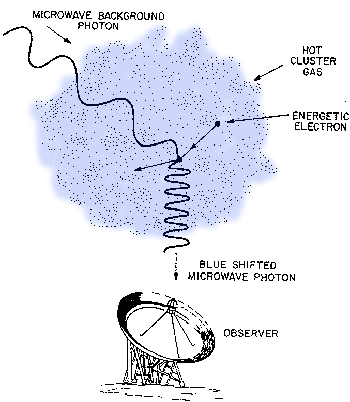}}
\caption{Image adapted from \citet{vanspeybroeck1999}.}
\label{fig:sze_schem}
\end{figure}

The Sunyaev-Zel'dovich effect (SZE) is a unique probe of the hot gas in clusters, since
it does not depend on emission processes, unlike X-ray emission or the broad majority of
luminous astrophysical processes routinely measured by astronomers.
The SZE arises by inverse Compton scattering of CMB photons off the hot electrons in the ICM
(depicted schematically in Fig.~\ref{fig:sze_schem}).  
This leaves a spectral signature on the CMB that is independent of redshift, 
since at higher redshifts the CMB is both denser in photon number and less redshifted in energy.\footnote{
For a given state of the electrons in the ICM at any redshift, the same number fraction of CMB photons are 
inverse Compton scattered by the same fraction of the photon energy.  Therefore, the $(1+z)^{4}$ dependence 
in how the energy of the CMB is redshifted, which also maintains its blackbody spectrum, ensures
the SZE spectral signature on the primary CMB is constant relative to the CMB.}

There are two separable components to the Sunyaev-Zel'dovich effect: the kinetic SZ effect,
or KSZ, due to the line-of-sight proper motions of clusters, and the thermal SZ effect
(TSZ, simply called `SZE' in the chapters that follow).  
I do not discuss the KSZ here, since we could not measure it with the Sunyaev-Zel'dovich Array 
(SZA, the instrument presented in this thesis; see Chapter~\ref{sza}).  
The thermal SZE is measurable as a spectral distortion of the CMB, computed
\begin{equation}
\label{eq:deltaT}
\frac{\Delta T}{\Tcmb}  = f(x) \, y,
\end{equation}
where  $\Delta T$ is the temperature distortion of the CMB, which has temperature \Tcmb, $y$ is called the ``Compton $y$ parameter''
and is integrated along the line of sight, and $f(x)$ (given in Eqs.~\ref{eq:fx} and \ref{eq:x_fx}) contains the SZE frequency dependence.

The line-of-sight Compton $y$ parameter is computed
\begin{equation}
\label{eq:compy}
y = \frac{k_B \, \sigT}{m_e c^2} \int \! n_e(\ell) \, T_e(\ell) \,d\ell,
\end{equation}
where $k_B$ is Boltzmann's constant, \sigT\ is the Thomson scattering cross-section of the electron,
$m_e$ is the mass of an electron, $c$ is the speed of light, and $n_e(\ell)$ and $T_e(\ell)$ are 
the electron density and temperature along sight line $\ell$.
The Compton $y$ parameter has a linear dependence on electron pressure $P_e$ when the ideal gas law, $P_e=k_B \, n_e \, T_e$, is assumed:
\begin{equation}
\label{eq:compy_Pe}
y = \frac{\sigT}{m_e c^2} \int \! P_e(\ell) \, d\ell.
\end{equation}

\begin{figure}
\begin{center}
\includegraphics[width=5in]{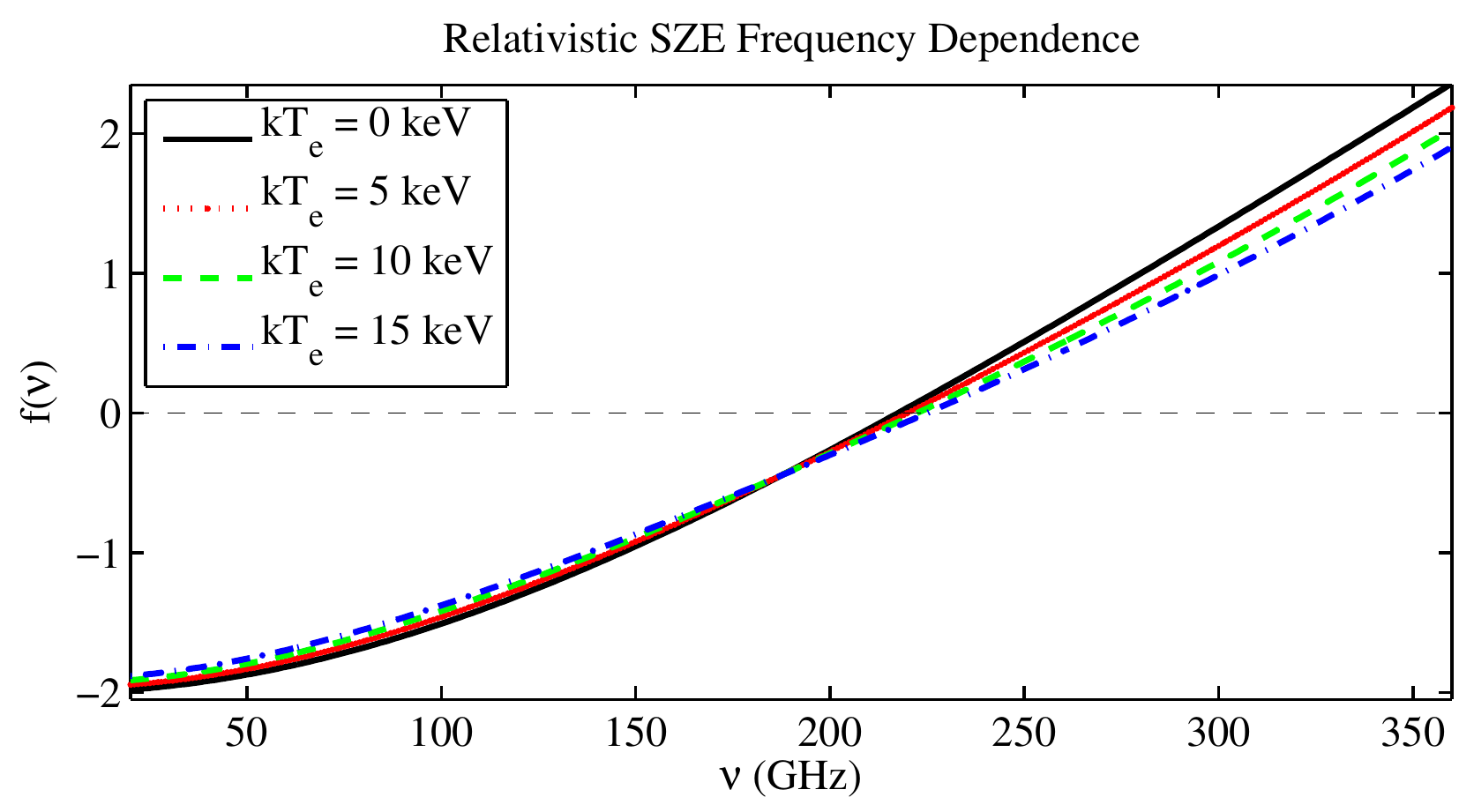}
\includegraphics[width=5.25in]{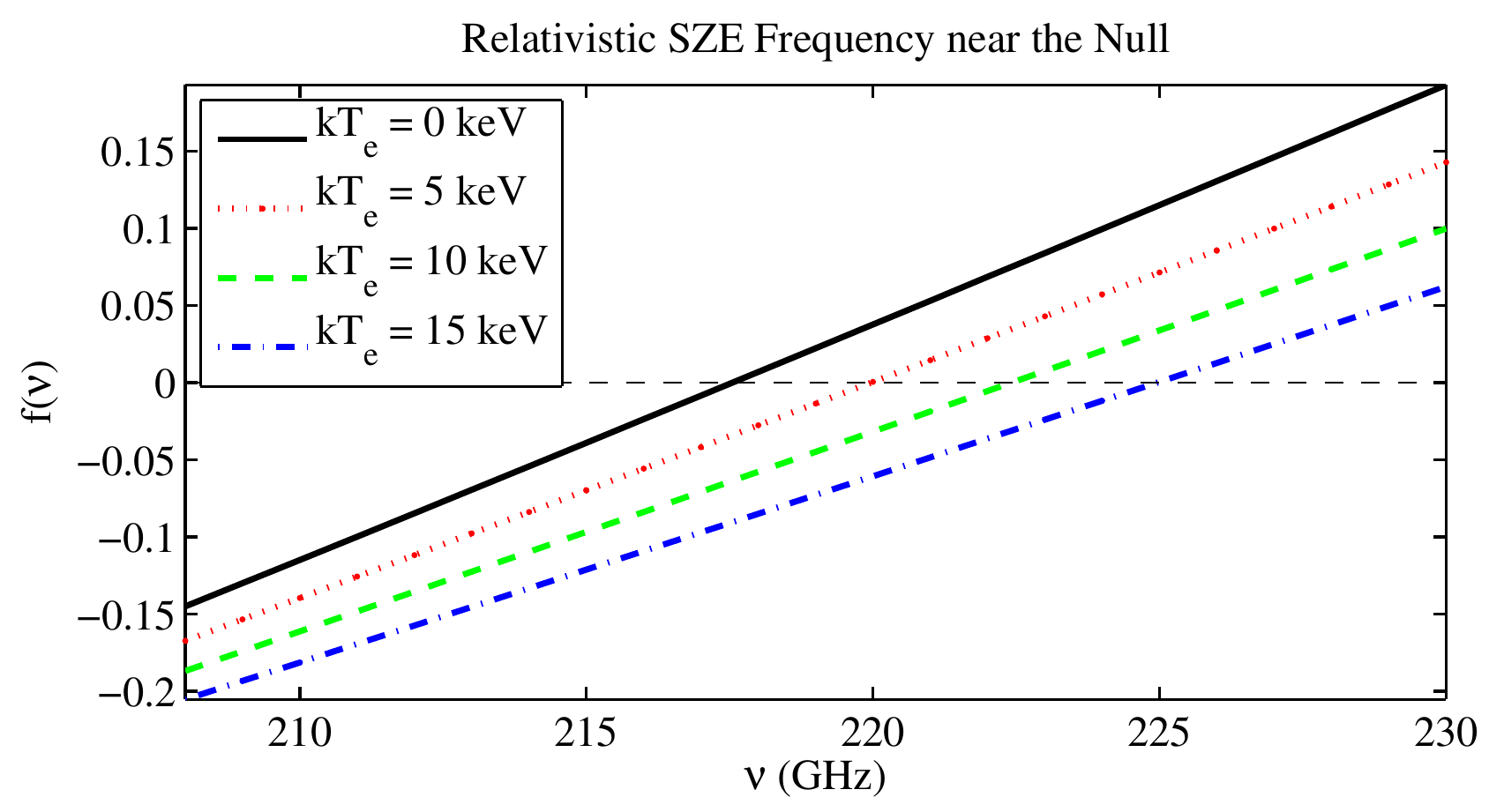}
\end{center}
\caption{SZE spectral dependence $f(\nu)$, plotted as a function of frequency (see Eq.~\ref{eq:x_fx}).  
The upper panel shows $f(\nu)$ over a broad range of frequencies, while
the lower panel shows a detail of the null in the SZE spectrum.
The relativistically-corrected $f(\nu)$ for a cluster with temperature $kT_e = 0~\rm keV$ (black, solid line) 
reduces precisely to the classical frequency dependence (Eq.~\ref{eq:fx}), since
electrons with no temperature are not moving at relativistic random velocities. The other lines
show how the relativistically-corrected $f(\nu)$ departs from the classical behavior for higher temperature electrons.
The relativistic corrections to $f(\nu)$ shown here are computed out to fifth-order using the equations provided in 
\citet{itoh1998}.  
The classical SZE spectrum has a null at $\nu \approx$ 217.5~GHz,
above which the SZE signal becomes an increment.
Higher temperature electrons require relativistic corrections \citep{itoh1998} to the classical
SZE frequency dependence, which shift the null to higher frequencies.  
A high temperature cluster would have a non-negligible
thermal SZ effect at the classical null (217.5~GHz). }
\label{fig:sz_fx}
\end{figure}

\begin{figure}
\centerline{\includegraphics[width=4in]{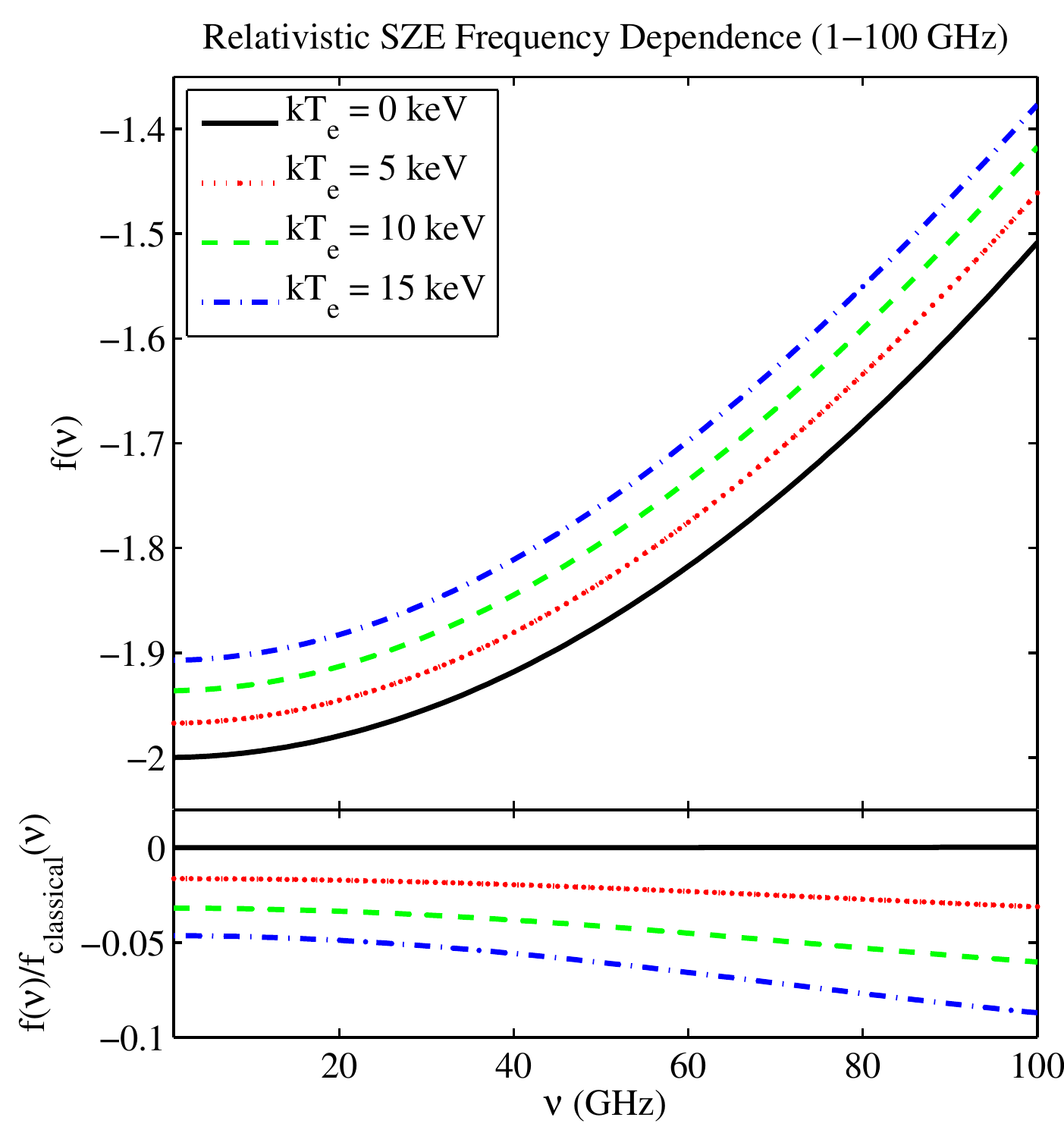}}
\caption{SZE frequency dependence below 100~GHz.  See Fig.~\ref{fig:sz_fx} for caption.
Below the null (see lower panel of Fig.~\ref{fig:sz_fx}), we often use the term ``SZE decrement'' 
to refer to the strength of the thermal SZ effect, which is negative at low frequencies.  
The lower panel shows the fractional deviation from the classical SZE for clusters at 
higher temperatures, due to the relativistic electron velocities in the ICM.  
Treating the thermal SZE at 30~GHz from a massive 10~keV cluster as classical introduces a 
$\approx - 3.6 \%$ bias in quantities derived from the SZE fits (i.e. the line-of-sight electron 
pressure in Eq.~\ref{eq:compy} would be underestimated by this amount, since the strength of the SZE 
would be overestimated by the classical calculation).}
\label{fig:sz_fx_100GHz}
\end{figure}

The classical, non-relativistic frequency dependence $f(x)$ of the SZE is
\begin{equation}
\label{eq:fx}
f(x) = x \left(\frac{e^x + 1}{e^x - 1}\right) - 4 
\end{equation}
where the dimensionless frequency $x$ is
\begin{equation}
\label{eq:x_fx}
x = h \nu / k_B \Tcmb.
\end{equation}
Here $h$ is Planck's constant, and $\nu$ is the frequency of the observation.
The classical spectral dependence $f(x)$ given in Eq.~\ref{eq:fx}, and relativistic corrections to it --
necessary due to the high thermal velocities of electrons in the ICM -- are plotted in the upper panel of Fig.~\ref{fig:sz_fx} 
over a broad range of frequencies.  
Here I used the calculations of \citet{itoh1998}, which correct the SZE $f(x)$ out to fifth-order.  
The SZE decrement $\lesssim 218~\rm GHz$ becomes an increment at 
$\gtrsim 218~\rm GHz$.  A detail of this crossover is shown in the lower panel of Fig.~\ref{fig:sz_fx}, which shows that
the relativistic velocities of the electrons, due to the cluster's temperature, shift the precise location of SZE null.
Finally, Fig.~\ref{fig:sz_fx_100GHz} shows a detail of the SZE decrement at the frequencies
the SZA can probe.

\section{Interferometry Overview}\label{interf}

\begin{figure}
\centerline{\includegraphics[width=4.5in]{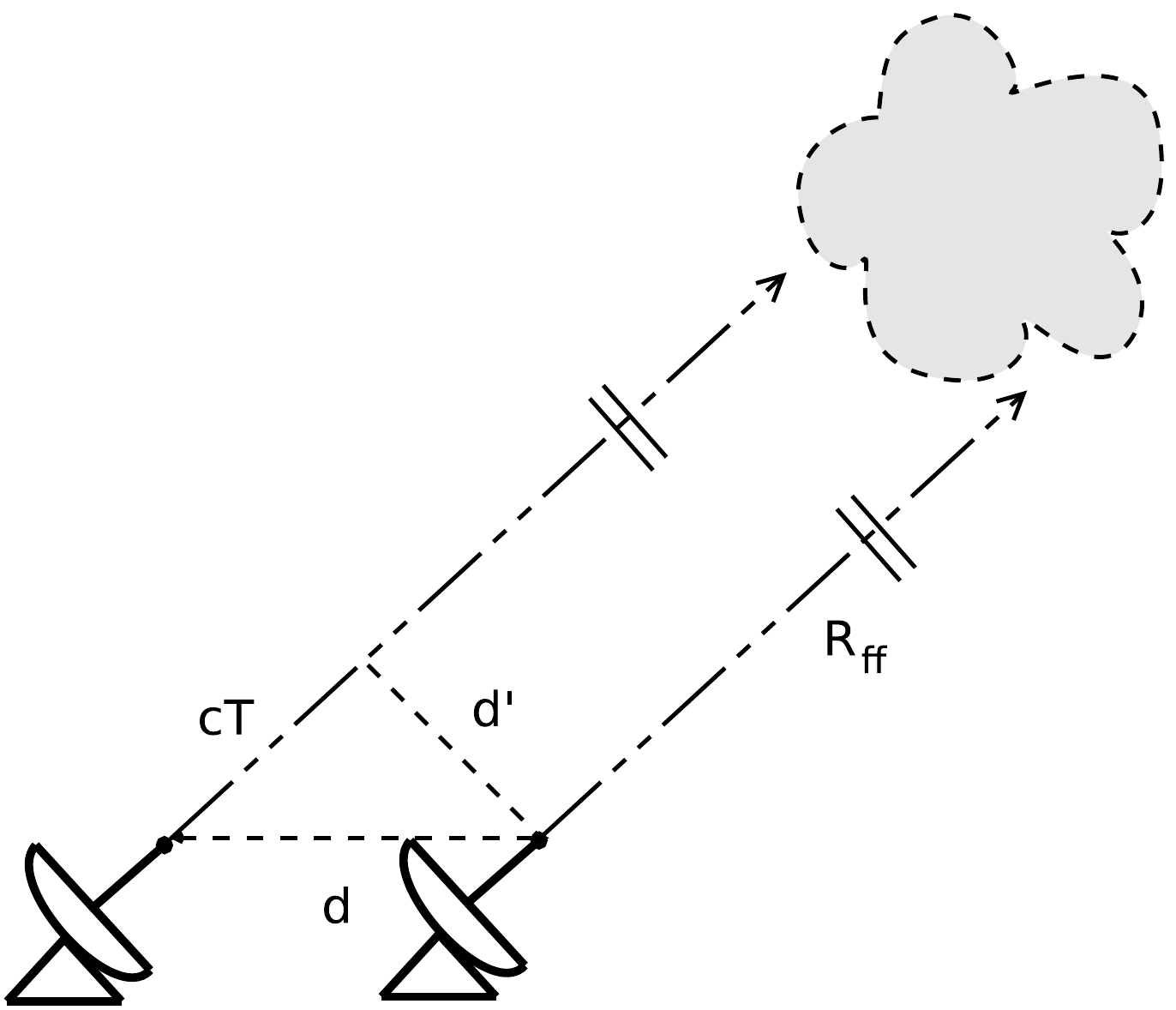}}
\caption{Example baseline formed by a pair of antennae.  The dish separation 
is $d$, while the projected baseline as seen by the source is $d^\prime$. 
The distance to the source (represented by a cloud) is labeled $R_{ff}$.
Note the scale of the broken lines, representing this astronomical distance $R_{ff}$,
is not accurate; in reality, the lines of sight are nearly parallel and point at the same 
location on the source. The time delay $T$ is equal to the difference in distances to the source,
from each antenna, divided by the speed of light $c$.  
By delaying the signal measured by the antenna on the right by time $T$, we ensure that the same astronomical 
wavefront is used in the correlation of the two signals.  Tracking an astronomical
source requires both the physical pointing of each antenna toward the source in the far
field, and the implementation of instrumental delays ($T$) that ensure
each antenna measures the same wave front as the Earth rotates.}
\label{fig:baseline3}
\end{figure}

Monochromatic light from an arbitrarily-shaped, spatially-incoherent aperture propagates via 
Fraunhofer diffraction to become the spatial Fourier transform of the light's intensity pattern, 
after traveling a distance many times its wavelength.
Radio astronomical interferometry takes advantage of this simple fact, where the astronomical source
serves as the arbitrary aperture.  An interferometric array measures the Fourier transform of the 
source's spatial intensity distribution, probing angular scales $\lambda/d^\prime$ (in radians) 
for wavelength $\lambda$ and projected baseline separation $d^\prime$ (see Fig.~\ref{fig:baseline3}).

\subsection{Criteria and Assumptions for Interferometry}

The conditions necessary to take advantage of astronomical interferometry constitute the
``van Cittert-Zernike theorem'' \citep[see][for a derivation and many further details]{thompson2001}.
I summarize the necessary criteria here:
\begin{itemize} 
\item The source must be in the far field, meaning that its distance $R_{ff} \gg (d^\prime)^2 / \lambda$ 
(illustrated in Fig.~\ref{fig:baseline3}),
where $d^\prime$ is the longest projected baseline in the array for a given observation,
and $\lambda$ is the wavelength at which the observation is performed.  
A large distance is essential for statistically independent emission (or scattering events, in the case
of the SZE) from the source to become coherent plane waves.
For the Sunyaev-Zel'dovich Array, this means we must observe
sources more than 360~km away, a condition met by any astronomically interesting source.\footnote{  
The Moon is $\sim 3.8 \times 10^5~\rm km$ away, and is much closer than the closest 
source we have observed, Mars.}  
In meeting the far field condition, for a sufficiently narrow band, we can treat incoming light from the 
source as a series of plane waves.  
\item As a corollary to the above, we assume there are no sources in the near field, so
we are only observing sources in the far field.
\item The source must be spatially incoherent, meaning that two arbitrary points 
within the source must not have statistically correlated emission. 
Since two points within a source are physically separated and entirely independent, 
emission from these two points will not in general be coherently emitted.
\end{itemize}

We also take advantage of one more approximation: the small angle approximation.  While this is 
not necessary for the van Cittert-Zernike theorem, it simplifies the analysis of interferometric 
observations, as it allow us to treat the data as a 2-D Fourier transform (discussed in the next section,
\S \ref{uvspace}). 

As I discuss later in this section, the largest radial scales probed by the SZA are on the order of 
$\sim 5\arcmin$ ($1.45 \times 10^{-3}$ radians).  
The small angle approximation is therefore well-justified for single, pointed observations made with the SZA
(as opposed to mosaicked observations).  
We therefore treat the spatial intensity pattern of the source as if it were truly in a plane perpendicular to the 
line of sight. We call this the ``image plane,'' and justify this by pointing out that the sources we
observe are far enough away that the distance to any part of the source is negligibly
different from the radial distance to the source's center.  

At a given instant in time, all antennae must measure a single, monochromatic wavefront for the 
astronomical signals to be correlated.  
The signals are measured within spectral channels (a ``channel'' is a smaller range of frequencies within
some band) that are small compared to the central frequency of each band, assuring each signal is 
approximately monochromatic.\footnote{The SZA observes at sky frequencies $\sim$ 30 and 90~GHz, 
and breaks each of its sixteen $\sim$ 500~MHz bands into fifteen usable 31.25~MHz channels (for more details, 
see \S \ref{dcon}).  Each channel is therefore $\sim 1/1000$ the sky frequency.}
To ensure the condition that we are measuring a single plane wave front, the 
differing distances from each antenna to the source are corrected by adding (computationally) an adjustable
time delay to each antenna's signal.  
This adjustable delay is the difference in path lengths from the astronomical source to each antenna,
divided by the speed of light (labeled $T$ in Fig.~\ref{fig:baseline3}).
The path length $c T$ changes throughout the course of an observation, as the source traverses the sky.

In radio interferometry, computing and applying the proper delays to the signal from each antenna, as the Earth moves, 
is called ``fringe tracking.'' 
The Cartesian location $(x,y)=(0,0)$ (typically north-south and east-west angular offsets) is assigned
to the point in the sky called the ``pointing center.'' Mechanical tracking keeps this point (e.g. a cluster's center) 
in the center of each antenna's primary beam; the pointing center is also the ``phase center'' for the observation, since the adjustable delays 
$T$ are computed in order to precisely compensate for the different distances to this point.  These differing distances
are based simply on the geometry illustrated in Fig.~\ref{fig:baseline3}.



\subsection{Probing Sources in Fourier Space (\emph{u,v}-space)}\label{uvspace}

\begin{figure}
\centerline{\includegraphics{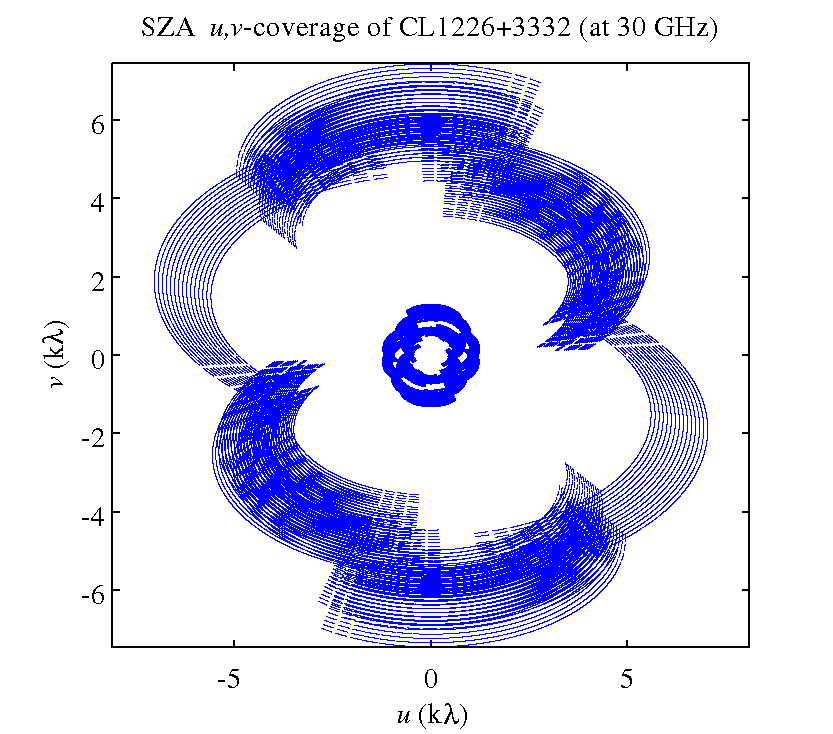}}
\caption{SZA \emph{u,v}-space coverage of CL1226.9+3332 (12:26:58.0, +33:32:45.0), a cluster 
that passes near zenith for the SZA (the SZA's latitude is $\sim 37^\circ$N).  Each blue point represents a data 
point's \emph{u,v}-space location,
using frequency independent units of $k\lambda$ (distance divided by wavelength, divided by 1000).
Note that there are 2 groupings of points: those $\sim$ 0.35--1.3~k$\lambda$ from the center, due
to the short baselines of the compact inner array, and those $\sim$ 3--7.5~k$\lambda$ from 
the center, due to the long baselines formed with the outer antennae (i.e. each baseline formed with
Antennae 6 or 7; the antenna layout of the SZA is shown in Fig.~\ref{fig:sza_ant_loc}).  
The SZA was designed to provide this broad, uniform coverage in \emph{u,v}-space on the shorter 
baselines, while simultaneously providing higher resolution coverage with longer baselines.
The data locations in \emph{u,v}-space exhibit inversion symmetry (i.e. $(u,v)=-(u,v)$) because the visibilities
are the transform of real data in image space.  The visibilities therefore exhibit Hermitian symmetry 
(i.e. $V_\nu(u,v)=V_\nu^*(-u,-v)$ in Eq.~\ref{eq:visibility}).}
\label{fig:uv_coverage_cl1226}
\end{figure}

\begin{figure}
\centerline{\includegraphics{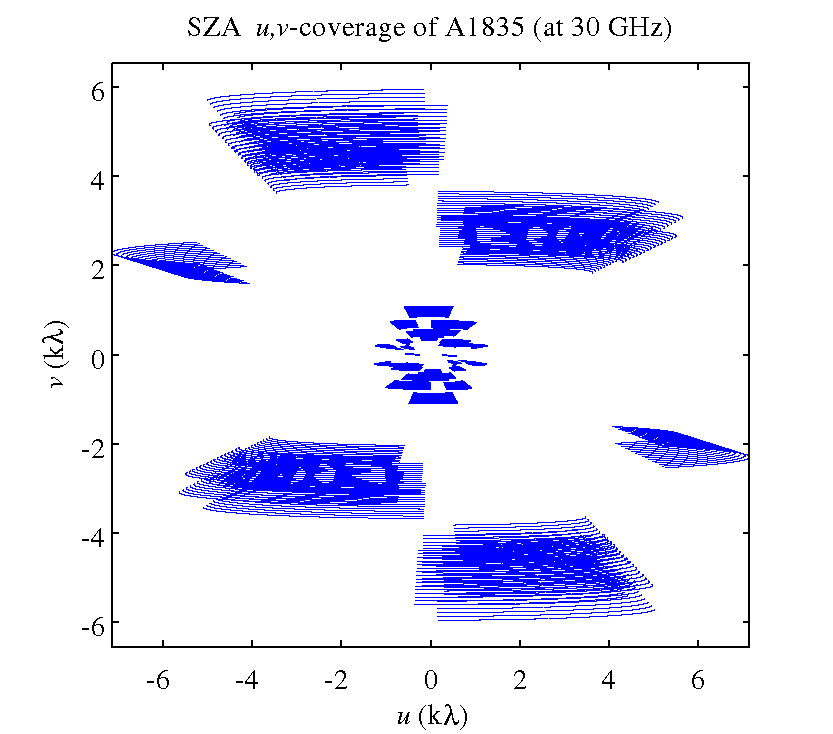}}
\caption{SZA \emph{u,v}-space coverage of A1835 (14:01:02.03, +02:52:41.71), a low declination cluster
(the SZA's latitude is $\sim 37^\circ$N).  
See Fig.~\ref{fig:uv_coverage_cl1226} for caption.  Note that the 
coverage provided by the inner baselines is nearly as complete as that in  Fig.~\ref{fig:uv_coverage_cl1226}.
The coverage on longer baselines, used to constrain the fluxes of sources known to be point-like 
(using independent radio surveys), is less complete.  
If a source is truly point-like, the \emph{u,v}-space coverage provided 
by the long baselines does not need to be complete to remove it, since 
a point source has the same magnitude of flux over all of \emph{u,v}-space 
(we generally do not use the SZA to determine whether a source is point-like, 
and need only constrain its flux.  See \S \ref{point_sources} for more details
about modeling unresolved radio sources.).}
\label{fig:uv_coverage_a1835}
\end{figure}

\begin{figure}
\centerline{\includegraphics[width=4.1in]{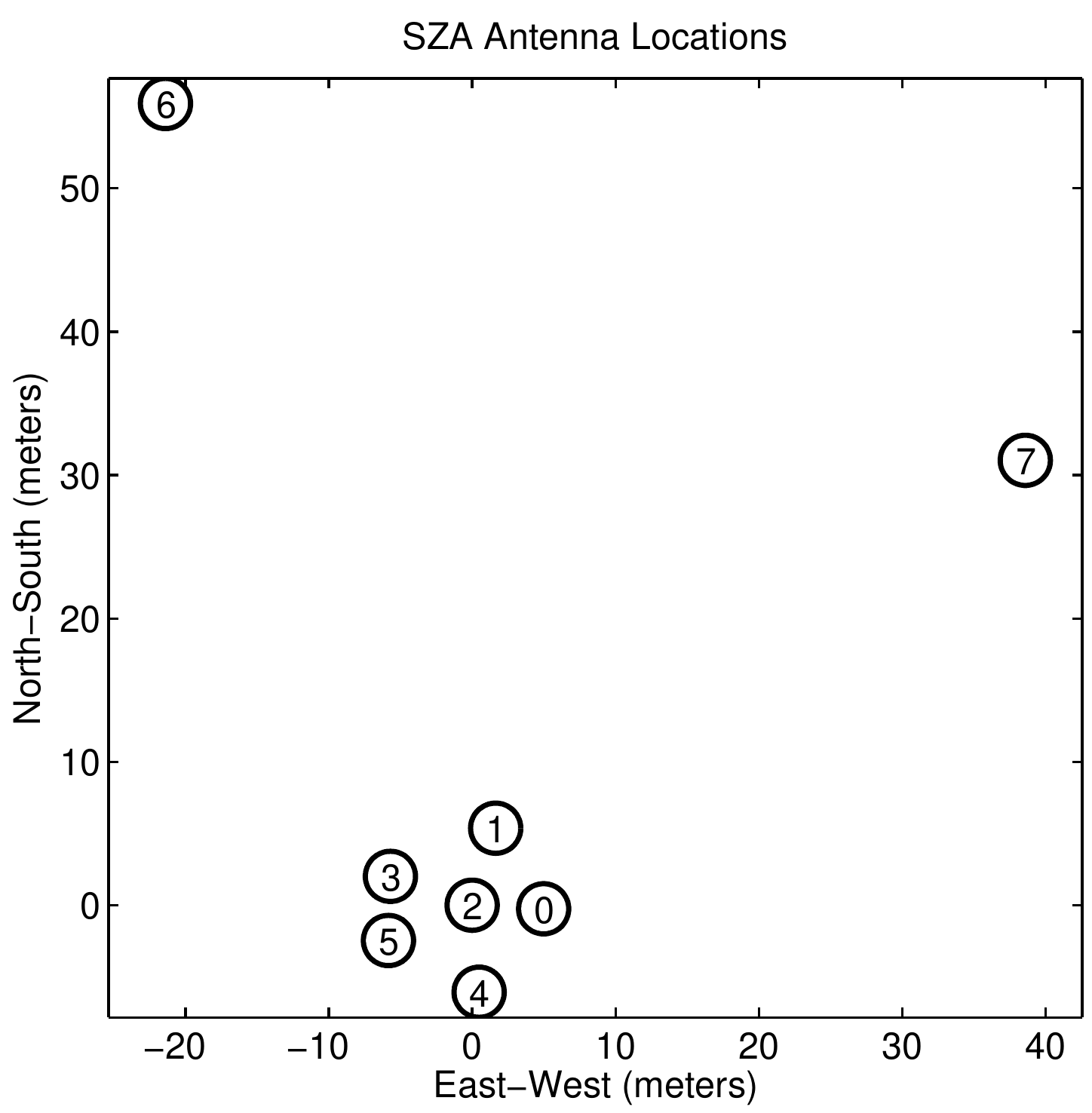}}
\caption{SZA Antenna Locations.  Antenna 2 is the reference antenna, and is therefore located 
at the origin.  Antennae 6 \& 7 provide 13 long baselines (between each other and paired with
each of the six inner antennae).  These long baselines probe small scales, thus aiding point source 
subtraction. Figures \ref{fig:uv_coverage_cl1226} \& \ref{fig:uv_coverage_a1835} show the \emph{u,v}-space
coverage of two sources observed with this array configuration.}
\label{fig:sza_ant_loc}
\end{figure}

\begin{figure}
\centerline{\includegraphics[width=6in]{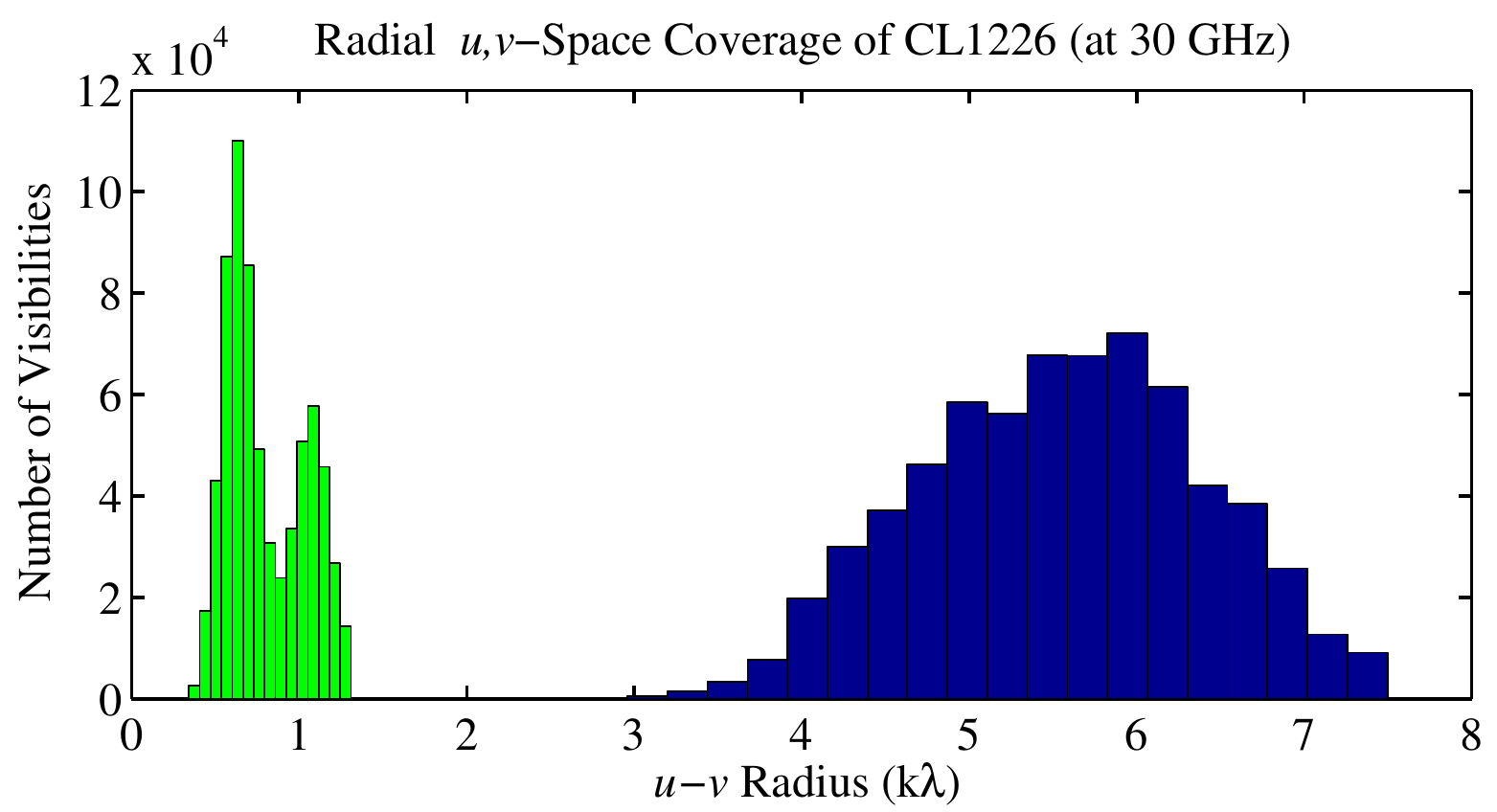}}
\caption{Radial distribution of scales probed in \emph{u,v}-space ($\sqrt{u^2 + v^2}$), 
for coverage shown in Fig.~\ref{fig:uv_coverage_cl1226}.
The distribution is extremely bimodal due to the short ($\sim$ 0.35--1.3~k$\lambda$, plotted in green) and long 
($\sim$ 3--7.5~k$\lambda$, plotted in blue) baselines of the SZA (shown in Fig.~\ref{fig:sza_ant_loc}).
}
\label{fig:uv_space_hist_CL1226}
\end{figure}

\begin{figure}
\centerline{\includegraphics[width=6in]{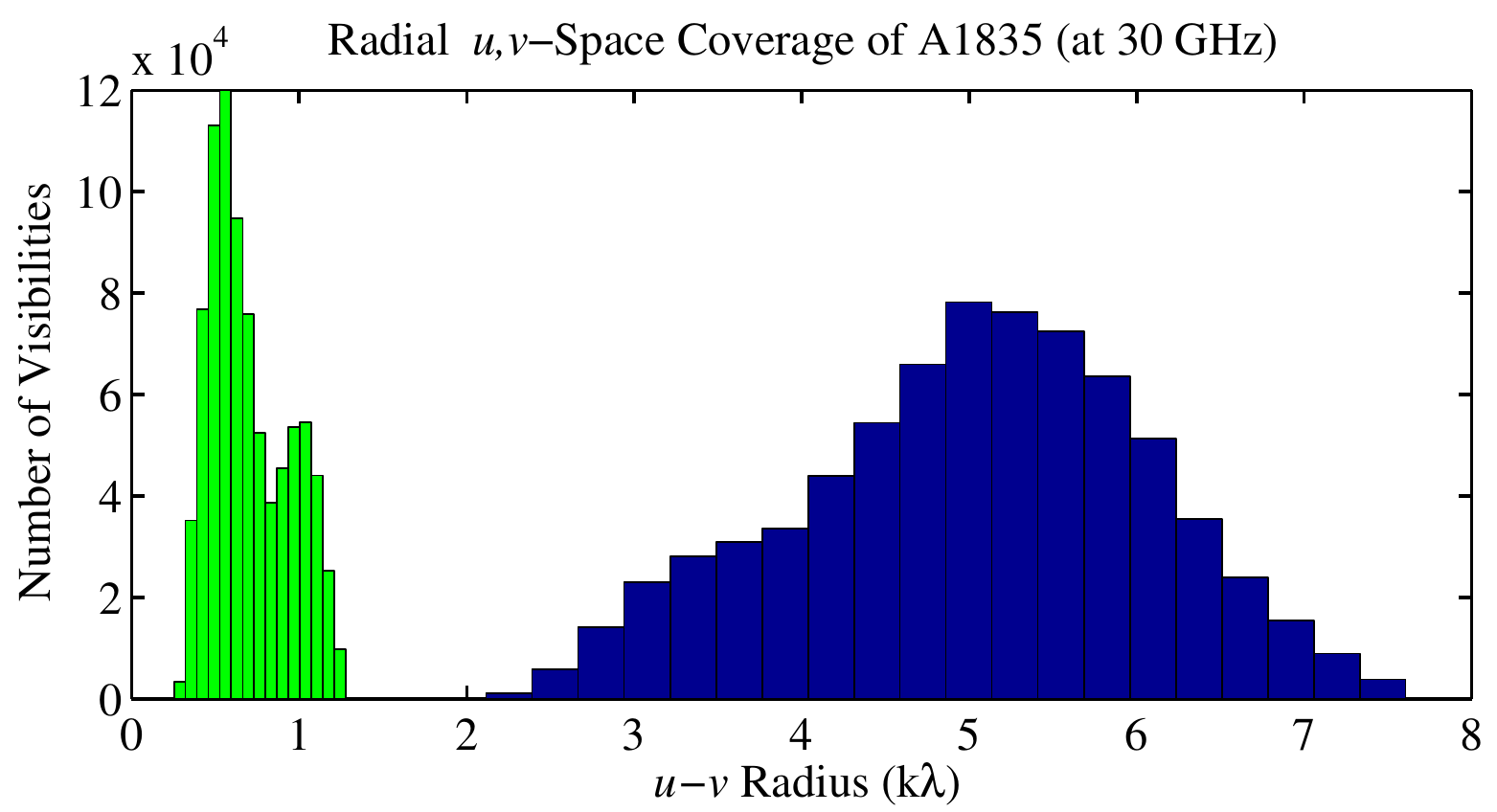}}
\caption{Radial distribution of scales probed in \emph{u,v}-space 
for coverage shown in Fig.~\ref{fig:uv_coverage_a1835}.
See Fig.~\ref{fig:uv_space_hist_CL1226} for details.
Because of the low declination of A1835, the projected baselines for this observation are shorter, on average,
than those for CL1226 (Fig.~\ref{fig:uv_space_hist_CL1226}).
}
\label{fig:uv_space_hist_A1835}
\end{figure}

In the context of astronomical interferometry, Fourier space is often referred to as ``\emph{u,v}-space,'' 
since $u$ and $v$ are the Fourier conjugates of image space coordinates $x$ and $y$.\footnote{Since we
are performing interferometry on sources that are far (compared to the observation's wavelength), 
we can ignore the $z$ spatial component along the line of sight, as well as its transform $w$.  
This is equivalent to stating that our sources lie in the image plane.}
When calibrated against an astronomical source of known flux, an interferometric array's output  
can be expressed as the flux (in Janskies\footnote{In S.I. units, $1~\rm Jy = 10^{-26}~\rm W \cdot m^{-2} \cdot Hz^{-1}$.}) 
at each point probed in \emph{u,v}-space, for each frequency probed by the instrument.  
For a given integration time (typically $\sim$ 20~sec for the SZA),
each baseline and band of the radio interferometric array produces one binned point in \emph{u,v}-space,
called a ``visibility.''\footnote{For 
the 28 baselines and 16 bands of the SZA, four hours of useful, 
on-source observation time produces $\approx 322,560$ independently measured visibilities. 
This is after the channels of each band are binned into one visibility per unit time.}
The visibilities behave collectively as a Fourier transform of the spatial intensity pattern $I_\nu(x,y)$ (flux per unit solid angle) 
at frequency $\nu$, and is calculated \citep{thompson2001}:
\begin{equation}
 V_\nu(u,v) = \int \! \int A_\nu(x,y) \, I_\nu(x,y) \, \frac{e^{- j 2 \pi (ux + vy)}}{\sqrt{1 - x^2 - y^2}} \, dx \, dy.
\label{eq:visibility}
\end{equation}
Here $A_\nu(x,y)$ is the spatial sensitivity of each antenna, the central lobe of which is called 
the primary beam (illustrated in Fig.~\ref{fig:baseline1}).  Each antenna's beam was mapped 
by Ryan Hennessy and Mike Loh, and was found to be well-approximated by a single, 
circularly-symmetric Gaussian with a half-power radius of $\sim 4.9\arcmin$. 
Because our observations are on the scales of arcminutes, we can make the approximation that 
$\sqrt{1 - x^2 - y^2} \approx 1$, where $x$ and $y$ in are radians.\footnote{For example, 
observations that probe a radial distance on the sky of 5\arcmin implies 
$\sqrt{x^2+y^2} \lesssim 0.0015$ radians, and therefore the term $\sqrt{1 - x^2 - y^2} \gtrsim 0.9985$. 
This justifies our use of the small angle approximation for single, targeted cluster 
observations used to probe arcminute scales.}  
Doing this, Eq.~\ref{eq:visibility} simplifies to the 2-D Fourier transform of the spatial intensity pattern
multiplied by the beam:
\begin{equation}
V_\nu(u,v) = \int \! \int A_\nu(x,y) \, I_\nu(x,y) \, {e^{- j 2 \pi (ux + vy)}} dx \, dy.
\label{eq:visibility2}
\end{equation}

\begin{figure}
\centerline{\includegraphics{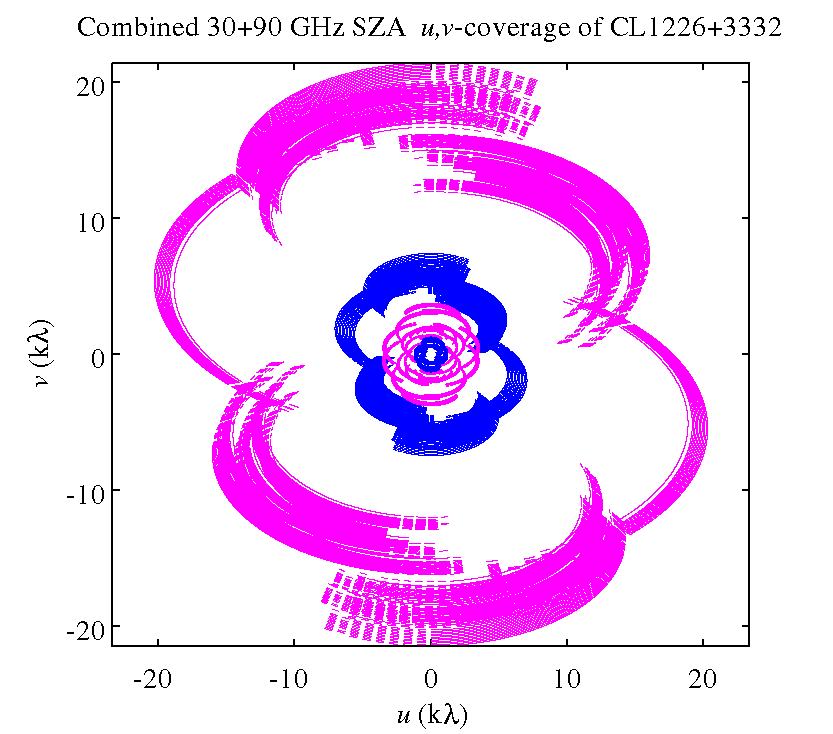}}
\caption{SZA \emph{u,v}-space coverage of CL1226.9+3332, combining observations from both the 30-GHz (blue) and
90-GHz (magenta) instruments. See Fig.~\ref{fig:uv_coverage_cl1226} for caption. Observations with the SZA using the 
90-GHz receivers (see \S \ref{sza_overview}), thus probe finer scales (larger \emph{u,v}-scales) than the 
30-GHz system for the same array configuration.}
\label{fig:uv_coverage_cl1226_30_90}
\end{figure}

\begin{figure}
\centerline{\includegraphics[width=6in]{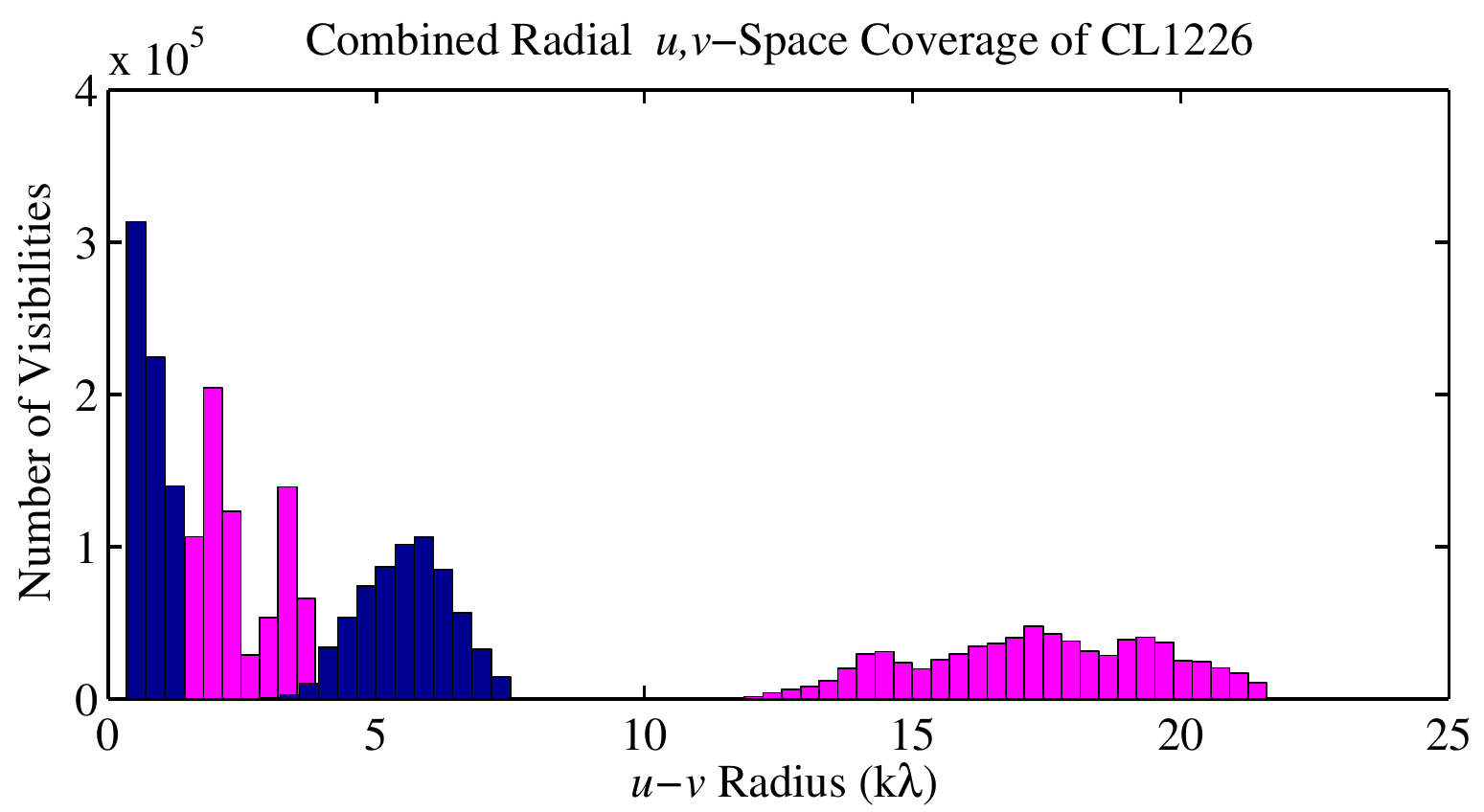}}
\caption{Radial distribution of scales probed in \emph{u,v}-space ($\sqrt{u^2 + v^2}$), 
for coverage shown in Fig.~\ref{fig:uv_coverage_cl1226_30_90}.
The $\sim$ 1.3--3~k$\lambda$ gap in the coverage at 30~GHz (shown in blue) is filled in by performing
complementary observations at 90~GHz (show in magenta).  For a given array configuration and observation length,
the 90-GHz \emph{u,v}-coverage can be obtained (approximately) by multiplying each \emph{u,v} coordinate 
in the 30-GHz \emph{u,v}-coverage by $\sim3$.
}
\label{fig:uv_space_hist_CL1226_30_90}
\end{figure}

Figures \ref{fig:uv_coverage_cl1226} \& \ref{fig:uv_coverage_a1835} show the \emph{u,v}-space
coverage of the SZA interferometer during two typical cluster observations.  This coverage
was provided using the array configuration shown in Fig.~\ref{fig:sza_ant_loc}, performing observations
with the 30-GHz receivers. Figures \ref{fig:uv_space_hist_CL1226} \& \ref{fig:uv_space_hist_A1835} respectively show the distributions 
of \emph{u,v}-radii probed by the \emph{u,v}-spacings plotted in Figs.~\ref{fig:uv_coverage_cl1226} \& \ref{fig:uv_coverage_a1835}.
The incompleteness of this coverage necessarily means information on those scales not probed is missing; an interferometer
filters information on scales it cannot access. In fact, the SZA was designed specifically to probe cluster 
(arcminute) scales.  Since the scales to which we are sensitive are proportional to $\lambda/d$ in radians,
long baselines of a 30-GHz SZA observation ($\sim$ 3--7.5~k$\lambda$) probe angular scales 
$\theta \propto \lambda/d = $ 0.46--1.15$\arcmin$, while short baselines of that same observation 
($\sim$ 0.35--1.3~k$\lambda$) probe angular scales $\theta \propto \lambda/d = $ 2.6--9.8$\arcmin$.
The SZA 30-GHz system is essentially not sensitive to radial scales larger than $\sim 5\arcmin$ (the 90-GHz
system, discussed below, probes scales one-third these sizes).  
In this way, the SZA isolates the small ($\Delta T$ in Eq.~\ref{eq:deltaT} typically peaks $\sim 10$~mK) 
cluster signal from the relatively large background (e.g. the atmospheric and instrumental noise 
discussed in \S \ref{tsys}, as well as the 2.73~K primary CMB).

By including SZA observations performed using the 90-GHz receivers, using the same array configuration shown in Fig.~\ref{fig:sza_ant_loc},
the gap at $\sim$ 1.3--3~k$\lambda$ in the \emph{u,v}-coverage at 30~GHz can be filled.  Since the wavelenth is
$\sim$ 3 times shorter at 90~GHz than it is at 30~GHz, each baseline at 90~GHz is effectively $\sim$ 3 times longer
in $\lambda$ (number of wavelengths). The combined 30+90~GHz coverage is shown in Fig.~\ref{fig:uv_coverage_cl1226_30_90}, 
while the corresponding histogram of radial \emph{u,v}-scales probed by combining the observations is shown
in Fig.~\ref{fig:uv_space_hist_CL1226_30_90}.
The short baselines of the 90-GHz system, ranging $\sim$ 1.2--3.8~k$\lambda$, constrain cluster signals in the intermediate 
angular scales $\theta \propto \lambda/d = $ 0.9--2.9$\arcmin$.

\section{X-ray imaging of Galaxy clusters}\label{xray}

In this thesis, I use X-ray imaging data taken with the \emph{Chandra} X-ray Observatory,\footnote{\url{http://cxc.harvard.edu/}}
extending the work of \citet{reese2002}, \citet{laroque2006}, \citet{bonamente2006}, and others to combine X-ray imaging data
with new SZE data taken with the SZA.

\begin{figure}
\centerline{\includegraphics[width=6in]{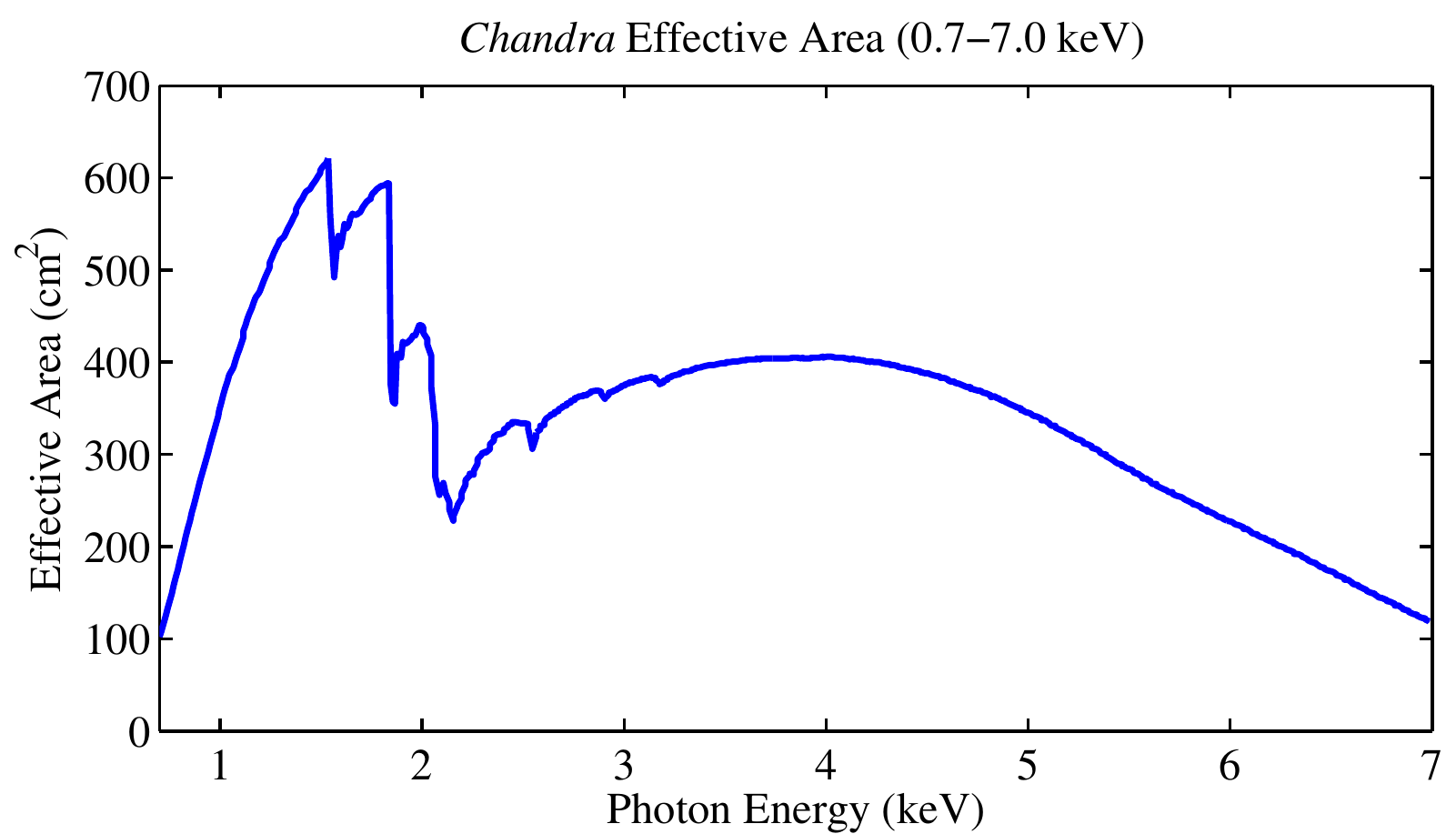}}
\caption{The effective collecting area of the \emph{Chandra} primary mirror, as a function of 
photon energy.  This response dominates the energy response of the instrument.  It also reduces the 
plasma emissivity that \emph{Chandra} effectively sees from a high temperature plasma (Fig.~\ref{fig:lambda}).
}
\label{fig:effarea}
\end{figure}

At X-ray wavelengths, emission from a cluster is predominantly due to thermal processes that occur in the 
same gas that produces the SZE.  
The two thermal processes by which the ICM emits X-rays are bremsstrahlung\footnote{Free-free 
emission due to electron collisions.} and line emission \citep[see, e.g.][]{longair1998, sarazin1988}, both of which depend on
collisions between pairs of particles within the gas (i.e. two particles are involved in the emission process).  
The X-ray emission therefore scales as number density-squared. In contrast, the SZE depends linearly upon electron 
pressure (Eq.~\ref{eq:compy}).  These two ways of probing the ICM therefore complement each other.

X-ray imaging data are sensitive to the surface brightness $S_X$ (in $\rm cts ~ arcmin^{-2} ~ s^{-1}$):
\begin{equation}
\label{eq:xray_sb}
S_X = \frac{1}{4\pi (1+z)^4} \! \int \!\! n_e(\ell)^2 \Lamee(T_e(\ell),Z) \,d\ell 
\end{equation}
where $n_e(\ell)$ and $T_e(\ell)$ are the electron density and temperature along sight line $\ell$, 
$\Lambda_{ee}(T_e,Z)$ (in $\rm cts ~ cm^{5} ~ s^{-1}$) is the X-ray emissivity measured by the 
instrument within the energy band used for the observation, 
$z$ is the cluster's redshift, and $Z$ is the plasma's metallicity.
Metallicity accounts for elements heavier than helium, where $Z_\odot \equiv 1$ is defined to be the elemental abundance 
measured in the solar atmosphere, and the abundances in a cluster are measured relative to the solar abundance (see \S \ref{mus} 
for a more detailed discussion of elemental abundances and metallicity).  For $T \gtrsim 3~\rm keV$, 
the X-ray emissivity of the plasma depends weakly on temperature.  Before accounting for the instrument's response (discussed below),
$\Lambda_{ee}(T_e) \propto T^{1/2}$ since the X-ray emission is bremsstrahlung-dominated at these energies.  

\begin{figure}
\centerline{\includegraphics[width=6in]{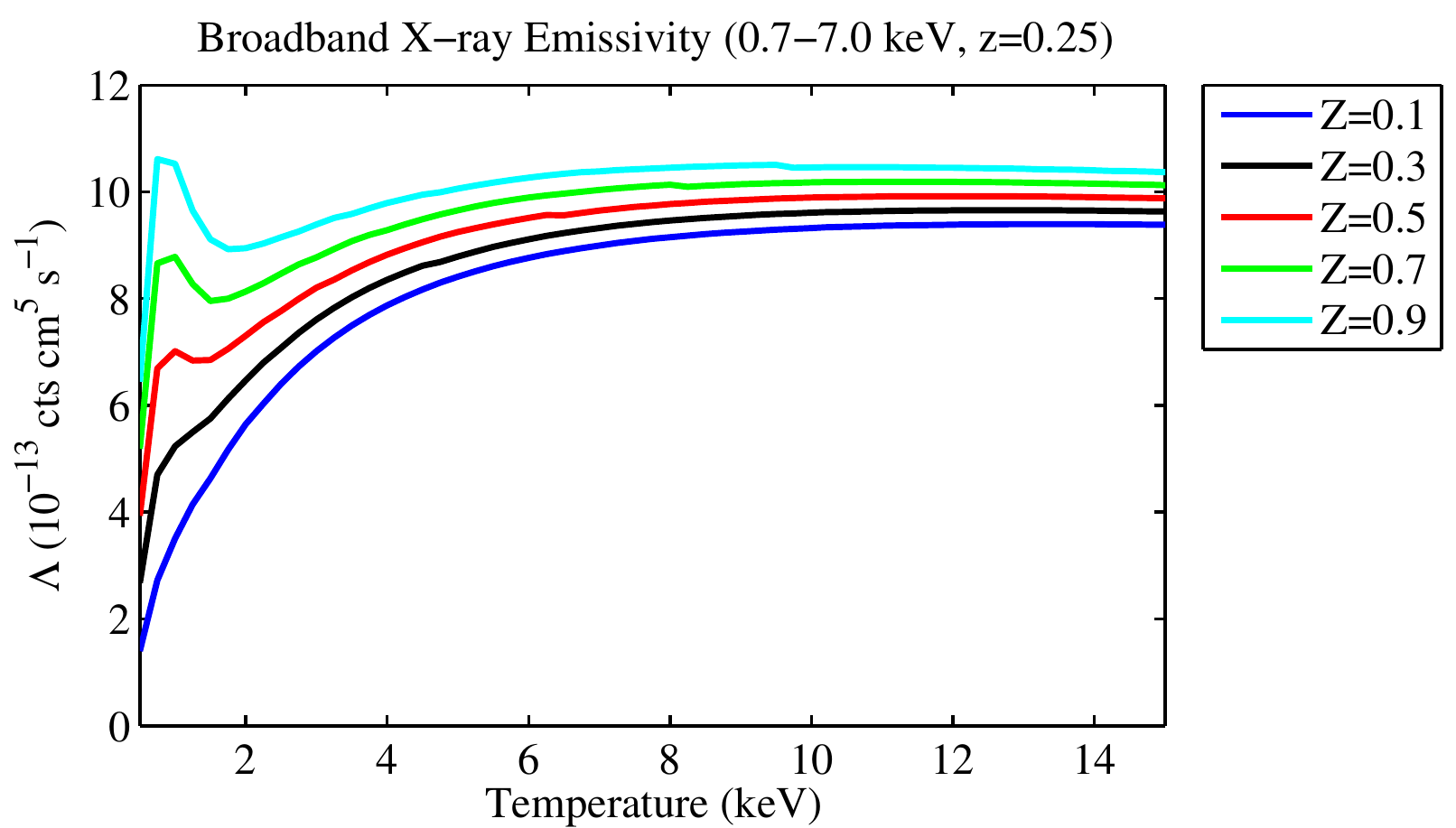}}
\caption{The X-ray emissivity of a cluster plasma at redshift $z=0.25$, redshifted to local
photon energy range 0.7--7.0~keV, as measured by \emph{Chandra} (using the instrument's effective
area shown in Figure~\ref{fig:effarea}).  
The plasma emissivity was computed using the plasma model of \cite{raymond1977} for a range
of cluster temperatures and metallicities (see \S \ref{mus}).  
The effective emissivity we measure is reduced by \emph{Chandra}'s efficiency, which declines 
for photon energies $> 4~\rm keV$  (see Fig.~\ref{fig:effarea}).
For plasma temperatures $\lesssim 2~\rm keV$, the cluster X-ray emission is dominated by lines produced by elements 
heavier than helium.  Since \emph{Chandra}'s sensitivity peaks at energies between 1--2~keV,
the effective emissivity of a plasma with temperature $\sim 1~\rm keV$ has a strong metallicity dependence
(compare at $\sim 1~\rm keV$ the effective emissivity of a plasma with metallicity 
$Z=0.9$, versus that with $Z=0.1$).
}
\label{fig:lambda}
\end{figure}

In addition to the already-weak temperature dependence of the plasma emissivity, the ``effective emissivity'' 
of the plasma as measured by \emph{Chandra} is reduced by the instrument efficiency, which declines for photon energies $> 4~\rm keV$. 
\emph{Chandra}'s efficiency is dominated by the ``effective area'' of the primary mirror for a 
photon of a given wavelength; the typical energy response of the \emph{Chandra} ACIS-I CCD is plotted in Fig.~\ref{fig:effarea}.
The effective emissivity measured by \emph{Chandra} in a given energy band accounts for the instrument's 
efficiency at that energy, and is plotted in Figure \ref{fig:lambda} for a range of plasma temperatures 
and metallicities for a plasma at $z = 0.25$.\footnote{
The effective area $A_{eff}(E) = \epsilon(E) \times A$ for the efficiency $\epsilon(E)$ for photon energy $E$ (local to the instrument) 
of an aperture with physical area $A$.
The ``effective emissivity'' could be computed, then, as $\Lambda_{ee,eff} = \int_{E_1}^{E_2} \Lambda_{ee,E} \, A_{eff}(E) \, dE$,
for an X-ray exposure in the local energy range $E_1$--$E_2$, given a physical plasma emissivity of $\Lambda_{ee,E}$ redshifted 
to local photon energy $E$.}
To obtain the total number of counts per pixel in an exposure, the result of Eq.~\ref{eq:xray_sb} must 
be multiplied by the exposure time and the field of view (in arcmin$^2$) of a pixel.\footnote{There are additional
complications. First, the effective area changes over the field of view of an X-ray instrument.  We use an 
exposure map, which gives the effective area each pixel sees at 1~keV (near the peak in \emph{Chandra}'s 
effective area) to correct for off-axis effects.   We also correct the X-ray image for the quantum efficiency of the 
CCD. See \citet{bonamente2004} for more details on the X-ray analysis.}

\section{Structure of the Thesis}\label{overview}

I describe the SZA instrumentation in Chapter~\ref{sza}, and discuss how we addressed challenges encountered 
during SZA commissioning observations in Chapter~\ref{sza_debug}.  In Chapter~\ref{data_calib}, I discuss SZA data calibration 
and the final data product used in cluster analyses.

In Chapter~\ref{modeling}, I discuss the joint modeling of SZE+X-ray data, and present the new models used.  I also
discuss our data fitting routine and how we derive cluster parameters of interest from the models.
In Chapter~\ref{model_application}, I apply the models to real observations and discuss the results.  Finally,
in Chapter~\ref{conclusions}, I present my conclusions and some ideas for extensions to these modeling
techniques.

\part{Instrumentation and Data Reduction}

\chapter{The Sunyaev-Zel'dovich Array}\label{sza}

\section{Overview of the Sunyaev-Zel'dovich Array \label{sza_overview}}

The Sunyaev-Zel'dovich Array (SZA), an interferometer composed of eight 3.5 
meter telescopes, is located at the Owens Valley Radio Observatory, in 
Big Pine, California.  
Its coordinates are $37^{\circ}14\arcsec02\arcsec$ latitude and 
$118^{\circ}16\arcsec56\arcsec$ longitude, and it is at an altitude of
1222 meters above sea level.  
The Owens Valley is a desert, and therefore provides suitable atmospheric conditions 
for performing centimeter and millimeter-wave observations for most of the year.

\begin{figure}
\centerline{\includegraphics[width=5.5in]{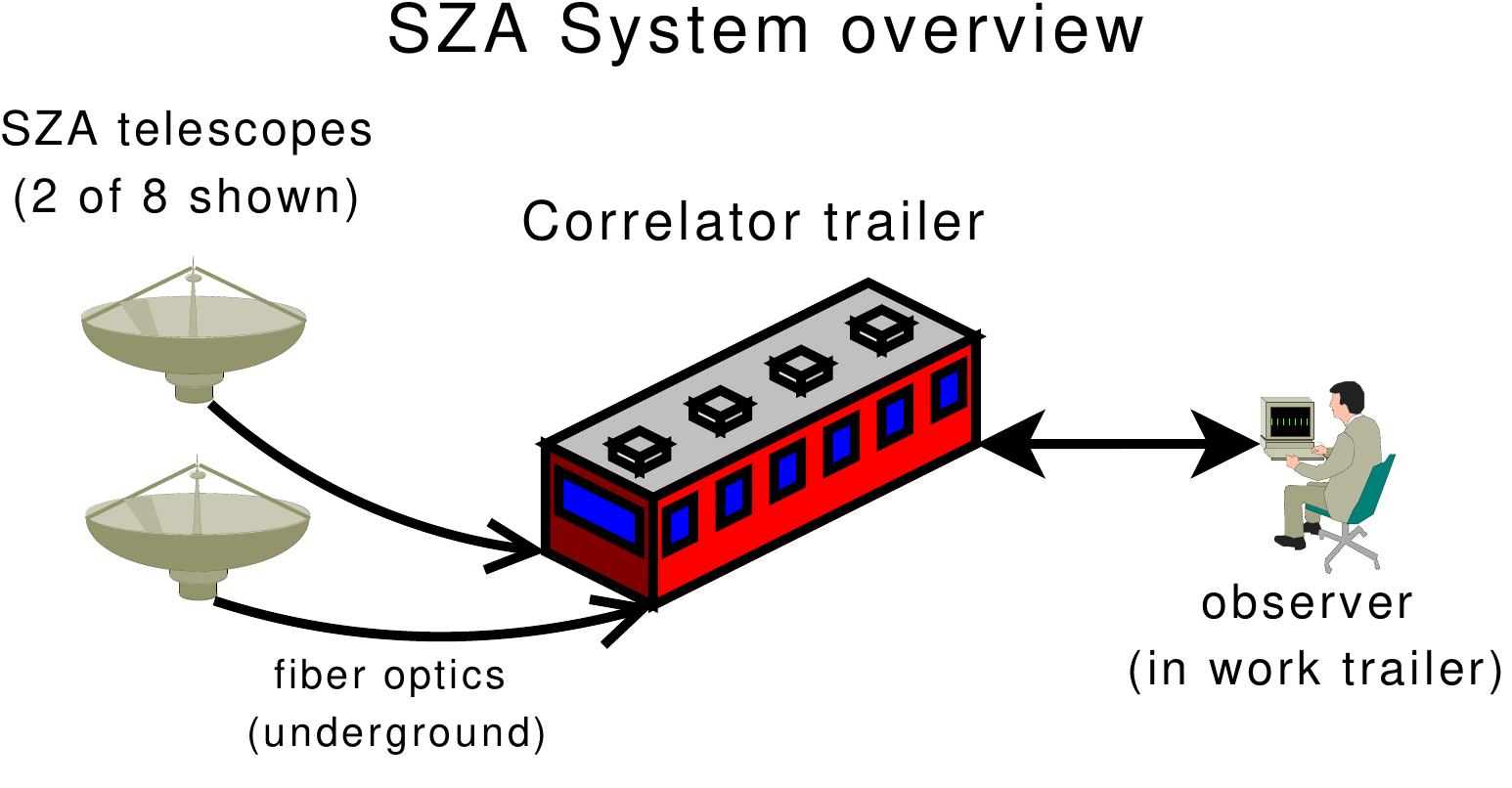}}
\caption{SZA System Overview.  The SZA has eight 3.5~m antennae that communicate 
via fiber optic connections to equipment in a general-purpose utility trailer, 
which is referred to as the ``Correlator Trailer.''
This trailer houses the downconverter, correlator, control system computer, and other electronics
common to the system.  The observer commands the control system using an interface
programmed by Erik Leitch that utilizes the SSH (secure shell) protocol.
}
\label{fig:system_overview}
\end{figure}

The SZA has a digital correlator with 8~GHz of bandwidth \citep[see][for
details on the correlator]{hawkins2004}, and each antenna is equipped with
two wideband receiver systems, capable of observing from 27-36~GHz (in the Ka band, 
hereafter referred to as the ``30-GHz'' band) and from 85-115~GHz (in the W band,
referred to as the ``90-GHz'' band).  
See Fig.~\ref{fig:system_overview} for a broad overview of the SZA system.
The large, 8-GHz receiver and correlator bandwidth provides the SZA with
the high sensitivity required to detect rapidly cluster signals.  
This wide bandwidth also provides the 
ability to probe a wide range of {\em u,v}-space (Fourier conjugate of image space) 
simultaneously, meaning it is sensitive to a wider range of angular scales than 
a comparable instrument with a smaller bandwidth.
 
The SZA receivers contain high electron mobility transistors (HEMT) for high gain,
low noise amplification of incoming signals.  The 30-GHz receivers
contain the same HEMT amplifiers that were in the Degree Arcminute-Scale
Interferometer (DASI) \citep[see][]{leitch2005}, in the same configuration
used for the OVRO/BIMA SZE receivers \citep[see][]{carlstrom1998,carlstrom2000}.  
The 90-GHz receivers utilize HEMTs implemented through monolithic microwave integrated 
circuit (MMIC) technology, which integrates four HEMTs into each MMIC block.

\begin{figure}
\centerline{\includegraphics[width=4in]{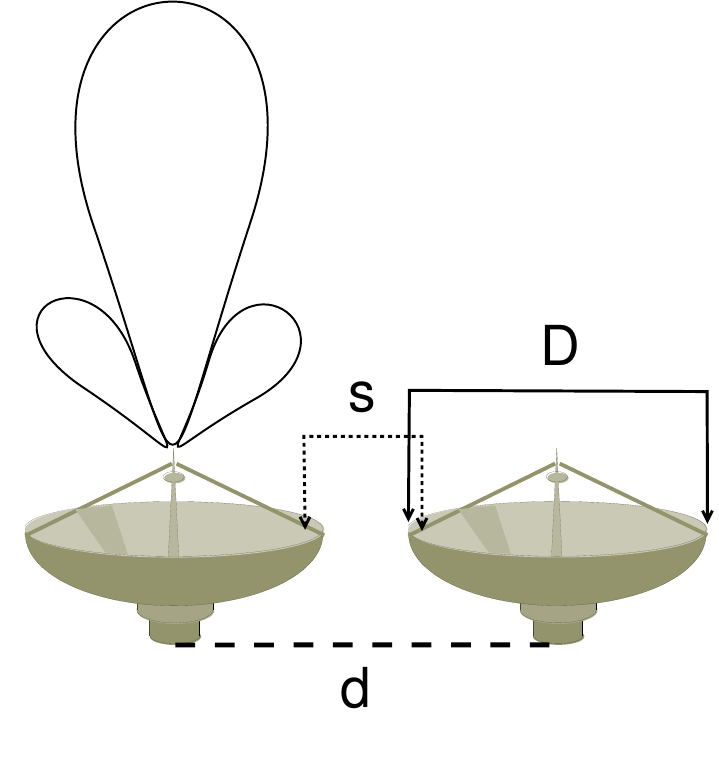}}
\caption{Detail of a pair of antennae.  The primary mirror diameter is $D$ (solid line).  
The center-to-center antenna separation (long dashed line) is $d$, which when looking at zenith
(as depicted here) is also the projected baseline length.
Separation $s$ (short dashed line) is a distance between two arbitrary points on the primaries 
separated by less than $D$. A cross-sectional representation of the main and secondary 
lobes of the antenna sensitivity pattern is shown above the left antenna, where the first 
sidelobe (secondary lobe) is greatly exaggerated in scale; It was, in reality, measured (by James Lamb) 
to be -25~dB ($10^{-2.5}\approx 1/316^\text{th}$) less than the sensitivity at the center of the primary 
beam (main lobe).}
\label{fig:baseline1}
\end{figure}

The SZA was designed to detect and probe clusters at intermediate 
and higher ($z\gtrsim0.2$) redshifts.  The relatively small (3.5~m) primary 
mirrors and their short focal lengths (see \S \ref{optics}) provide 
two advantages for SZE observations of clusters: a large primary beam, which 
scales as $\lambda/D$ for observational wavelength $\lambda$ and primary mirror
size $D$, and the ability to closely pack the antennae without resulting
in inter-antenna collisions.
The full width half maximum (FWHM) size of the truncated Gaussian primary beam, where the sensitivity 
falls to -3 dB of the peak, is $\approx10.7\arcmin$ at the center of the 30~GHz band.  This is necessary
so that objects on arcminute scales, such as clusters are intermediate redshifts, are relatively unattenuated
by the primary beam.  
The scales probed by an interferometer are determined by the projected separation 
distance $d$ (the distance between the centers of two antennae as seen from the 
source), and scale as $\lambda/d$.
Because the SZA routinely observes clusters all the way to the antenna shadowing
limit of 3.7 meters (i.e. the projected antenna spacing is nearly as small as the
dish size), it can in principle probe radial angular scales larger
than $\sim 4.8\arcmin$ at the central frequency of 30.938~GHz.
This is because, for a projected baseline of length $d$, there are points on the primary 
mirrors separated by distances $s \in [d-D,d+D]$ that are not entirely 
attenuated by the primary beam (see Fig.~\ref{fig:baseline1}).

\begin{figure}
\centerline{\includegraphics[width=6in]{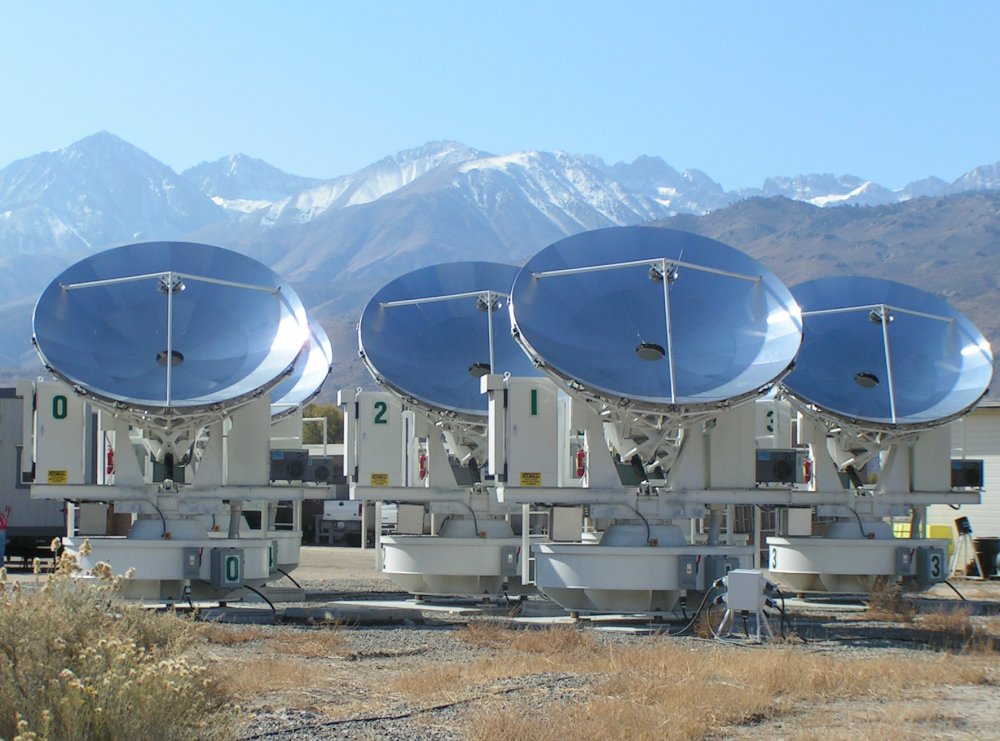}}
\caption{Photo of the inner 6 telescopes (Antennae 0--5)
 of the SZA.  Photo is taken from the northeast (roughly along Baseline 2-7; see Fig.~\ref{fig:sza_ant_loc}). 
Antenna 5 is behind Antennae 1 and 3, and Antenna 4 is behind Antenna 0.}
\label{fig:innerarray_pict}
\end{figure}

\begin{figure}
\centerline{\includegraphics[width=3in]{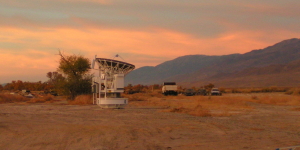} \includegraphics[width=3in]{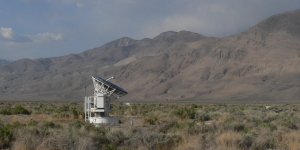}}
\caption{Photo of the outer 2 telescopes of the SZA (Antennae 6 \& 7, see \ref{fig:sza_ant_loc}), taken while 
standing near the inner array. Antenna 6 is shown in the left panel, and Antenna 7 is shown in the right panel.}
\label{fig:outerarray_pict}
\end{figure}

Six of the eight SZA antennae, forming fifteen baselines,\footnote{The number
of baselines $N_\text{bl}$ is computed: $N_\text{bl} = N (N-1)/2$}
comprise the closely-packed ``inner array,'' which is sensitive to
$\approx$ 1-5\arcmin scales.  The inner array is shown in Fig.~\ref{fig:innerarray_pict}. 
Two outer antennae (Fig.~\ref{fig:outerarray_pict}), which are identical to the inner six,
form thirteen more baselines (i.e. one baseline between the two outer antennae, 
and six baselines formed between each outer antenna with each of 
the inner six; see Fig.~\ref{fig:sza_ant_loc}).  These outer antennae provide the 
ability to fit simultaneously any unresolved ($\lesssim 20\arcsec$) radio sources
(hereafter ``point sources'') in the cluster field, which could
otherwise mask the SZE decrement at 30~GHz.


As discussed in \S \ref{interf}, when observing an astronomical source, in addition to the physical pointing 
of each antenna that keeps the source in the main lobe of the antenna beam pattern (see Fig.~\ref{fig:baseline1}), 
the delays necessary for fringe tracking must be computed.
Fringe tracking corrects the phases for the instantaneous projected baseline 
changes due to geometry, as the source moves through the sky, relative to the 
array (see Fig.~\ref{fig:baseline3}).
Phase changes can also be due to effects that are more difficult to calculate. For example,
 variations in the lengths of the fiber optics and cables that carry the signals
can change the phases of the signals from each antenna.  
Phase shifts due to properties of the electronics themselves can also act as 
path length differences (at a particular frequency).  
These instrumental phase effects are particularly sensitive to temperature fluctuations.

In this chapter, I provide an overview of the components that comprise the SZA.
Emphasis is placed on thermal and mechanical aspects of the instrument, as this was  
the focus of the instrumentation component of this thesis.  
Thermal stability and the consequences of poor thermal regulation are discussed as appropriate.

\section{Telescope Optics}\label{optics}

\begin{figure}
\centerline{\includegraphics[width=5.5in]{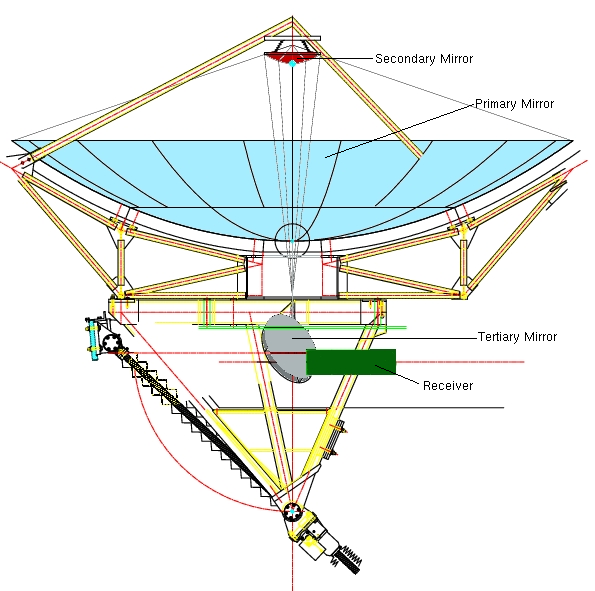}}
\caption{Overview of the SZA antenna optical design.  The primary (light blue) and secondary (red) mirrors are on-axis 
reflectors.  The tertiary mirror (silver) selects the 30 or 90-GHz receiver within the receiver cryostat (dark green).
The receiver cryostat is located within the larger receiver enclosure box (not shown).
Image adapted from David Woody.}
\label{fig:antenna_design}
\end{figure}

The antennae of the SZA are designed as on-axis, altitude-azimuth telescopes with small 
primary mirrors that have short focal lengths (see Fig.~\ref{fig:antenna_design}).  
Their design allows for a compact array configuration with the antennae spaced as 
close as $1.2 D$, where $D$ is the diameter of the primary, without ever colliding.
The panels of the primary mirrors were machined by Jerry Forcier of Forcier Mechanics 
using a conventional, computer numeric controlled (CNC) mill, programmable using standard 
computer-aided design/machining (CAD/CAM) code.  Since the mill operates in standard 
Cartesian coordinates, the machining process provided the mirrors with a scalloped surface.  
The tool and toolpath used to machine the primaries were chosen so that the scalloping 
scatters higher frequency light (e.g. IR and optical), which would otherwise damage 
the instrument if, for example, the sun were focused on the receiver.\footnote{Not all 
telescopes can point directly at the sun without damaging the receivers; the SZA, however, can.}
The surface was measured by Marshall Joy to have an \emph{rms} roughness of $25~{\rm \mu m}$, meeting the 
design requirements necessary for the 1.3~mm upgrade the SZA may see as part of CARMA \citep[see e.g.][]{woody2004,scott2006}.

\begin{figure}
\centerline{\includegraphics[width=6in]{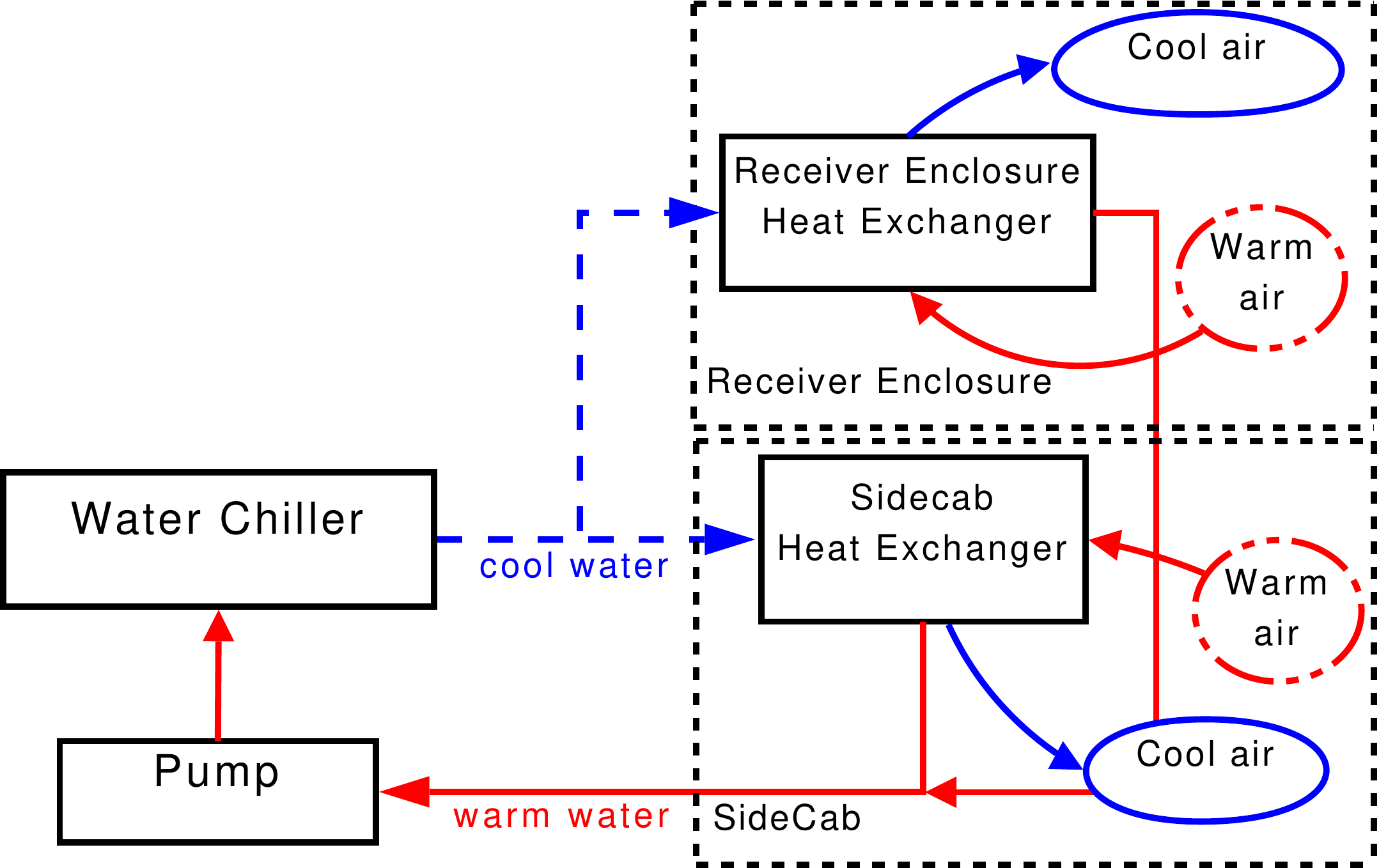}}
\caption{Flow chart of the closed-loop chiller system.  Warm water is pumped through a cool chiller reservoir.
The chilled water continues on to the sidecab, which is a weather-tight equipment rack containing the 
antenna computers, the motion control system and servos, and electronics power supplies, and to the 
receiver enclosure (described in the text).
The warm air is blown across the fins of the heat-exchangers, through which the cool water is 
circulated, removing heat.  The chilled air returned to the surrounding sidecab or receiver 
box, as appropriate, where it cools the electronics.
The water, now warm, returns to the pump input, where it is recirculated through
the system.}
\label{fig:chillers}
\end{figure}

The secondary mirrors are convex paraboloids which slow the focus\footnote{See Fig.~\ref{fig:antenna_design}, which shows 
that the rays are being quickly focused by the primary; the secondary mirror extends the focal length.} 
before it enters the receiver enclosure.
The receiver enclosure is simply a large, weather-proof, thermally-regulated box 
that holds the receiver cryostats, the back-end electronics (which are further enclosed
in a smaller, thermally-regulated electronics box),  the tertiary mirror, an ambient 
calibrator load, and various support electronics.  
In the fall and spring, the receiver enclosure is cooled by air circulating through 
closed-cycle air/air heat-exchangers. In the summer, the enclosure is cooled by a 
refrigerated water/antifreeze solution, which is pumped through a closed liquid/air 
heat-exchanger (similar to a car radiator, containing refrigerated fluid) between 
the air in the box and the chilled fluid (see Figure \ref{fig:chillers}).  
In the winter, the box is heated by ceramic, AC-powered heaters.
In addition to protecting equipment from the elements (e.g. dust, harsh sunlight, and rain), 
the receiver enclosure serves as the first step in achieving thermal stability, and its stability helps
reduce expansion and contraction of the section of fiber optics closest to the receiver.

The signal enters the receiver enclosure through a microwave transparent (has a very 
small absorption/reflection coefficient -- less than a few percent), weather-proof 
window made of $75~{\rm \mu m}$ mylar and 1-inch thick zotefoam (a nitrogen-filled, polyethylene foam).
Zotefoam, a commercially-available foam used in housing construction, was chosen because it
is highly transparent at microwave frequencies.  It is a closed-cell foam that does 
not permit air or water diffusion, and is very robust both chemically and mechanically.
This mylar layer in front of the zotefoam of the receiver enclosure's window provides an additional 
-- and easily replaceable -- layer
of protection from dust and water, which cling to the porous surface of the zotefoam.\footnote{A porous
surface develops as the zotefoam weathers and degrades in sunlight.}
Thin mylar, such as that used for the windows, is also microwave-transparent.

The next mirror in the optical path -- directly after the window to the receiver enclosure -- 
is an adjustable tertiary mirror: a concave, off-axis paraboloid 
that bends the optical path by $90^{\circ}$ into the receiver.
This programmable tertiary selects the receiver into which the signal is focused.  

\section{Receivers}\label{sza_receivers}

\subsection{Receiver Noise Considerations}\label{tsys}

The primary goals of receiver design are the attainment of high gain and low
noise.  We characterize noise in terms of the ``receiver noise temperature'' 
$T_{\rm rx}$, which is derived in this section. 

The power emitted in the Rayleigh-Jeans tail of a blackbody is approximately linearly 
proportional to its temperature. For two thermal sources at temperatures $T_{\rm hot}$ and $T_{\rm cold}$, 
we measure output powers $P_{\rm hot}$ and $P_{\rm cold}$, which are 
linearly proportional to the source temperatures as long as the gain of 
the receiver system also remains linear over that range. 
Our measurement includes an additional, constant amount of power\footnote{This is the 
zero-intercept in a hypothetical graph of measured output power versus input blackbody 
temperature.} due to the noise in the receiver, which can be attributed to a hypothetical 
thermal source at temperature $T_{\rm rx}$.  
This receiver noise temperature therefore characterizes
the output power that we would measure in the absence of any input power.
We define the receiver $y$ factor (the ratio of the measured output powers):
\begin{equation}
y \equiv \frac{P_{\rm hot}}{P_{\rm cold}}
= \frac{T_{\rm hot}+T_{\rm rx}}{T_{\rm cold}+T_{\rm rx}} 
\label{eq:yfact}
\end{equation}
Solving for $T_{\rm rx}$, we have:
\begin{equation}
T_{\rm rx} = \frac{T_{\rm hot} - y~\! T_{\rm cold}}{y - 1}
\label{eq:tsys}
\end{equation}
Contributions to $T_{\rm rx}$ can include noise from the receiver cryostat windows, amplifiers, mixers, 
waveguides, and the feed horns.  These contributions are all affected by the physical temperatures 
of the components.  

The total contribution $T_{\rm rx}$ from the receiver to the noise is calculated 
in terms of each component's gain $G$ (which can be less than unity) for $n$ stages:
\begin{equation}
T_{\rm rx} = T_1 + \frac{T_2}{G_1} + \frac{T_3}{G_1 G_2} + \ldots + \frac{T_n}{G_1 G_2 \ldots G_{n-1}}
\label{eq:t_rx}
\end{equation}
In Eq.\ \ref{eq:t_rx}, the noise due to each successive stage is reduced by the 
combination of gains before it ($G_1 \cdot G_2 \cdot \ldots \cdot G_{n-1}$).
The first components in a receiver -- the window and the feed horn -- are chosen to
have low loss and be well-matched to the incoming signal, respectively, in order to reduce their
noise contributions.  In addition to being well-matched to the incoming signal, the feed horns 
are kept at low temperature, which reduces their noise contribution.
The first stage amplifier typically dominates the overall receiver noise temperature $T_{\rm rx}$.
We therefore optimize the system to have a high-gain, low-noise amplifier at the first (and at every, if possible) 
stage of amplification.
Higher current, higher noise amplifier stages follow this stage, providing most of the 
overall amplification of each receiver system.

For the overall noise contribution to a signal from a telescope -- including contributions 
from the back-end electronics, the atmosphere, optical elements in the signal path to the receiver,
and of course the noise from the receiver itself -- we define the system temperature $T_{\rm sys}$,
where the $y$ factor in Eq.~\ref{eq:yfact} is instead measured at the output of the entire
system, and $T_{\rm sys}$ is scaled to above the Earth's atmosphere (i.e. we include the noise
contribution of the atmosphere with the instrumental noise in one term; see \S \ref{noise}.).
The dominant contributions -- in order of importance -- to the total system temperature $T_{\rm sys}$ 
are therefore the receiver noise $T_{\rm rx}$, the sky ($\sim$ 10--20~K at zenith), 
noise scattered from the ground\footnote{Dirt is essentially a 
blackbody at $\sim 300$~K.  Power emitted by the ground can scatter off the feed legs holding 
the secondary mirrors.  This small amount of noise was measured to be $\sim 4$~K at elevations 
$\lesssim 12^\circ$.  We typically limit our observations to sources above a $30^\circ$ horizon.}
(which is called ``spillover,'' and is $\sim 4$~K at very low elevations, which we avoid when observing),
loss due to the windows in the optical path, poor coupling with the feed horn, 
and the noise contribution from first stage of the HEMT or MMIC amplifier.  

\subsection{Measuring $T_{\rm rx}$}
 
In the laboratory, we used a piece of Eccosorb$^\circledR$\footnote{Eccosorb$^\circledR$ 
is a microwave blackbody made by Emerson \& Cuming Microwave Products. 
Details about this product can be found here: 
\url{http://www.eccosorb.com/europe/english/page/63/eccosorb}} as a ``hot load'' (a thermal load with $T=T_\text{hot}$ in Eq.~\ref{eq:tsys}). 
The ``cold load'' (a thermal load with $T=T_\text{cold}$ in Eq.~\ref{eq:tsys}) was another piece of Eccosorb$^\circledR$
at liquid nitrogen's (abbreviated $LN_2$) boiling point ($\approx$ 77~K at 1 atmosphere of pressure).
This is achieved by keeping the blackbody material in a bath of $LN_2$, and removing it for
the measurement (for only a few seconds, so it retains evaporating $LN_2$ throughout the measurement).  

On the SZA telescopes, the hot load is a mechanized version of the same ambient blackbody used in the lab, 
mounted to the front of the receiver cryostat.  However, a true cold load with an $LN_2$ 
bath would be impractical in our system.  
Instead,  careful tracking of atmospheric temperature and humidity allow a model to be 
developed for the power contribution from the sky, and therefore the sky serves as 
the ``cold load'' for calibration purposes.  

Under ideal observational conditions 
the atmospheric contribution to the noise scales as the optical depth.  
The optical depth scales roughly as $\sec(\theta)$ for angle $\theta$ 
measured from zenith. This $\sec(\theta)$ dependence only holds if 
assuming a plane-parallel atmosphere; this is a gross simplification, and the true calculation 
used in data calibration does not make this assumption.  
To reduce the noise contribution from the atmosphere, we typically observe 
sources with elevations $\gtrsim 30^\circ$ above the horizon.

\subsection{Receiver RF Components}\label{rx_rf_components}

\begin{figure}
\centerline{\includegraphics[width=5.45in]{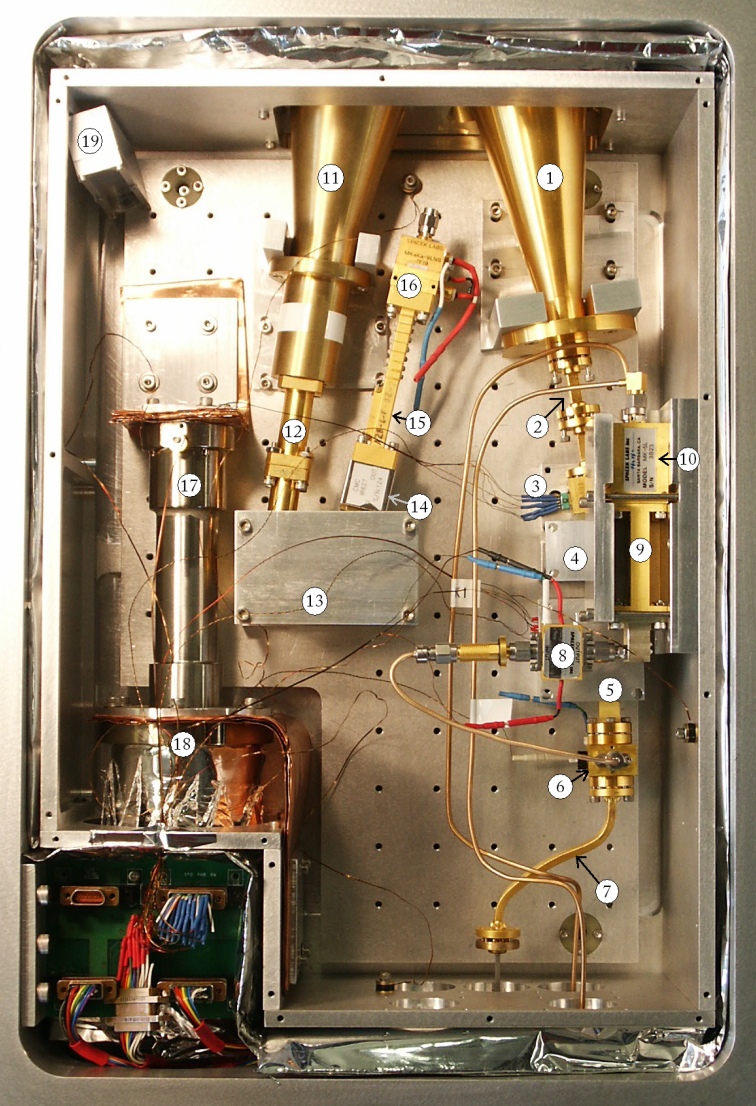}}
\caption{Photo of Receiver Cryostat Electronics test setup.  Items {\bf 1}--{\bf 10} are components
of the 90-GHz receiver, and {\bf 11}--{\bf 16} are components of the 30-GHz receiver.
See \S \ref{rx_rf_components} for details on each RF component.
The stages of the refrigerator ({\bf 17} \& {\bf 18}) are discussed in \S \ref{rx_thermal}.
The copper strapping attached to {\bf 17} is not the final version.
}
\label{fig:rxphoto}
\end{figure}

\begin{figure}
\centerline{\includegraphics[width=6in]{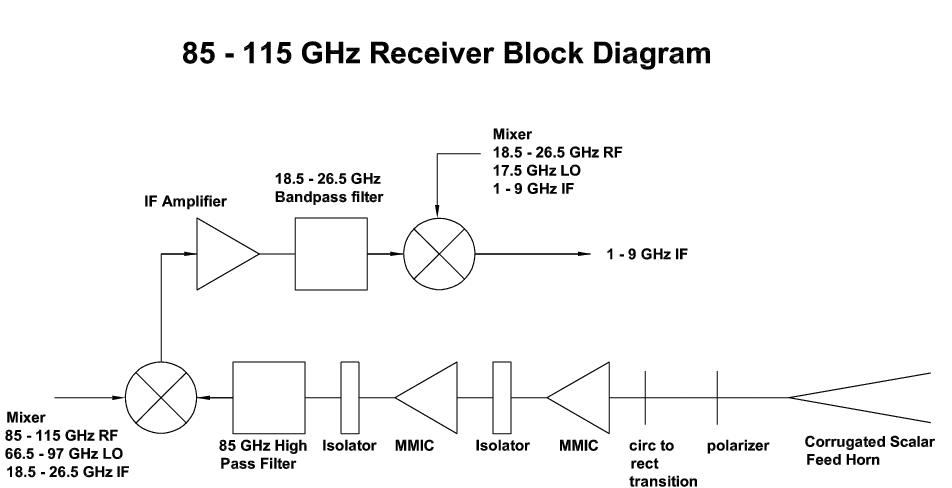}}
\caption{90-GHz receiver block diagram, courtesy of Amber Miller. 
See the photo in Fig.~\ref{fig:rxphoto}, which shows these components:
{\bf 1.} 90-GHz Feedhorn. 
{\bf 2.} Circular Polarizer \& Circular-to-Rectangular Transition.
{\bf 3.} 1$^{\rm st}$ MMIC HEMT Amplifier.
{\bf 4.} Isolator.
{\bf 5.} 2$^{\rm nd}$ MMIC HEMT Amplifier, Isolator, \& High Pass Filter (under bracket).
{\bf 6.} W-band Mixer.
{\bf 7.} Waveguide for the incoming signal from 90~GHz tunable LO, the bias-tuned Gunn.
{\bf 8.} IF Amplifier.
{\bf 9.} Bandpass Filter.
{\bf 10.} K-band Mixer.
}
\label{fig:90block}
\end{figure}

\begin{figure}
\centerline{\includegraphics[width=6in]{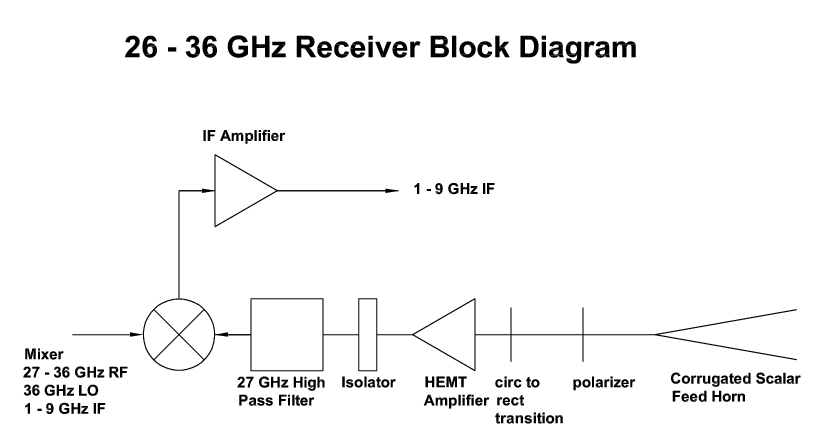}}
\caption{30-GHz receiver block diagram, courtesy of Amber Miller.
See the photo in Fig.~\ref{fig:rxphoto}, which shows these components:
{\bf 11.} 30-GHz Feedhorn.
{\bf 12.} Circular Polarizer \& Circular-to-Rectangular Transition.
{\bf 13.} HEMT amplifier (under a mount).
{\bf 14.} Isolator.
{\bf 15.} 27~GHz High Pass Filter.
{\bf 16.} Mixer.
The IF Amplifier in this figure was later moved outside the receiver.
}
\label{fig:30block}
\end{figure}

Within the cryostat there are two complete, independent receiver systems
(see photo in Fig.~\ref{fig:rxphoto}).  These receivers are capable of observing at sky frequencies 27--36~GHz 
(the ``30-GHz receiver'') and 85--115~GHz (the ``90-GHz receiver'').  
Each receiver (see Figures \ref{fig:30block} and \ref{fig:90block}) begins with a corrugated feed horn
(\#1 and \#11 in Fig.~\ref{fig:rxphoto}) that couples the incoming signal to a circular polarizer
(\#2 and \#12 in Fig.~\ref{fig:rxphoto}).  A circular-to-rectangular transition, which unfortunately
rejects one linear polarization, couples the circular polarizer to a rectangular (linearly polarized) 
waveguide.  The components that follow -- starting with the HEMT and MMIC amplifiers\footnote{Their
definitions are repeated here, for convenience.  HEMT is ``high electronic mobility transistor,'' 
and MMIC is ``monolithic microwave integrated circuit.''} (\#3 and \#13 in Fig.~\ref{fig:rxphoto}) -- 
are specific to each receiver system.
I present the 30-GHz receiver first, but note that it is similar in design to the 90-GHz system.

\subsubsection{30-GHz Receiver RF Components}

In the 30-GHz receiver, a single HEMT amplifier with four stages (i.e. four discrete HEMTs) amplifies the signal.
An isolator (\#14 in Fig.~\ref{fig:rxphoto}), which reduces reflections coming back from the filter, 
follows the HEMT amplifier.  
Next, a high-pass filter (\#15 in Fig.~\ref{fig:rxphoto}) eliminates unwanted signals below 27~GHz, 
and helps to determine the range
of the intermediate frequency (IF) signal, which is the sky signal mixed down to 1--9~GHz (see \S \ref{dcon}
for a description of the SZA downconversion scheme).  
We refer to the 1--9~GHz band as the ``IF band,'' and provide the corresponding sky frequencies of each band within
the IF band (see Table~\ref{table:band_scheme}, \S \ref{dcon}).
The IF band is the band in which all the back-end electronics (\S \ref{backend}) common to both 
the 30 and 90-GHz receivers work.  
A mixer (\#16 in Fig.~\ref{fig:rxphoto}), using a 35.938~GHz local oscillator (LO) as its reference, 
mixes sky frequencies 26.938--34.938~GHz down to the 1--9~GHz IF band (the LO is discussed
in \S \ref{ebox_layout}, while the full downconversion from sky to IF bands is 
discussed in \S \ref{dcon}).

The high pass filter determines the upper end of the IF band by attenuating signals below 
27~GHz.\footnote{Note that the channel at 26.938~GHz is not used; See Table~\ref{table:channel_scheme}, 
\S \ref{corr}, noting that none of the channels at the band edges are used.}
The high filter attenuates unwanted harmonics related to the LO, since the LO is actually the 4th harmonic of 
an 8.972~GHz oscillator (discussed in \S \ref{backend}) mixed with a 50~MHz phase reference frequency, 
and other harmonics of the 8.972~GHz oscillator can be present in the system.  
I discuss RF contamination -- and how we eliminated it -- in \S \ref{rfi}.

\subsubsection{90-GHz Receiver RF Components}

The 90-GHz receiver (see Fig.\ \ref{fig:90block}) is similar to the 30-GHz receiver, with a few exceptions.  
Notably, it has two MMIC amplifiers (\#3 and \#5 in Fig.~\ref{fig:rxphoto}), each of which contains four HEMTs 
(i.e. four stages), since these higher-frequency MMIC HEMTs have less gain than their 30-GHz counterparts.
The MMICs are separated by an isolator (\#4 in Fig.~\ref{fig:rxphoto}), which reduces coupling between the 
gain stages. Feedback between the stages could otherwise lead to oscillations, since 
coupling from reflections is a form of positive feedback (which is intrinsically unstable).

There are two mixers in the 90-GHz system.  The first (\#6 in Fig.~\ref{fig:rxphoto}) mixes the sky signal with a tunable LO that, along
with an 18.5--26.5~GHz bandpass filter (\#9 in Fig.~\ref{fig:rxphoto}), selects the range of observable sky frequencies for the 
90-GHz receiver.  The second mixer (\#10 in Fig.~\ref{fig:rxphoto}), using a fixed-frequency, 17.5~GHz 
dielectronic resonant oscillator (DRO), brings the signal down to the 1--9~GHz IF band.  
An additional IF amplifier (\#8 in Fig.~\ref{fig:rxphoto}) is present in the 90-GHz system for two 
reasons -- the MMIC HEMTs have less overall gain than their 30-GHz counterparts, and mixers are lossy components 
that split the input signal power among the output mixing products. Since the output of the 90-GHz receiver
is handled by the same electronics setup as that which handles the 30-GHz receiver's output, the two systems
need to have comparable output power levels.

\subsection{Thermal Considerations for the Receivers}\label{rx_thermal}

The signal enters the receiver cryostat through a zotefoam and mylar window.  The RF components in the receiver
are sensitive to temperature,\footnote{The charge carriers in a semiconductor are less likely to be in the 
conduction band due to thermal motion as $T\rightarrow 0{\rm~K}$, which means most of the current
in the amplifier is due to the applied field from the incoming signal.} and therefore must be kept 
cool (11-20 K) to reduce noise.  This necessitates the placement of the amplifiers and other 
noise-sensitive components in a vacuum-sealed cryostat.

\begin{figure}
\centerline{\includegraphics[width=6in]{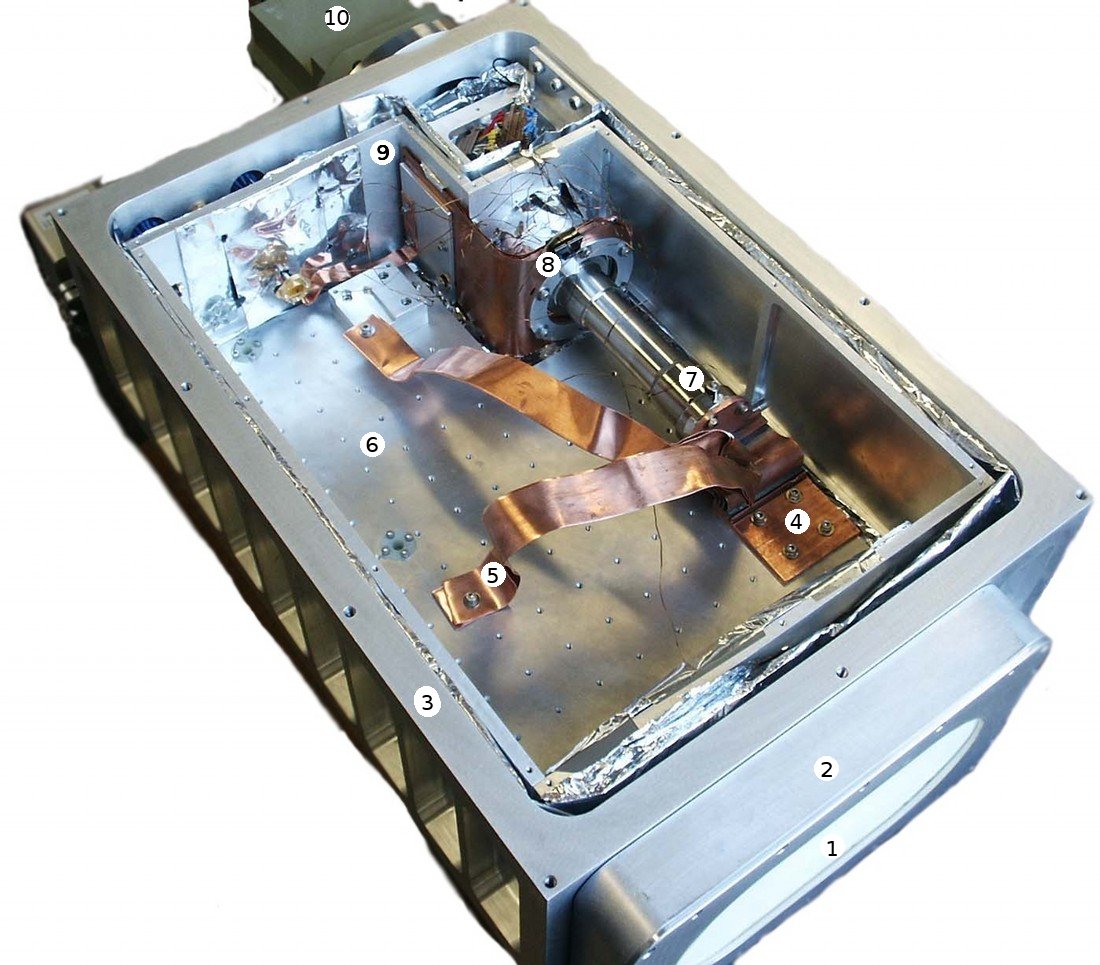}}
\caption{Receiver Cryostat Thermal test setup.  Test hard-soldered strapping ({\bf 4}) on the
second stage of the cold head ({\bf 7}) is shown.  The long copper shims ({\bf 5}), used to increase
the thermal conductivity between the RF components and the cold head, were replaced with more 
flexible, nickel-plated, braided-copper straps. Also in the photo: {\bf 1} is the zotefoam window,
within the window holder ({\bf 2}). The cryostat case is labeled {\bf 3} (the upper lid, which is not shown,
mates here to make a vacuum seal). The cold plate, to which most
of the RF components are mounted, is {\bf 6}.  The first stage of the refrigerator head is
{\bf 8}.  The radiation shield is {\bf 9}, and the exposed part of the refrigerator head is {\bf 10}.
Mylar blanketing can be seen filling the space between the cryostat case and the radiation shield.}
\label{fig:rxstrapping}
\end{figure}

\begin{figure}
\centerline{\includegraphics[width=5.5in]{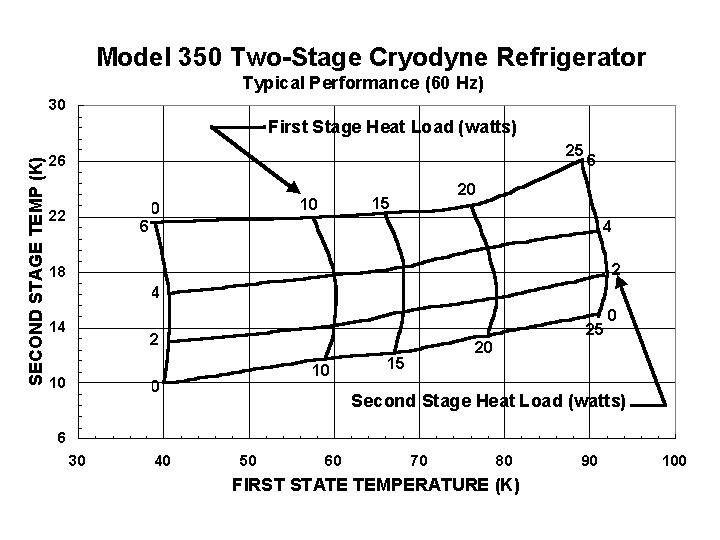}}
\caption{The load curve for the CTI-cryogenics Model 350 Cryodyne refrigerator.  
Figure from Brooks Automation, the company that now owns Helix Technology CTI-Cryogenics.}
\label{fig:cryodyne_load}
\end{figure}

The cryostat is cooled by a CTI-cryogenics Model 350 Cryodyne refrigerator, which offers 
two stages of cooling:
the first (\#8 in Fig.~\ref{fig:rxstrapping}) is a higher power, warmer stage that runs at 
$\sim$ 55~K under a load of about 10~W, and the second (\#7 in Fig.~\ref{fig:rxstrapping}) is a 
lower power, cooler stage that runs at $\sim$ 12~K when the load
is about 1~W 
(see Fig.\ \ref{fig:cryodyne_load}).\footnote{Figure from \url{http://www.brooks.com/documents.cfm?documentID=4932}}  
The cryostat is therefore built in two stages as well:  A warmer radiation shield, made of 
aluminum ($\gtrsim 90 \%$ reflectivity in the microwave and far infrared, where the thermal
emission at $\lesssim 100$ K peaks), and a cooler stage within the shield, where the most sensitive
components of the system are mounted.  

\begin{figure}
\includegraphics[width=5in]{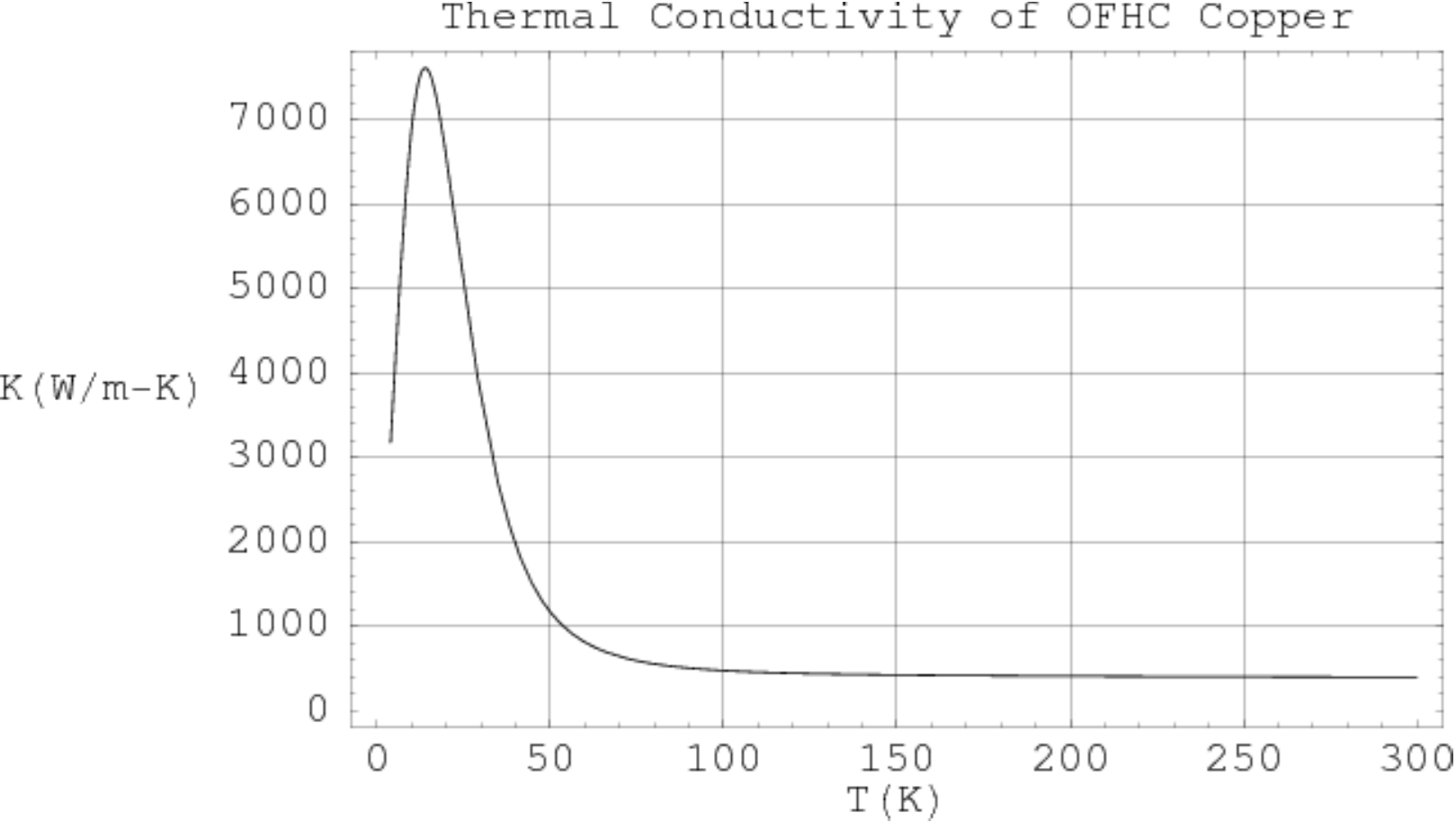}
\caption{Thermal conductivity of OFHC Copper, computed from tables available on the NIST website \citep{nistweb}.
These values have been measured in the temperature range $T=4-300~\rm K$.}
\label{fig:kappa_ofhc_cu}
\end{figure}

The cold head of the refrigerator interfaces with the two stages using copper strapping.  
Copper was chosen for its high thermal conductivity (see Fig.~\ref{fig:kappa_ofhc_cu}).
For the coldest stage, it was necessary to machine and hard solder (using silver solder) a 
piece that ensured good thermal coupling on each end without mechanically stressing the 
refrigerator as the system is warmed and cooled (labeled \#4 in Fig.~\ref{fig:rxstrapping}).  
The thermal link from the second stage of the refrigerator cold head to the cold plate 
is made by four thin copper shims, each of which has a small kink (zig-zag)
that helps provide strain relief.
The thermal power $P$ conducted across a thermal link between temperatures $T_1$ and $T_2$
can be calculated,
\begin{equation}
P = \frac{A}{L} \int_{T_1}^{T_2}\! \kappa(T) dT
\label{eq:Pthermal}
\end{equation}
where $\kappa$ is the thermal conductivity in ${\rm (W m^{-1} K^{-1})}$.

\begin{figure}
\includegraphics[width=5in]{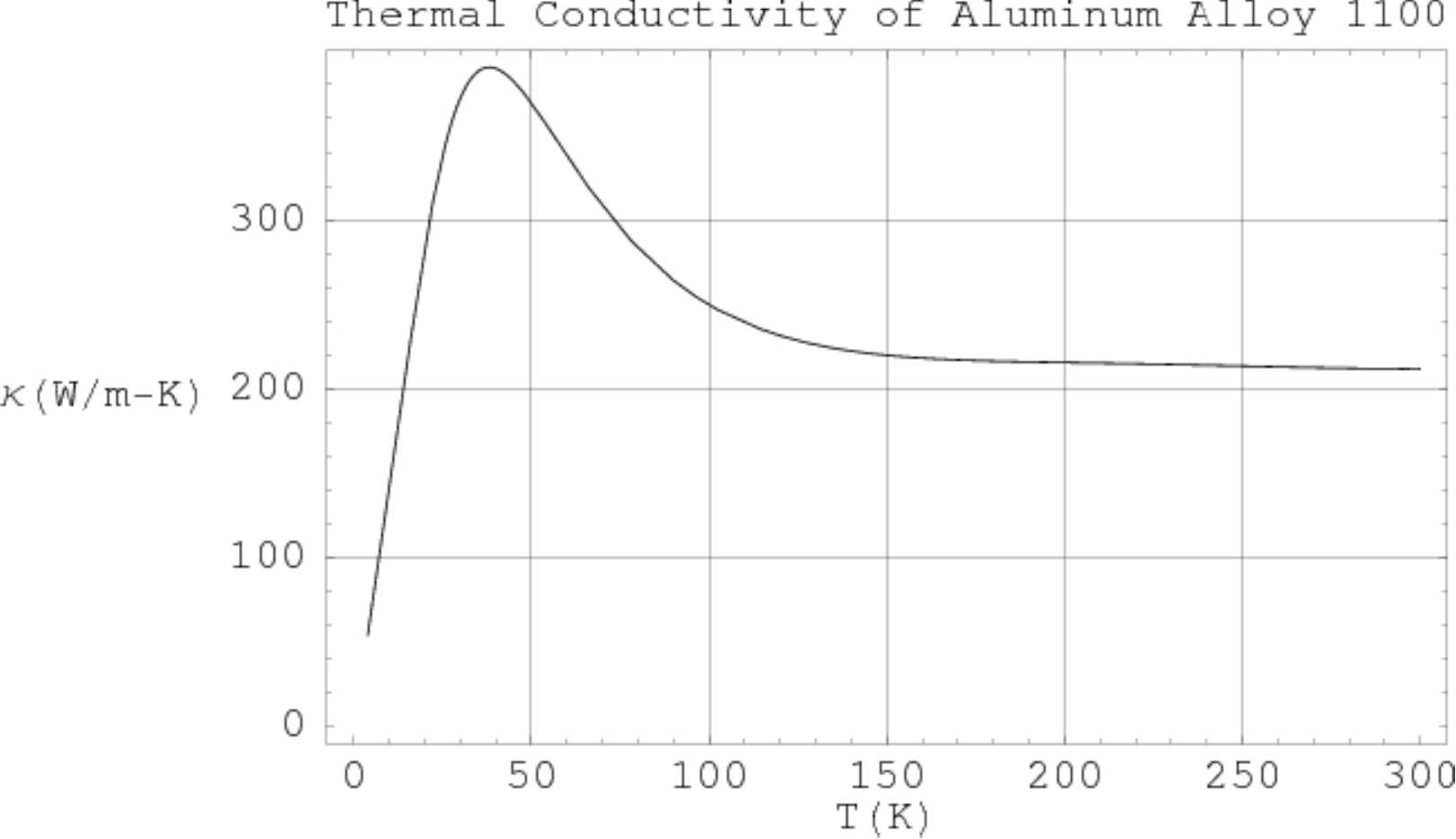}
\caption{Thermal conductivity of Aluminum Alloy 1100, computed from tables available on the NIST website \citep{nistweb}.
These values have been measured in the temperature range $T=4-300~\rm K$.}
\label{fig:kappa_al1100}
\end{figure}

Indium film (0.01$\arcsec$ thick) provides a soft, conductive layer between the 
nickel-plated copper refrigerator head and the copper strapping.
Indium  is also used in the interface between the strapping and the cold plate.
The four oxygen-free high-purity (OFHC) copper shims that conduct heat from the cold plate to the second stage of the refrigerator head
are each 0.05$\arcsec$ thick, 2$\arcsec$ wide,  and about 2$\arcsec$ long, yielding an area-to-length ratio 
$A/L = 4 \times 0.05\arcsec \simeq 0.005~{\rm m}$ (see Eq.~\ref{eq:Pthermal}).
At 12~K, OFHC copper has a high thermal conductivity\footnote{See 
NIST website, \url{http://www.cryogenics.nist.gov/MPropsMAY/material}~\url{properties.htm}.} 
of $\sim 7500~{\rm (W m^{-1} K^{-1})}$, so a load of 1 W produces a temperature differential 
of only $\sim 0.03~{\rm K}$ along the length of the strapping to the cold plate.  The measured differential
across this copper strapping agreed with this calculation to within the resolution\footnote{The temperature
resolution of the readout was 0.125~K, and the time resolution was $\sim$ 1~s. Given this temperature resolution and the
fact that the refrigerator cycles $\pm 0.5~$K in $\sim$ 2~s, the temperature at each end of the strapping
changes too rapidly to measure precisely this differential.} of the Lakeshore temperature sensors we used.
Aluminum alloy 1100, the common alloy of which the receiver parts are machined -- including the cold plate where 
the sensitive receiver electronics are mounted -- has a far lower conductivity at low temperatures 
than copper (aluminum alloy 1100 has a conductivity $\kappa \approx 10-100~{\rm (W m^{-1} K^{-1})}$; 
see \cite{woodcraft2005,nistweb} and Fig.\ \ref{fig:kappa_al1100}).
Even with a thickness of 0.25$\arcsec$, the aluminum cold plate exhibits a $\sim 3~\rm K$ 
thermal differential.
Braided copper strapping was therefore used to link the most sensitive components 
-- the 30-GHz HEMT and two 90-GHz MMIC HEMT blocks -- directly to the refrigerator head.

\begin{figure}
\includegraphics[width=5in]{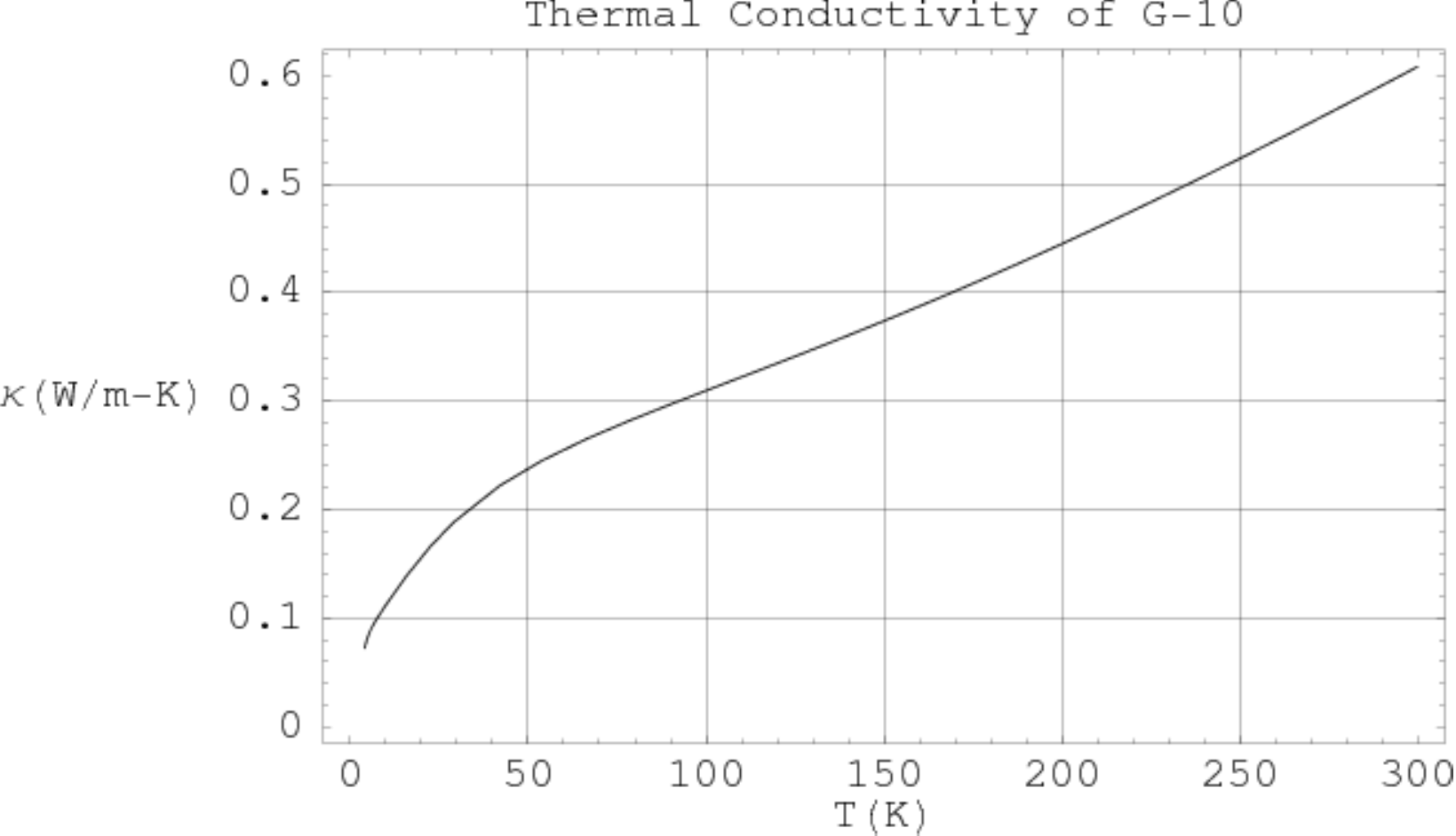}
\caption{Thermal conductivity of G-10, computed from tables available on the NIST website \citep{nistweb}.
These values have been measured in the temperature range $T=4-300~\rm K$.}
\label{fig:kappa_g10}
\end{figure}

The stages of the cryostat are separated by stand-offs made of a 
machinable fiberglass/epoxy composite known as ``G-10'' or ``garolite.''  
It has a thermal conductivity $\sim$ 0.1--0.2~${\rm (W m^{-1} K^{-1})}$ at temperatures from
10--100 K, and $\sim 0.6~{\rm (W m^{-1} K^{-1})}$ at room temperature, making it a good insulator 
(see Fig.\ \ref{fig:kappa_g10}).
The six stand-offs for the radiation shield, each 0.813~in (2.06~cm) long and with a 
cross-section of 0.062 in$^2$ (0.40~cm$^2$), contribute about 0.8 W to the total thermal 
load on the radiation shield.  Five stand-offs are used to separate the cold plate from
the radiation shield.  These stand-offs are each 0.5~in (1.25~cm) long and have the same cross-sectional 
area as the first set of stand-offs; they contribute roughly 50~mW to the cold plate.

A multi-layered blanket of aluminized mylar fills the space between the inner wall of the 
cryostat outer case (which is, of course, $\sim$ 300~K) and the radiation shield (see Fig.~\ref{fig:rxstrapping}).
The radiation shield equilibrates at $\sim$ 80~K, even though the first stage of the refrigerator is at $\sim$ 55~K, 
due to the poor conductivity of the aluminum of which the shield is made.
This blanket reduces thermal loading due to emission, as each layer is highly reflective
and will equilibrate to some temperature between 90-300 K.  Thermal emission
can be computed using the Stefan-Boltzmann Law ($P = b \sigma T^4 A$),
where $b$ is the emissivity coefficient of the material ($b \approx 0.10$ for aluminum), 
$\sigma$ is the Stefan-Boltzmann constant, $T$ is temperature (in K), and $A$ is the area of 
the absorbing/emitting surface. 
The strong temperature dependence encapsulated in the Stefan-Boltzmann Law implies that the innermost blanketing layer, 
which equilibrates at $\sim$ 80~K, radiates $\sim$ 200 times less than the 300~K cryostat inner wall.

In Fig.~\ref{fig:rxphoto}, the item labeled \#19 is called a ``getter.'' The getter contains activated charcoal, 
which acts as a trap for contaminants in the cryostat. 
The charcoal, which is cold, provides a large surface area onto which contaminants are adsorbed. 
Note that the final receiver configuration had the getter mounted to the cold plate.

Zotefoam was chosen as the material out of which to build the receiver windows, as it
also can hold vacuum (in addition to the properties discussed in the previous section). 
However, zotefoam is very elastic, and a window made of zotefoam alone 
bows under vacuum enough to be compressed against the corrugated feed horns.  Since these feed horns 
couple directly via waveguide to the HEMT amplifiers, and are mounted on the electronics plate,
they are ideally kept at the same temperature as the amplifiers.  Even given the 
poor conductivity of zotefoam, this strong contact between it and the feed horns 
is enough to dominate the thermal load on the cold stage.

The bowing problem was solved by introducing a layer of mylar as a backing 
for the zotefoam. 
Thin, microwave transparent mylar, like that in the receiver windows, is strong enough to support the zotefoam, 
but is not robust to sharp impacts. The zotefoam layer acts as a mechanical
protection for the mylar, and makes the vacuum seal.

\begin{figure}
\centerline{\includegraphics[width=5.5in]{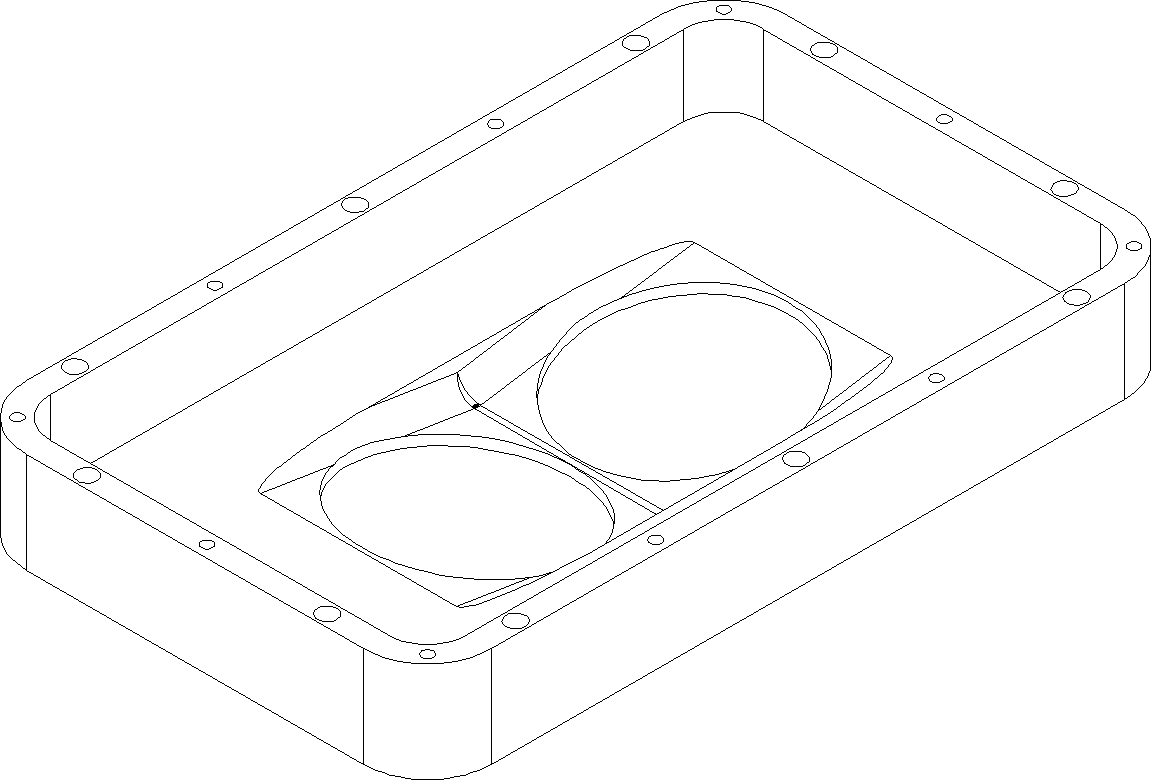}}
\caption{The window holder, as seen looking at the bed where the window is epoxied.}
\label{fig:windowholder}
\end{figure}

\begin{figure}
\centerline{\includegraphics[width=5.5in]{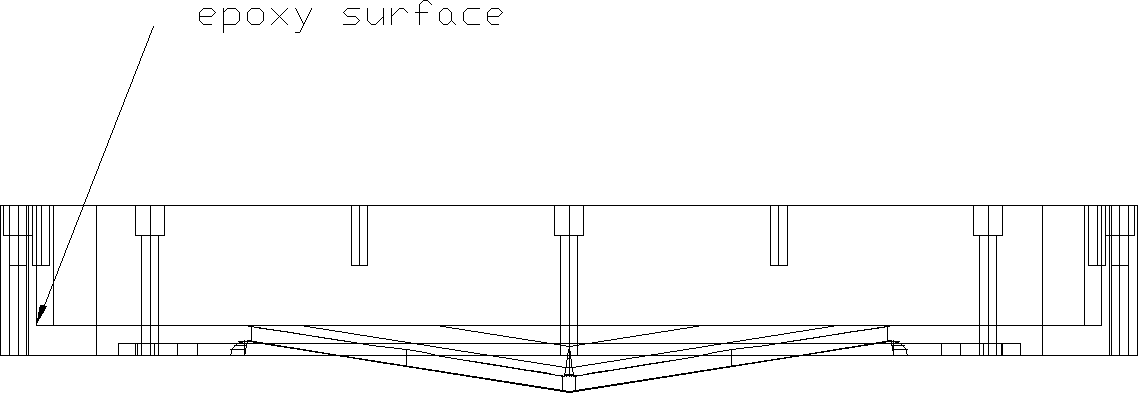}}
\caption{Cross-section of the window holder, depicting where the first layer of epoxy is applied.}
\label{fig:windowholder_crosssect}
\end{figure}

The windows are held by a custom-designed piece machined from a solid aluminum block 
(see Fig.\ \ref{fig:windowholder}).  
Two holes, each about 1.2 times larger than the outer diameter of the feed horns, allow the signal
to enter.  The bed of the holder provides a large, flat surface which was sandblasted to further increase
the surface area.  This surface was coated with a thin layer of epoxy (specifically, a clear 
Emerson \& Cuming Eccobond$^\circledR$ 24, two-part epoxy with high viscosity, which avoids getting 
epoxy in unwanted places), and a single layer of mylar is joined here, carefully removing bubbles 
by hand from the interface (see Fig.\ \ref{fig:windowholder_crosssect} for an illustration of this interface).  
After the first layer sets, the zotefoam is joined here by applying another layer of epoxy directly above 
the first layer.  This is done to avoid having epoxy in the optical path.  After this layer of epoxy sets, one last 
layer of epoxy is used to fill in the $\sim$ 1/10$^{\rm th}$ inch
gap around the zotefoam's edge, and a protective plate with a large elliptical gap for the signal is mounted
on the front.  The final step in this process is to add a pinhole in the mylar (on the vacuum side), 
allowing any air/moisture to escape from between the mylar/zotefoam interface when vacuum is drawn.
The final product is a zotefoam window that bows gently into the mylar without making
contact with any of the components inside the cryostat.

\section{Back-end Electronics}\label{backend}

\begin{figure}
\centerline{\includegraphics[width=6in]{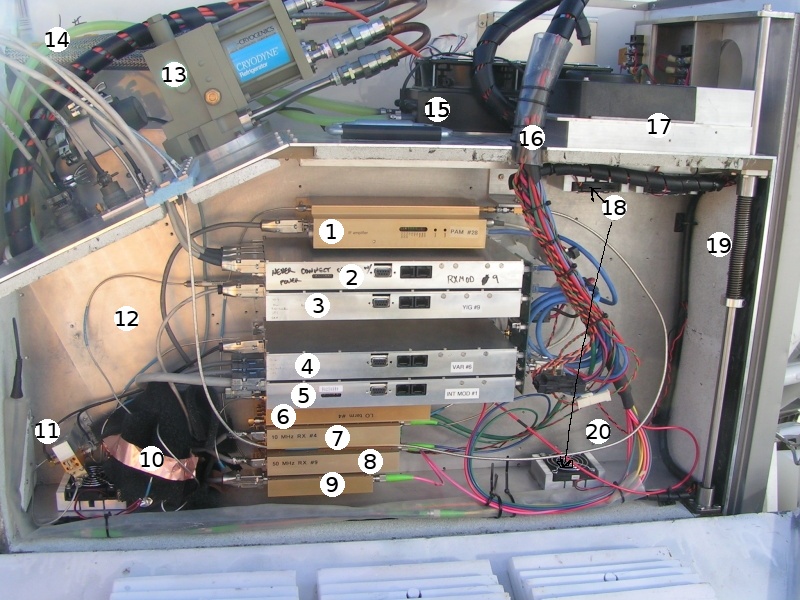}}
\caption{Photo of Ebox (lid removed), set up for 30-GHz observations.  
The Ebox houses the modules used to control and process signals from the receiver. 
See text in \S \ref{ebox_layout} for a description of the electronics.
Note that the bias-tuned Gunn (BTG), the module to control the bias-tuned Gunn (the ``BTG Mod'' in 
Fig.~\ref{fig:ebox_modules}), and the IF switch had not been installed at the time of 
this photo, as they were only required for 90~GHz observations.
They were later installed in locations {\bf 12} and {\bf 20}, respectively.
The fiber bundle and power cables enter the Ebox at location {\bf 16}.
The refrigerator head ({\bf 13}) for the receiver (located behind the Ebox), the receiver enclosure's 
chiller line ({\bf 14}), the chiller heat exchanger ({\bf 15}), and the TEC cooling fans ({\bf 17}) 
can all be seen above and outside the Ebox. The internal Ebox air circulation fans at label {\bf 18}.  A spring
that assists in lifting the Ebox is labeled {\bf 19}.
The walls of the Ebox are lined with an open-cell PVC foam for insulation, which
was painted white to prevent degradation and flaking due to weathering.
}
\label{fig:ebox_30GHz}
\end{figure}

\begin{figure}
\centerline{\includegraphics[width=6in]{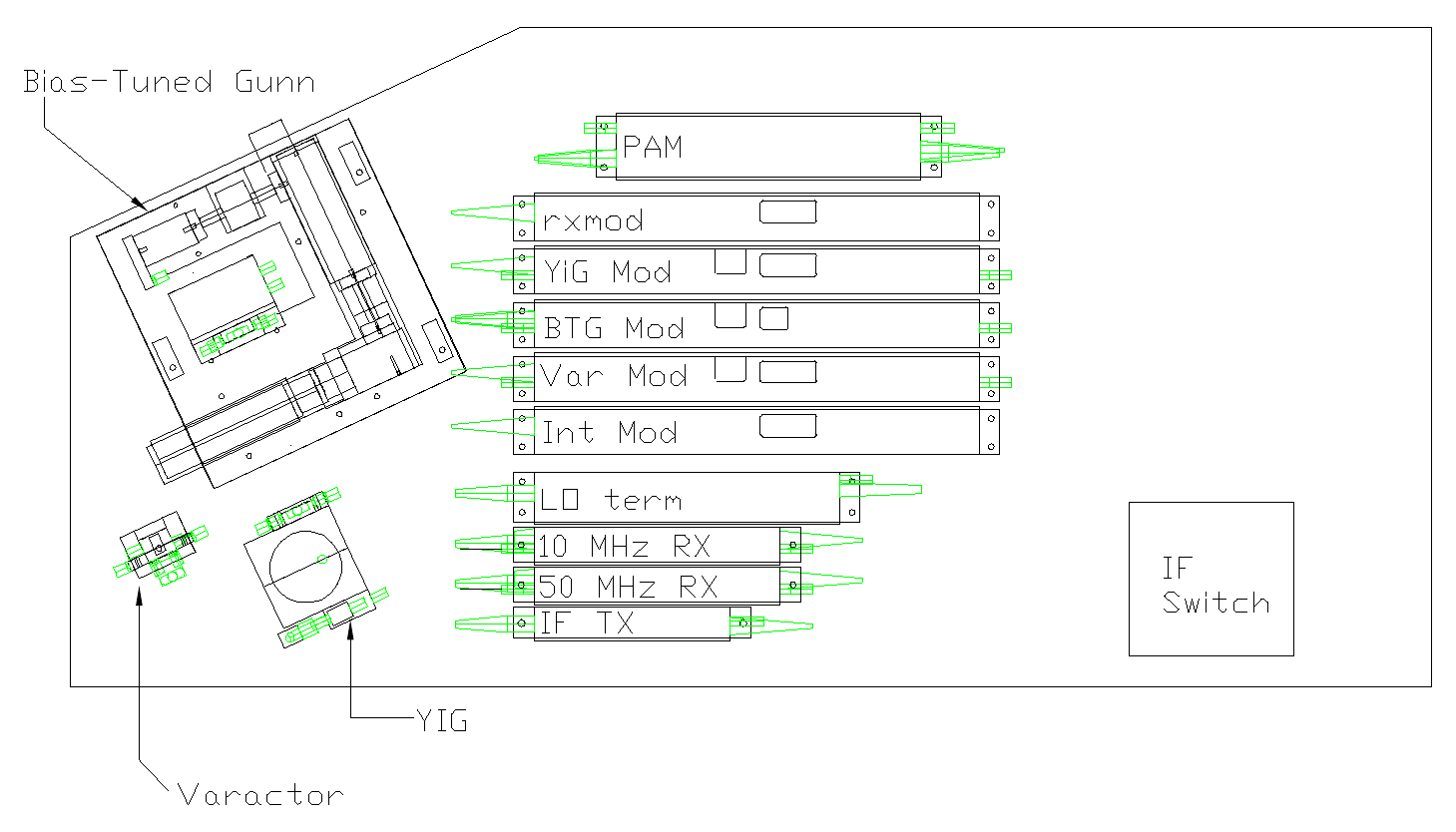}}
\caption{Layout of Modules in the Ebox.  Connectors are shown in green.}
\label{fig:ebox_modules}
\end{figure}

\subsection{Overview of the back-end electronics}\label{ebox_layout}

All support electronics for the receiver are located in a large, thermally-regulated, 
insulated box, called the ``Ebox'' (see Fig.~\ref{fig:ebox_30GHz}). 
The Ebox is placed within the receiver enclosure, just outside the cryostat.  
The primary components in this box (shown schematically in Fig.~\ref{fig:ebox_modules}) are:
 
\begin{itemize}

\item An IF switch, used to select the receiver output appropriate for 30 or 90-GHz observations.
The IF switch is mounted in location \#20 in Fig.~\ref{fig:ebox_30GHz}.

\item A pre-amplifier module (PAM in Fig.~\ref{fig:ebox_modules}, \#1 in Fig.~\ref{fig:ebox_30GHz}), 
which is primarily an IF amplifier, is common to both receivers
as it follows the IF switch.  The PAM also includes a set of precisely-calibrated attenuators 
(in 0.5~dB steps, up to a total attenuation of -30~dB), both on the input and output of the module.  
These attenuators are useful in avoiding compression effects when the power becomes high enough to
saturate electronics further down the signal chain.
Since the hot load calibrator (a thermal blackbody at ambient temperature) 
increases the receiver output power by a factor $\sim (300~{\rm K})/T_{\rm sys}\sim 4-6$,
where $T_{\rm sys}\sim$ 40--50~K, these attenuators are also inserted during 
calibrations to prevent compression of the signal (for more information, see \S \ref{data_calib}).

\item A power supply for the receiver (labeled ``rxmod'' in Fig.~\ref{fig:ebox_modules}
\#2 in Fig.~\ref{fig:ebox_30GHz}).  This powers both the 30 and 90-GHz receivers.

\item An high-frequency oscillator, based on a Yttrium Indium Gallium (YIG in Fig.~\ref{fig:ebox_modules},
\#10 in Fig.~\ref{fig:ebox_30GHz})  diode in a cavity, that operates at 8.972~GHz.
The module that monitors the YIG's status is labeled \#3.

\item The varactor, a ``variable capacitance'' component (\#11 in Fig.~\ref{fig:ebox_30GHz}) 
that is used to produce 
harmonics of the YIG frequency as well as lock the phase of the LO.  A phase-lock 
loop (PLL), controlled by electronics in the correlator trailer, locks the phases 
of all antennae to a single reference.
The module that controls the varactor is \#3.

\item A bias-tuned Gunn (BTG) oscillator, which is the 90-GHz system's equivalent of the YIG.  
This uses a Gunn diode forced into oscillation within an adjustable waveguide cavity.  
This was later installed in location \#12 in Fig.~\ref{fig:ebox_30GHz}, while the module
to control it was installed between \#3 and \#4 (``BTG Mod'' in Fig.~\ref{fig:ebox_modules}).

\item A dielectric resonance oscillator (DRO), which supplies the fixed 17.5~GHz secondary LO used by 
the 90-GHz system's secondary mixer.  This is mounted on the same plate as the BTG, so it was also 
installed in location \#12 in Fig.~\ref{fig:ebox_30GHz}.

\item Various CANBUS (a protocol like TCP/IP that allows daisy-chaining) controlled electronics, used to 
control the various oscillators and reference frequencies of the PLL and lobe rotator (which accounts
for the part of fringe tracking due to movement of the source across the sky), regulate the temperature,
etc.  Among these is the interface module (``Int Mod'', \#5 in Fig.~\ref{fig:ebox_30GHz}), 
which monitors the phase reference and lobe rotator. These references are received
by the 10 and 50~MHz receivers (\#7 and \#8 in Fig.~\ref{fig:ebox_30GHz}).

\item The local oscillator terminator (``LO Term'' in Fig.~\ref{fig:ebox_modules}, \#6 in Fig.~\ref{fig:ebox_30GHz}), 
which locks the YIG to the ninth harmonic of a central, synthesized reference frequency with the 10~MHz signal mentioned above.  
This phase lock loop is separate from, but analogous to, the one which locks the varactor to the YIG using 
the 50~MHz reference.

\item An optical IF band transmitter (called the ``OTX''; labeled ``IF Tx'' in Fig.~\ref{fig:ebox_modules},
\#9 in Fig.~\ref{fig:ebox_30GHz}), 
which takes as its  input the output of the PAM. This sends the signal over fiber optics 
(\#16 in Fig.~\ref{fig:ebox_30GHz}) to the downconverter and correlator (which are discussed in 
\S \ref{dcon} \& \ref{corr}, respectively).

\end{itemize}

\subsection{Thermal considerations for the electronics box}

Thermal stability is the primary motivation for the implementation of this insulated electronics box.  
The thermal sensitivity of the amplifiers, local oscillators, and fiber optics (whose lengths 
change with temperature) leads to phase variations during periods of poor thermal regulation.  
While these effects, discovered during commissioning operations, varied from antenna 
to antenna, one extreme example exhibited $\pm45^{\circ}$ 
phase oscillations as the temperature varied sinusoidally about its nominal regulation temperature
by $\pm2.5^{\circ}$C (which could happen even with the final thermal regulation in place). 
Since these oscillations had periods on the order of 15-20 minutes, the same frequency 
as that of our calibration observations, the sampling of these phase changes is not frequent
enough to remove the effects through data calibration.  This problem was eventually solved
with better regulation algorithms and the occasional manual intervention by the observer.

Thermal regulation of the electronics box, utilizing thermo-electric coolers (TECs, or Peltier junctions), 
is set up so that the temperature at a single point on the electronics mounting plate is kept within $\pm 0.125 
^{\circ}$C of the 32$^{\circ}$C (28$^{\circ}$C) set point in the summer (winter).  
The thermal differentials within this box are reduced by fans, and by the large thermal mass and 
high thermal conductivity\footnote{Note, as in Figs.\ \ref{fig:kappa_al1100} 
and \ref{fig:kappa_ofhc_cu}, that -- at room temperature -- aluminum is only a factor of $\sim 2.5$ 
less conductive than copper, but is also lighter by an equivalent factor. 
Weight was an important consideration, as the Ebox has to be lifted from the receiver enclosure 
to access the receivers.} of the 1/4-inch aluminum plate to which all the electronics are mounted.
A small amount of silicone grease (heatsink compound) is used to ensure decent conductivity 
between the electronics modules and the mounting plate.

\begin{figure}
\centerline{\includegraphics[width=6.5in]{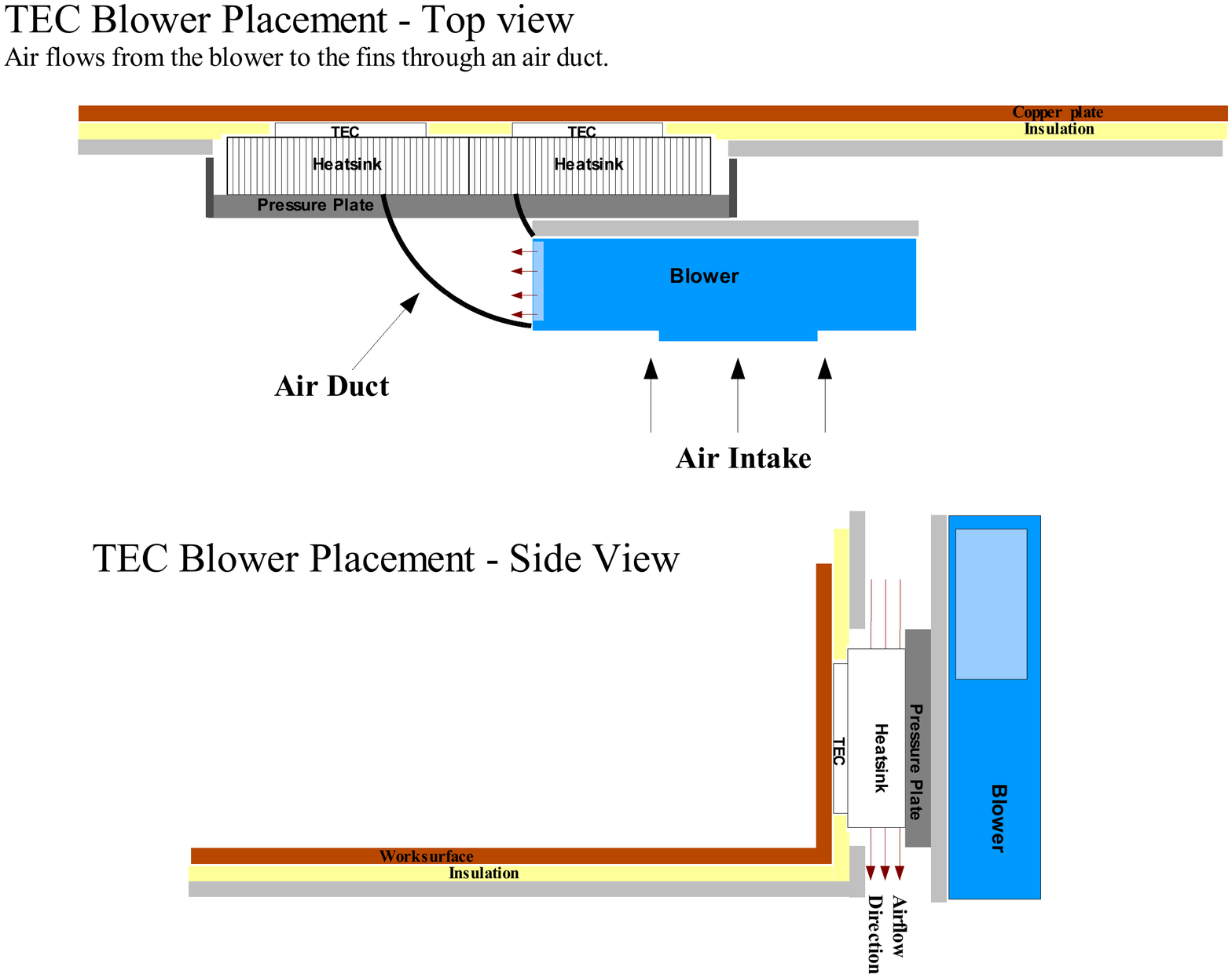}}
\caption{TEC Blower Conceptual Diagram, courtesy Marshall Joy and Georgia Richardson.  Ambient air from the receiver
enclosure is forced through the heatsink fins by the blower (see Fig.~\ref{fig:ebox_30GHz}, \#17).  
These copper heatsinks are thermally coupled to the hot sides of the TECs
(which are $\gtrsim 20^\circ \rm C$ hotter than the worksurface side).  The TECs diffuse more heat away from the 
worksurface than toward it, thus removing heat from the Ebox.}
\label{fig:tec_concept}
\end{figure}

A conceptual overview of the TEC setup for the electronics box is show in Fig.~\ref{fig:tec_concept}.
Ambient air is forced, by a large impeller fan (\#17 in Fig.~\ref{fig:ebox_30GHz}), through the fins 
of the heatsink attached to the hot side of each TEC, thereby carrying away the heat.  
Good thermal contact between the heatsink and the TEC is ensured by the pressure plate, which
is held down by five springy Belleville washers, each providing 20-30 pounds of force.
This setup also forces the cold side of the TECs onto a copper plate that makes thermal
contact with the electronics mounting plate, labeled ``worksurface'' in Fig~\ref{fig:tec_concept}.  
Since the TECs are ceramic, their porous surfaces are coated with a layer of heatsink compound,
which fills the pores and provides more surface area for thermal contact.
Open-cell foam insulation isolates the temperature-regulated copper plate and worksurface from
the warmer outside air.

The TECs can be operated at a maximum power of $P_\text{max} = 300$~W (10~A at 30~V, DC), removing 
$P_\text{max,remove} = \xi P_\text{max}$ for TEC efficiency $\xi$.  With cooling efficiencies of 
$\xi \approx 30 \%$, they can keep a thermal load of up to 90~W regulated at a desired set temperature.  
The electronics within the Ebox draw $\sim$3~A at 24~V (DC), and therefore dissipate $\sim$ 72~W of heat.  
There is a remaining thermal load that is due to both the finite thermal conductivity of the 
G-10 supports and foam insulation of the Ebox, and to the imperfect thermal isolation provided by the Ebox.  

The Ebox insulation is a rigid PVC foam, chosen because it is the best insulating foam offered by 
McMaster-Carr\footnote{\url{http://www.mcmaster.com}}.  The conductivity of the foam is $90~{\rm mW \cdot in/ft^2/K}$
($0.17~\rm btu \cdot in/ft^2/^\circ F$).  
With $\sim 11~\rm ft^2$ of 0.375~in foam insulation, 2.6~W of heat flow through this insulation
for each degree of differential between the outside and inside of the box (see Eq.~\ref{eq:Pthermal}),
In practice the electronics box cannot maintain a temperature more than $\sim 1^\circ$C less than the 
ambient temperature of the receiver enclosure, due to leaks that allow some air exchange between
the inside and outside of the box.

\section{The Correlator Trailer}

The signal from each telescope is sent over fiber optic cables, located in conduits 
underground (as illustrated in Fig.~\ref{fig:system_overview}), to an air-conditioned trailer 
-- referred to as the ``correlator trailer'' -- containing the instrumentation necessary 
to downconvert (\S \ref{dcon}) and correlate (\S \ref{corr}) the signals.  
The correlator trailer also houses the computer that controls the array and collects all data
taken during an observation.

\subsection{Downconverter}\label{dcon}

\begin{figure}
\centerline{\includegraphics[width=6in]{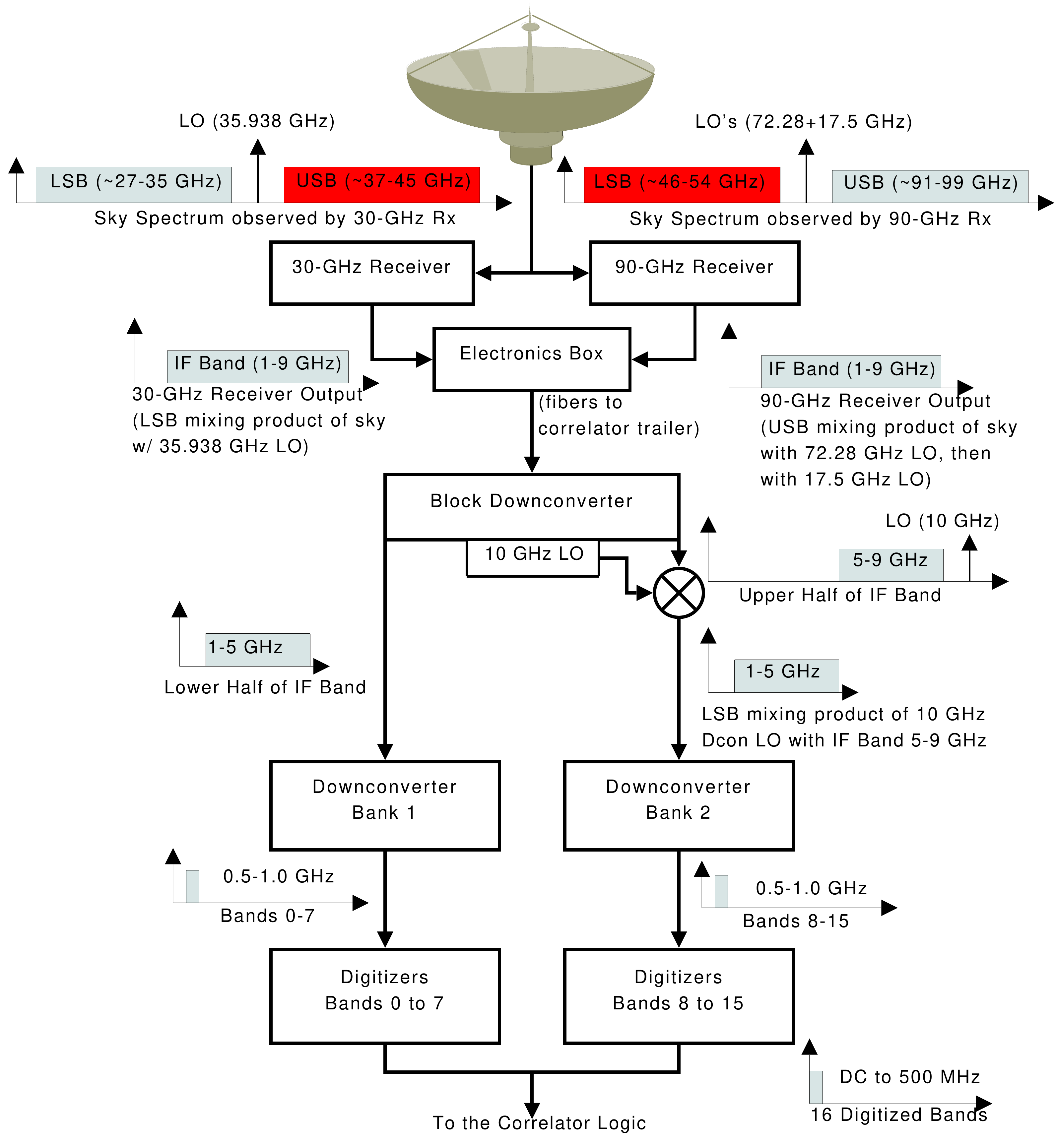}}
\caption{Block Diagram of the SZA Band Downconversion, showing how the sky observation
frequencies 26.938--34.938 and 90.78--98.78~GHz are split into the sixteen digitized bands, each 500~MHz
in bandwidth, that are the input to the correlator.  See description in the text (\S \ref{dcon}).
Bands shown in red are the side bands that are not used.}
\label{fig:sza_bands}
\end{figure}

The 1--9~GHz IF is processed by an array of mixers, amplifiers, and bandpass filters, 
known collectively as the ``downconverter,'' 
which splits the IF into sixteen 500~MHz bands running from 0.5--1~GHz (see Fig.~\ref{fig:sza_bands}).
The reason for using this range is that electronics that can handle 0.5~GHz and 1~GHz are similar, while
those that could handle a signal from 0--0.5~GHz; for instance,
DC signals cannot be capacitively coupled.  

Here I describe the full signal downconversion chain, starting from the sky frequencies 
26.938--34.938 and 90.78--98.78~GHz, and ending with the sixteen digitized 500~MHz bands that constitute
the input to the correlator. Figure \ref{fig:sza_bands} illustrates this signal chain.

\begin{figure}
\centerline{\includegraphics[width=6in]{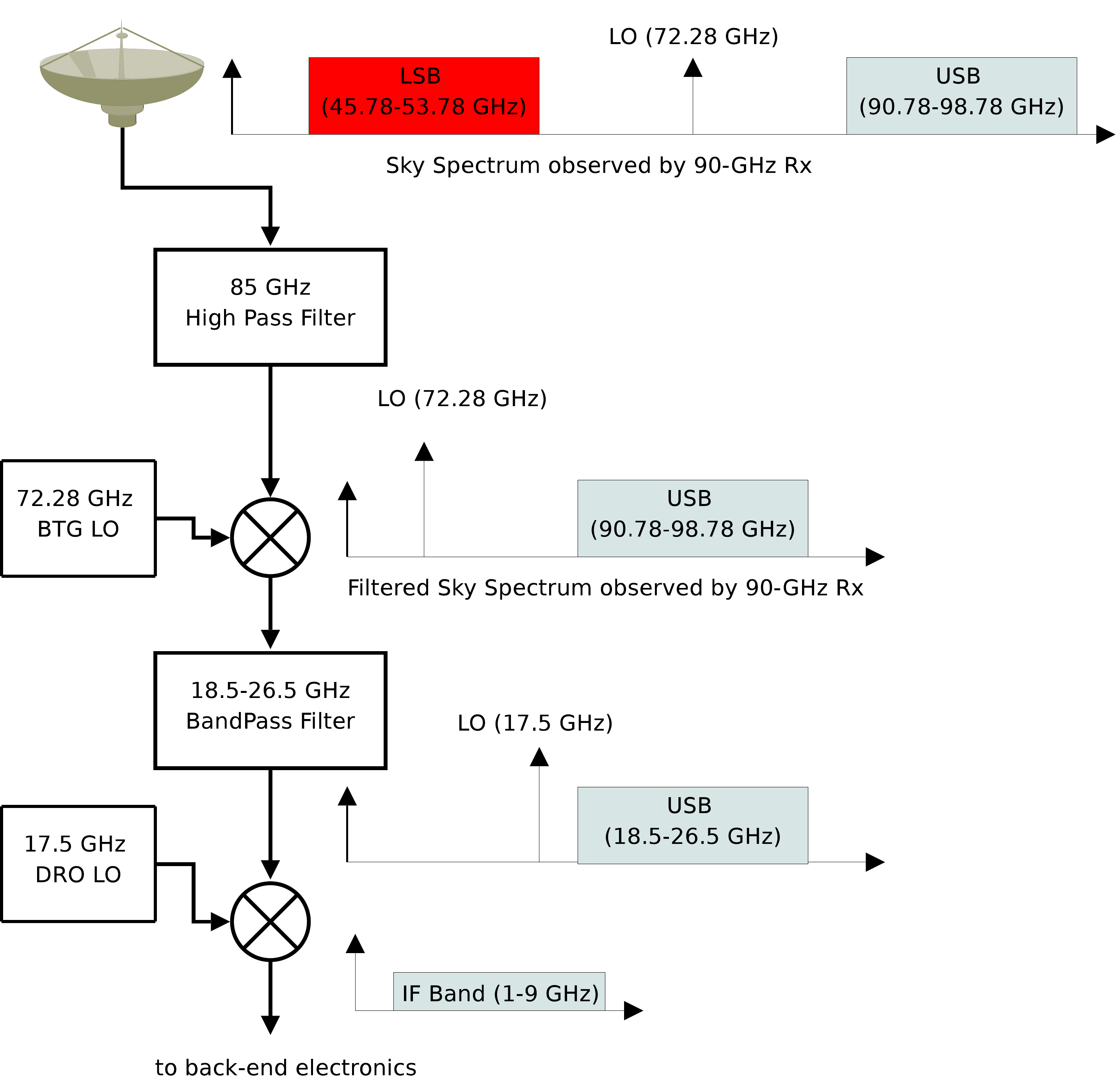}}
\caption{Detailed Diagram of the 90-GHz Receiver Downconversion, showing how the sky observation
frequencies are brought down to the 1--9~GHz IF band (see Fig.~\ref{fig:90block} for details on the 
receiver RF components).  This figure accompanies the broader downconversion scheme illustrated in
Fig.~\ref{fig:sza_bands}.
In the current tuning scheme, sky frequencies 90.78--98.78~GHz are mixed ($\otimes$) with
a 72.28~GHz LO (the bias-tuned Gunn), producing an 18.5--26.5~GHz USB product.  
Sky frequencies $< 85~\rm GHz$ are blocked by the high pass filter, so no LSB
mixing product is produced (e.g. the band shown in red).
The bandpass filter passes the 18.5--26.5~GHz USB product to a second mixer.  
The second LO, at 17.5~GHz, mixes with this 18.5--26.5~GHz USB product to 
place the 90-GHz receiver output in the 1--9~GHz IF band.  This 1--9~GHz IF signal is the input
to the back-end electronics common to both the 30 and 90-GHz systems.  See \S \ref{dcon} 
for more details.}
\label{fig:sza_90ghz_bands}
\end{figure}

The tertiary mirror of each telescope selects either the 30 or the 90-GHz receiver.  In the 
30-GHz receiver system, the lower side band (LSB) mixing product of the sky with a 35.938~GHz local 
oscillator (LO) places sky frequencies 26.938--34.938~GHz in the 1--9~GHz IF band.  
In the 90-GHz receiver, a 72.28~GHz LO (in the current tuning scheme) mixes with sky 
frequencies 90.78--98.78~GHz, producing a 18.5--26.5~GHz upper side band (USB) product;  
a second LO at 17.5~GHz mixes with this to place the 90-GHz receiver output (also USB) 
in the 1--9~GHz IF band (see Fig.~\ref{fig:sza_90ghz_bands} for more details on the downconversion
in the 90-GHz receiver).  Note that the 30-GHz receivers have a full bandwidth of 10~GHz, while
the 90-GHz receivers have a much broader bandwidth that allows observations from 85--115~GHz.  
For both systems, the back-end electronics determine what range of sky frequencies are observed.

\begin{table}[t]
\centerline{
 \begin{tabular}{|c|c|c|c|}
  \hline
  Band & IF Band Range & 30-GHz Sky Frequencies & 90-GHz Sky Frequencies\\
  (\#) & (GHz) & (GHz) & (GHz) \\
  \hline
  0  & 1.0--1.5 & 34.688 & 91.030 \\
  1  & 1.5--2.0 & 34.188 & 91.530 \\
  2  & 2.0--2.5 & 33.688 & 92.030 \\
  3  & 2.5--3.0 & 33.188 & 92.530 \\
  4  & 3.0--3.5 & 32.688 & 93.030 \\
  5  & 3.5--4.0 & 32.188 & 93.530 \\
  6  & 4.0--4.5 & 31.688 & 94.030 \\
  7  & 4.5--5.0 & 31.188 & 94.530 \\
  8  & 5.0--5.5 & 30.688 & 95.030 \\
  9  & 5.5--6.0 & 30.188 & 95.530 \\
  10 & 6.0--6.5 & 29.688 & 96.030 \\
  11 & 6.5--7.0 & 29.188 & 96.530 \\
  12 & 7.0--7.5 & 28.688 & 97.030 \\
  13 & 7.5--8.0 & 28.188 & 97.530 \\
  14 & 8.0--8.5 & 27.688 & 98.030 \\
  15 & 8.5--9.0 & 27.188 & 98.530 \\
  \hline
 \end{tabular}
}
\caption{The 1--9~GHz IF band is separated into 16 bands, as outlined in Fig.~\ref{fig:sza_bands}. 
These are the central sky frequencies of both the 30 and 90-GHz SZA systems that correspond to 
the downconverted IF bands, discussed in \S \ref{dcon}.  Note that higher band number corresponds to
a lower sky frequency for the 30-GHz system because it uses the lower side band.
}
\label{table:band_scheme}
\end{table}

Back-end electronics in the electronics box select the proper receiver output 
and send the signal over fibers to the correlator trailer (see Fig.~\ref{fig:system_overview}).  
The IF is then split into 1--5 and 5--9~GHz bands by the block downconverter.
The lower half of the IF, from 1--5~GHz, is passed directly to the second stage downconverter.  
The 5--9~GHz band is mixed with a 10~GHz LO, also in the block downconverter, 
and the LSB mixing product (from 1--5~GHz) continues to the second stage downconverter.
I use the $\otimes$ symbol to represent the mixer, which is located within the block downconverter, to 
clearly denote how the 1--9~GHz IF band is first split into 1--5 and 5--9~GHz.

The second stage downconverters mix and filter each 1--5~GHz band
into sixteen separate 0.5--1.0~GHz bands, listed in Table~\ref{table:band_scheme}.  
Each 0.5--1.0~GHz band is then digitized at a sampling frequency $f_s=1$~GHz. This sampling
operation digitally aliases each 0.5--1.0~GHz band to a 0--500~MHz 
baseband, digitized signal.\footnote{Recall that the Nyquist theorem states 
$f_s \geq 2 f_\text{bw}$, where $f_s$ is the sampling frequency and
 $f_\text{bw}$ is the bandwidth of the signal.
A sampling frequency of 1~GHz limits the sampled signals bandwidth to 0.5~GHz.}
This sampled signal is the input to the SZA correlator, discussed in the next section. 

\subsection{The SZA Correlator}\label{corr}

\begin{figure}
\centerline{\includegraphics[width=6in]{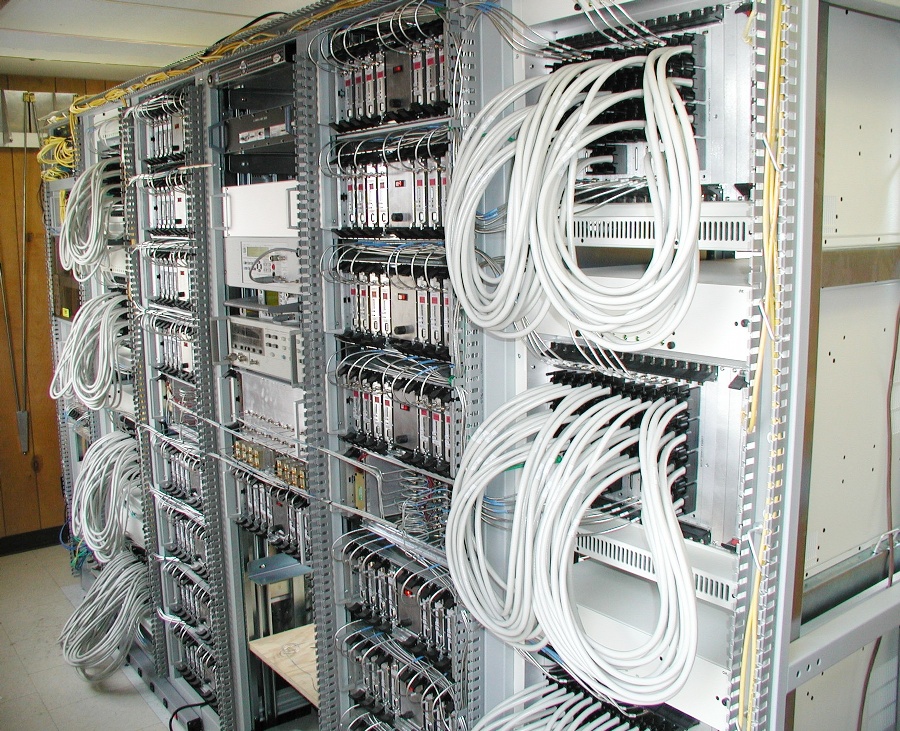}}
\caption{Photo of the SZA Correlator and Downconverter, courtesy David Hawkins.}
\label{fig:corrphoto}
\end{figure}

In the discussion of interferometry presented in \S \ref{interf}, details about the process
of interfering the signals from a pair of antennae, and the instrument that performs this 
\emph{cross-correlation}, were for simplicity ignored.
I now discuss the correlator, which is the single instrument most crucial for performing
interferometry.  Fig.~\ref{fig:corrphoto} shows a photo of the SZA correlator.

The cross-correlation $r_{xy}(m)$ for digital time delay $m$ of two discrete-time signals
$x(n)$ and $y(m+n)$ (delayed by $m$ samples with respect to $x(n)$), such as the digitized 
signals from a pair of antennae, is defined
\begin{equation}
r_{xy}(m) \equiv \sum_{n=-\infty}^\infty x[n]~y[m+n].
\label{cross_corr}
\end{equation}
Here, $n$ is a sample taken at time $t=n\tau$ for sampling period $\tau \equiv1/f_s$, and $m$ a whole number
\emph{lag} -- a discrete time delay -- applied to one antenna's signal.  
Since the SZA bands are sampled at $f_s=1$~GHz, $\tau = 1$~ns.
In a real instrument, we cannot sum over the infinite number of samples indicated in Eq.~\ref{cross_corr}.
For the SZA, correlation is performed for 32 lags of each signal, 
after having been properly delayed by fringe tracking and downconverted to 
0.5--1~GHz bands (as discussed in \S~\ref{dcon}).
The overall effect of the correlator is to multiply the signals from two antennae,
sliding one signal past the other.\footnote{For comparison, note that convolution, 
a similar operation to correlation, is the flipping and sliding of one signal past the other.} 
Since the signals $x(n)$ and $y(n)$ scale as voltage, the correlator output 
scales as voltage-squared; cross-correlation is therefore proportional to the power
received by an antenna.

We are accustomed to dealing with visibilities as a function of frequency.  
Visibilities, when calibrated against an astronomical source of 
known flux, give the flux at each data point in \emph{u,v}-space, for each frequency in 
a spectrum (see \S \ref{interf}).
The spectrum is simply obtained by Fourier transforming the correlated data.

The SZA correlator is an \emph{XF} or lag-correlator,
where \emph{X} stands for the cross-correlation, and \emph{F} stands for the Fast 
Fourier Transform (FFT). 
The SZA correlator performs cross-correlation in the time domain, but note that \emph{FX}
correlators are also common, since correlation has a simple Fourier relation
\begin{equation}
r_{xy}(m) = \mathcal{F}\{ X[k]^* Y[k] \},
\label{cross_corr_Fourier}
\end{equation}
where $X[k]$ and $Y[k]$ are the Fourier transforms of $x(n)$ and $y(n+m)$ for discrete frequencies
$k$, $\mathcal{F}$ is the Fourier transform operation, and $^*$ denotes the complex 
conjugate.\footnote{Note for comparison that convolution in the time domain is simply 
multiplication in the frequency domain (i.e. neither signal is complex-conjugated, unlike in correlation).}

The FFT is a computationally efficient implementation of the Discrete Fourier Transform (DFT),
which in turn is a truncation of the Discrete Time Fourier Transform (DTFT).
The DTFT is mathematically complete \citep[e.g. it satisfies Parseval's Theorem;
see][for further information]{oppenheim1999,bracewell2000},
and becomes the familiar (continuous time) Fourier Transform in the limit of infinitesimally
small steps in time.

The DTFT is:
\begin{equation}
X(\omega) = \sum_{\scriptscriptstyle n=-\infty}^{\scriptscriptstyle \infty} 
x(n)~e^{\scriptscriptstyle -j \omega n}
\label{eq:dtft}
\end{equation} 
where
\begin{equation}
x(n) = (1/2\pi) ~ \int_{\scriptscriptstyle -\pi}^{\scriptscriptstyle \pi} 
X(\omega)~e^{\scriptscriptstyle j \omega n} d\omega.
\label{eq:dtft2}
\end{equation} 
The DFT is a special case of the DTFT, where the 
number of samples is limited to a finite number $N$. This has the effect
of discreting frequency at $\omega = 2 \pi k/N$, where $k=N(f/f_s)$ is the discretized
wavenumber.   
In the DTFT, $X(\omega)$ wraps outside the range $\omega = [-\pi, \pi]$;
in the DFT, $X(\omega)$ is sampled at only the discrete values of $k$, 
and $X(k)$ wraps outside the range $k=[0,N-1]$.  
The DFT is:
\begin{equation}
X(k) = \sum_{\scriptscriptstyle n=0}^{\scriptscriptstyle N-1} 
x(n)~e^{\scriptscriptstyle -j 2 \pi k n / N}
\label{eq:dft}
\end{equation} 
where
\begin{equation}
x(n) = (1/N) ~ \sum_{\scriptscriptstyle k=0}^{\scriptscriptstyle N-1} 
X(k)~e^{\scriptscriptstyle j 2 \pi k n / N}
\label{eq:dft2}
\end{equation} 
The substitutions that take us from the DTFT to the DFT have the effect
of limiting the frequency resolution to $N/2$ elements (in the Nyquist
bandwidth of $f_s/2$; there are $N$ discrete $k$'s in the range 
$f=[-f_s/2,f_s/2]$).
Notice that the wrapping in the DTFT and DFT (Eqs.~\ref{eq:dtft2} \& \ref{eq:dft2}) is the same
as wrapping every $f_s$ in frequency, or every harmonic of the sampling frequency; 
this aliasing is simply that due to Nyquist sampling at $f_s$, mentioned in
\S \ref{dcon}.  Only the range $k=[0,N/2-1]$, 
corresponding to frequencies $\omega = [0, \pi]$, contains unique spectral information.
Since the time-domain signal is real, its transform is Hermitian, meaning those frequencies 
above $f_s/2$ are the complex conjugate of the positive frequencies.
Table \ref{table:channel_scheme} lists the 17 channels produced by FFT-ing the 32 samples.

\begin{table}[t]
 \centerline{
  \begin{tabular}{|c|c|}
   \hline
   Channel & Frequency \\
   (\#) & (MHz) \\
   \hline
   0  & ~~0.00 \\
   1  & ~31.25 \\
   2  & ~62.50 \\
   3  & ~93.75 \\
   4  & 125.00 \\
   5  & 156.25 \\
   6  & 187.50 \\
   7  & 218.75 \\
   8  & 250.00 \\
   \hline
  \end{tabular}
  \begin{tabular}{|c|c|}
   \hline
   Channel & Frequency \\
   (\#) & (MHz) \\
   cont'd & cont'd \\
   \hline
   9  & 281.25 \\
   10 & 312.50 \\
   11 & 343.75 \\
   12 & 375.00 \\
   13 & 406.25 \\
   14 & 437.50 \\
   15 & 468.75 \\
   16 & 500.00 \\
   \hline
  \end{tabular}
 }
\caption{Each 500~MHz band has 17 channels, numbered 0--16.  
Channels 0 and 16 are both attenuated by the 500~MHz bandpass filters used in the 
downconverter \S \ref{dcon}, and are not used for data.  Note that these are discrete
frequencies that arise from the DFT of a discrete-time signal (Eqs.~\ref{eq:dft}), rather
than continuous bands.}
\label{table:channel_scheme}
\end{table}

In the time domain, the limited number of correlator lags $m$ used in the DFT is equivalent 
to rectangular windowing -- the one-dimensional equivalent to the top hat function --  
of the correlated signal $r_{xy}(m)$ (Eq.~\ref{cross_corr}). 
Outside of the range $n=[0,N-1]$ (in discrete time), the signal is zero, 
and this has important implications when interpreting the DFT as a power spectrum
(which are discussed in \S \ref{spectral_leakage}).  

The mathematics of the cross-correlation and Fast Fourier Transform (FFT) are programmed 
into field-programmable gate arrays (FPGAs), which can be faster than using 
general-purpose processors to perform digital correlation.
The raw visibilities produced by this process are averaged for 0.5 second intervals and stored
to disk as raw visibilities in correlator units.\footnote{The correlator electronics are set
up such that the auto-correlation of each antenna's signal -- applying Eq.~\ref{cross_corr} with 
$x(n)=y(n)$ to find how a signal correlates with itself -- is 1.  
Deviations from this are used later, in calibration (\S \ref{data_calib}),
to diagnose problems in the correlator.}
The reader is encouraged to see \citet{hawkins2004} for further details, where the
SZA correlator is described in much greater detail.

\subsection{Thermal Considerations for the correlator trailer}

\begin{figure}
\centerline{\includegraphics[width=5.75in]{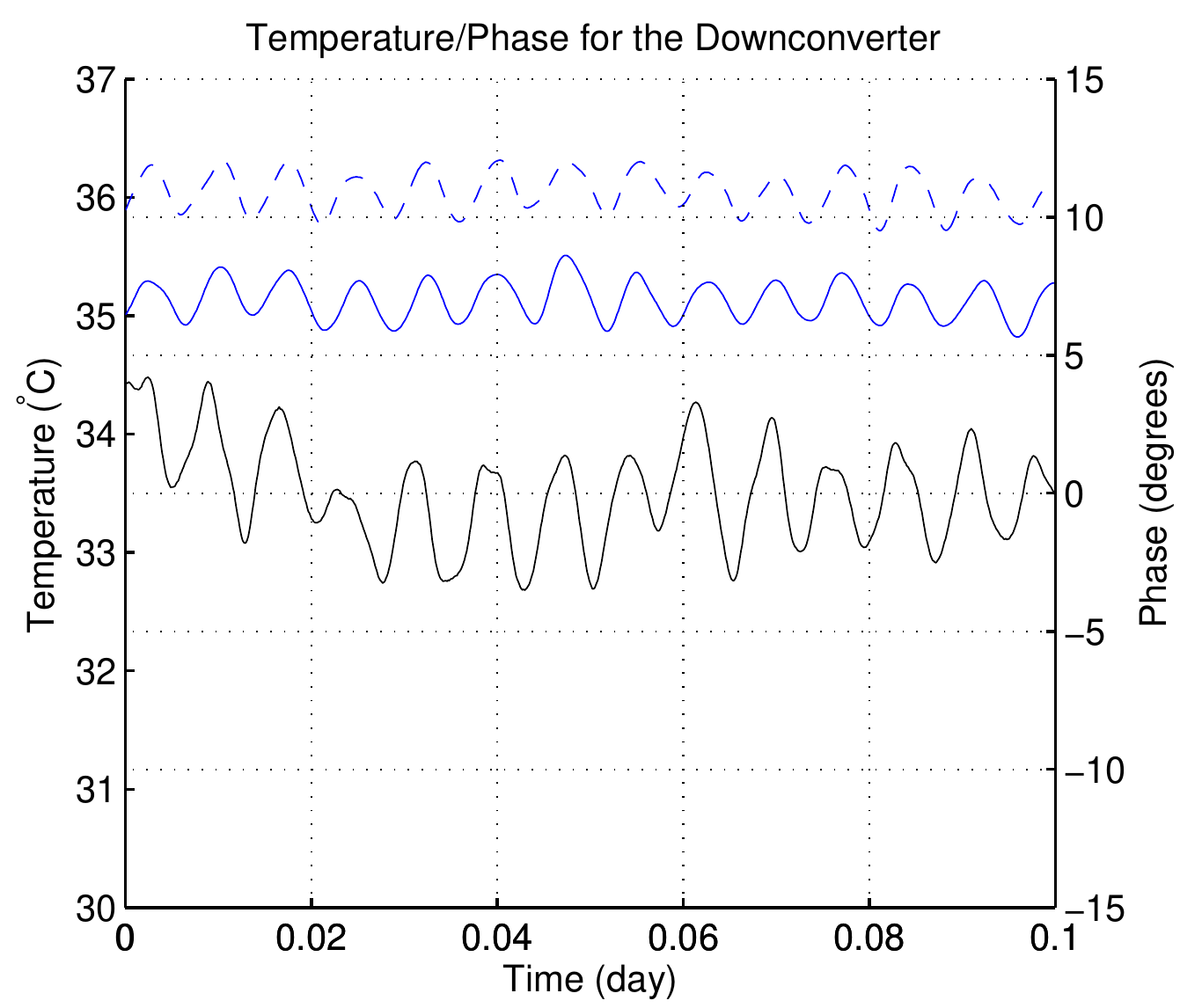}}
\caption{Plot of phases on a single baseline (Baseline 0-1) and the corresponding downconverter 
temperatures during the same period.  
The blue lines are the temperatures measured in the downconverters for Antennae 0 (dashed line) 
and 1 (solid line).  The black line is the phase of the raw data on baseline 0-1, taken
while staring at a strong point source, which ideally has flat, constant phase 
(see \S \ref{phasecal} for details).
Note that thermal variations in adjacent digitizers, due to cycling of the A/C, of 
0.5$^\circ$C (peak to peak) produced 6$^\circ$ phase variations.  
Modifications to the A/C removed these thermal oscillations on short time scales
(see Fig~\ref{fig:thermal_phases2}).}
\label{fig:thermal_phases}
\end{figure}

The correlator trailer draws $\sim 1$~MW of power, and thus
requires an industrial-strength air conditioner (A/C) with two independent compressor
units (though much of this power is drawn by the A/C itself, and its heat is dissipated
outside).   
At the beginning of SZA commissioning, we used a state-of-the-art air conditioner with 
a commercially-available governor that can keep the temperature within $\pm 1^{\circ}$C 
of a desired set point.  It was soon discovered, due to the temperature sensitivity of the 
digitizers, that the 20 minute air 
conditioner cycling of $\pm 0.25^{\circ}$C led to $\pm3^{\circ}$ phase variations in 
the data (see Fig.\ \ref{fig:thermal_phases}), and a more robust solution had to be found.

\begin{figure}
\centerline{\includegraphics[width=5.75in]{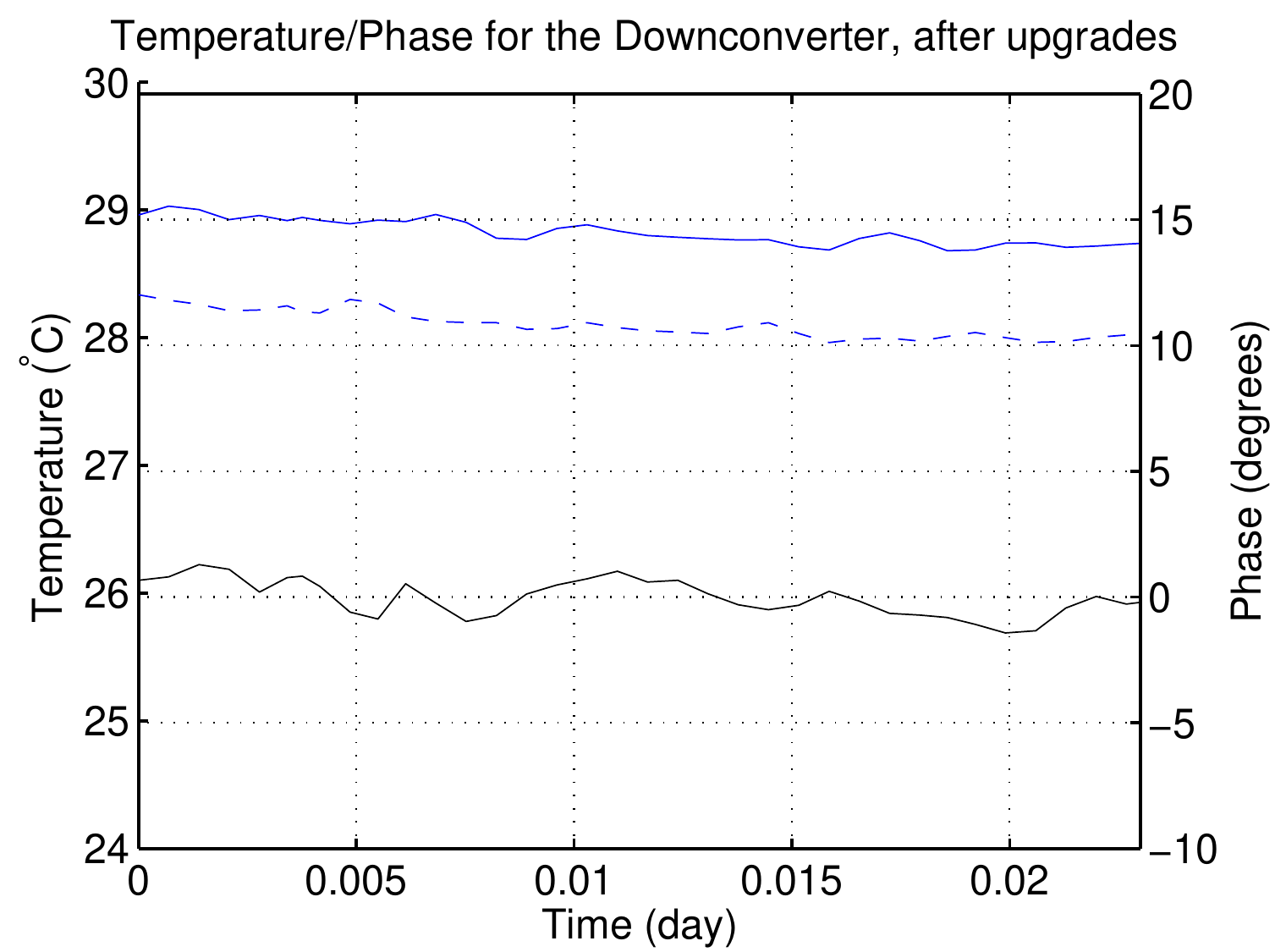}}
\caption{Plot of phases on a single baseline (Baseline 0-1) and the corresponding downconverter 
temperatures during the same period.  
The blue lines are the temperatures measured in the downconverters for Antennae 0 (dashed line) 
and 1 (solid line).  The black line is the phase of the raw, uncalibrated data on baseline 0-1, 
taken while staring at a strong point source, with the linear slope and mean value removed.
Note that this is on a much shorted time scale than that of Fig.~\ref{fig:thermal_phases}, but
is long enough to have shown 3 full cycles of the A/C if they had persisted.  Any residual
phase error is both negligble and is removed by calibration (see \S \ref{data_calib})}
\label{fig:thermal_phases2}
\end{figure}

The resolution to this problem was largely worked out by the SZA engineer, Ben Reddall.
He removed the governor and set the A/C to be continuously on.  He then modified the unit
so that louvers could push the air through the heat exchanger of one compressor at a time,
so that the other compressor could be shut off when not necessary.  This strong, continuous 
air flow in a closed cycle ensures good mixing and stability within the trailer;  it also
prevents the active compressor from icing up, as it essentially doubles the airflow through it.
Various layers of insulation further contributed to the thermal stability of the trailer.\footnote{Note 
that the original OVRO correlator was housed in a basement underground, and the newer CARMA correlator 
-- for which the SZA correlator served as prototype -- is in a large, heavily-insulated room with a concrete
foundation, rather than in a wooden trailer located above pavement.}  
The final steps implemented by Ben Reddall were the addition of water bottles 
(water has a high thermal capacity) to the air supply duct, and of three 1~kW space 
heaters used during the winter.  The heaters help prevent the A/C, which must be run continuously,
from freezing on cold winter nights.  
Figure \ref{fig:thermal_phases2} demonstrates the continuing success of the correlator trailer
thermal solution, three years after it was implemented.

The temperature in the correlator trailer typically varies by 20$^\circ$C over the day,
but does so in a way that these slow, diurnal phase changes are common mode to the fibers 
and electronics. The slowly-changing phase drifts produced by these temperature changes 
can be tracked using our standard astronomical calibration (see \S \ref{data_calib}) 
and removed from the data.




\chapter{Sources of Contamination}\label{sza_debug}

This chapter addresses two sources of contamination found during the SZA commissioning
phase: radio interference from external sources and antenna cross-talk.
While I have attempted to make this palatable for those who are not engineers 
or radio astronomers, the information is best suited to an audience interested 
in building and debugging a radio interferometer.

\section{Radio Interference}\label{rfi}

Here we consider the problem of radio frequency interference (RFI), which we found in our early 
observations to be contaminating frequency channels or even entire bands.  
A ``birdie,'' the colloquial name for RFI, is in general some contaminating, 
non-astronomical signal, which can increase the scatter in the radio visibilities
at the birdie frequency, corrupting data in a given channel or even a whole band (discussed
in \S \ref{spectral_leakage}).
The strongest birdies are generally from sources local to the antenna electronics, 
and could be the mixing product of the local oscillator (LO),
the LO reference, and/or the YIG.  Birdies can also come from 
external transmissions, generally ground-based sources (e.g. military communications
or satellite uplinks) or geostationary satellites.

In this section, we examine ways of lessening the impact of birdies, since they could
potentially contaminate maps made from observations with the SZA.
For simplicity, we consider a birdie to be an ideal sinusoidal signal.  
We calculate how much fringe tracking helps to reduce
birdies, since they are stationary with respect to the antennae\footnote{As mentioned
in the first paragraph of \S \ref{rfi}, birdie sources are generally local to the antennae, 
from ground-based transmissions, or from geostationary satellites; all of these sources
are stationary with respect fringe tracking.  Other possible birdie sources
include satellites in low earth orbits, which would move through the fringes on a given 
baseline much more rapidly than the astronomical sources we track (e.g. 45~minute orbits versus
the sidereal day that a celestial object takes).  
We found no indication of contamination from any satellites with an orbital period on 
the order of a day, which -- if they exist -- could produce coherent, cross-correlated contamination.} 
and lose coherence in the cross-correlated data.  

Early observations with the SZA revealed the presence of a strong birdie, just outside our
IF band, at the receiver output frequency 9.022~GHz. This birdie arose in the output of 
the receivers as the mixing product of the YIG 3$^\text{rd}$ harmonic 
($3 \times 8.972~\rm{GHz} = 26.916~\rm{GHz}$) and 
the 35.938~GHz LO ($35.938~\rm{GHz}-26.916~\rm{GHz}=9.022~\rm{GHz}$).  
The 30-GHz receivers detected both the YIG 3$^\text{rd}$ harmonic 
emitted within their own Eboxes and those from adjacent antennae (\S \ref{rx_rf_components}
for information on the YIG oscillator).   
This birdie showed up in Band 15, which is the lowest sky frequency band (26.938--27.438~GHz),
despite the fact that this band runs from 8.5--9~GHz, and the birdie was 9.022~GHz.

We demonstrate in this section that an additional 23~dB of attenuation, at the birdie's source,
was required to bring the birdie to less than 10\% of the \emph{rms} noise level via fringe tracking.
We then proceed to consider how a birdie can corrupt all the channels 
in a band, even if the birdie is a pure tone, singular in frequency.  
Finally, we demonstrate that a clever retuning of local oscillators can avoid 
band corruption by insuring all the birdie's power appears in one channel, 
which can readily be flagged and excised from the data.

\subsection{The Effect of Fringe Tracking}

Since birdies come from sources not associated with the field being observed, 
they can be treated as an untracked source.  Since the SZA birdie was local to the 
antenna, it was effectivley moving across the sky in the opposite direction as 
the astronomical source being observed.
The instantaneous correlations of the birdie signal affecting a pair of antennae 
do not sum up coherently over time (see \S \ref{interf}), and the birdie is 
therefore attenuated by this loss of coherence.  
The factor by which the birdie is attenuated, $F_b$, 
can be calculated as a sinc function of the fringe frequency $\nu_f$ 
(the rate at which an untracked source moves across the sky) and the integration
time $\tau_a$.  The attenuation factor is
\begin{equation}
F_b = \frac{\sin(\pi \nu_f \tau_a)}{\pi \nu_f \tau_a},
\label{eq:birdie_atten}
\end{equation}
where $\nu_f$ is the fringe frequency
\begin{equation}
\nu_f = u ~\! \omega_{\scriptscriptstyle e} \cos(\delta),
\label{eq:fringe_freq}
\end{equation}
$\omega_{\scriptscriptstyle e}$ is the earth's angular velocity, 
($\omega_{\scriptscriptstyle e} = 7.29115 \times 10^{-5}~\text{rad s}^{-1}$), $\delta$ is the source's declination, and
$u$ is the Fourier conjugate (in \emph{u,v}-space, the standard notation used by radio astronomers) 
of the baseline along the east-west direction, normalized by the wavelength of observation:
\begin{equation}
u = X_{\scriptscriptstyle \lambda} \sin H + Y_{\scriptscriptstyle \lambda} \cos H.
\label{eq:ucoord}
\end{equation} 
Here $H$ is the hour angle of the source, where $H=0^h$ corresponds 
to the meridian, and each hour increment is 15$^{\circ}$.
$X_\lambda$ and $Y_\lambda$ are respectively the x (north-south) and y (east-west) components 
of the baselines divided by the wavelength at which the observation is made.  
A longer baseline increases the fringe frequency, since it yields a tighter, more 
highly-resolved fringe pattern on the sky. 
A source at $\delta=0^{\circ}$ (in the equatorial plane) moves most quickly across the sky, 
so the fringe frequency $\nu_f$ is highest there. 
A point source at the pole, on the other hand, would not move through fringes at all.

The $1/{\pi \nu_f \tau_a}$ envelope of the birdie attenuation factor 
(denominator of Eq.~\ref{eq:birdie_atten}) results from the time averaging
of the measured visibities, the smallest binning of which is a 1/2 second frame 
(see \S \ref{corr}).
We are interested in how much the birdie is attenuated during an observation, 
and how this compares to the Gaussian noise floor of the same observation.
The noise floor decreases according to \citep[see, e.g.][]{rohlfsw96,thompson2001}:
\begin{equation}
\frac{\Delta T_A}{T_{\rm sys}} = \frac{M}{\sqrt{\Delta \nu \times \tau_{a}}}.
\label{eq:integration}
\end{equation}
In the above equation, $M$ is a factor -- less than or equal to unity -- that accounts for the 
quantization noise in the correlator system,\footnote{We do not need to know $M$ here.  However, 
$M\approx 0.88$ for the SZA correlator \citep{hawkins2004}.} $\tau_{a}$ is the integration
time, $\Delta \nu$ is the bandwidth, and $T_A$ is the antenna temperature.  

\subsection{Can Fringe Tracking Beat Down the Birdie?}\label{fringe_beating}

We consider here if the SZA 9.022~GHz (IF band) birdie that plagued Band 15 of the SZA 30-GHz system
(see Table~\ref{table:band_scheme}) in the first few months of operation could
be sufficiently attenuated by fringe tracking.  
We assume our source is at declination $\delta=0^{\circ}$, which is our best case scenario;
this yields the highest fringe frequency, since the source moves most rapidly across
the sky.  
If our source were at the north celestial pole, on the other hand, fringe tracking would
not attenuate the birdie at all.  Clearly, we cannot rely in all cases on
fringe tracking to eliminate birdies.
We will also for simplicity assume our source is close to the meridian (i.e. near transit). 
With $H \simeq 0$, the $X_\lambda$ component is negligible (see Eq.~\ref{eq:ucoord}),
simplifying the calculation.

Let us set $Y_\lambda = 1000$ (see Eq.~\ref{eq:ucoord}).  For the SZA, this would be among the longest
of the inner array baselines, with a 10 meter, east-west baseline component while observing at 30~GHz.  
If we chose a purely north-south baseline ($Y_\lambda = 0$), a source near 
transit would give $u=0$, and fringe tracking would not attenuate the birdie
at all (see Eq.~\ref{eq:fringe_freq}).  Again, this indicates that fringe tracking
is not a panacea for reducing birdies, but we proceed to compute how much it helps in 
our best case scenario for the inner array.

Calculating the fringe frequency (Eq.~\ref{eq:birdie_atten}) for these conditions, 
$\nu_f \simeq 7.3 \times 10^{-2} \rm~Hz$.
We can see that $F_b$ hits its first null at $\tau_a= 1/\nu_f \simeq 14 \rm~sec$, 
and we might be tempted to choose this as our integration time so that 
no stationary sources could affect us.  
Since this frequency is different for each baseline, it is clear that this scheme will not work.

We can proceed in calculating the birdie's attenuation without needing to know 
$M$ and $T_{\rm sys}$ (in Eq.\ \ref{eq:integration}) by measuring the ratio of the 
birdie signal power with fringe-tracking off (which is therefore unattenuated by integration) to the 
noise level in a known time interval.
Given the functional behavior of the birdie's attenuation (Eq.\ \ref{eq:birdie_atten})  
we only measure the birdie's unattenuated amplitude $A_{b,0}$ and the
noise's \emph{rms} amplitude $A_{n}(\tau_m)$ in a given time interval $\tau_m$, with fringe
tracking off.  
It should be noted that the noise integrates down as $1/\sqrt{\tau}$ (as in Eq.\ \ref{eq:integration}) 
without regard to fringe tracking, so it is reasonable to compare the noise levels
for given integration times, regardless of whether fringe tracking is on.
The birdie and fringe amplitudes as functions of time are given by:
\begin{equation}
A_{b}(\tau_a) = A_{b,0}~F_b(\tau_a)
\end{equation}
and
\begin{equation}
A_{n}(\tau_a) \sqrt{\tau_a} = A_n(\tau_m) \sqrt{\tau_m}
\end{equation}
so that the ratio of the birdie to the noise level for integration time $\tau_a$ is:
\begin{equation}
\frac{A_{b}(\tau_a)} {A_n(\tau_a)} 
= \frac{A_{b,0}~F_b(\tau_a)} {A_n(\tau_m)\sqrt{\tau_m/\tau_a}}
\label{eq:birdie}
\end{equation}

With fringe tracking off, the level of the 9.022~GHz birdie was measured to be 
$\sim 250$ times the noise in a 20~second, uncontaminated integration.  
Since it's more convenient to express integration times in seconds, we solve for
$A_{b,0}$ normalized to the noise in a 1~s integration:
\begin{equation}
A_{b,0} =  250~A_{n}(20~\rm s) = \frac{250}{\sqrt{20}} A_{n}(1~\rm s).
\end{equation}
Plugging the ratio $A_{b,0}/A_n(1~\rm s)$ into Eq.~\ref{eq:birdie}, we find the ratio of the birdie to the noise 
level after time $\tau_a$ (in seconds):
\begin{equation}
\frac{A_{b}(\tau_a)}{A_{n}(\tau_a)} = \frac{250}{\sqrt{20}} ~\! F_b(\tau_a)\sqrt{\tau_a} 
\end{equation}
Taking the local maxima of $F_b$ (defined in Eq.\ \ref{eq:birdie_atten}) as our worst case scenario
for the birdie's amplitude, 
we ignore the sinusoidal component of the sinc function and keep the $1/(\pi \nu_f \tau_a)$ 
envelope.
This yields:
\begin{equation} 
\frac{A_{b}(\tau_a)}{A_{n}(\tau_a)} = \frac{250}{\sqrt{20}} \, \frac{1}{\pi \nu_f \sqrt{\tau_a}} = \frac{17.8}{\nu_f \sqrt{\tau_a}}.	
\label{eq:birdie2}
\end{equation}

Now, we can solve for the time when the envelope for the birdie amplitude falls 
below the noise floor.  For our fringe rate $\nu_f$, we find that it takes
about 16.5 hours for this birdie to integrate down to the same level as the noise.
This of course violates the assumption that the source is near transit.  
Furthermore, this time scale is much longer than any amount of time 
we would sensibly bin as a single data point. 

We would like the birdie to be attenuated to $< 10\%$ of the noise 
level in each binned data point, so that maps are relatively unaffected by the birdie (with a $\sim$ 5-$\sigma$
cluster detection, the birdie would affect the SZE flux on the $2\%$ level, which is
less than our calibration uncertainty).  
To get this birdie down to the 10\% level in a 5 minute integration, the maximum time which we might reasonably
bin into a single data point, we need about 22~dB of attenuation on each antenna.\footnote{
As noted in \S \ref{corr}, cross-correlation scales as power. Therefore, each antenna-based birdie's 
power has to be attenuated by the same amount as the total desired birdie attenuation.}  
When observing sources at higher declinations, additional attenuation would be 
required to compensate for the lower fringe frequency $\nu_f$.  A source at $\delta = 37^{\circ}$ 
takes 25\% longer to beat the birdie down to the noise level of a $\delta = 0^{\circ}$,
since its fringe frequency is 80\% of the $\delta = 0^{\circ}$ value.  Conversely, an extra dB of 
attenuation would be required to bring the birdie down to the same
level in the same time.  Therefore, we would need 23~dB of attenuation of the birdies
at their sources in order for this measured birdie not to affect a typical 
SZA observation.

In Eq.~\ref{eq:birdie2}, we ignored the fact that $F_b$ varies sinusoidally (as in Eq.~\ref{eq:birdie_atten}).  
In a typical SZA observation, many integrations are taken, and are often binned
into intervals ranging from 20--300~s.  For each binned data point, the birdie level would vary randomly (since $\nu_f$
changes throughout the observation), with both positive and negative variations that are several times larger than the noise
(e.g. 250 times larger for a 20~s observation, as measured before attenuation).
Thus, as long as the fringe rate $\nu_f \neq 0$ a birdie would increase the scatter in the data taken at that frequency, 
rather than integrating to a coherent signal.

\subsection{Spectral Leakage}\label{spectral_leakage}

As noted in \S \ref{corr}, the limited number of samples used in correlation
has the effect of rectangular windowing\footnote{A rectangular window is the one-dimensional
equivalent to the top hat function.} of the cross-correlated data.  
In this section we examine how a sinusoidal signal -- such as an idealized birdie --
can contaminate an entire band of data, rather than only appearing at its frequency.  
This phenomenon is called ``spectral leakage.''

\begin{figure}
\includegraphics[width=6in]{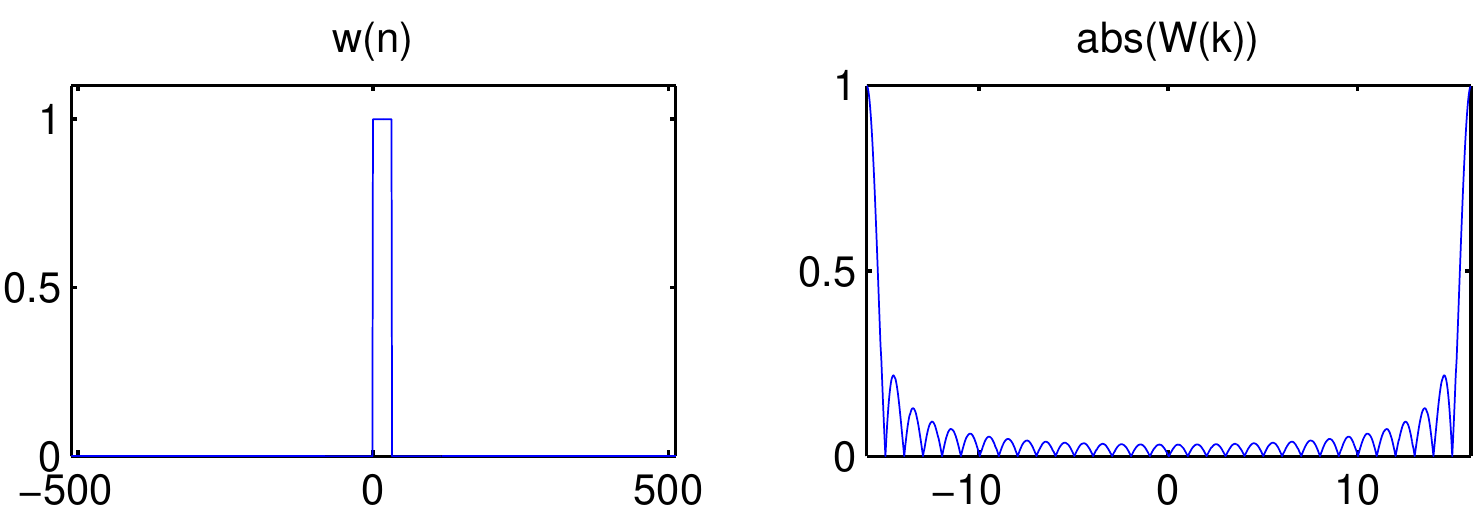}
\caption{{The rectangular window function and its transform.
\bf Left:} The rectangular window function $w(n)$ in the time domain.
{\bf Right:} Magnitude of the Fourier transform of $w(n)$, $|W(k)|$.
$w(n)$ is unity for $n=[0,31]$, corresponding to 32 discrete samples
in time (i.e. the digitized waveform), and zero outside this range. 
$|W(k)|$ is unity for $k=16$ and $k=-15$, and zero at all other
integer values of $k$. See discussion in text.}
\label{fig:window_func}
\end{figure}

Consider a birdie at frequency $f_0 = k_0 f_s/N$, where the signal is sampled $N$ 
discrete times at a sampling frequency $f_s$, and $k_0$ is the birdie's wavenumber. 
Let $x(n)$ be the discrete-time representation of the continuous, sinusoidal birdie 
signal (i.e. the digitally-sampled birdie, cross-correlated with a similar birdie 
from another antenna, idealized to a sine wave).  
\begin{equation}
x(n) = A \sin \left(\frac{2 \pi n k_0}{N}\right) 
= \frac{A}{2}(e^{j 2 \pi n k_0 / N}+e^{-j 2 \pi n k_0 / N})
\end{equation}
The DTFT (Eq.~\ref{eq:dtft}) of $x(n)$ is:
\begin{equation}
X(k) = \frac{A}{2}\left[\delta(k-k_0) + \delta(k+k_0)\right].
\end{equation}

The signal $x(n)$ is limited to the range $n=[0,N-1]$ because we only 
have $N$ samples, and is zero elsewhere.  Therefore, the signal is rectangular
windowed by the limited number of samples.
The transform $W(k)$ of a rectangular window function in the time
domain, $w(n)$, which is equal to 1 in the range $n=[0,N-1]$ and is zero 
elsewhere (see See Fig.~\ref{fig:window_func}), is
\begin{equation}
W(k)= \sum_{\scriptscriptstyle n=0}^{\scriptscriptstyle N-1} e^{-j 2 \pi k n / N}
= e^{-j \pi k (N-1)/N} \left[\frac{\sin(\pi k)}{\sin(\pi k/N)}\right].
\end{equation}
Since multiplication in the time domain is equivalent to convolution in the 
frequency domain, the DTFT of $x(n)$ multiplied by the window $w(n)$ is
\begin{equation}
Y(k) = \sum_{\scriptscriptstyle n=-\infty}^{\scriptscriptstyle \infty} 
W(k) X(k) = \frac{A}{2}\left[W(k-k_0)+W(k+k_0)\right].
\label{eq:W_k}
\end{equation} 
The convolution of $W(k)$ with the birdie $X(k)$ simply shifts the $W(k)$ 
to be centered on $\pm k_0$ (hence $W(k \pm k_0)$ in Eq.\ \ref{eq:W_k}).  
In the DFT, which only samples the DTFT at discrete values of $k$, if $k_0$
is a multiple of $f_s/N$, we have an integer value of $k_0$, meaning we
sample $W(k)$ only at values where it is equal to one or zero.
For $k = k_0$, $Y(k)=1$, while $Y(k)=0$ for $k \neq k_0$.

For non-integer $k_0$, $W(k)$ is shifted so that it is sampled
at values other than zero and one. A birdie not centered on a spectral channel thus leaks into 
all channels within the band, with the strongest contamination appearing in the channels nearest 
the birdie frequency.\footnote{Note that, 
in the XF/lag correlator scheme, the FFT is taken on the cross-correlated data, hence the 
rectangular windowing of the signal only happens once, and frequency response goes as sinc, 
not sinc$^2$.}  
It is easiest to remove birdie contamination from an observation if it is
centered on a frequency channel; in this case, all the birdie power appears in 
one frequency channel, which can be excised during data reduction (see \S \ref{autoflag}). 

Since spectral leakage arises from the DFT itself, this leakage will not show 
up in any band other than that in which the birdie lies.  This is simply because 
each band is sampled and FFT'd individually (see Fig.~\ref{fig:sza_bands}).
Furthermore, increasing the number of channels (which could be done by increasing the 
number of correlator lags) would localize the leakage; in the limit of infinite 
channels, a birdie would be measured at its true frequency.  
This is the case where the DFT limits to the full DTFT (Eqs.~\ref{eq:dtft} and \ref{eq:dtft2}), 
which has infinite spectral resolution within a band $f_s/2$ in size.

\begin{figure}
\includegraphics[width=5.5in]{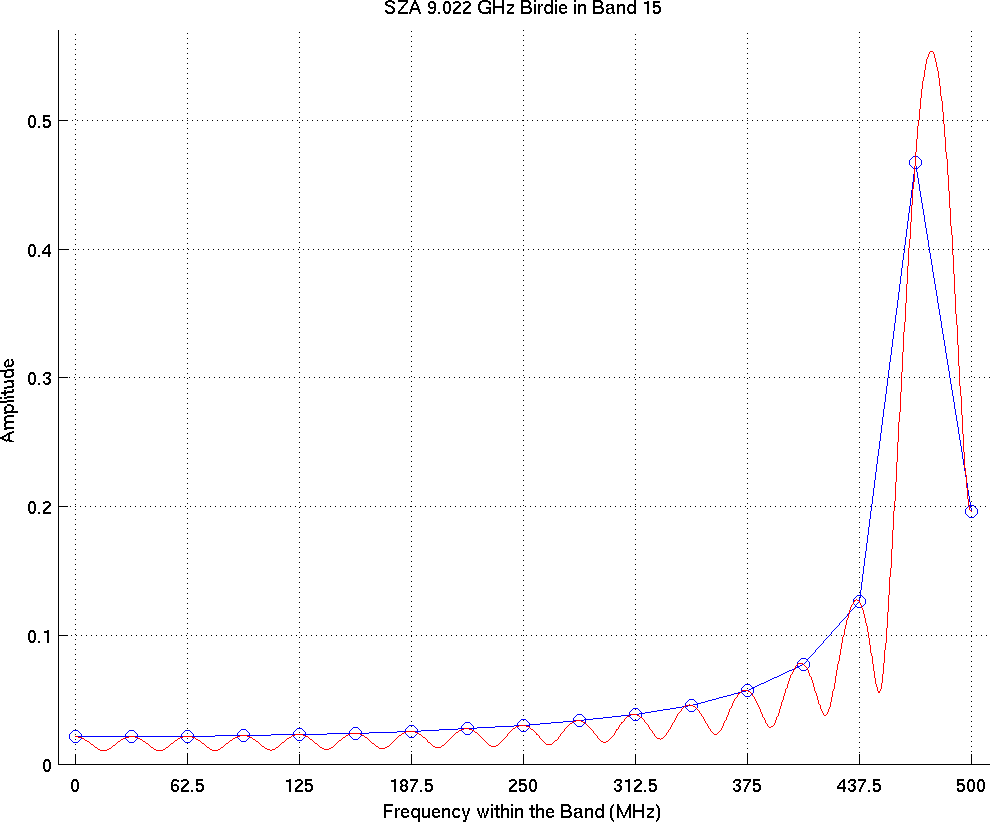}
\caption{9.022~GHz birdie in the IF, downconverted to 1.022~GHz, 
	sampled in time and FFT'd.  The blue line is the amplitude of the 
	bandpass that would be determined from the SZA spectral data.  
	The birdie does not alias cleanly to the center of any channel, 
	and therefore leaks into every other channel in the band.  
	The DTFT of the birdie (red curve) is the sum of two sinc functions, 
	centered at 478 and 522~MHz, (or at $\pm 478 {\rm~MHz}$, since the transform 
	of a the windowed sinusoid is $W(\pm k_0)$). 
	Values of $k = [0,16]$ from the DFT are plotted here, corresponding to 
	the positive frequencies of Channels 0 through 16~in the SZA correlator (see Table~\ref{table:channel_scheme}).
 	The band wraps outside $k = [0,31]$ (or $k = [-15,16]$), which corresponds to
	frequencies of 0-1000~MHz (or -500-500~MHz).}
\label{fig:birdie_leakage}
\end{figure}

\begin{figure}
\includegraphics[width=5.4in]{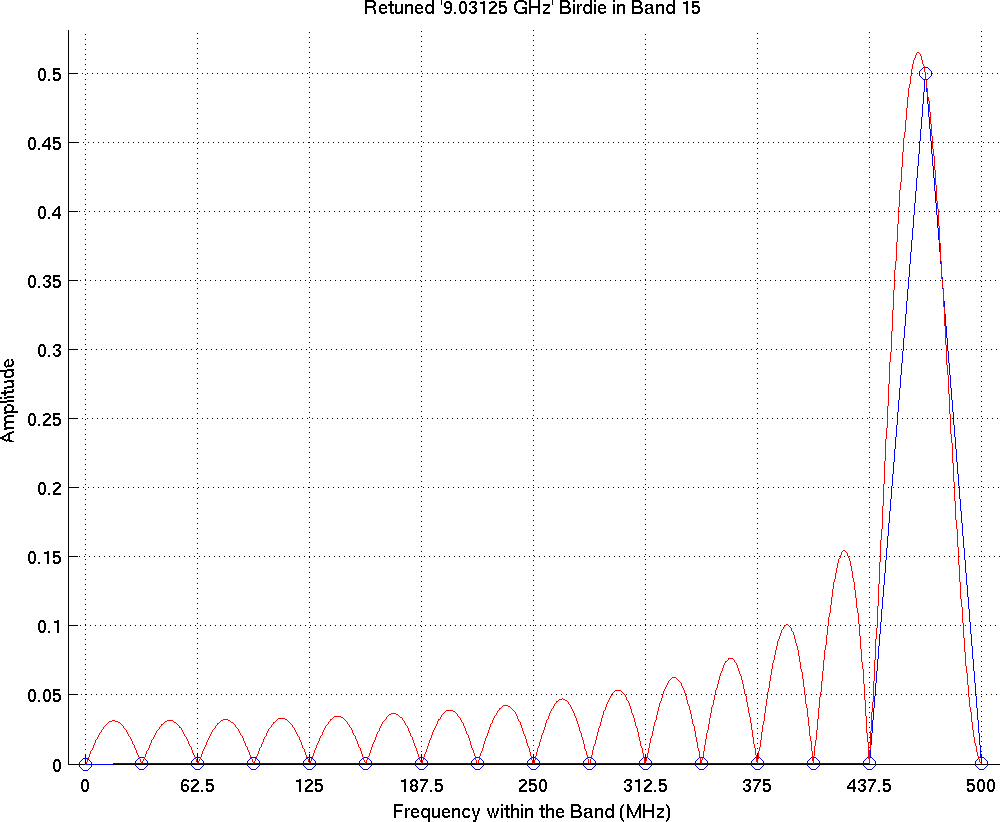}
\caption{Retuned ``9.03125~GHz'' birdie (placing it at an integer $k$ frequency), downconverted to 1.03125~GHz,
	then sampled and FFT'd.  This birdie aliases cleanly to the center of 
	the Channel 16 (see Table~\ref{table:channel_scheme}), which can be excised 
	from the data.
        The other channels' centers sample the nulls of the rectangular
 	window's transform (red curve), which is a summation of sincs located at 
 	$\pm 468.75 {\rm~MHz}$.
	See Fig.\ \ref{fig:birdie_leakage} for more details.}
\label{fig:birdie_centered}
\end{figure}

\subsection{Example of Spectral Leakage Corrupting a Band}

The 9.022~GHz birdie, arising from the 3$^\text{rd}$ YIG harmonic 
(detailed at the beginning of this section, \S \ref{rfi}), corrupted all of Band 15
during initial testing of the SZA (see Table~\ref{table:band_scheme}).  
Just before digitization, this birdie is at 1.022~GHz; 
this is not a discrete multiple of a 500/16~MHz = 31.25~MHz.
The birdie's presence in the final data product (before the problem was solved) indicates that 
the bandpass filters of the block downconverter could not completely eliminate strong signals from 
outside the band.

Sampling at 1~GHz aliases the birdie to 522~MHz.
Due to the Hermitian property of a real signal's FFT, frequency is reflected about
500~MHz. The 522~MHz birdie therefore appears in the 0--500~MHz band, between Channels 
15 and 16 (468.75 and 500~MHz). 
The effect of spectral leakage, shown in Fig.~\ref{fig:birdie_leakage}, is calculated 
by evaluating the DFT expression for $X(k)$ (Eq.\ \ref{eq:dft}), where
$x(n)$ was a sinusoid at 522~MHz.  
While the birdie is strongest in the channels adjacent to its real frequency, 
it remains above 3\% of its full power throughout the entire band, rendering the band
useless for observations.  

A clever retuning of the last downconverter LO (\S \ref{dcon}) seen by Band 15 
avoids contamination of the entire band, localizing the birdie's power to one channel.
The details of this are as follows:
The 9.022~GHz birdie first ends up at 978~MHz after mixing with the 10~GHz LO of the block 
downconverter; it is then mixed with a 2.0~GHz LO (Downconverter Bank 2, in 
Fig.~\ref{fig:sza_bands}), which places Band 15 at in the 0.5--1.0~GHz band that is
input to the digiters.  The digitizers therefore see the birdie at 1.022~GHz.
We retuned the last LO to 2.00925~GHz, shifting the birdie to 1.03125~GHz.
Fig.~\ref{fig:birdie_centered} shows the spectrum that results from this retuning.
The result is that birdie's power only shows up in Channel 16, which is set 
in the data reduction pipeline to be automatically flagged (\S \ref{autoflag}).

The final solution to the birdie question was to eliminate the birdies at
their sources: the YIG oscillators. 
By wrapping the YIG in Eccosorb$^\circledR$ (see Fig.~\ref{fig:ebox_30GHz}), we found
the signal was sufficiently attenuated; this was a temporary solution, since the 
form of Eccosorb foam used in this test has a tendency to flake and degrade in the field.
A more permanent solution was implemented by simply shielding the YIGs in small aluminum boxes.  
While the retuning of the 2.0~GHz downconverter LO was retained, periodic measurements 
continue to show that the YIG harmonics do not escape their aluminum housings (though
we still flag Channel 16 of Band 15, as this precaution only costs us one of our $16\times15$
spectral channels).

\section{Antenna Cross-talk}\label{crosstalk}

In the survey fields, which were all selected to pass close to zenith in order to
minimize atmospheric noise, an apparent excess in noise\footnote{The \emph{rms} noise of the data
when observing blank sky would slowly rise and fall while tracking a field through
certain antenna configurations.  This excess in noise was not correlated with changes in the
atmosphere.} was discovered between adjacent antennae when observing at antenna elevations 
$\gtrsim 80^{\circ}$.  
It was recognized that all the antenna positions that allowed this ``excess noise'' 
on these short baselines corresponded to positions where the secondary mirror feedlegs 
on one antenna had direct lines of sight to the feedlegs on the adjacent antenna.  

It was further discovered that, with fringe tracking off
and the antennae in a position known to produce ``excess noise,'' 
the measured phase versus frequency on that baseline (which for blank, untracked sky should be random noise) 
was strongly wrapping every two channels; this observed ``excess noise'' was actually a signal due to ``cross-talk''
between elements of the antennae, and the wrapping was due to the pathlength between the elements involved
in the antenna cross-talk.
Since each channel is $31.25~\text{MHz}$, the measured wrapping corresponded 
to a delay of 16 nanoseconds (i.e. $2\times31.25~\text{MHz}$), and the position
of the antennae was such that the astronomical delay ($T$ in \S \ref{interf})
for that pair of antennae was also 16 nanoseconds.
This is because projected baseline at the cross-talk maximum position was roughly 
$480~\text{cm}$, or $c T = (3\times10^{10}~\text{cm s}^{-1}) \times (16~\text{ns})$.
Thus the delay used to track astronomical fringes on the baseline formed by
the two antennae also flattened the phase of the cross-talk signal, providing
a coherent, correlated signal.\footnote{Flattening the phase of a signal
across a band is the spectral equivalent to sending the proper, fringe-tracking
delay necessary to correlate data.}

Several ways of eliminating this antenna cross-talk were explored.  While
the precise cause is unknown, it is hypothesized that the feed horn from one
antenna could launch a signal that scattered coherently into the other antenna, and
the other antenna would do the same in reverse.  With fringe tracking on, the 
delay being sent to an antenna sometimes precisely flattened this phase 
wrapping, producing strong, correlated amplitudes in the data.

In order to test that the scattering came from the feedlegs of the secondary,
a wire mesh was added to the side of one antenna from a pair that exhibited cross-talk.
This shifted the cross-talk to higher antenna elevations.  We considered building collars
for some or all of the antennae in the inner array, but this was determined to be 
too costly, and the high winds in Owen Valley would make it difficult to maintain.
Furthermore, if the collars extended too far, this would increase the risk of 
collisions between close-spaced antennae, something the SZA was designed not
to have.

\begin{figure}
\centerline{\includegraphics[width=5in]{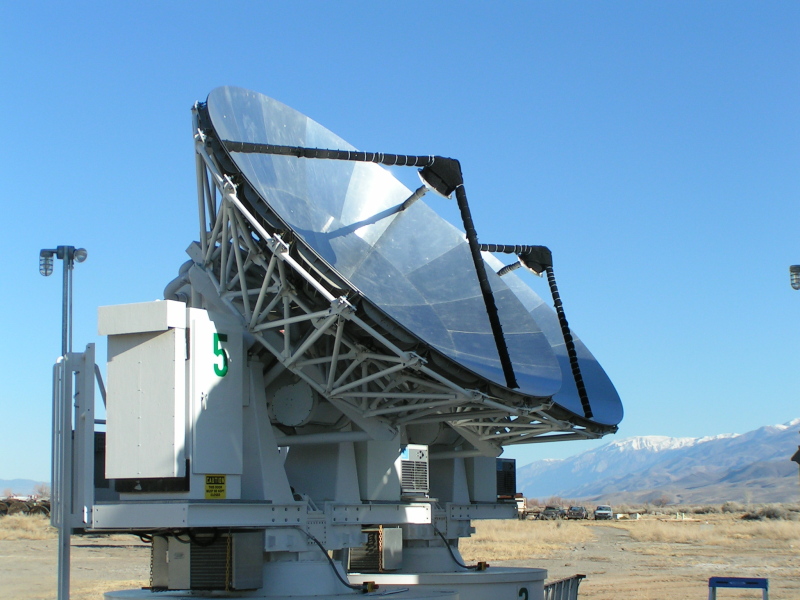}}
\caption{Photo of the closest pair of antennae (Antennae 3 \& 5, see Fig.~\ref{fig:sza_ant_loc}), 
with the test Eccosorb$^\circledR$ wrapped around the feedlegs.
Ultimately, the Eccosorb$^\circledR$ was replaced by crinkled aluminum foil.}
\label{fig:ant35_eccosorb}
\end{figure}

We then tried adding Eccosorb$^\circledR$ to the feedlegs (see Figure~\ref{fig:ant35_eccosorb}).  
While the absorber did eliminate cross-talk, it also increased the
noise in the system, as each feedleg was now radiating at $\sim 300~\text{K}$. 
For comparison, the cluster signals 
we observe peak at $\sim 10~\text{mK}$.  This increase in system noise was determined 
to be unacceptable, as it increased the required integration times.

The final solution to the cross-talk problem was to wrap the feedlegs in crinkled,
heavy-duty aluminum foil ($\sim$50 mil), available at most grocery stores.  The crinkles
were carefully constructed to produce large, 3--10~cm facets of arbitrary orientations
on the side of the feedleg facing the primary mirror.
These facets have the effect of randomly changing the pathlength and angle between 
corresponding parts of each antenna.

\chapter{Data Reduction}\label{data_calib}

In this chapter, I provide a brief overview of the steps necessary to calibrate
SZA data.  For a more detailed treatment of the SZA data analysis routines, I
refer the reader to the thesis of Stephen Muchovej \citep{muchovej2008}.  

The SZA data calibration routine was written by students in the SZA collaboration 
(primarily Mike Loh, Stephen Muchovej, Matthew Sharp, and Chris Greer) in 
MATLAB$^\circledR$\footnote{\url{http://www.mathworks.com}} code.  
Since MATLAB$^\circledR$ is a scripting language optimized for matrix manipulation,
it is well-suited to manipulate large data sets when performing uniform mathematical operations.
Some of the more computationally-intensive routines -- in particular, those that required loops --
were implemented in external, compiled C/C++ functions that are called from within MATLAB$^\circledR$.
In the following sections, I describe each step in the data reduction pipeline in 
order of operation

\section{System Temperature Computation}\label{tsys_calib}

The first step in the data calibration routine is a calculation of the system temperature 
$T_{\rm sys}$ (see Eq.~\ref{eq:tsys}).
As described in \S \ref{tsys}, $T_{\rm sys}$ is defined similarly to $T_{\rm rx}$, 
but accounts for the noise
of the entire system scaled to above the Earth's atmosphere.  
We therefore use $T_{\rm sys}$ to compute the \emph{rms} noise level in an observation,
\begin{equation}
\sigma_{rms} \propto \frac{T_{\rm sys}}{\sqrt{N(N-1)\, \tau \, \Delta \nu}}.
\label{sigma_rms}
\end{equation}
Here $N$ is the number of antennae in the array, $\tau$ is the integration time, and $\Delta \nu$
is the bandwidth of the system. For a given integration time, on a given instrument,
$T_{\rm sys}$ characterizes the noise in an observation. 

\begin{figure}
\centerline{\includegraphics[width=6in]{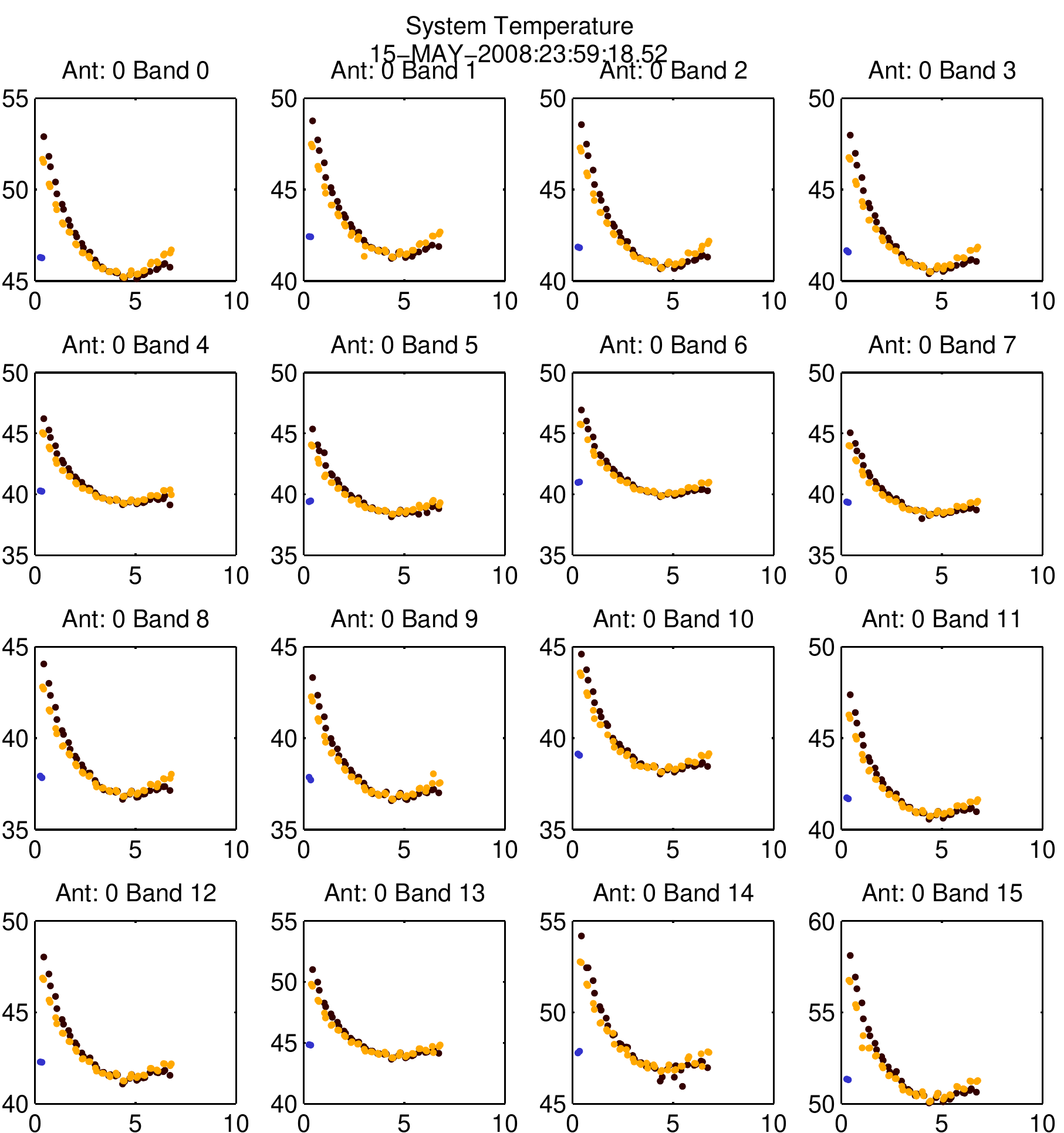}}
\caption{System temperatures of a typical SZA cluster observation, computed for each band.
One representative antenna is shown, since the other seven all have similar system temperatures.  
The $y$-axis shows $T_{\rm sys}$ in K, and the $x$-axis is time in hours from the start of the
observation.  The dark blue point at the start of the observation is the $T_{\rm sys}$ of the bandpass 
calibrator (\S \ref{bandpass}).
The black points are the $T_{\rm sys}$ for the target data (usually a cluster), and the
orange points are $T_{\rm sys}$ for the phase calibrator (\S \ref{phasecal}).  
The points differ due to the differing columns of atmosphere 
to each source. The minimum in each curve occurs when the source transits (reaches its highest elevation in
the sky, and therefore the optical depth reaches its minimum for the observation).}
\label{fig:tsys}
\end{figure}

During a typical cluster observation, we compute the noise temperature while observing the
phase calibrator and while observing the target (usually a cluster) separately, since each 
is observed through a different column of atmosphere.  
See Fig.~\ref{fig:tsys} for a plot of $T_{\rm sys}$ in an observation.

Since $T_{\rm sys}$ is used to compute the theoretical \emph{rms} noise in the observation at a later
stage of the reduction pipeline (see \S \ref{noise}), we interpolate our $T_{\rm sys}$ measurements 
past large deviations from ideal behavior.\footnote{As discussed in \S \ref{tsys}, 
determining $T_{\rm sys}$ 
requires a measurement of total output power (i.e. not correlated visibilities, but
an actual measurement of power from each antenna) while observing two thermal loads with known temperatures.  
Therefore, $T_{\rm sys}$ cannot be determined while observing the source, and must be interpolated 
between measurements, which are taken every $\sim$ 10 minutes.  
This interpolation provides an estimate of $T_{\rm sys}$ for each data point.} 
This ideal noise is used in flagging data that show large deviations from the theoretical \emph{rms} noise.
A jump in $T_{\rm sys}$ of more than $\sim$ 1.5~K can indicate a sudden change in weather or an instrumental effect,
so we flag and interpolate stable $T_{\rm sys}$ measurements past these jumps.  Additionally, data are flagged if no $T_{\rm sys}$ measurements 
are available for over $\sim$ 30--35 minutes.  This was determined during pipeline testing to be the longest
reasonable time to go without a $T_{\rm sys}$ measurement, as we are effectively interpolating the behavior
of the atmosphere between actual measurements.


\section{Automatic Flagging}\label{autoflag}

Some bad data are next automatically flagged, without any user intervention, on a number of conditions.
These conditions can depend on the state of the array,  in which case they are flagged by the control 
system (i.e. for these instances, the reduction routine simply propagates the 
flags set by the control system).  
Examples of the common flags set by the control system are:

\begin{itemize}

\item {\bf Tracking (physical):} If a telescope is not mechanically tracking a source for 
any reason (hardware malfunction, an antenna is off-line for repair, etc.), data from that 
antenna are flagged.

\item {\bf Fringe Tracking:} If the software that implements fringe tracking (\S \ref{interf}) is not 
functioning properly while observing an astronomical source, there can be no correlated data.  
Therefore, these data are flagged.

\item {\bf Correlator Bands not received:} If the correlator software is not running 
for a given band, no correlated data are returned in that band; that band will be flagged.
\end{itemize}

Other automatic flags can be due to properties of the array that can either be computed or have been
measured.  These are:

\begin{itemize}

\item {\bf Antenna Shadowing:} For given antenna positions and primary mirror sizes,
in a homogeneous array like the SZA, it is straightforward to compute whether one 
antenna obstructs the view of another.  We use a projected antenna spacing 
of 3.6 meters, which treats the effective primary diameter also as 3.6~m in this 
calculation (recall that the SZA primary diameter is 3.5~m). This buffer is used to 
avoid diffraction effects at the edges of the primaries, which are due to the sidelobes of the 
antenna beam pattern (see Fig.~\ref{fig:baseline1}).  
If an antenna is shadowed, all data from that antenna are flagged during that period (i.e. we flag
each baseline involving the shadowed antenna for the entire time it is shadowed).

\item {\bf Birdies:} As discussed in \S \ref{rfi}, radio interference was found to contaminate
some of the channels in our observations.  A list of bad channels was kept with the rest of the data
reduction code, in a Concurrent Versions System \footnote{\url{http://www.nongnu.org/cvs/}}
(CVS) archive. 

\item {\bf Cross-talk:} As discussed in \S \ref{crosstalk}, coherent signals between
closely-spaced antennae could corrupt baselines.  Baselines measured to have significant
amounts of cross-talk were noted in the CVS, and those corrupted baselines were excluded. 

\end{itemize}

After the automatic flagging, a baseline correction is applied to the data.    
By applying small corrections to the antenna locations used by the control system, slightly different 
astronomical delays are computed for each antenna as the source traverses the sky.  
The set of correct antenna locations, which produce a constant phase versus time on each 
baseline, is called the ``baseline solution.''  

We typically measure the baselines every two weeks by observing many point sources across the sky
over the course of a few hours, and permuting the old baseline solution to find the 
flattest phases for the set of point sources.  Common reasons for changes to the baseline solution, 
which are typically less than a millimeter, involved expansions and contraction of the ground due to 
rain, or freezing and thawing of moisture in the soil between the concrete antenna pads.
No flagging is performed when the baseline solution is applied, as this is simply a mathematical
correction that is applied to the data.


\section{Interactive Data Calibration}\label{interactive_flagging}

The next step in the pipeline is a series of calibrations that can be run interactively
or by specifying scripts with desired parameters on which to flag.
While the following steps are now generally run in an automated fashion using scripts supplied by 
the user, we retain the flexibility to flag data by hand.  
This manual mode of flagging was particularly important when testing and debugging the 
pipeline, and was necessary to establish reasonable limits to use in script-driven data 
calibration.


\subsection{Bandpass calibration}\label{bandpass}

Any component -- an amplifier, a cable, a filter, or a mismatch between components or sections of 
waveguide\footnote{A waveguide mismatch produces standing waves in frequency across a spectrum, 
due to reflections.} -- 
can add a spectral shape to the signal that is non-astronomical in origin.  
By observing an astronomical source known to have a simple power law spectrum, 
instrumental phase and amplitude features can be calibrated out of the data.  
Bandpass calibration is done by solving for the corrections necessary to flatten both the 
amplitude and phase of each band's spectrum, on a \emph{per antenna} basis.  
Since these instrumental features do not change on the timescales of the observation (and typically only
do change when a component in the receiver or back-end electronics is changed), the bandpass calibration 
is performed once per observation, and can often be shared between adjacent observations.  

The first interactive flagging step is therefore the bandpass calibration, in which the shape of the 
instrument's frequency response is removed from the channel-based data before averaging the bands.  
Recall from \S \ref{dcon} \& \ref{corr} that there are 15 channels in each of the 16 bands. 
Data are bandpass calibrated using an $\approx 10$ minute integration on a strong point source 
($\gtrsim 10~\rm Jy$).  This short bandpass calibration observation is taken once per observation, 
at either the start or finish of an observation (``track'').  

\begin{figure}
\centerline{\includegraphics[width=6in]{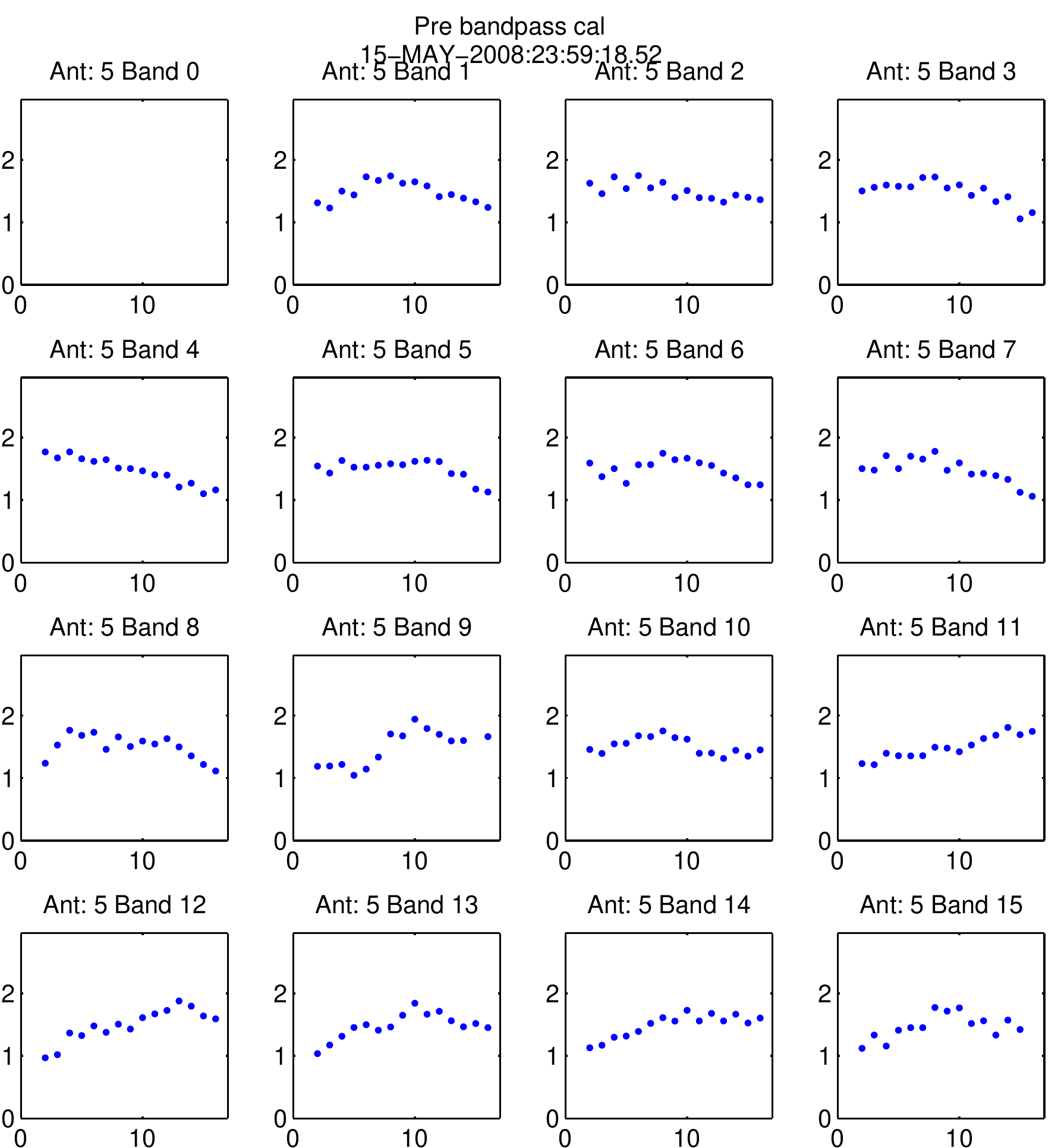}}
\caption{Amplitude vs frequency channel of the bandpass calibrator, measured for each band of Antenna 5, 
before calibration. The $y$-axis is the amplitude in Jy, and the $x$-axis is the channel
number.  Fig.~\ref{fig:postbp5} shows the calibrated amplitude of the bandpass.}
\label{fig:prebp5}
\end{figure}

\begin{figure}
\centerline{\includegraphics[width=6in]{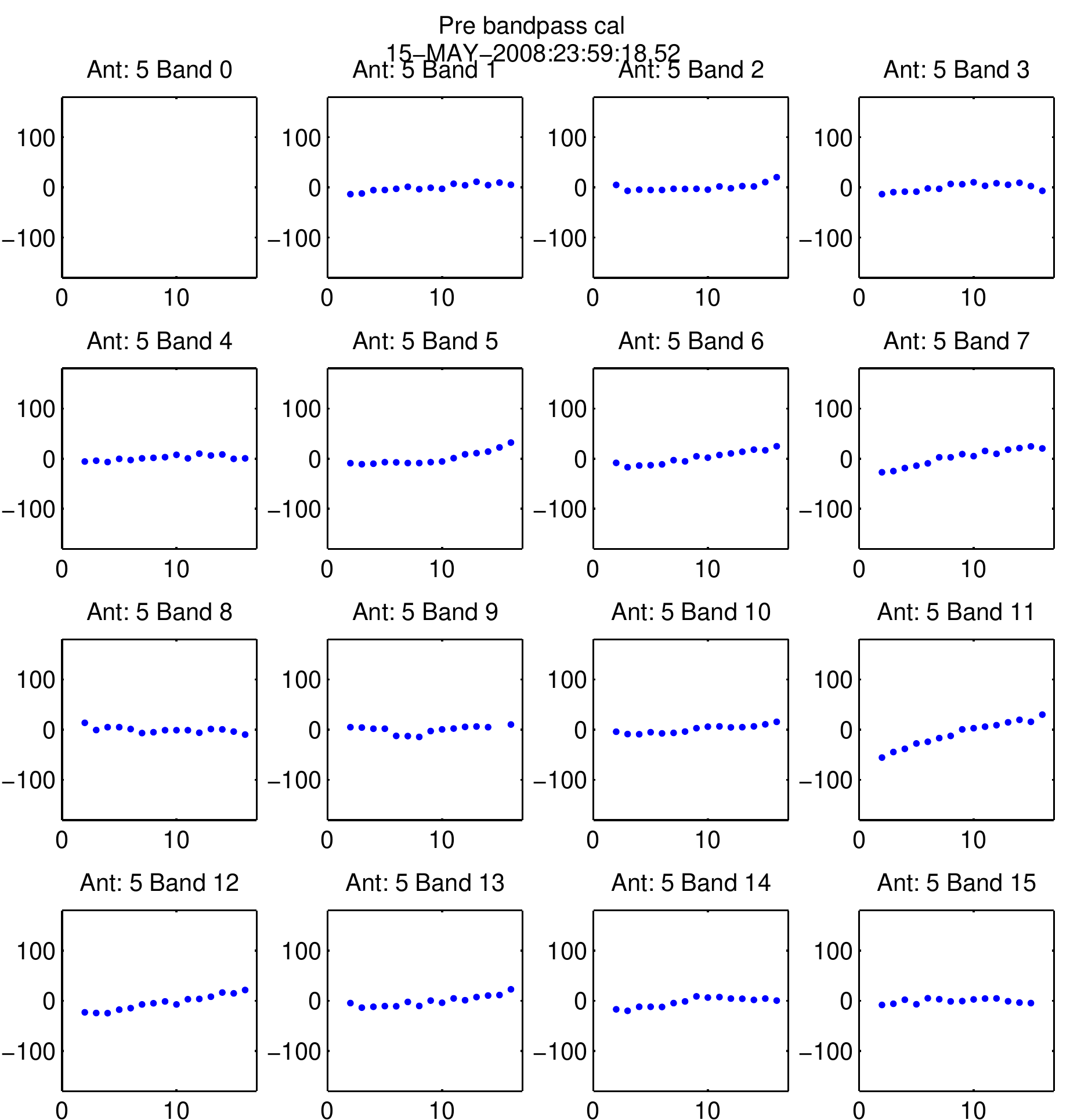}}
\caption{Phase vs frequency channel for the bandpass calibrator, measured for each band of Antenna 5, 
before calibration.
The $y$-axis is the phase in degrees, and the $x$-axis is the channel number.
Fig.~\ref{fig:postbpphase5} shows the calibrated amplitude of the bandpass.}
\label{fig:prebpphase5}
\end{figure}

\begin{figure}
\centerline{\includegraphics[width=6in]{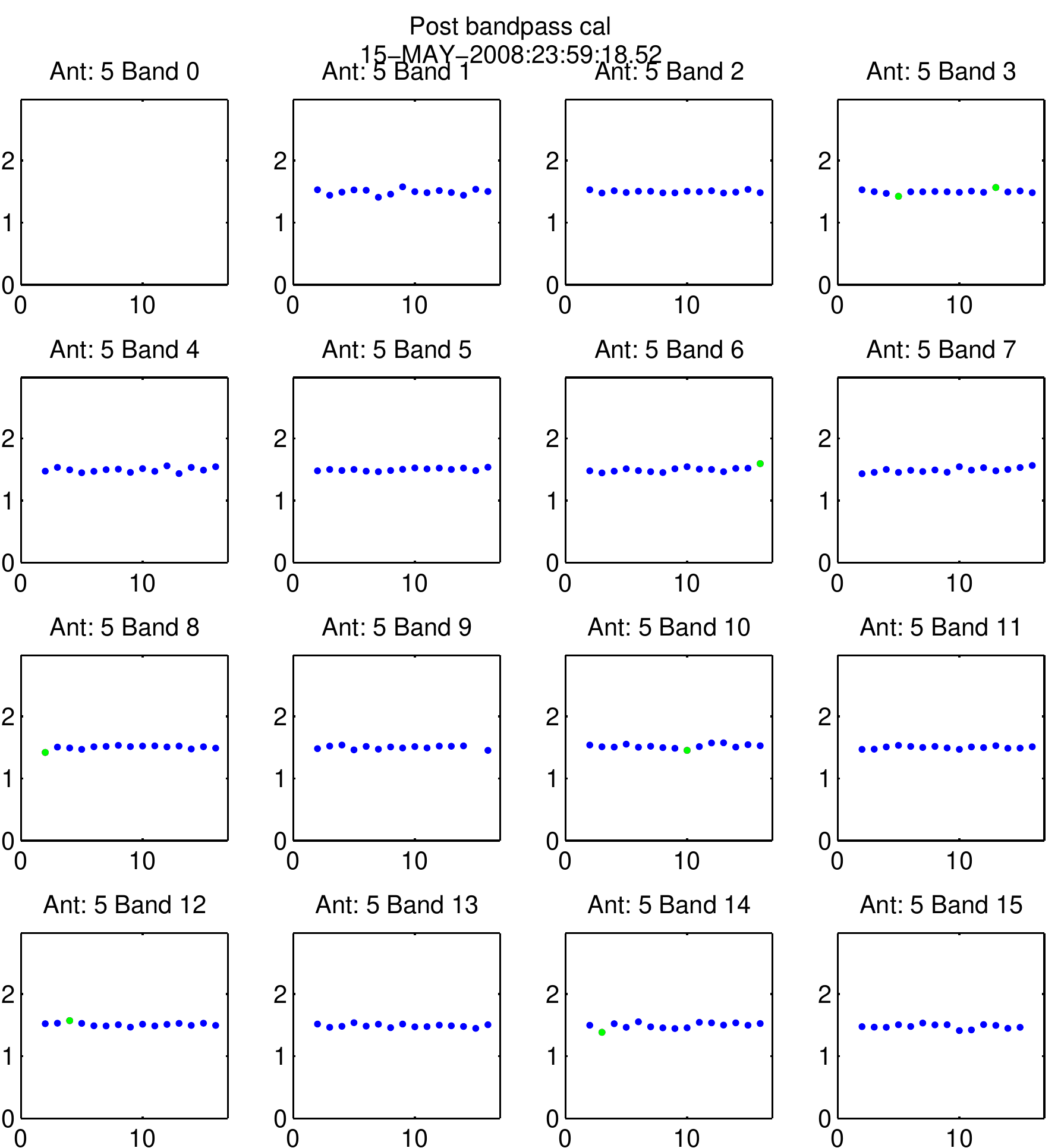}}
\caption{Amplitude (vs frequency channel) of the bandpass calibrator for each band of Antenna 5, after calibration.
See Fig.~\ref{fig:prebp5}, which shows these bandpass amplitudes pre-calibration, for caption.}
\label{fig:postbp5}
\end{figure}

\begin{figure}
\centerline{\includegraphics[width=6in]{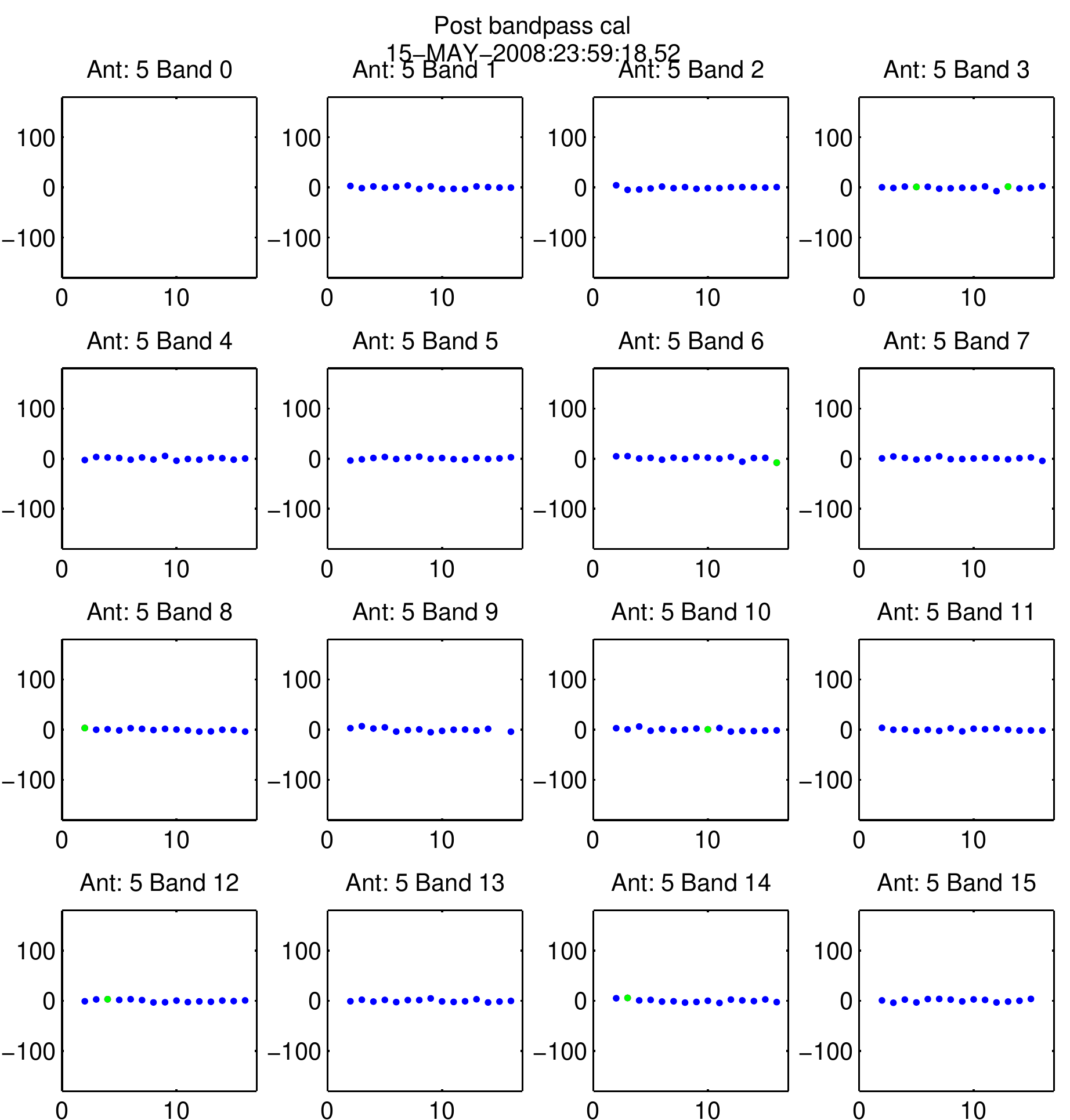}}
\caption{Phase (vs frequency channel) of the bandpass calibrator for each band of Antenna 5, after calibration.
See Fig.~\ref{fig:prebpphase5}, which shows these bandpass phases pre-calibration, for caption.}
\label{fig:postbpphase5}
\end{figure}

Figures \ref{fig:prebp5} \& \ref{fig:prebpphase5} show the bandpasses for each band of 
Antenna 5's amplitude and phase, respectively, before calibration.
Figures \ref{fig:postbp5} \& \ref{fig:postbpphase5} show these bandpasses after calibration.
After the bandpass calibration has been applied, the channels of each band are binned, and we no longer
work with the channel-based data.  This binning reduces the size of the dataset by a factor of 17
(see Table~\ref{table:channel_scheme}). 


\subsection{Phase/Amplitude calibration}\label{phasecal}

The next step in the calibration corrects for drifts in the overall amplitude
and phase of each band.
Every $\sim 15$ minutes, a moderately strong ($\gtrsim 2~\rm Jy$) point
source is observed for 3--5 minutes.  An unresolved source at the array's pointing center 
(i.e. at the spatial location $(0,0)$) is mathematically a 2-dimensional Dirac delta function, 
$\delta(0,0)$.\footnote{The integral of the point source's flux density over space 
is the total flux $f$ of the point source (i.e. $f = \int \delta(x,y) \, dx \, dy$).  
The general equation for a point source in \emph{u,v}-space with a flux $f_{0}$, normalized at
frequency $\nu_0=30.938~\rm GHz$ (center of the SZA 30-GHz band), and with a spectral index $\alpha$, 
at image location $(x,y)$ (in radians from the phase center), is 
\begin{equation}
f(u,v) = f_{0} \left(\frac{\nu}{\nu_{0}}\right)^\alpha 
	[\cos(2 \pi (ux+vy)) + j \sin(2 \pi (ux+vy))].
\end{equation}
For $(x,y)=(0,0)$, the point source is purely real, and has a constant flux at all locations
in \emph{u,v}-space.  The simplicity of this solution motivates the use of point sources in 
tracking the system's response, as each baseline will ideally measure the same flux.
Another way of stating this is that a point source is smaller than the beam formed by any
baseline pair, so the flux probed by any baseline, in Jy/beam, equals the total flux in Jy.} 
Therefore, its spatial Fourier transform is a purely real constant, which implies that
each baseline ideally would measure the same amplitude, with zero phase.  Deviations
from zero phase, therefore, are tracked and corrected using this source as a ``phase calibrator.''  

\begin{figure}
\centerline{\includegraphics[width=6in]{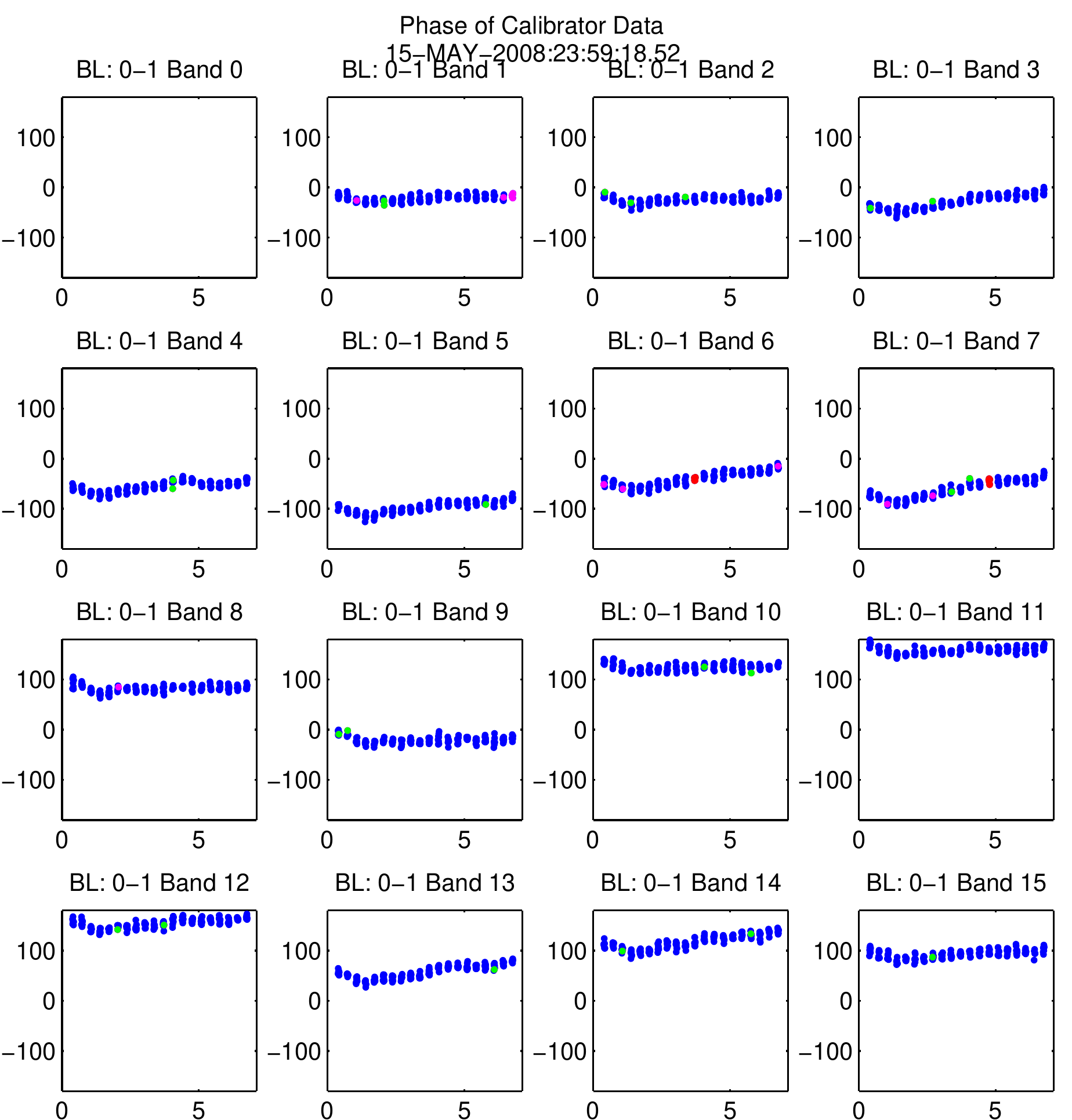}}
\caption{Phase of the point source calibrator versus time, as measured by a short baseline (Baseline 0-1).
The $y$-axis is the phase in degrees, and the $x$-axis is time in hours since the start of the track.
Band 0 was entirely flagged for this track, due to broken digitizers and a 
lack of spares at the time.  The magenta points are those flagged due to the automatic flagging, 
and the green points are those flagged by a user-specified script's limits,
which catches 30$^\circ$ outliers from the underlying, interpolated phase.
}
\label{fig:phase01}
\end{figure}

\begin{figure}
\centerline{\includegraphics[width=6in]{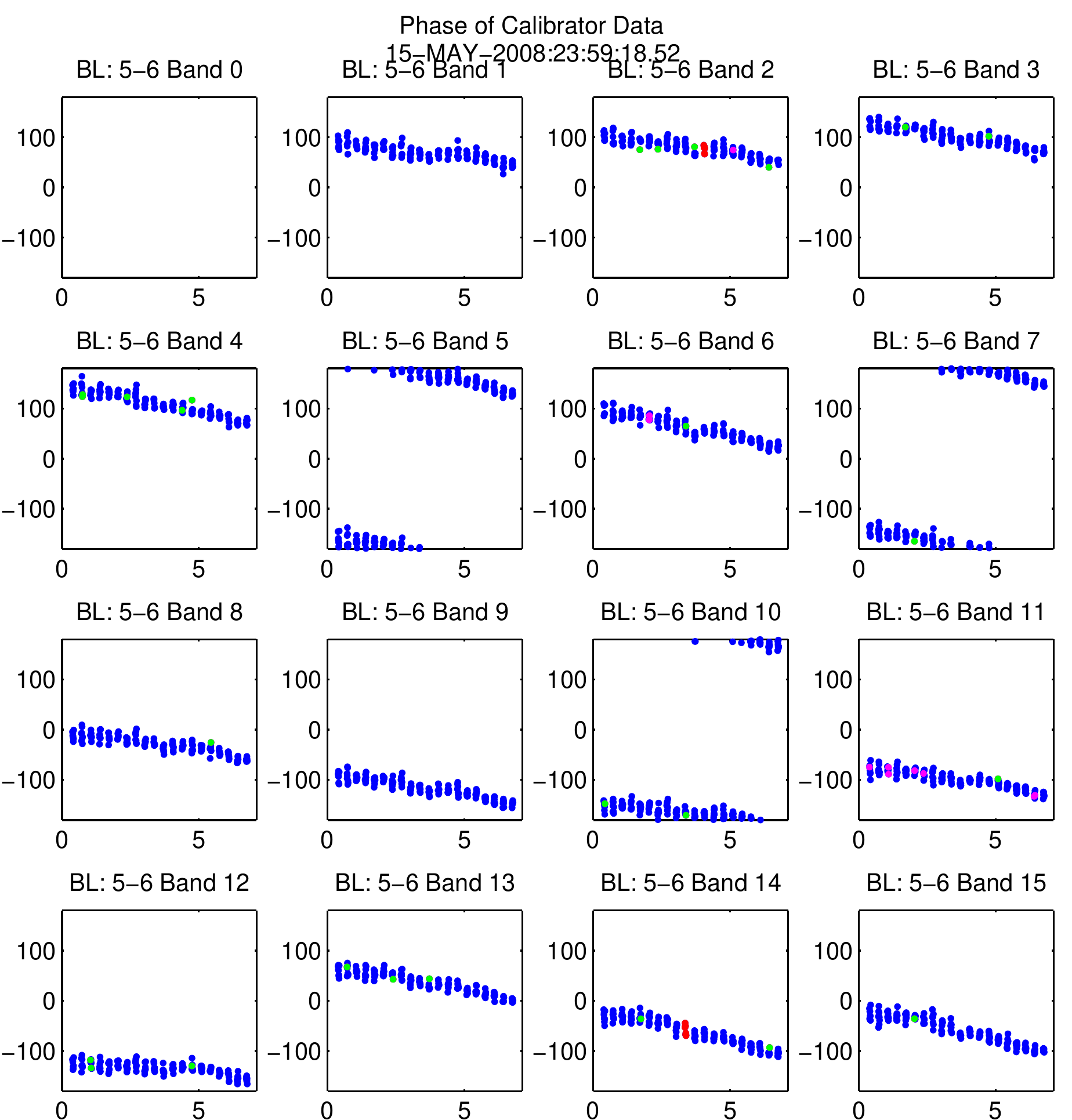}}
\caption{Phase of the point source calibrator versus time, as measured by a long baseline (Baseline 5-6).
See Fig.~\ref{fig:phase01} for further details.  
Note the slightly larger scatter in the phase on this baseline than on Baseline 0-1 (Fig.~\ref{fig:phase01}).  
This is due to atmospheric coherence being slightly poorer on long baselines (the antennae are 
looking through different columns of air, and atmospheric turbulence has a scale size on the order of tens of meters). 
The underlying slope in the phase calibrator is easily determined in this observation, as the atmospheric coherence
for the example track was typical of a clear spring day.}
\label{fig:phase56}
\end{figure}

\begin{figure}
\centerline{\includegraphics[width=6in]{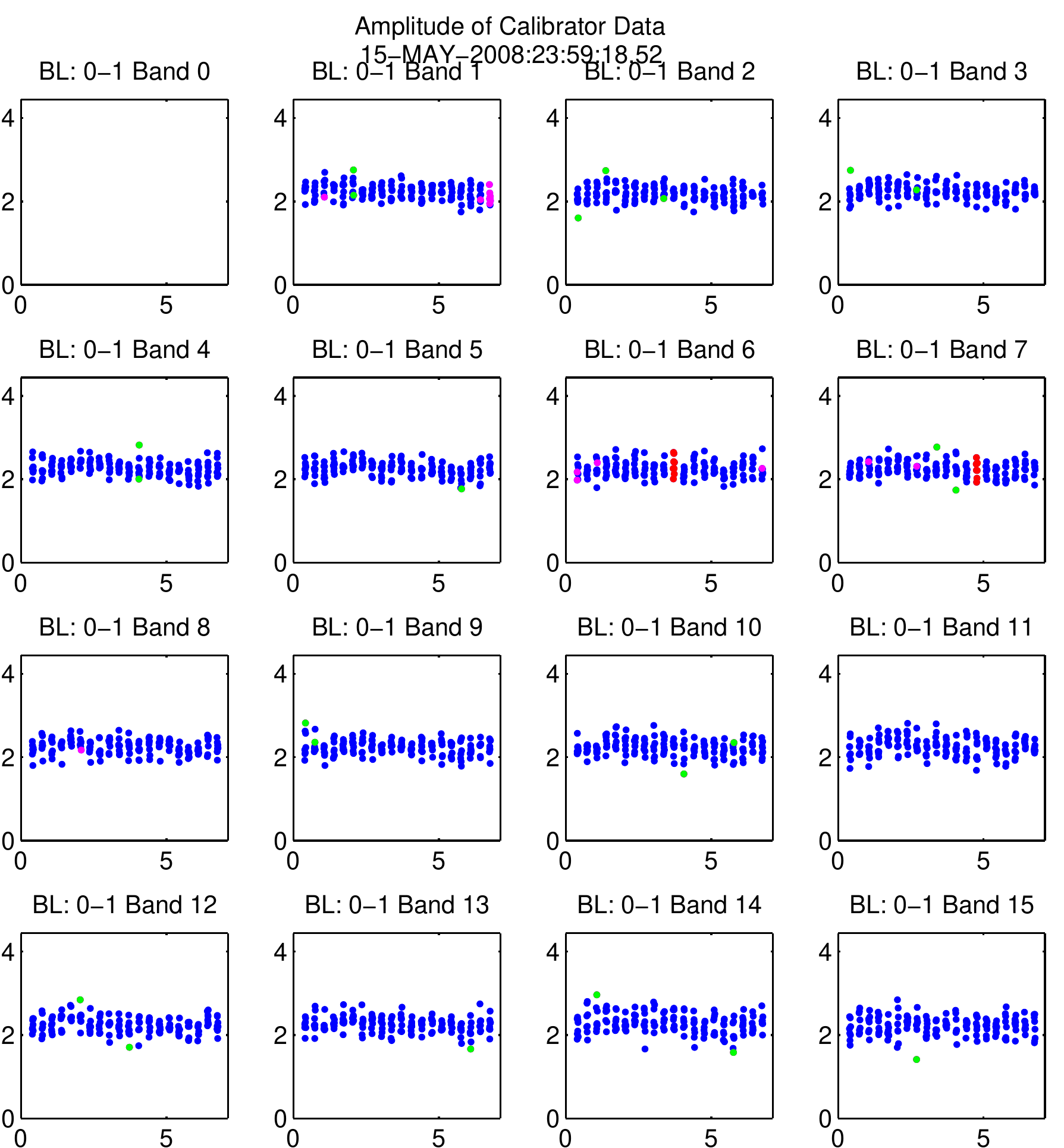}}
\caption{Amplitude of the point source calibrator versus time, as measured by a short baseline (Baseline 0-1).
The $y$-axis is the amplitude in Jy, and the $x$-axis is time in hours since the start of the track. 
Since a  point source is unresolved (the beam is larger than the point source), the amplitude in Jy equals 
that in Jy/beam.
This is the amplitude of the calibrator corresponding to the phase shown in Fig.~\ref{fig:phase01}.}
\label{fig:amp01}
\end{figure}

\begin{figure}
\centerline{\includegraphics[width=6in]{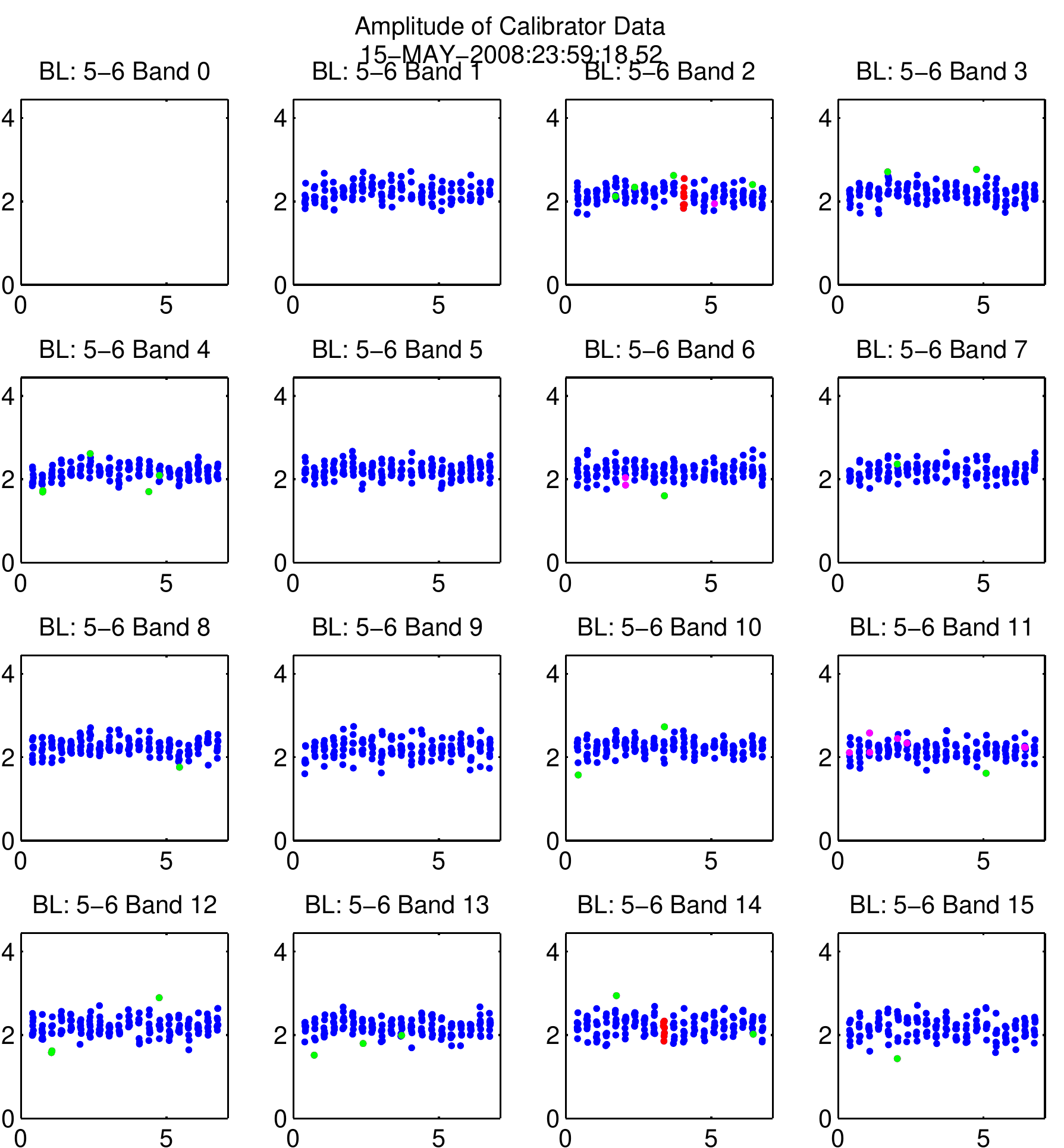}}
\caption{Amplitude of the point source calibrator versus time, as measured by a long baseline (Baseline 5-6).
This is the amplitude of the calibrator corresponding to the phase shown in Fig.~\ref{fig:phase56}.
See Fig.~\ref{fig:amp01} for further details.}
\label{fig:amp56}
\end{figure}

\begin{figure}
\centerline{\includegraphics[width=6in]{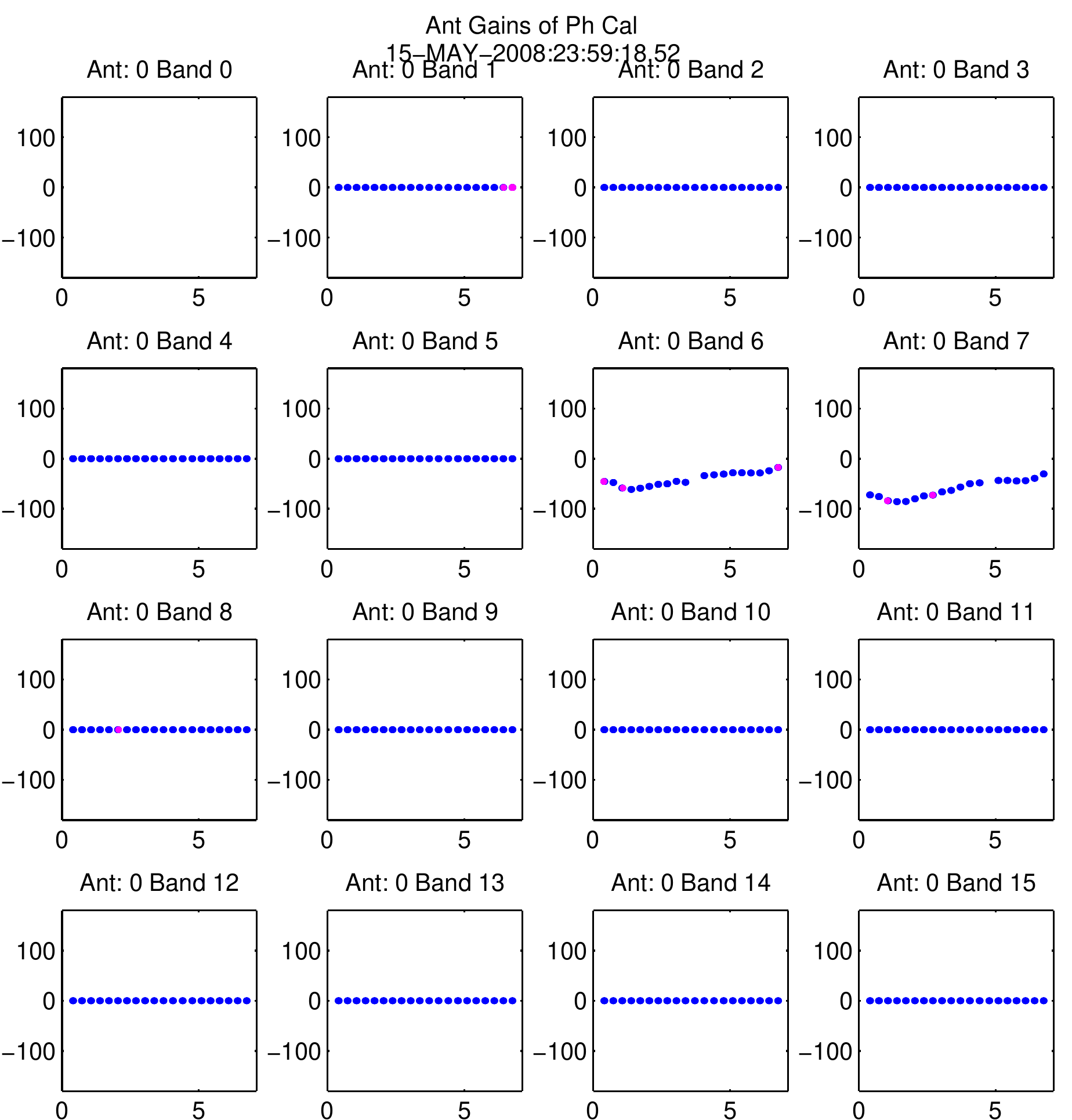}}
\caption{Antenna-based phase of the point source calibrator versus time.
One antenna must be chosen as a reference in order to compute the antenna-based
phase of the other antennae.  In this plot, Antenna 0 served as that reference for all
but Bands 6 \& 7, which had flagged calibration observations (see red points in Fig.~\ref{fig:phase01})}
\label{fig:antgains0phase}
\end{figure}

Figures \ref{fig:phase01}--\ref{fig:amp56} show the baseline-based phases and amplitudes of the 
phase calibrator, measured by the example Baselines 0-1 and 5-6.
By setting one antenna per band as the ``reference antenna,'' which for that band is assumed to 
have zero phase, the calibrator phase for every other antenna can be computed.  
The reference antenna for each band is simply chosen to be the first antenna with
no flagged calibrator data.  In Fig.~\ref{fig:antgains0phase}, it is clear -- from the flat phase that 
is identically equal to zero -- that Antenna 0 was used as the reference for all bands except 
Bands 6 \& 7, which had flagged phase calibrations mid-track
(see red points in Fig.~\ref{fig:phase01}).

\begin{figure}
\centerline{\includegraphics[width=6in]{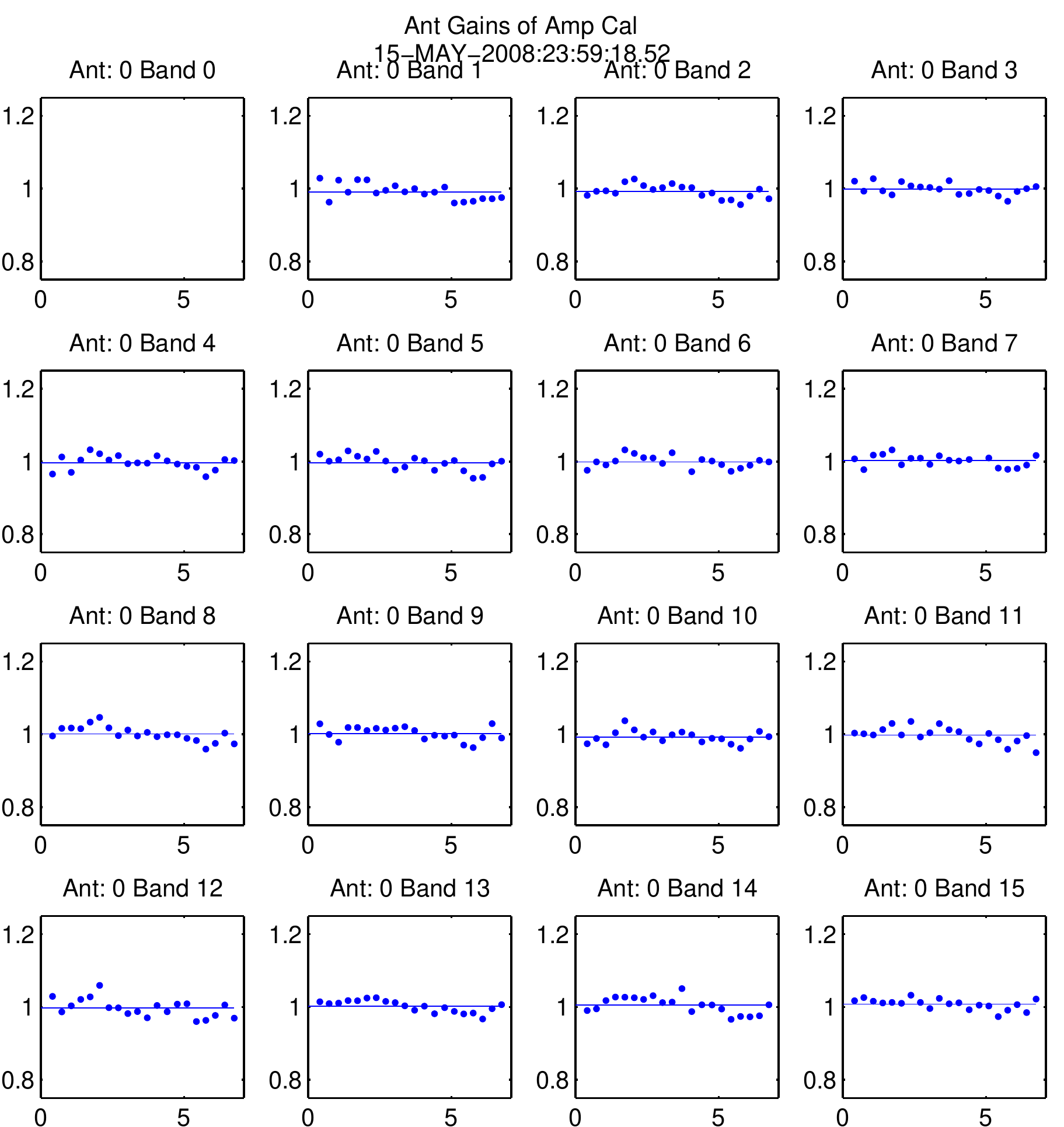}}
\caption{Antenna-based gain of the point source calibrator versus time, for Antenna 0.
The $y$-axis is the gain (ideally 1), and the $x$-axis is time in hours since the start 
of the track.  The scatter seen here, which is simply due to noise in the measurement, 
is smaller than our absolute calibration uncertainty ($\sim 5\%$, see \cite{muchovej2007}).}
\label{fig:antgains0}
\end{figure}

The phase calibrator is used to track changes in the complex antenna-based gains throughout
the track , which can happen for a number of reasons (see Figures \ref{fig:antgains0} \& 
\ref{fig:antgains0phase}).  
The SZA receivers are well-characterized and have stable gains under 
normal operating conditions;  changes in the complex gain of the actual instrument therefore usually 
arise when equipment is thermally
unstable (failed refrigerator, poor regulation of the electronics box, rapid heating/cooling
of the correlator trailer during sunrise/sunset, etc.).  Sharp changes in phase,
which lead to degradation of the measured amplitude, are used to flag data.

Scatter in the phase due to poor atmospheric coherence can lead to a lower measured
amplitude in the correlated data.
Clouds and atmospheric turbulence during inclement weather are generally of
a small enough scale that a baseline in the inner array measures the full amplitude of the
calibrator, as the phase scatter is common-mode on short baselines (i.e. they see
the same fluctuations); during such
a period, long baselines measure a diminished signal, since the atmosphere
is essentially adding random, uncorrelated phase fluctuations to each signal.  
The example observation presented throughout this chapter was taken on a clear spring day 
with good atmospheric coherence.


\begin{figure}
\centerline{\includegraphics[width=6in]{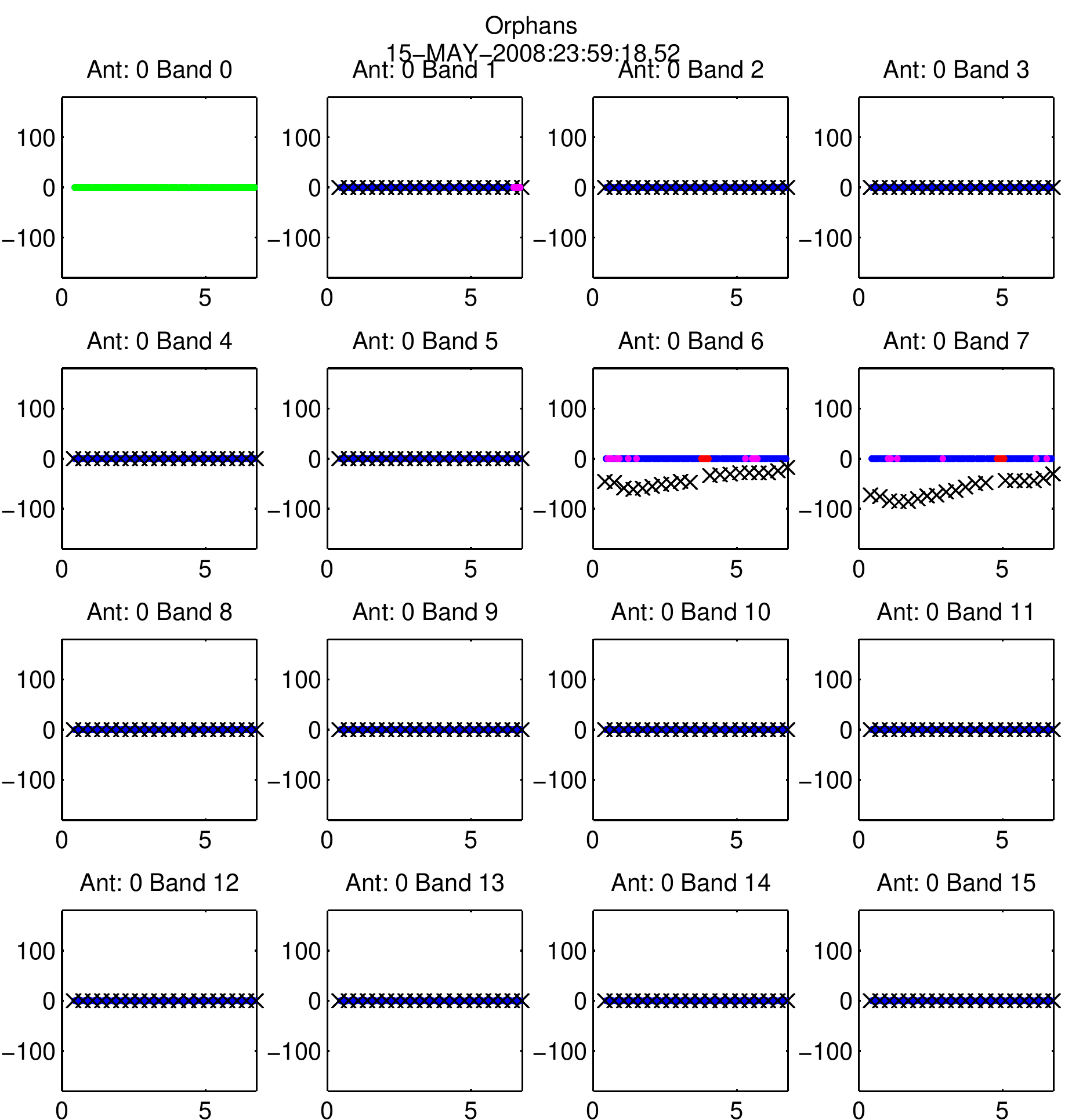}}
\caption{Orphaned data from Antenna 0.  Automatically flagged data in Band 0 
are shown as green points, and were due to a digitizer hardware problem (and a temporary 
lack of replacement digitizer boards).  
Data orphaned by gaps in the calibrator data in Bands 1, 6, \& 7 are shown in red (see Fig.~\ref{fig:antgains0phase}).
Useful data are represented as blue points.
For simplicity, target data are plotted with zero phase (rather than plotting the noisy 
distribution of raw target data phases, since we are only trying to determine which data
are not bracketed by calibrator observations). 
The antenna-based phases of bracketing calibration observations are shown as black X's
(which are identically zero in bands where Antenna 0 was the reference; see Fig.~\ref{fig:antgains0phase}).
}
\label{fig:orphans0}
\end{figure}

\begin{figure}
\centerline{\includegraphics[width=6in]{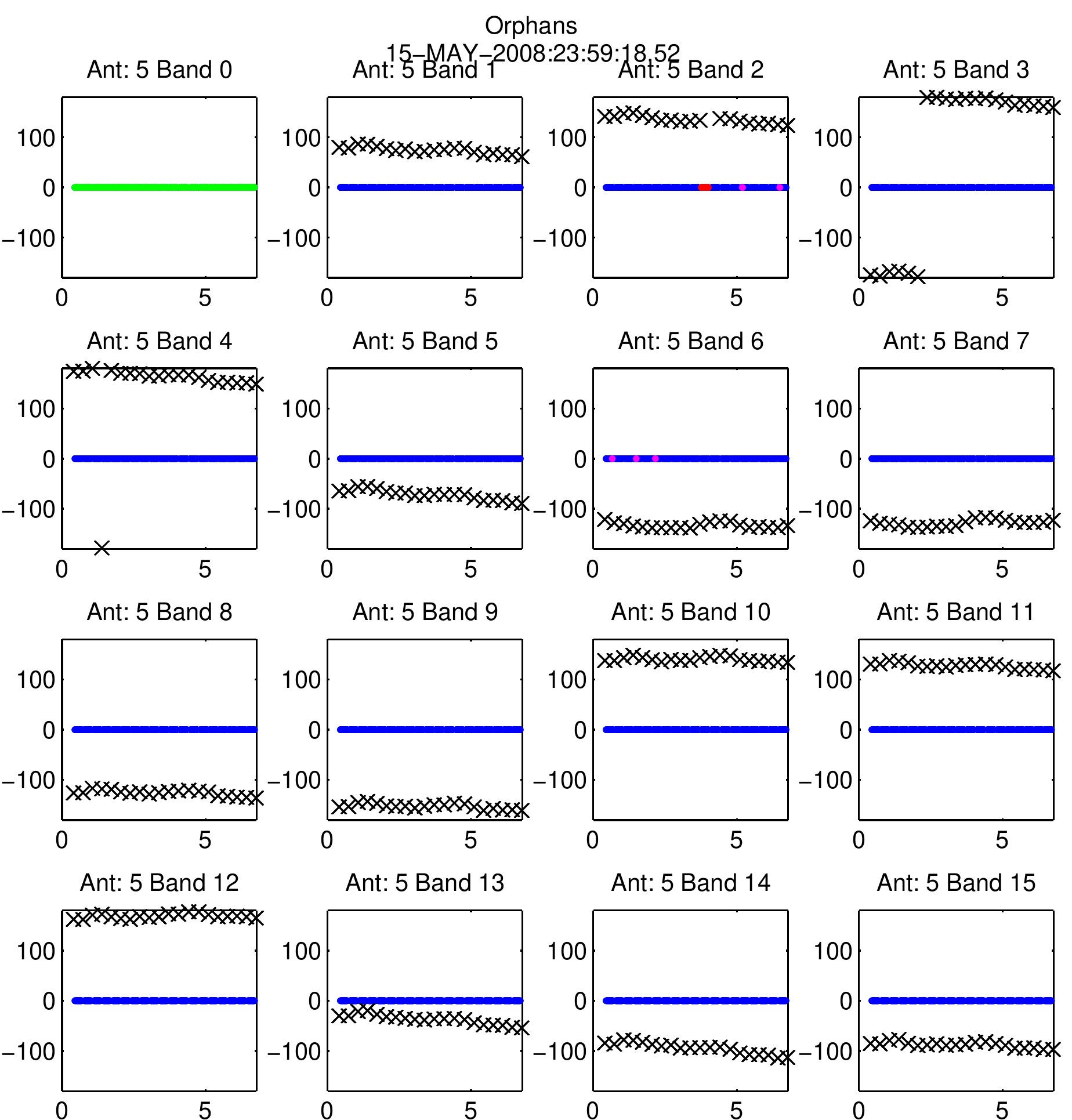}}
\caption{Orphaned data from Antenna 5.  
Data orphaned by gaps in the calibrator data in Band 2 are shown in red (see Fig.~\ref{fig:phase56}).
See Figure~\ref{fig:orphans0} for details.
Note that Antenna 5 was not the reference antenna for any band in the antenna-based phase 
calculation, so none of its phases are identically zero.}
\label{fig:orphans5}
\end{figure}

\subsection{Orphaned data}

After phase calibration is complete, periods when data are not bracketed by $T_{\rm sys}$
or phase calibrator observations are flagged, as those data cannot be calibrated.  
We call these data points ``orphans'' and discard them.  This includes 
data orphaned by flagged calibrator observations (such as the red points in Figures 
\ref{fig:phase01} and \ref{fig:phase56}).
Note that this step is antenna-based, since data are typically orphaned by antenna-specific problems
(e.g. an antenna was shadowed, or was not tracking and missed a phase calibrator observation).
See Figures \ref{fig:orphans0} \& \ref{fig:orphans5} for plots that show the flagging of
orphaned data.


\subsection{Noise and variance of the data}\label{noise}
The next step in the user-guided calibration is to flag target data with a high or biased 
level of \emph{rms} noise. The real \emph{rms} noise of the target data is compared with the 
expected theoretical value, computed using the measured values of $T_{\rm sys}$ 
(see \S \ref{tsys_calib}, Eq.~\ref{sigma_rms}).
If the variance of the data does not agree with the theoretical value on the $\sim 35\%$ level, 
the data are flagged.  This threshold was determined by Matthew Sharp in a series of jackknife tests
(differencing of essentially blank fields) on many CMB fields.  
Since $T_{\rm sys}$ data have been cleaned of outlier measurements, this flags periods with
higher than expected noise.

\begin{figure}
\centerline{\includegraphics[width=6in]{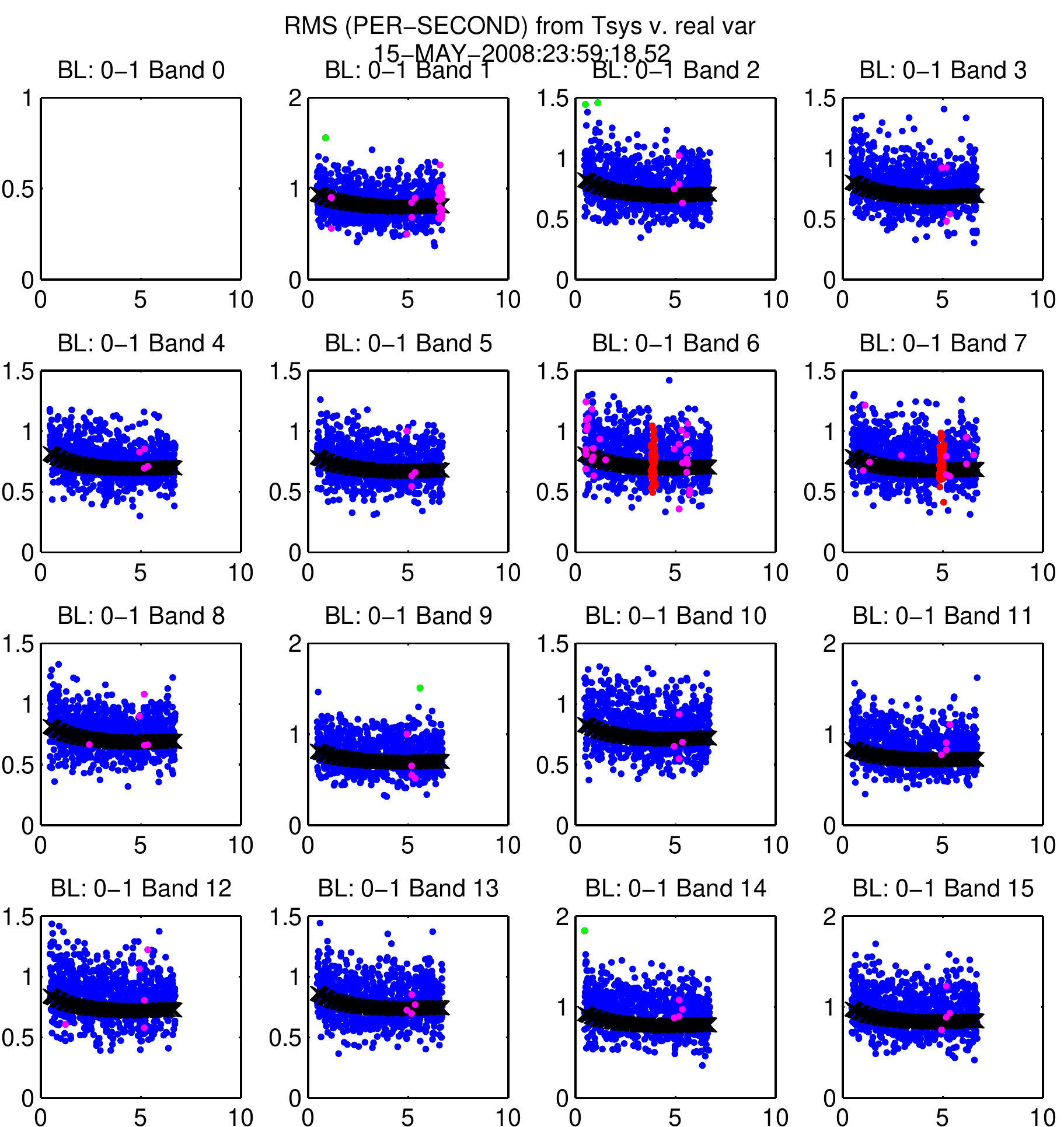}}
\caption{\emph{rms} of target data, taken on short Baseline 0-1.  
The theoretical prediction, based on the measured $T_{\rm sys}$, is shown in black. 
Flagged data -- already caught by other steps (see Figures \ref{fig:phase01} \& \ref{fig:orphans0}
in particular) -- are plotted in red and magenta.  Newly flagged data are in green.  
Useful, unflagged data are in blue.}
\label{fig:rms01}
\end{figure}

\begin{figure}
\centerline{\includegraphics[width=6in]{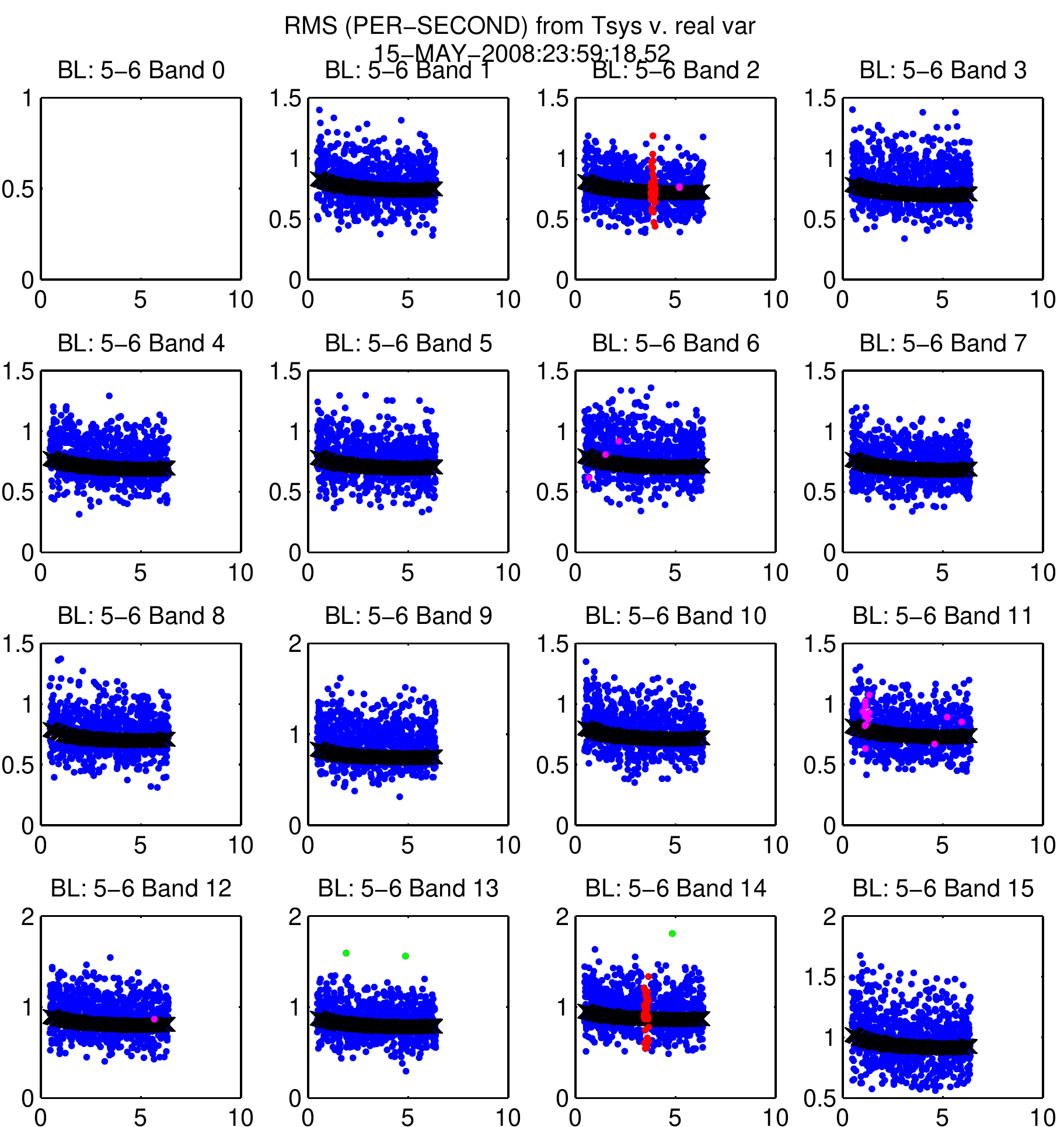}}
\caption{\emph{rms} of target data, taken on long Baseline 5-6.  
Flagged data -- already caught by other steps (see Figures \ref{fig:phase56} \& \ref{fig:orphans5}) 
-- are plotted in red and magenta.
See Figure~\ref{fig:rms01} for more details.}
\label{fig:rms56}
\end{figure}

Additionally, any data points with more than a 4-$\sigma$ deviation from the distribution are flagged. While this
filter is useful in principle, data with such large error bars are typically caught in the previous steps.
See Figures \ref{fig:rms01} \& \ref{fig:rms56} for plots of the \emph{rms} noise of target data for a long
and a short example baseline.

The expected \emph{rms} noise is computed using cleaned $T_{\rm sys}$ data, where any jumps in $T_{\rm sys}$
have been interpolated past.  However, if the jump was due to a real instrumental or atmospheric effect,
the data will show a similar jump in measured \emph{rms} noise. This is why large deviations from the 
expected theoretical noise are flagged.  
An example of an instrumental effect that leads to a higher than expected \emph{rms} 
level (without affecting the measured $T_{\rm sys}$) is antenna cross-talk, discussed in \S \ref{crosstalk}.


\subsection{Amplitude of the target data}
\begin{figure}
\centerline{\includegraphics[width=6in]{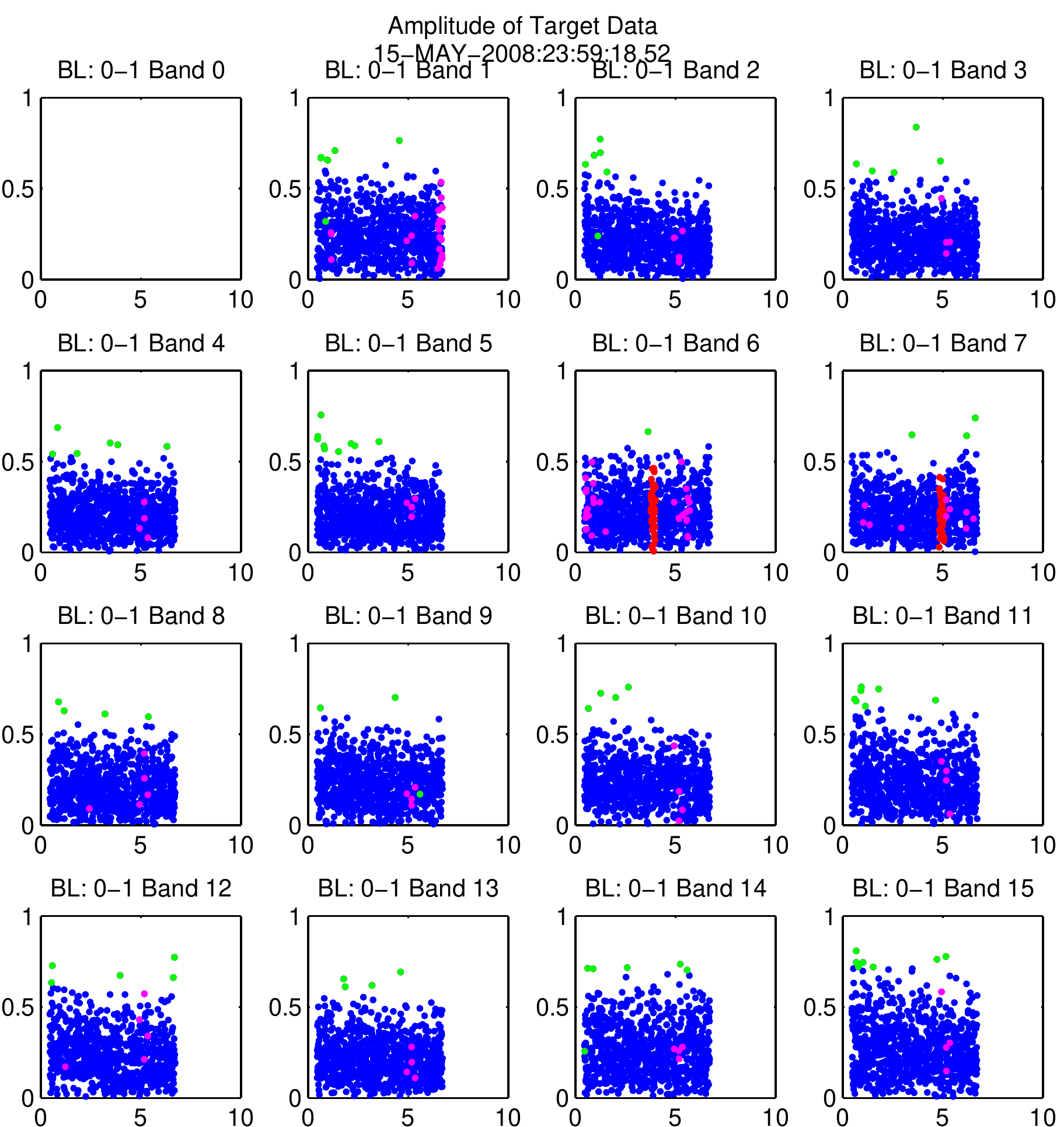}}
\caption{Amplitude of visibility data (in Jy, see \S \ref{interf}) for a short baseline (0-1),
taken on the target source  (typically a cluster). Blue points are unflagged data.  
Red and magenta points are flagged data caught by previous steps in the data calibration 
(see Figures \ref{fig:phase01} \& \ref{fig:orphans0}), while green points may indicate
data flagged in this step as outliers or from previous steps (e.g. Bands 1, 2, \& 14 contain
some green points that are not outliers, flagged in Fig.~\ref{fig:phase01}).
The SZE flux from the cluster and the fluxes of point sources in the cluster field are on the 
$\sim$ mJy level; they are therefore not noticeable in the raw data plotted here (which are 
noise dominated).}
\label{fig:target_amp01}
\end{figure}

\begin{figure}
\centerline{\includegraphics[width=6in]{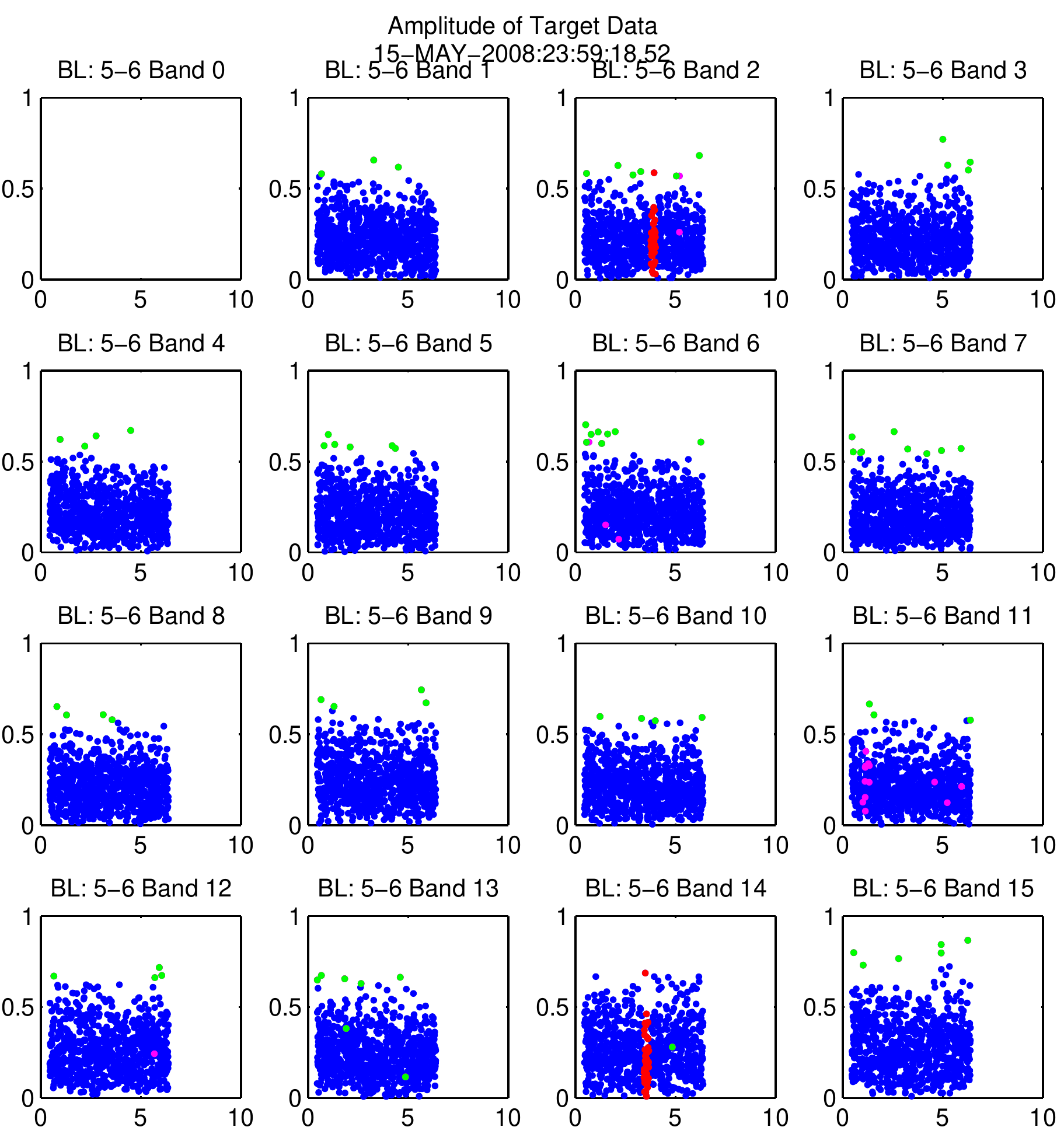}}
\caption{Amplitude of visibility data in Jy (see \S \ref{interf}) for a long baseline (5-6) 
taken on the target source. See Fig.~\ref{fig:target_amp01}.  Note that long baselines do
not typically measure cluster scales, but the raw target data here and in Fig.~\ref{fig:target_amp01}
have similar amplitudes.  Also note that some green, flagged points (e.g. in Band 13) were flagged
in Fig.~\ref{fig:phase56}, not because they are outliers.}
\label{fig:target_amp56}
\end{figure}

The last step of user-guided calibration flags for anomalies in the amplitude 
of the raw target data.   For the example cluster observation (shown in Figs.~\ref{fig:target_amp01} 
\& \ref{fig:target_amp56}), there are no strong ($\gtrsim 1~\rm Jy$) 
point sources in the field.  Individual data points are dominated by noise, since the cluster 
signal is on the order of ten mJy (total, integrated over the sky), and the \emph{rms} noise per second is 
$\sim 0.5~\rm Jy/beam$.\footnote{This is for typical binning 
of the visibilities -- of course the signal-to-noise ratio ($S/N$) increases when data are binned, revealing a
high-significance detection of the cluster's SZE signal, as well as the accompanying point sources in the field.}
Data points with more than a 4-$\sigma$ deviation from the mean -- which could be due to high
atmospheric noise -- are typically flagged.\footnote{While data with larger, random \emph{rms} noise would
have a lower weight (see \S \ref{final_data_product}), and thus would not bias datasets, including them
increases the size of the calibrated dataset. Since noisy data points are the exception, it is
easier to exclude them.}  
Also, any jumps in target amplitude are immediately suspect, as they are indications of 
undesired instrumental/atmospheric effects (since the cluster SZE signals and point source fluxes
are not time-dependent on the scale of hours).

Though we never attempted observations of clusters with ($\gtrsim 1~\rm Jy$) point sources in the field, 
the CMB and cluster survey data often do contain such sources.  
Their presence simply raises the mean value of the amplitude
of the target data.  Since we only flag on outliers from the distribution, note 
that this step does not prevent us from performing observations of fields with 
strong point sources.\footnote{Since the SZA's dynamic range is $\sim 100$ the 
presence of a 1~Jy point source in a field would prevent us from obtaining good 
constraints on a $\sim$ 10~mJy cluster SZE signal.}


\section{Dirty Maps}

\begin{figure}
\centerline{\includegraphics[width=4in]{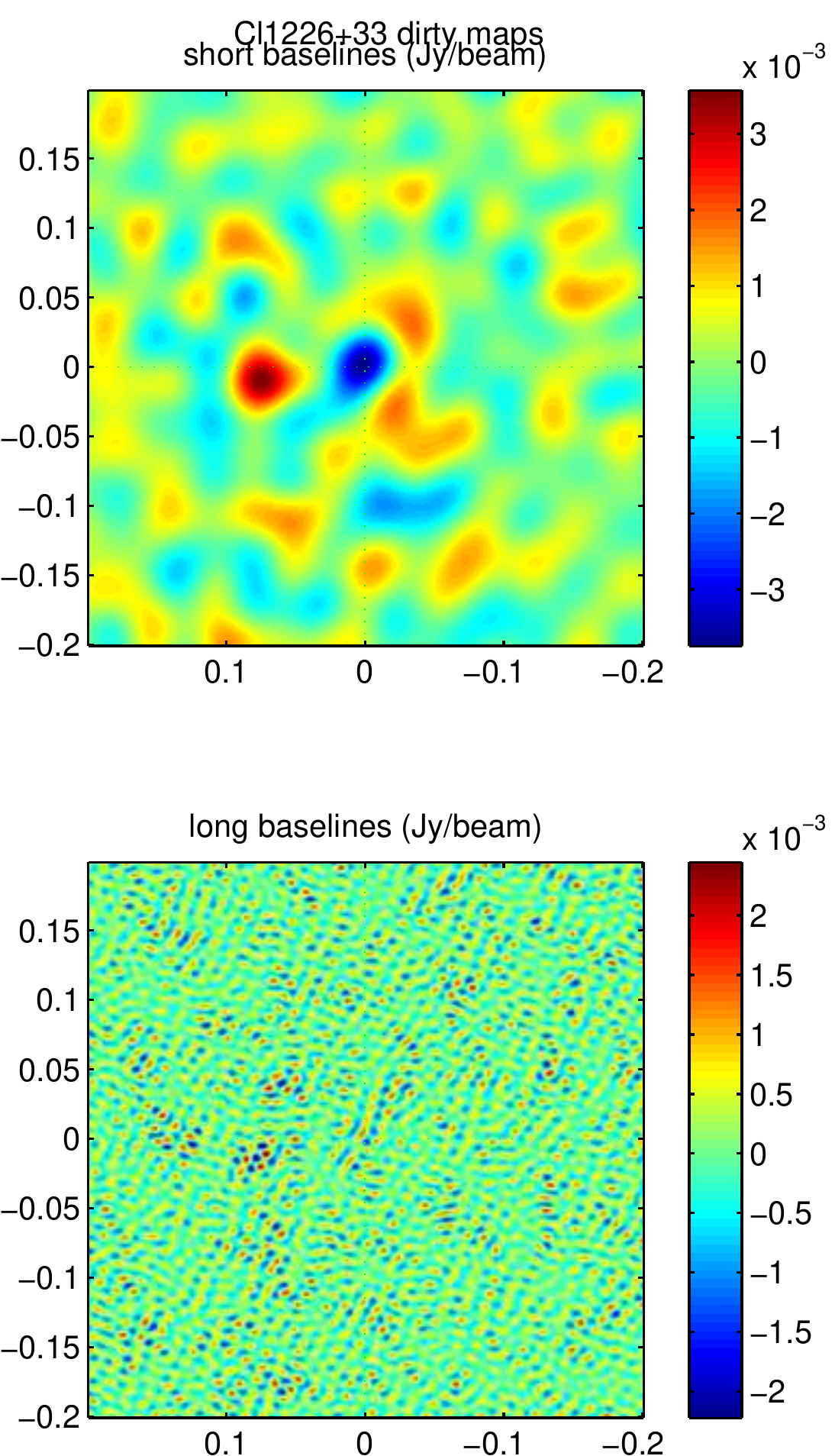}}
\caption{Dirty maps of the short (upper panel, $\lesssim 2~\rm k\lambda$) and long (lower panel, 
$\gtrsim 2~\rm k\lambda$). $x$ and $y$ axes are the map coordinates in degrees.  The colors represent 
the signal in Jy/beam (the flux detected within the beam formed by each baseline).  
See text for more details.}
\label{fig:dirty_maps}
\end{figure}

After all steps of the calibration are complete, the machinery of \emph{Difmap} \citep{shepherd1997}, imported into
MATLAB$^\circledR$, is used to produce rudimentary maps of the observation (shown in 
Fig.~\ref{fig:dirty_maps}).  These maps are largely illustrative, as all quantitative results of the SZE cluster 
observations are determined by fitting the visibilities in \emph{u,v}-space (see \S \ref{markov}).
We do this because the noise properties of the visibilities in \emph{u,v}-space are well understood.  

In Fig.~\ref{fig:dirty_maps}, the short baseline data (upper panel) show a central, dark blue ``blob,'' 
which is the flux decrement due to the cluster CL~J1226.9+3332.  To the left of the cluster, 
which is east on the sky, a $\sim 4~\rm Jy$ source is seen as a red ``blob'' 
(attenuated by the primary beam to $\sim 2.5~\rm Jy$, as it is $4.5\arcmin$ from the center).  
The resolution of the upper panel map is determined by the scale probed by
the short baselines, and not the size of the source in the sky (if larger, the baseline would
not be sensitive to the source). 
The lower panel shows a slight excess in flux at the point source position.
The long baselines do not probe cluster scales.
For well-calibrated data, only the very strongest sources show up in a 
single track, and a typical cluster observation requires more than a single track to constrain a 
cluster's SZE signal (as well as the weak $\sim$ few mJy point sources in the field).  
We therefore expect a single track to be dominated by noise.


\section{Final Data Product}\label{final_data_product}

Calibrated data are output, using a C++ function written by Erik Leitch, in 
the standard Flexible Image Transport System (FITS)\footnote{See \url{http://fits.gsfc.nasa.gov/}.} 
format for astronomical data.  Since these are radio interferometric data in \emph{u,v}-space,
we use the UVFITS specification of the FITS format.
These files contain the real and imaginary components of the visibility data (in Jy, see \S \ref{interf}) 
from all 16 SZA bands, stored in binary format.  They also contain the \emph{u,v}-space coordinates and 
weights (inverse variance, 1/$\sigma_\text{rms}^2$) for each data point\footnote{Interferometric maps of 
the \emph{u,v}-data typically weight the data by inverse variance (1/$\sigma_\text{rms}^2$).  
Information about the noise is stored in this form to avoid redundant computation steps.}.  

Further useful information is also stored in the UVFITS file.
An ASCII header provides observation specifics, such as the date of the observation,
name of the observatory, name of the target observed, the pointing center used for
the target, and the frequency at which it was observed, while a binary table includes
each antenna's physical location.

\part{Modeling and Analysis of Clusters}
\chapter{Modeling the Cluster Signal}\label{modeling}

\section{Introduction to Cluster Models}

As detailed in \S \ref{uvspace}, the SZA interferometer probes a cluster's SZE signal
in \emph{u,v}-space  (see Figures \ref{fig:uv_coverage_cl1226}--\ref{fig:uv_space_hist_CL1226_30_90}).  
In order to extract physical parameters from an interferometric cluster observation,
we need to fill in information on the scales we do not probe.
A simple model to extrapolate through the individually noisy, sparsely-sampled visibilities\footnote{There are typically
$\sim$10-100 thousand visibilities in an SZA cluster observation, yielding high signal-to-noise SZE
observations when the data are binned.} (Eq.~\ref{eq:visibility}), is used to extract meaningful constraints on physical 
cluster parameters.
Given the large gaps in the \emph{u,v}-coverage, a simple, 2-parameter model was found to work best.

When fitting data in \emph{u,v}-space, it is helpful to consider two extreme cases:
scales below the resolution of the instrument (i.e. smaller than the scales probed by
the longest baseline), and scales larger than the shortest baseline probes.  The first case
defines what we consider to be a point source, where in \emph{u,v}-space each antenna ideally
measures constant, uniform power.  

The second case implies that we are not sensitive to backgrounds, such as the primary
CMB (most of its power is on much larger angular scales than we probe) or large-scale 
galactic emission (from our galaxy).  
This makes the SZA a good spatial filter for clusters at high redshift, since it interferometrically
probes Fourier modes of the spatial intensity pattern corresponding to the bulk of a cluster's signal. 
It also means we cannot constrain cluster features on scales larger than the shortest 
baseline can probe.
A cluster model can predict an arbitrary signal at scales larger than we can probe; 
a poorly-motivated model can have Fourier modes in its transform that agree with the 
measured SZE signal, but 
might predict too much or too little flux on scales not accessible to an interferometer.  
At the longest wavelength of the SZA, observing down to the shadowing limit, this upper
limit is $\approx 12.7\arcmin$; above this scale, information is lost due to lack of \emph{u,v}-coverage.

Modeling of any signal, including that due to the SZ effect in clusters,
introduces priors on the resulting fit and its derived quantities.
I discuss below the SZE models I tested, as well as the X-ray models used to complement
them.  I then discuss the Markov chain Monte Carlo (MCMC) fitting procedure we use to
fit these models jointly to SZE+X-ray data. Finally, I show how we derive galaxy cluster 
properties from the resulting model parameters.

Throughout this section, we use several common cluster overdensity radii that 
appear often in the literature.\footnote{The ``overdensity'' is simply the factor by which
the average density in a given volume is higher than the critical density of the Universe at that
redshift, which is the density
it takes for an object to begin to collapse, rather than expand with the Universe.} 
These overdensity radii are defined with respect to the critical 
density of the Universe $\rho_c(z)$ at the cluster's redshift (see Eq.~\ref{eq:rho_c}).
We define $r_\Delta$ as the radius within which the average (gas+dark matter) density 
is $\Delta \, \rho_c(z)$  (e.g. $r_{2500}$ is the radius containing an overdensity 
$2500 \rho_c(z)$, and $r_{500}$ is the radius containing $500 \rho_c(z)$).  
Since the mean cluster density drops with radius, a smaller overdensity $\Delta$ corresponds to
a larger radius.  A good rule of thumb is that $r_{500} \approx 2 \, r_{2500}$ for most massive clusters.
We show how we solve for these radii for real observations in \S \ref{r_delta}.

\section{$\beta$-Model SZE Profiles}

\subsection{The $\beta$-model \label{beta_model}}

The $\beta$-model, also called the King profile, has a long history \citep{cavaliere1976,cavaliere1978} 
of performing remarkably well at parameterizing cluster radial gas density profiles.  Part of its appeal
is its simplicity.  Here I show how to derive the line of sight Compton $y$ parameter 
when a $\beta$-model is used to describe the radial electron pressure profile, 
\begin{equation}
P_e(r) = P_{e,0} \left[1+(r/r_c)^2\right]^{-3\beta/2}.
\label{eq:beta}
\end{equation}
Here $P_{e,0}$ is the central electron pressure, $\beta$ is a single slope that describes how
pressure decreases with radius, and $r_c$ is the ``core radius,'' typically $\sim$ 100 kpc for 
a massive cluster. 
For $\beta=2/3$ -- a typical value when this model is used to describe the intra-cluster
medium's (ICM) density -- 
$r_c$ is the radius at which the electron pressure falls to half its central value.  

\begin{figure}
\centerline{\includegraphics[width=2in]{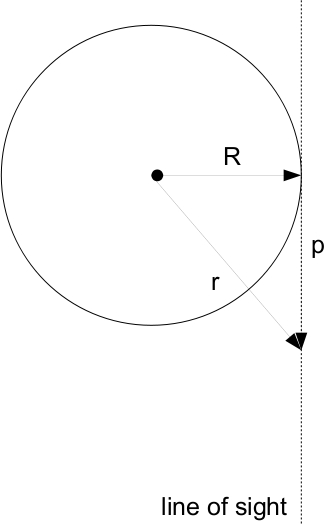}}
\caption{Geometry for the line of sight integral of a spherically-symmetric model.}
\label{fig:beta_calc}
\end{figure}

Figure \ref{fig:beta_calc} shows the geometry for the line of sight integral of a 
spherically-symmetric ICM model.
Here, $\vec{r}$ is the radius within the cluster, $R$ is the 
radius on the sky (from the projected cluster center on the sky), and 
${p}=\sqrt{\vec{r}\,^2-R^2}$ is the distance along the line of sight.  
The radius on the sky, $R$, is a constant in the integration, equal to the minimum 
in $\vec{r}$ (see Figure \ref{fig:beta_calc}). 
The Compton $y$ parameter as a function of $R$ is
\begin{equation}
y(R) = \frac{\sigma_T}{m_e \, c^2}
\int_{-\infty}^\infty 
P_{e,0} \left[1+({r}/r_c)^2\right]^{-3\beta/2}
\frac{{r}}{\sqrt{{r}^{2}-R^2}} \, d{r}.
\label{eq:y_beta}
\end{equation}
Substituting in ${p}$ and $d{p}/d{r} = {r}/\sqrt{{r}^{2}-R^2}$, 
and taking advantage of the symmetry, we have 
\begin{equation}
y(R)  = 2 P_{e,0} \frac{\sigma_T}{m_e \, c^2} \int_{0}^\infty 
\left[1+({p}/r_c)^2+(R/r_c)^2\right]^{-3\beta/2} d{p}.
\label{eq:y_beta2}
\end{equation}
A rearrangement of terms yields
\begin{equation}
y(R) = 2 P_{e,0} \, \frac{\sigma_T}{m_e c^2} \, \left[1+(R/r_c)^2\right]^{-3\beta/2}
\int_{0}^\infty \left\{1+\frac{{p}^{2}}{r_c^2[1+(R/r_c)^2]}\right\}^{-3\beta/2} \!\! dp.
\label{eq:y_beta3}
\end{equation}
Defining the quantity $p_c = \sqrt{r_c^2 + R^2}$, the integral in Eq.\ \ref{eq:y_beta3} 
becomes integrable in terms of the hypergeometric Eulerian function $\Gamma$.
\begin{equation}
\begin{split}
\int_{0}^\infty \left\{1+\frac{{p}^{2}}{r_c^2[1+(R/r_c)^2]}\right\}^{-3\beta/2} \!\! d{p}
= \int_{0}^\infty \! \left[1+(p/p_c)^2\right]^{-3\beta/2} dp \\
= \sqrt{\pi} \, p_c  \, \frac{\Gamma(3\beta/2 -1/2)}{\Gamma(3\beta/2)} \\
= \sqrt{\pi (r_c^2 + R^2)} \, \frac{\Gamma(3\beta/2 -1/2)}{\Gamma(3\beta/2)} \\
= \sqrt{\pi} \, r_c [1 + (R/r_c)^2]^{1/2}  \,\frac{\Gamma(3\beta/2 -1/2)}{\Gamma(3\beta/2)}
\end{split}
\label{eq:beta_integral}
\end{equation}
Putting the result of Eq.\ \ref{eq:beta_integral} back into Eq.\ \ref{eq:y_beta3}, we finally have
\begin{equation}
y(R) = 2 P_{e,0} \, \frac{\sigma_T}{m_e c^2} \, \sqrt{\pi}\, r_c \, [1 + (R/r_c)^2]^{-3\beta/2+1/2} \, 
\frac{\Gamma(3\beta/2 -1/2)}{\Gamma(3\beta/2)}
\label{eq:y_beta_final}
\end{equation}
Note that Eq.~\ref{eq:y_beta_final} has the same form as the integral of the line of sight through 
the center of the cluster.  We therefore can define the central Compton $y$ parameter $y_0$:
\begin{equation}
y_0 = 2 P_{e,0} \, \frac{\sigma_T}{m_e c^2} \, \sqrt{\pi} \, r_c \, \frac{\Gamma(3\beta/2 -1/2)}{\Gamma(3\beta/2)}.
\label{eq:y_0}
\end{equation}
This yields the expression for the SZE $\beta$-model commonly fit to observations:
\begin{equation}
y(R) = y_0\, [1 + (R/r_c)^2]^{-3\beta/2+1/2}.
\label{eq:y_beta_sky}
\end{equation}
In terms of temperature decrement\footnote{Below the null in the SZE spectrum, of course.  Above $\sim$218~GHz, we would 
refer to the ``temperature increment'' instead.  See \S \ref{sze}.} $\Delta T/\Tcmb = f(x)\,y$ (as in Eq.~\ref{eq:deltaT}), 
we also define the central decrement
\begin{equation}
\Delta T_0 \equiv f(x) \, y_0 \, \Tcmb,
\label{eq:dT0}
\end{equation}
where $f(x)$ contains the frequency dependence of the SZE (see Eq.~\ref{eq:fx}). 

The $\beta$-model thus provides a simple, semi-analytic formula for fitting SZE data.

\subsection{The Isothermal $\beta$-model \label{iso_beta_model}}

In \S \ref{beta_model}, we made no assumptions about the temperature, density, or metallicity 
distribution within the ICM; we only assumed the electron pressure can be described by
a $\beta$-profile.  

For the following, we make the common assumption of isothermality.  We will also assume the 
X-ray-determined isothermal spectroscopic temperature $T_X$ is the temperature of the
ICM gas ($T_e(r)=T_X$).  Using Eq.~\ref{eq:beta} and applying the ideal gas law,
\begin{equation}
P_{e} = k_B\,n_{e}\,T_e,
\label{eq:idealgaslaw}
\end{equation}
we arrive at this expression for the electron density:
\begin{equation}
n_e(r) = n_{e,0} \left[1+(r/r_c)^2\right]^{-3\beta/2}.
\label{eq:ne_beta}
\end{equation}

A number of joint SZE+X-ray cluster studies have been performed using the isothermal $\beta$-model 
\citep[see, for example][]{grego1999a, laroque2006, bonamente2008}.
Throughout these studies, the isothermal $\beta$-model has been applied in various ways; 
for simplicity we only discuss the most recent.
In \cite{laroque2006} and \cite{bonamente2008}, the jointly-fit SZE and X-ray data were used to constrain an SZE-determined central 
electron density, $n_{e,0,SZ}$.  This comes from applying Eq.~\ref{eq:idealgaslaw} to the SZE-constrained 
central pressure $P_{e,0}$, and using the X-ray-constrained temperature $T_X$ to solve for the central density
(i.e. They used $n_{e,0,SZ} = P_{e,0} / k_B T_X$). 
The analogous X-ray-determined central electron density, which does not require SZE data, is called $n_{e,0,X}$ 
in this context (and later is just called $n_{e,0}$, as we never use $n_{e,0,SZ}$ in this work).
By not requiring $n_{e,0,SZ}=n_{e,0,X}$ to hold when jointly fitting SZE+X-ray data,
the central decrement (Eq.~\ref{eq:dT0}) can be allowed to fit freely the amplitude of the SZE signal.  
These different normalizations for the electron density profile were used to provide somewhat independent
constraints on the hot gas fraction, using the X-ray-constrained estimate of the total cluster mass with
the X-ray or SZE-constrained gas mass estimates (we show how these quantities are derived from the fit
profiles in \S \ref{mass_deriv}).  An important drawback of this method is that, since X-ray imaging data
have a much higher spatial significance than SZE data, the jointly-fit shape parameters $r_c$ and $\beta$ are
driven to those values preferred by the X-ray data (as we will demonstrate in \S \ref{fit_params}).
Furthermore, it has been well-established by X-ray observations that isothermality is a poor 
approximation both within the cluster core and at large radii 
\citep[see e.g.][and references therein]{piffaretti2005,vikhlinin2005a,pratt2007}).

Since current X-ray imaging data have higher signal to noise than current SZE observations, and X-ray imaging 
has been shown to recover well cluster gas density \citep{nagai2007}, we will instead only consider, for comparison, what the SZE fit
yields for SZE-specific quantities, such as electron pressure and the integrated, intrinsic Compton-$y$ parameter,
\Yint\ (see \S \ref{yint_deriv}).  We use the X-ray-determined density and temperature for all other quantities 
derived from cluster observations that have been fit with the isothermal $\beta$-model, and do not attempt
to constrain electron density with SZE data.


As noted above, the assumption of isothermality does not remain valid over a broad radial extent of a cluster,
even when the core is excluded from the X-ray data (as was done in \cite{laroque2006} and 
\cite{bonamente2008}, for example).  Early joint SZE+X-ray data fitting only allowed the SZE data
to constrain one unique parameter -- $y_0$ -- to within $\sim 10\%$ error bars, while
X-ray data primarily constrained the cluster shape parameters ($\beta$ and $r_c$, used to fit 
both the SZE and X-ray data) and density normalization $n_{e,0}$.\footnote{Using X-ray data alone, temperature
and density can both be determined.  Assuming the ideal gas law, $y_0$ can be computed using only X-ray data. 
However, $y_0$ was allowed to be fit independently by SZE data in earlier works.  In order to assess
the performance of the $\beta$-model as it was applied previously, I did not change how this fit
was performed.}
Since the SZA can integrate to the OVRO/BIMA signal levels
$\sim 8$ times more quickly, the resulting higher quality data of a typical 30-GHz SZA 
observation allows two parameters to be uniquely determined by SZE data.  
An additional benefit of the SZA over OVRO/BIMA is in the shorter baselines, 
due to the close-packed inner array (discussed in \S \ref{sza_overview}).  
The short baselines of the SZA allow us to probe larger scales, where isothermality
is not expected to hold.

Note that we present the results of the isothermal $\beta$-model jointfits for comparison only, and use its limitations
to demonstrate how a new model, presented next, can move beyond the $\beta$-model.

\section{A New SZE Pressure Profile\label{genNFW}}

\subsection{Motivations for a New Pressure Profile\label{genNFW_motivation}}

\begin{figure}
\begin{center}
\includegraphics[width=4.5in]{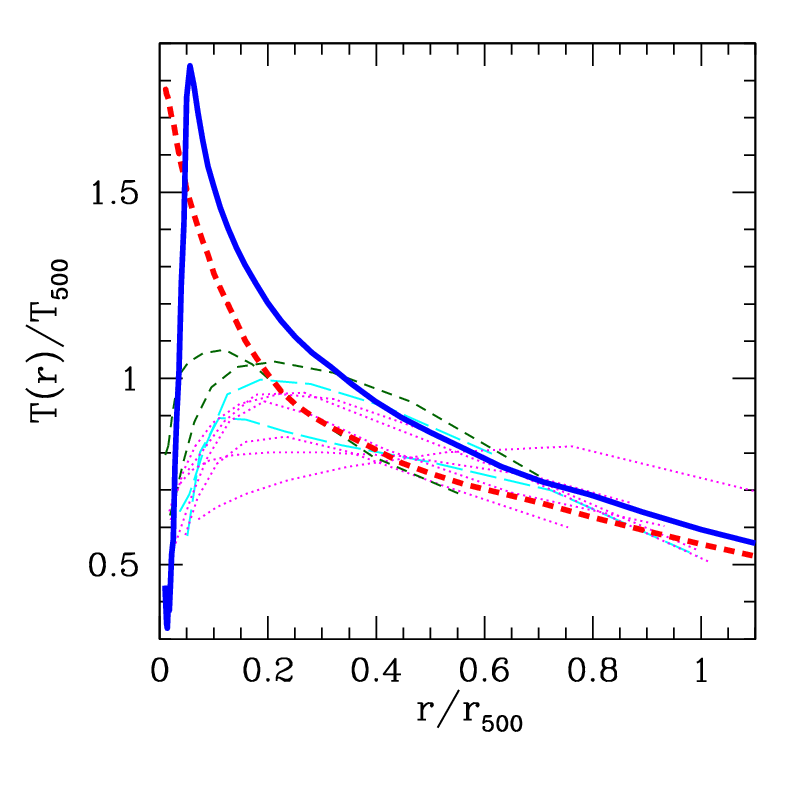}
\caption{The above figure shows fits to the temperature profiles of 11 nearby, relaxed real clusters 
(in green, magenta, and cyan), as well as to the average temperature profiles of 16 clusters 
simulated using adiabatic (red) and cooling + star-formation (blue) physics.  
All profiles are scaled to their best-fit $r_{500}$
values.  V06 used the eight-parameter temperature model (Eq.~\ref{eq:V06_tprof}) 
to capture the details of the real cluster temperature profiles.  Figure from N07.
}\label{fig:n07_temperature}
\end{center}
\end{figure}

We motivate here the use of a new pressure profile, chosen not to have 
shape parameters linked to those used in fitting the X-ray imaging data,
which primarily determine ICM density (note the $n_e^2(r)$ dependence in 
Eq.~\ref{eq:xray_sb} for the X-ray surface brightness implies a strong density 
dependence)\footnote{The X-ray surface brightness in Eq.~\ref{eq:xray_sb} behaves as 
$S_X \propto (1+z)^{-4} \int n_e^2 \, \Lambda_{ee}(T_e,Z) \, d\ell$, where the plasma 
emissivity $\Lambda(T) \propto T^{1/2}$. 
Therefore, the surface brightness is much more sensitive to density than it is to temperature.}.   
By chosing a new pressure profile, we explicitly allow the electron pressure and density 
profiles, $P_e(r)$ and $n_e(r)$, to have different shapes, consistent with a non-isothermal
electron temperature profile $T_e(r)$.
In order to motivate this new pressure profile, first presented in \citet[hereafter N07]{nagai2007b}, 
we examine the empirical evidence afforded by recent, detailed X-ray studies of galaxy clusters.
It is known that the isothermal assumption does not hold over a broad range of cluster radii 
(see, e.g., Figure~\ref{fig:n07_temperature}).  
In this section, we discuss some of the background work that provides the foundations
of the new models.

In looking at clusters over a broad range of masses, we hope to identify whether or not they exhibit 
\emph{self-similarity}, which (if self-similarity holds) implies that they tend toward the same shape or 
slope when scaled to a fiducial radius such as the virial radius or $r_{500}$.
While several works have demonstrated that temperature can exhibit a nearly self-similar decline at large radii 
($r>0.2 \, r_\text{vir}$, see e.g. \citet{loken2002,markevitch1998}), it is also clear in 
these works and  in recent detailed X-ray measurements that cluster temperature profiles are not 
universal or self-similar over all cluster radii (particularly within the cores;  
see \citet{vikhlinin2005a,vikhlinin2006} and Fig.~\ref{fig:n07_temperature}).  

\begin{figure}
\begin{center}
\includegraphics[width=4.5in]{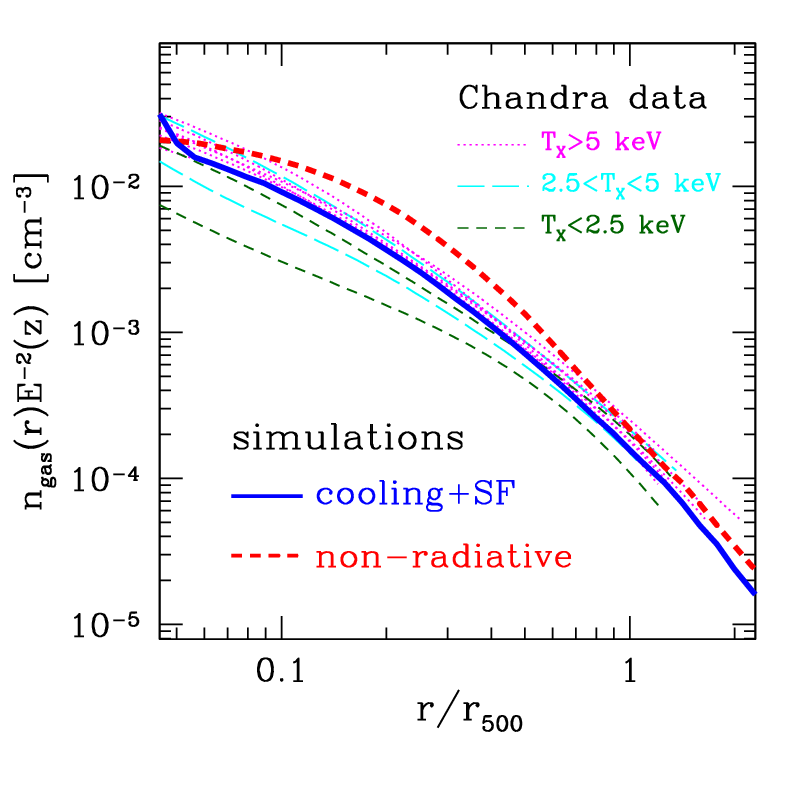}
\caption{The above figure shows fits to the density profiles of 11 nearby, relaxed real clusters, 
as well as to the average density profiles of 16 clusters simulated using adiabatic (red) and 
cooling + star-formation (blue) physics.  
V06 motivate the use of a nine-parameter density model (Eq.~\ref{eq:Vikhlinin}) 
to model these density profiles.  
Figure from N07.
}\label{fig:n07_dens}
\end{center}
\end{figure}

Recently, \citet[][hereafter V06]{vikhlinin2006} demonstrated a new density model, which they
used to fit the surface brightness profiles of 11 nearby, 
relaxed clusters of galaxies with deep \emph{Chandra} X-ray exposures.
The density profile V06 fit is
\begin{equation}
    n_e(r) = \sqrt{\frac{n_{e0}^2 \, (r/r_c)^{-\alpha}}{\left[1+(r/r_c)^2\right]^{3\beta-\alpha/2}}
                \frac{1}{\left[1+(r/r_s)^\gamma\right]^{\varepsilon/\gamma}}
              + \frac{n_{e02}^2}{\left[1+(r/r_{c2})^2\right]^{3\beta_2}}}.
\label{eq:Vikhlinin}
\end{equation}
We refer to this equation as the ``V06 density model.'' 
We note that the $\alpha$ component in the first term of Eq.~\ref{eq:Vikhlinin} was 
introduced by \cite{pratt2002} to fit the inner slope of cuspy cluster density profiles.  
For $\alpha=0$, the first half of the first term is a $\beta$-model.
The second part of the first term -- $\left[1+(r/r_s)^\gamma\right]^{-\varepsilon/\gamma}$ --
accounts for any steepening around $r_{500}$ observed in these clusters.
The additive second term -- $n_{e02}^2\left[1+(r/r_{c2})^2\right]^{-3\beta_2}$ -- is simply 
another, smaller $\beta$-model component, which is also present explicitly to fit 
the cluster core (and is degenerate with the $\alpha$ component of the first term).
Of the nine free parameters in the V06 density model, four are present explicitly 
to fit the inner regions of the cluster ($\alpha,
\beta_2, n_{e02}, \& ~  r_{c2}$).\footnote{V06 used 
this profile with $\varepsilon=3$, so it is not a free parameter.}

V06 coupled Eq.~\ref{eq:Vikhlinin} with a new radial temperature profile, which is
\begin{equation}
  T_{\mathrm{3D}}(r) =  T_0
	\left[\frac{(r/r_{\text{cool}})^{a_{\text{cool}}}+T_{\text{min}}/T_0}
	     {(r/r_{\text{cool}})^{a_{\text{cool}}}+1}\right]
	\left[\frac{(r/r_t)^{-a}}{(1+(r/r_t)^b)^{c/b}}\right].
\label{eq:V06_tprof}
\end{equation}
This profile has eight free parameters. The term in the first set of square brackets describes any cool core present in the cluster,
where the temperature approaches $T_\text{min}$ as $r \rightarrow 0$.  
The term in the second set of brackets describes
the decline in temperature in the cluster outskirts, as $r \rightarrow \infty$.

Figure \ref{fig:n07_dens} compares the density profiles of the 11 real clusters presented in V06
to those fit to mock X-ray observations of the \citet{kravtsov2005} sample of clusters; 
those clusters were simulated separately with adiabatic (`non-radiative') and 
with gas cooling and stellar formation (`cooling+SF' or CSF) feedback physics.
The figure and analysis of the mock X-ray observations are presented in N07 (I simply
use their results here).
Note that outside the core ($r>0.15 \, r_{500}$), the density profiles 
of these relaxed clusters have slopes that are more similar than within $r<0.15 \, r_{500}$. 
This can be seen as the gas density profiles of the real clusters with $T_X>5~\rm keV$ (magenta)
and the CSF simulated clusters tend toward roughly the same slopes beyond $r>0.15 \, r_{500}$.

Figure \ref{fig:n07_temperature} compares the radially-averaged temperature profiles fit to 
\emph{Chandra} X-ray spectroscopic data of these same 11 relaxed clusters from V06.  
It also shows the average temperature profiles of the \citet{kravtsov2005} simulated clusters.
Note in Fig.~\ref{fig:n07_temperature} that the large spike in the temperature profiles 
of the simulated clusters, accompanied by a sudden drop for the CSF simulations, is generally 
accepted as evidence that one or more cluster feedback mechanisms is missing from the 
simulations.  The CSF simulations in particular, which exhibit an exaggerated cusp in density 
(Fig.~\ref{fig:n07_dens}),
suffer from this ``over-cooling problem.''  

\subsection{Generalized NFW Model for ICM Pressure \label{genNFW_model}}

\begin{figure}
\begin{center}
\includegraphics[width=4.5in]{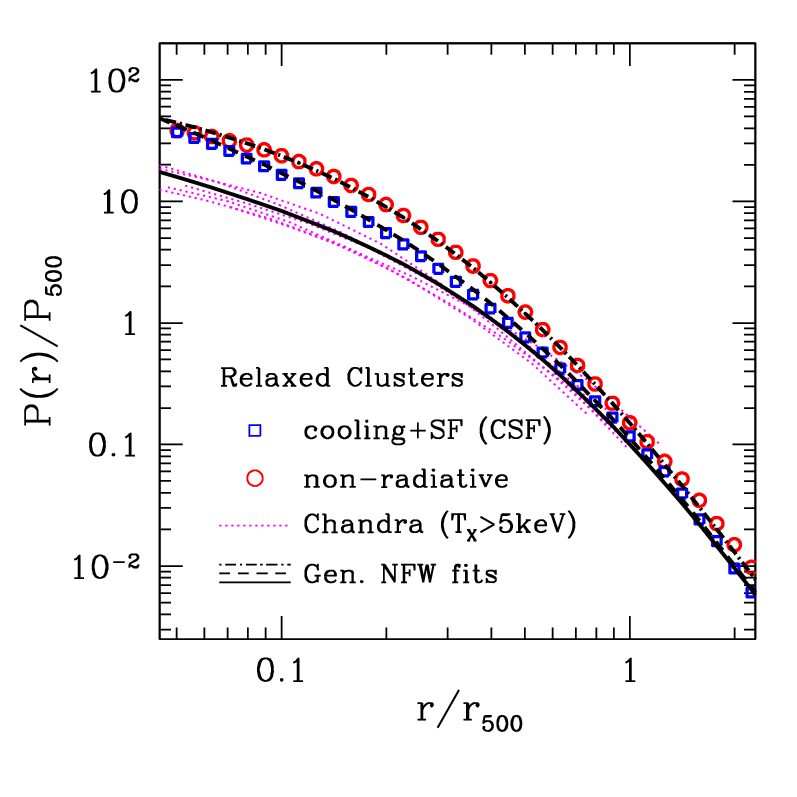}
\caption{The above figure shows the X-ray-derived pressure profiles of 5 nearby, relaxed real clusters
with $T_X > 5~\rm keV$, as well as to the average 3-D, radially-averaged pressure profiles of 16 clusters simulated 
using adiabatic (red) and cooling + star-formation (blue) physics.  
The black lines are the best fit generalized NFW profiles for each type of cluster plotted.
Figure from N07.
}\label{fig:n07_pressure}
\end{center}
\end{figure}

Applying the ideal gas law (Eq.~\ref{eq:idealgaslaw}) to the density and temperature fits presented 
in V06,  N07 derived the ICM pressure profiles of the V06 relaxed clusters.  
Additionally, N07 analyzed the 3-D, radially-averaged pressure profiles of their 
simulated cluster sample (the \cite{kravtsov2005} sample).  N07 found that the pressure profiles of 
both simulated and real clusters can be accurately described by similar generalizations of the 
Navarro, Frenk, and White \cite[][NFW]{navarro1997} model used to fit dark matter halos 
of simulated clusters.
Note that the cuspy core was excluded from the analysis in N07, so the over-cooling in the simulations
does not affect the conclusions they draw about the derived pressure profile.

The generalized NFW profile (abbreviated ``genNFW'' for convenience)
applied to the electron pressure has the form:
\begin{equation}
P_e(r) = \frac{P_{e,i}}{(r/r_p)^c 
\left[1+(r/r_p)^a\right]^{(b-c)/a}}.
\label{eq:press}
\end{equation}
where $P_{e,i}$ is a scalar normalization of the pressure profile,\footnote{$P_{e,i}$
is not a central pressure, as this profile does not become flat in the core, unless
$c = 0$.}
$r_p$ is a scale radius (typically $r_p \approx r_{500}/1.8$), and the
parameters $(a,b,c)$ respectively describe the slopes at intermediate
($r\approx r_p$), outer ($r>r_p$), and inner ($r \ll r_p$) radii;
we refer to $(a,b,c)$ as the genNFW ``slope parameters.''
Note that choosing a pressure profile similar to the dark matter 
halo's density profile is reasonable because the gas pressure distribution is 
primarily determined by the gravitationally dominant dark matter component.  
As we show in \S \ref{yint_deriv}, the integral of thermal pressure (ergs/cm$^3$) can be directly related
to the (thermal) energy content of a cluster.  As the dominant form of kinetic energy in the ICM, 
thermal energy tracks the underlying gravitational potential -- and ultimately the dark
matter halo -- to the extent that the cluster is in pressure equilibrium (i.e. pressure
balances the gravitational pull).  It is therefore
unsurprising that the pressure profiles of relaxed clusters should be self-similar and 
have a profile resembling that of dark matter.

It is worth noting that, to derive pressure from the V06 profiles, seventeen parameters (many degenerate) --
eight for $T(r)$ and nine for $n_e(r)$ -- have been fit, while the X-ray derived pressures
of all clusters in the V06 sample were well-fit by the five parameters of the genNFW profile (on the percent level).
N07 provide best-fit, fixed slopes $(a,b,c)$ for the genNFW profile, which they obtained by
fitting both types of simulated clusters as well as the real clusters with spectroscopic temperatures 
above $5~\rm keV$, presented in V06. 

The implications of this self-similarity in ICM pressure profile can be understood further by
the analysis of cluster dynamical timescales.
Since inhomogeneities in pressure propagate as sound waves in the ICM, the sound-crossing 
time $t_{\rm sc}$ is relevant in determining how rapidly a disturbance in pressure equilibrates.  
\citet{ettori2001} provides the sound-crossing time:
\begin{equation}
t_{\rm sc} = 1.2 \, \left(\frac{10~{\rm keV}}{kT_{\rm gas}}\right)^{\! 1/2} \left(\frac{R}{1~\rm Mpc}\right) \rm Gyrs.
\end{equation}
Here $R$ is the scale of a feature out of pressure equilibrium, and $kT_{\rm gas}$ is the feature's
temperature in keV.  As an example, a hot subcluster with a temperature of 10~keV and size of 0.1~Mpc
merging subsonically with a larger cluster would reach pressure equilibrium with its immediate 
surroundings in only $\sim 0.12~\rm Gyrs$.  This is much less than the age of a typical cluster.
For such a merging cluster not to be near pressure equilibrium, the merger would have to be 
very recent, or the subclump would have to be traveling close to the speed of sound.   

\subsubsection{Some Common Simplifications of the Generalized NFW Model}\label{genNFW_cases}

For different choices of slopes, the genNFW profile reduces to several interesting or familiar cases:

\begin{itemize}

\item \underline{Dark Matter Halo Profiles}: The slopes $(a,b,c) = (1,3,1)$ yield
 the NFW profile for dark matter halos.  Similarly, the genNFW profile reduces to the 
\citet{moore1999} profile by setting $(a,b,c) = (1.5,3,1)$, and to the
\citet{jing2000} profile for galaxy clusters when $(a,b,c) = (1.1,3,1)$.

\item \underline{The $\beta$-model:}: Setting the slopes to $(a,b,c) = (2,3\beta,0)$ 
reproduces the $\beta$-model, where $b = 3 \beta$.  
In this case $P_{e,i}=P_{e,0}$ and $r_p=r_c$.

\item \underline{The N07 Pressure Profile}: The slopes $(a,b,c) = (1.3,4.3,0.7)$ yield the profile we use 
in fitting real clusters of galaxies (see the discussion below, in \S \ref{N07}).

\item \underline{The N07 Pressure Profile for Simulated Clusters}: 
A slightly cuspier inner slope, $c=1.1$, yields the best-fit 
pressure profile clusters simulated with CSF physics.
The full set of slopes is therefore $(a,b,c) = (1.3,4.3,1.1)$.
\end{itemize}

\subsubsection{Slope of the Generalized NFW Pressure Profile ($dP/dr$)}

We provide here the derivative, with respect to $r$, of the genNFW pressure profile.
This will be useful when estimating the cluster total mass, \Mtot, using the equation
of hydrostatic equilibrium (Eq.~\ref{eq:hse}, \S \ref{total_mass}).
\begin{equation}
\frac{dP(r)}{dr} = -\frac{P_{e,i}}{r} 
	  \left(\frac{r}{r_p}\right)^{\! -c} 
	  \left[1 + \left(\frac{r}{r_p}\right)^{\! a}\right]^{-(a+b-c)/a}
	  \left[b \left(\frac{r}{r_p}\right)^{\! a} + c\right].
\label{eq:N07_dPdr}
\end{equation}

\subsection{The N07 Pressure Profile}\label{N07}

In order to reduce the number of free parameters to those that can be constrained by
existing SZE observations of clusters, we fix the slopes of the genNFW profile to the best-fit 
slopes provided by N07, and
test their ability to extract cluster physical parameters in Chapter \ref{model_application}.
By using these fixed slopes to fit the SZE, we take advantage of the self-similarity 
of cluster pressure profiles across a broad range of masses.  
N07 found that the pressure profiles of all sixteen clusters simulated using 
cooling and star formation physics could be well fit with the slopes of 
the profile fixed to the same values (namely $(a,b,c) = (1.3,4.3,1.1)$), while 
the pressure profiles of several relaxed, real clusters, studied in detail using 
{\em Chandra} and presented in V06, could all be fit using a slightly different value 
for the inner slope, $c=0.7$.
Therefore, the pressure profile we test -- Eq.~\ref{eq:press} with slopes fixed at 
$(a,b,c) = (1.3,4.3,0.7)$ -- becomes
\begin{equation}
P_e(r) = \frac{P_{e,i}}{(r/r_p)^{0.7} 
\left[1+(r/r_p)^{1.3}\right]^{3.6/1.3}}.
\label{eq:N07}
\end{equation}
The clusters simulated with non-radiative, adiabatic physics required quite different 
slopes, which we do not consider here, due to the simplified physics used in their simulation.
As demonstrated by N07, we note that the same outer slope used to fit the V06 (real) cluster
and the CSF simulated clusters, $b=4.3$, also fit well the clusters simulated using adiabatic
physics.

\subsection{Simplified Vikhlinin Density Model -- An X-ray Density Model 
to Complement the N07 Pressure Profile}

The density model we chose for fitting X-ray imaging data is a simplified, 
core-cut form of the V06 density model.
We justify this core cut because we are primarily interested in recovering cluster parameters 
at $r_{2500}$ and $r_{500}$, where the contribution from the core is negligible.  
We refer to our simplification as the ``Simplified Vikhlinin Model'' or ``SVM'' hereafter.  
Many have demonstrated \citep[see, e.g.][]{laroque2006} that a core-cut isothermal
$\beta$-model recovers \Mgas\ accurately outside the inner 100~kpc, out to $r_{2500}$.  
We chose this simple modification as a suitable description of intermediate-quality 
X-ray data outside the core, out to, and possibly beyond, $r_{500}$.  We note
that V06 presented this profile as an intermediate step to obtaining 
the full 9-parameter density model (Eq.~\ref{eq:Vikhlinin}), but never used the SVM
to fit cluster data.

We begin with the standard $\beta$-model.  Following V06, we add a scale radius $r_s$ 
at which the density starts to decline more rapidly, with a slope $\varepsilon$.
The SVM is therefore
\begin{equation}
\rho(r) = \rho_0 \left[1+(r/r_c)^2\right]^{-3\beta/2}
\left[1+(r/r_s)^3\right]^{-\varepsilon/6}.
\label{eq:svm}
\end{equation}
The two extra degrees of freedom, in $r_s$ and $\varepsilon$, attempt to overcome the 
limitations of the $\beta$-model in fitting a cluster out to $r_{500}$ 
\citep[for more detail, see][]{mroczkowski2009}.  Several authors 
\citep[see e.g][]{neumann2006,vikhlinin2006,maughan2007} have noted the $\beta$-model
is insufficient for describing clusters at $r \gtrsim r_{500}$, where their density profiles 
of steepen.
Thus we have developed the N07+SVM pair of models -- Eqs.~\ref{eq:N07} and \ref{eq:svm} -- 
primarily to fit clusters used in determing the SZE+X-ray bulk cluster scaling relations.

\begin{figure}
\begin{center}
\includegraphics[width=5in]{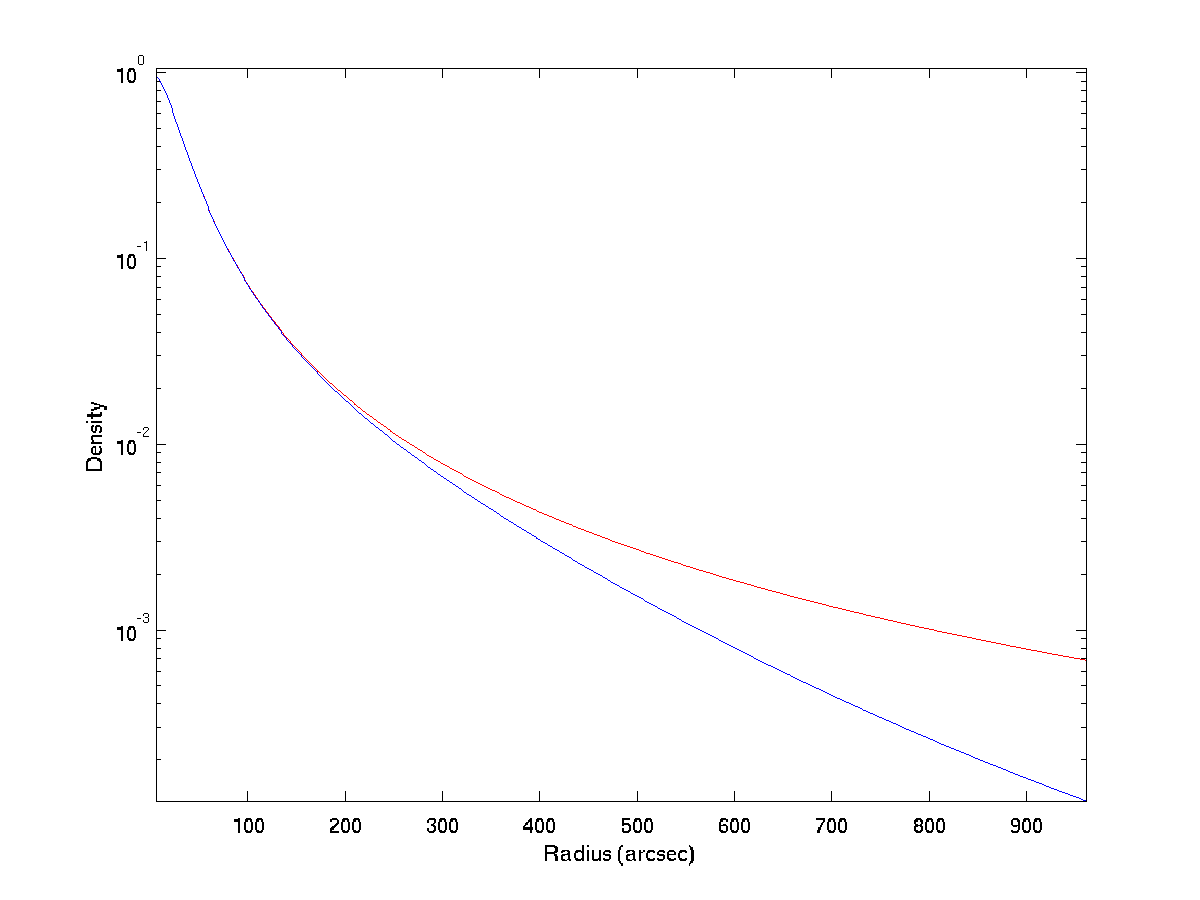}
\caption{This figure shows the density of the standard $\beta$-model (red) and SVM 
(blue) using the steepest outer slope V06 allowed, $\varepsilon=5$.  This was
chosen to illustrate the largest disparity the radially-average outer cluster slope 
V06 considered ``realistic.''
We chose typical parameters for the components common to the two profiles (namely, we chose 
$r_c=100\arcsec$ and $\beta=0.7$, and set the normalization 
$\rho_0 = 1$ for simplicity).  For the SVM, we show
a typical massive cluster's scale radius for the slope to steepen, $r_s=500\arcsec$.
}
\label{fig:vikh_dens}
\end{center}
\end{figure}

The SVM simplifies to a $\beta$-model when $\varepsilon=0$.
Figure \ref{fig:vikh_dens} shows the standard $\beta$-model (red) and SVM 
(blue) using $\varepsilon=5$, the steepest outer slope V06 allowed in their density fits.
Here, we chose typical parameters, found in cluster fits, for the components common to the 
two profiles (namely, we chose $r_c=100\arcsec$ and $\beta=0.7$, and set the normalization 
$\rho_0 = 1$ for simplicity).  For the SVM, we chose
a scale radius of $r_s=500\arcsec$, corresponding to the radius at which the density of 
a typical cluster at low redshift begins to steepen.
Note that since X-ray surface brightness scales as $\rho_{\rm gas}^2$, the signal will
be $<0.1\%$ that of the core for the regions where the profiles diverge ($r>200\arcsec$).

\subsection{Combining the N07 and SVM Profiles}\label{n07+svm}

We apply the N07 pressure profile (Eq.~\ref{eq:N07}) to fit the SZA data presented in 
Chapter~\ref{model_application}, combining it with the SVM (Eq.~\ref{eq:svm}) to fit 
complementary \emph{Chandra} X-ray imaging data. 
As discussed in \S \ref{xray}, fitting X-ray surface brightness (Eq.~\ref{eq:xray_sb}) requires
both density and temperature information.
Having assumed the ideal gas law (Eq.~\ref{eq:idealgaslaw}), the electron temperature profile 
$T_e(r)$ occupies a unique position while jointly fitting the N07+SVM profiles to X-ray+SZE 
cluster observations: it can be derived from the pressure and density fits without relying 
on X-ray spectroscopy.
For each iteration of the MCMC fitting routine, the temperature profile necessary to fit the X-ray surface 
brightness is derived for each set of test parameters of the density and pressure profiles.
The weak temperature dependence of the X-ray imaging data (discussed in \S \ref{xray}) has the 
ability to exclude some of the pressure model's parameters that fit well the SZE data alone, 
as will be shown in Chapter~\ref{model_application}.

\section{Markov chain Monte Carlo Analysis}\label{markov}

A Markov chain Monte Carlo (MCMC) method, implemented in a 
program called \emph{markov}, is used to determine the distribution of models that fit the data, 
as described by \citet{bonamente2004,laroque2006,bonamente2006,bonamente2008}. 
Iterations are accepted in the MCMC according to the Metropolis-Hastings algorithm, which
ensures that a fully-converged MCMC reflects the true probability distribution
of the resulting model fits. Assuming the model itself as a prior, the 
output chain of parameters -- the ``Markov chain'' itself -- can be used to determine the 
probability density of the model fits.

The SVM has five free parameters to describe the gas density (see Eq.~\ref{eq:svm}),
and the N07 profile has two free parameters to describe the electron pressure 
(see Eq.~\ref{eq:N07}).  Additional parameters such as the
cluster centroid,\footnote{Since they are separate observations, and the coordinates used in fitting
are simply in offset from the pointing center of the observations, the SZE and X-ray centroids are 
fit independently.  They are, however, later checked for consistency.} the X-ray background level, 
and the positions, fluxes, and spectral indices of field radio point sources are also included
as necessary.

The model fitting code was originally called {\em jointfit}, and used a downhill simplex method 
\citep[see, for example,][and references therein]{bonamente2004,joy2001,grego2000,grego2001,
laroque2003,patel2000,reese2000,reese2002}.  {\em markov} and {\em jointfit} have been used 
extensively to fit {\em Chandra} and {\em ROSAT} X-ray observations jointly with 
SZE observations made by OVRO and BIMA.  
Because the UVFITS files produced from SZA data contain sixteen bands, as opposed to 
the single-banded UVFITS files from OVRO/BIMA,\footnote{OVRO had two bands, but each 
band was output as a separate UVFITS file.} {\em markov}'s UVFITS reader required
an upgrade in order to read and properly account for the frequency of each band.  Furthermore,
in the transition from 32-bit to 64-bit desktop processors, the standard CFITSIO
libraries did not work, and I upgraded the reader to the well-documented UVFITS
reader Martin Shepherd wrote for \emph{Difmap} \citep{shepherd1997}. 

For each iteration in the MCMC process, model test parameters are generated by stepping
randomly from the last accepted link in the Markov chain.  The size of the step is determined
using a top hat distribution of adjustable width.
These test parameters are used to compute the model image over
a regular grid, with sampling defined to be less than half the
smallest scale the SZA can probe (i.e. Nyquist sampling of the scale probed by the longest
baseline in the observation).  This image is multiplied by the 
primary beam of the SZA, transformed via FFT to Fourier space (which is where the data 
are directly sampled by a radio interferometer), and interpolated to the Fourier-space
coordinates of the SZE data.  The likelihood calculation for the SZE data is
then performed directly in Fourier space, where the noise properties of 
the interferometric data are well-characterized.

The MCMC routine determines which trial sets of parameters are accepted into the 
Markov chain by computing the joint likelihood $\mathcal{L}$ of the models' fits 
to the X-ray and SZE data. The SZE likelihood is given by
\begin{equation}
\ln(\mathcal{L}_\text{SZE})=\displaystyle \sum_i \left[-\frac{1}{2} \left(\Delta R_i^2+\Delta I_i^2\right)\right] W_i
\end{equation}
where $\Delta R_i$ and $\Delta I_i$ are the difference between model
and data for the real and imaginary components at each point $i$ in
the Fourier plane, and $W_i$ is a measure of the Gaussian noise
($1/\sigma^2$).
Since the X-ray counts, treated in image space, are distributed
according to Poisson statistics, the likelihood of the surface brightness model fit is given by
\begin{equation}
\ln(\mathcal{L}_\text{Xray})=\displaystyle \sum_i \left[ D_i \ln(M_i) - M_i -\ln(D_i!) \right]
\label{eq:xray_L}
\end{equation}
where $M_i$ is the model prediction (including cluster and background
components), and $D_i$ is the number of counts detected in pixel
$i$. 
Spatial regions of the X-ray image, containing for example the cluster core, gas clumps, or 
X-ray point sources, can be excluded from the X-ray analysis simply by excluding data in those regions
from the X-ray likelihood calculation (Eq.~\ref{eq:xray_L}).\footnote{An alternative approach
might involve, for example, interpolation past the region of the X-ray image that we wish to exclude.  
We choose not to make any such assumptions about the cluster in the excluded regions.}

The joint likelihood of the X-ray and SZE models is given by
$\mathcal{L}=\mathcal{L}_\text{SZE}\cdot\mathcal{L}_\text{Xray}$.  
For the models that include X-ray spectroscopic temperature constraints, 
the likelihood is $\mathcal{L}=\mathcal{L}_\text{SZE}\cdot\mathcal{L}_\text{Xray}\cdot\mathcal{L}_\text{XSPEC}$,
where $\mathcal{L}_\text{XSPEC}$ comes from a table of $\chi^2$ values computed using
a plasma emissivity model fit to the X-ray spectroscopic data, using the software package XSPEC \citep{arnaud1996}.
The relationship between $\chi^2$ and $\mathcal{L}_\text{XSPEC}$ is
\begin{equation}
\mathcal{L}_\text{XSPEC} = \exp\left(-\frac{\chi^2}{2}\right).
\end{equation}

Throughout a {\em markov} run, the probability distribution of fits to the data 
is explored. Some number of iterations at the beginning of the Markov chain, however, will depend on the 
starting parameters {\em markov} chooses;  We refer to this as the ``burn-in period'' 
-- when the MCMC still depends on the starting conditions -- and ignore iterations accepted during this period.
After a sufficiently large number of total iterations, the allowed range of parameters 
is fully explored. The number of required iterations depends on the number of free
parameters, the size of the top-hat distribution out of which the test parameters are chosen, and the noise 
in the data. This is why we use convergence tests, described below, to determine the total number of required 
iterations.

The likeliest parameters will appear multiple times in the 
Markov chain.  Their distribution after the initial burn-in period is 
referred to as the ``stationary distribution,'' since further iterations do not
yield better constraints on the model.
We follow the methods outlined in \cite{bonamente2004}, first exploring large
regions of parameter space with test runs to ensure the solution is not stuck
in a local minimum.  Further runs are used to narrow in on the likeliest parameters
by limiting the parameter range to explore and tuning the top hat distribution width.
As in \cite{bonamente2004}, we use the Raftery-Lewis test to determine convergence
to a stationary distribution \citep{raftery1992,best1995,gilks1996,plummer2006}, 
This is an efficient way to determine how many iterations the initial burn-in 
should contain, and how many further iterations are required to reach the stationary 
distribution.
Once the MCMC routine has found the stationary distribution, we use the set
of accepted model parameters to derive cluster parameters such as \Mtot\ and their error bars.

\section{Mass analysis \label{mass_deriv}}

\subsection{Weighting factors for the cluster gas\label{mus}}

The first quantity we consider is the cluster gas density.  X-ray spectroscopic data have the ability to constrain 
the metallicity $Z$ of a gas (the mass fraction of elements heavier than helium) by fixing the abundance ratios to 
those observed in the solar atmosphere and fitting for the few prominent emission lines --notably 
those from iron -- that can be found in X-ray spectra. 
Since the lighter elements (e.g. hydrogen, helium) are fully collisionally-ionized 
at cluster temperatures ($\gtrsim 1 {\rm keV}$), no emission lines are observed from these,
and their abundances are assumed to be those produced by Big Bang Nucleosynthesis (a.k.a. ``cosmic abundance''), 
plus the amount produced by stellar nucleosynthesis, for the X-ray spectral fit to the metals.

When comparing the SZE-derived temperature (see \S~\ref{n07+svm}) to the X-ray spectroscopic 
temperature, we use the program XSPEC \citep{arnaud1996} to fit a plasma emissivity 
model to the X-ray data.  
The plasma spectroscopic emission models we most commonly fit to these data are 
the Raymond-Smith \citep{raymond1977} and MEKAL \citep{mewe1985,kaastra1993,liedahl1995} models, which
can have free parameters of temperature, redshift, and metallicity (redshift is often fixed to that
found using complementary optical observations, though typical X-ray data can
constrain a cluster's redshift to better than $\pm 1\%$).
Fitting simulated cluster emission with X-ray spectra generated using the MEKAL model, we found that the 
metallicity recovered in XSPEC from fits of the Raymond-Smith model was within $\gtrsim 80\%$ accuracy 
of the metallicity in the input MEKAL spectrum. 

Using the fit metallicity, we tested the X-ray surface brightness fitting procedure -- which relies
on the Raymond-Smith model when fitting X-ray emissivity -- against mock X-ray observations of 
these simulated clusters.  
While these mock X-ray exposures were generated using the MEKAL model, 
the emissivity $\Lambda_{ee}(T_e,Z)$ used to fit the mock observations (in Eq.~\ref{eq:xray_sb}) was 
computed using the Raymond-Smith model.  The recovered gas density was accurate to better than 98\%.  
We conclude that the continuum emission in an X-ray image is insensitive to the plasma 
emissivity model assumed in the fit, and continue to use the Raymond-Smith model to fit 
X-ray surface brightness.

\begin{table}[t]
 \centerline{
  \begin{tabular}{|c|ccc|}
   \hline
   Element & Mass Fraction & Number of Species & Mass \\
   \hline
   H  & $X = n_{\rm H} / (n_{\rm H} + 4 n_{\rm He})$    & $1 e^- + 1 p^+$         & $1 m_p$ \\
   He & $Y = 4 n_{\rm He} / (n_{\rm H} + 4 n_{\rm He})$ & $2 e^- + 1 \alpha^{2+}$ & $4 m_p$ \\
   \hline
  \end{tabular}
 }
\caption{Common parameters used to compute the weighting factors for a pure H and He plasma.}
\label{mu_table}
\end{table}

After determining the metallicity of the ICM (for either real or mock observations), several useful ``weighting factors'' 
can be determined by carefully accounting for the species of which the gas is composed.  
The weight per particle $\mu$ is determined by summing the masses of all the species 
(i.e. the electrons, protons, and atomic nuclei) in the plasma and dividing by one proton mass 
per particle.  
\begin{equation}
\mu = \frac{n_{\rm H}\times1 m_p + n_{\rm He} \times4 m_p}
{\left[ n_{\rm H}(1 e^- + 1 p^+) + n_{\rm He}(1 \alpha^{2+}+2 e^-)\right] m_p}.
\label{eq:mu}
\end{equation}
For example, a fully-ionized plasma consisting of 10\% helium and 90\% hydrogen
by number yields 
\begin{equation}
\mu = \frac{0.9 + 0.1\times4}{0.9(1 e^- + 1 p^+) +0.1(1 \alpha^{2+}+2 e^-)} = 0.62,
\label{eq:mu_example}
\end{equation}
where we use the nomenclature $e^-$ for an electron, $p^+$ for a proton, and $\alpha^{2+}$
for a helium nucleus, and we have assumed a hydrogen atom is exactly one proton mass $m_p$
and a helium atom is $4 m_p$ (see Table~\ref{mu_table}).

Similarly, we can define $\mu_e$, the weight per electron, by summing the masses of the species
and dividing by $1 m_p$ for each electron. 
\begin{equation}
\mu_e = \frac{n_{\rm H}\times1 m_p + n_{\rm He}\times4 m_p}
{\left[n_{\rm H}(1 e^-)  +n_{\rm He}(2 e^-)\right] m_p},
\label{eq:mue}
\end{equation}
For the example in Eq.~\ref{eq:mu_example}, this means
\begin{equation}
\mu_e = \frac{0.9 + 0.1\times4}{0.9(1 e^-)  + 0.1(2 e^-)} = 1.18,
\label{eq:mue_example}
\end{equation}

In practice, the detailed calculation of weighting factors requires accounting for all the species
in the gas. Having fit spectroscopically a metallicity $Z$, we use the abundance calculation of 
\citet{anders1989} to determine the relevant weighting factors.  We assume this abundance model
for all results presented in this thesis (though we consider the effects of a non-universal, non-constant
$Y/X$ ratio in \S \ref{Tsl_extension}).


\subsection{Relating the fit $n_e(r)$ and $P_e(r)$ to ICM gas properties\label{dens_press}}

The quantities $\mu_e$ can now be used to relate the electron number density 
profile $n_e(r)$ to the gas (mass) density $\rho_{\rm gas}(r)$, which is
\begin{equation}
\rho_{\rm gas}(r) = \mu_e m_p n_e(r).
\label{eq:gasdens}
\end{equation}
Similarly, the number density of all species in the gas $n(r)$ relates to the gas density through $\mu$:
\begin{equation}
\rho_{\rm gas}(r) = \mu m_p n(r).
\label{eq:gasdens2}
\end{equation}
Assuming the electron temperature equals the gas temperature ($T_e = T_{\rm gas}$),\footnote{This is because $\mu$ and $\mu_e$ 
are mass weighting factors, so we have to relate pressure to density to take advantage of them.
Hence we have $\mu/\mu_e = n_e(r)/n(r) = P_e(r)/P_{\rm gas}(r)$.} 
we combine Eqs.~\ref{eq:gasdens} \& \ref{eq:gasdens2} and use the ideal gas law to find
the electron pressure profile relates to the overall gas pressure via
\begin{equation}
P_{\rm gas}(r) = \left(\frac{\mu_e}{\mu}\right) P_e(r).
\label{eq:gaspress}
\end{equation}


\subsection{Mass of the X-ray-Emitting ICM Gas (\Mgas) \label{gas_mass}}

For a spherically-symmetric density profile, we now simply integrate  $\rho_{\rm gas}(r)$ to obtain
the gas mass \Mgas\ within a spherical volume with radius $r$: 
\begin{equation}
\Mgas(r) = 4 \pi \int_{0}^{r} \! \rho_{\rm gas}(r') r'^2 dr'.
\label{eq:gasmass}
\end{equation}

\subsection{The Total Mass of the Cluster (\Mtot) \label{total_mass}}

The total gravitational mass within a spherically-symmetric system in pressure equilibrium 
can be found by considering the balance between the inward gravitational force $f_{in}$ and the
outward pressure $f_{out}$.  This scenario of pressure support and negligible net mass transport
across the surface of the sphere is known as hydrostatic equilibrium (HSE).

Consider the forces on a spherical shell of thickness $dr$, at radius $r$, with gas mass 
$m = 4 \pi r^2 \, \rho_{\rm gas}(r) \, dr$. The inward force of gravity is
\begin{equation}
f_{in} = -\frac{G \, m \, \Mtot(r)}{r^2} =   -\frac{G \, 4 \pi r^2 \, \rho_{\rm gas}(r) \, \Mtot(r) dr}{r^2} 
\label{eq:f_in}
\end{equation}
where $G$ is Newton's gravitational constant and \Mtot(r) is the total gravitating 
mass within radius $r$.
The outward force of pressure is
\begin{equation}
f_{out} = 4 \pi r^2 \left(\frac{dP_{\rm gas}}{dr}\right) dr  
\label{eq:f_out}
\end{equation}
Setting $f_{in} = f_{out}$ and solving for \Mtot(r) yields
\begin{equation}
\Mtot(r) = - \frac{r^2}{G \rho_{\rm gas}(r)} \left(\frac{dP_{\rm gas}}{dr}\right) 
\label{eq:hse}
\end{equation}

Notice that this is a powerful way of using the observable gas distribution in a cluster of galaxies
to estimate its total matter content (i.e. dark + baryonic).  
By assuming further the pressure is entirely thermal\footnote{
Turbulence and magnetic fields are possible sources of additional, non-thermal pressure support.
The standard HSE total mass estimate could, for example, be systematically low by $\sim$5-20\% 
because it does not account for subsonic motions of the ICM, which can contribute to the total ICM pressure
but will not affect the temperature and density (see e.g. \cite{nagai2006} for details).  Additionally,
there are indications that cosmic rays could contribute to radiative support (see e.g. \cite{pfrommer2007}).} and that the ideal
gas law holds, one only has to determine two of the three quantities \{$P_{\rm gas}(r)$, $\rho_{\rm gas}(r)$, 
and $T(r)$\} in order to estimate a cluster's total mass.

\subsection{Hot Gas Mass Fraction (\fgas) \label{fgas}}

For each accepted iteration in the MCMC, we have a set of model parameters fit
to the available SZE+X-ray data, from which we can compute \Mgas\ and \Mtot.  
Solving for \Mgas\ and \Mtot\ using Eqs.~\ref{eq:gasmass} \& \ref{eq:hse}, we 
simply solve for the profile of the gas mass fraction individually
for each set of MCMC parameters:
\begin{equation}
\langle \fgas(r) \rangle = \langle \Mgas(r)/\Mtot(r) \rangle.
\label{eq:fgas}
\end{equation}
Doing this, the mean gas fractions we present in this thesis take advantage of the fact that
MCMC explores the probability density of model fits to the data.  Furthermore,
the confidence intervals presented also directly reflect the range of models fit (rather
than assuming the derived \Mgas\ and \Mtot\ are statistically independent).


\subsection{Overdensity Radius \label{r_delta}}

In order to compute global cluster properties such as \Mgas, \Mtot, and \fgas, one needs to
define a radius out to which all quantities will be calculated.
Following \citet{laroque2006,bonamente2006}, we compute global
properties of clusters enclosed within the overdensity radius $r_{\Delta}$,
within which the average density of the cluster is a specified
fraction $\Delta$ of the critical density, via
\begin{equation}
\frac{4}{3} \pi \, \rho_c(z) \, \Delta \, r_{\Delta}^3 = \Mtot(r_{\Delta}),
\label{eq:r_delta}
\end{equation}
where $\rho_c(z)$ is the critical density of the Universe at redshift
$z$, and is computed (with respect to $\rho_c= 3 H_0^2/8 \pi G$, the critical density at redshift
$z=0$):
\begin{equation}
\rho_c(z) = \rho_c \, [\Omega_M (1+z)^3 + \Omega_\Lambda].
\label{eq:rho_c}
\end{equation}

Throughout this work, we evaluate cluster properties at density
contrasts of $\Delta=2500$ and $\Delta=500$, corresponding to average
densities of 2500 and 500 times the critical density at the redshift
of the cluster.  The overdensity radius $r_{2500}$ has often been used
in previous SZE+X-ray studies, since that was the maximum radius
attainable in many X-ray observations of intermediate redshift clusters 
(see, e.g. \cite{laroque2006}).  The overdensity $r_{500}$ is now
reachable in deep SZA and \emph{Chandra} X-ray data,
without extrapolating the model beyond the image fitting region
or into regions with signal-to-noise ratio $S/N \lesssim 1$.

\section{SZE-Specific Quantities \label{sz_quants}}

\subsection{The Integrated, Intrinsic Compton $y$ Parameter (\Yint) \label{yint_deriv}}

Since the SZE signal arises via Compton scattering, the integrated 
SZ effect, when scaled to the distance of the cluster, can be converted to \Yint, 
the intrinsic, integrated Compton $y$ parameter (see e.g.~\cite{dasilva2004, nagai2006}).
The intrinsic line-of-sight (abbreviated ``los'' in the subscripts to follow) 
integrated Compton $y$ parameter $\Ylos$ is obtained by integrating $y$ 
(Eq.~\ref{eq:compy}) over the solid angle $\Omega$ subtended by the cluster, 
and scaling the area for the angular diameter distance of the cluster,
\begin{equation}
\Ylos \equiv {D_A^2} \int_{\Omega} y~d\Omega = \frac{\sigma_T}{m_e \, c^2} 
\int_{-\infty}^{\infty} \!\! d\ell \int_A \! P_e \, dA. \\
\label{eq:Yint}
\end{equation}
Here $A$ is the area of the cluster in the plane of the sky.  

In addition, SZE observations constrain the intrinsic volumetric (abbreviated 
``vol'' in  the subscripts to follow) integrated Compton $y$ parameter
$\Yvol$, obtained by integrating $y$ within a spherical volume of radius $r$:
{\begin{equation}
\Yvol = \frac{4 \pi \,\sigma_T}{m_e \, c^2} \int_{0}^{r} \!\! P_e (r') \, r'^2 dr. \\
\label{eq:Yvol}
\end{equation}

Since \Yint\ -- in the form of either \Yvol\ and \Ylos -- scales as pressure (ergs/cm$^3$) 
integrated over volume (cm$^3$), it is proportional to the thermal energy content
of the ICM within that volume.  
As the dominant form of kinetic energy in the ICM, thermal energy tracks the 
underlying gravitational potential, and ultimately the dark matter halo of the cluster, to
the extent that HSE holds. SZE flux thereby provides a robust, low scatter proxy, 
\Yint, for total cluster mass, \Mtot.

\subsection{The Total Thermal Energy Content: \newline An SZE-only Scaling Quantity \label{thermal_energy}}

A problem becomes apparent when one considers how the N07 pressure profile -- or any profile fit to 
SZE data alone -- could be used in an SZE survey to constrain cosmology.  
To maximize the utility of a cluster survey, it is crucial to 
understand how to relate a cluster's observable properties to its total mass.

When using joint SZE+X-ray data, one has the luxury of solving for \Mtot\ by assuming
HSE (Eq.\ \ref{eq:hse}).  Joint SZE+X-ray fits therefore allow one to solve for $r_\Delta$, the radius that 
encloses matter overdensity $\Delta$.  We can then compute \Mtot$(r_\Delta)$ and \Yint$(r_\Delta)$ (e.g. 
either \Ylos$(r_\Delta)$ or \Yvol$(r_\Delta)$, provided in Eqs.\ \ref{eq:Yint} \& \ref{eq:Yvol}), which, for example,
could be used to establish the \Yint$(r_\Delta)$--\Mtot$(r_\Delta)$ scaling relation.\footnote{One could also use the derived
$r_\Delta$ to obtain \Mtot$(r_\Delta)$ from a previously-established scaling relation, though
if one has $r_\Delta$, one necessarily has \Mtot$(r_\Delta)$ already.  This is simply because
$\Mtot(r_\Delta)=\Delta \rho_c r_\Delta^3$, where $\rho_c$ is the critical density of the Universe 
at that redshift.}

For an SZE-detected cluster with no complementary X-ray observation (or some independent
estimate of \Mtot, such as from lensing or galaxy velocity dispersion), the problem arises 
that we do not have a way to determine $r_\Delta$, and therefore do not know 
within which radius to compute \Yint$(r_\Delta)$.
There is no direct way to find $r_\Delta$ from an SZE-only observation of a cluster,
since we do not know $\Mtot(r)$ (see Eq.~\ref{eq:r_delta}).  
Unlike the isothermal $\beta$-model, the N07 pressure profile can be integrated over an infinite
volume, yielding a finite estimate for the total thermal energy content of the cluster. 
The thermal energy content should scale directly with the total mass of a cluster -- to the 
extent that thermal pressure supports the cluster against gravitational collapse --
and has been studied before by others \citep[e.g.][]{afshordi2007}.
The total thermal energy content $E_\text{tot}$ of a cluster that can be described by 
the spherically-symmetric pressure profile $P_{\rm gas}(r)$ is 
\begin{equation}
E_\text{tot} = \frac{3}{2} \int_0^\infty \!\! 4 \pi P_{\rm gas}(r) r^2 dr.
\label{eq:thermalEnergy}
\end{equation}
Using Eq.~\ref{eq:gaspress}, where the SZE-constrained electron pressure profile is 
parameterized in the form of the generalized NFW pressure profile (Eq.~\ref{eq:press}), 
a tool such as \emph{Mathematica$^\circledR$}\footnote{\url{http://www.wolfram.com}} 
can handily solve this integral.\footnote{Physically, this integral should not
be carried out to infinity.  However, the generalized NFW profile drops rapidly enough
that any contribution from outside the virial radius is negligible.}
For $a\!>\!0$, $b\!>\!3$, and $c\!<\!3$, the integral converges, yielding
\begin{equation}
E_\text{tot,genNFW} = 6\pi r_p^3 \, P_{e,i} \, \left(\frac{\mu_e}{\mu}\right)
\frac{\Gamma(\frac{b-3}{a}) \, \Gamma(\frac{3-c}{a})}{a \, \Gamma(\frac{b-c}{a})}.
\label{eq:thermalEnergy_genNFW}
\end{equation}
Here $\Gamma$ is the Euler gamma function.  
For fixed slopes $(a,b,c)$, the integral is only dependent upon profile parameters $P_{e,i}$ 
and $r_p$, and scales as \Yvol.
The parameters recommended by Daisuke Nagai (personal communication) for his simulated clusters 
all had $a \in [0.5,2]$, $c \in [0.5,1.5]$, and $b \in [4,5]$, which covers a range consistent 
with a finite thermal energy content.  The N07 profile provides $(a,b,c)=(1.3,4.3,0.7)$,\footnote{
These are the original values for $(a,b,c)$, published in N07.  These were updated after 
the completion of this work, however, and will be published in an erratum to N07.  
Fits utilizing the newer values are presented in \citet{mroczkowski2009}.}
which also yields finite thermal energy.
In Figure \ref{fig:A1835_thermalE}, we show the N07 profile predicts a thermal 
energy content within a given radius that converges to a finite total value.

\begin{figure}
\centerline{\includegraphics[width=5.25in]{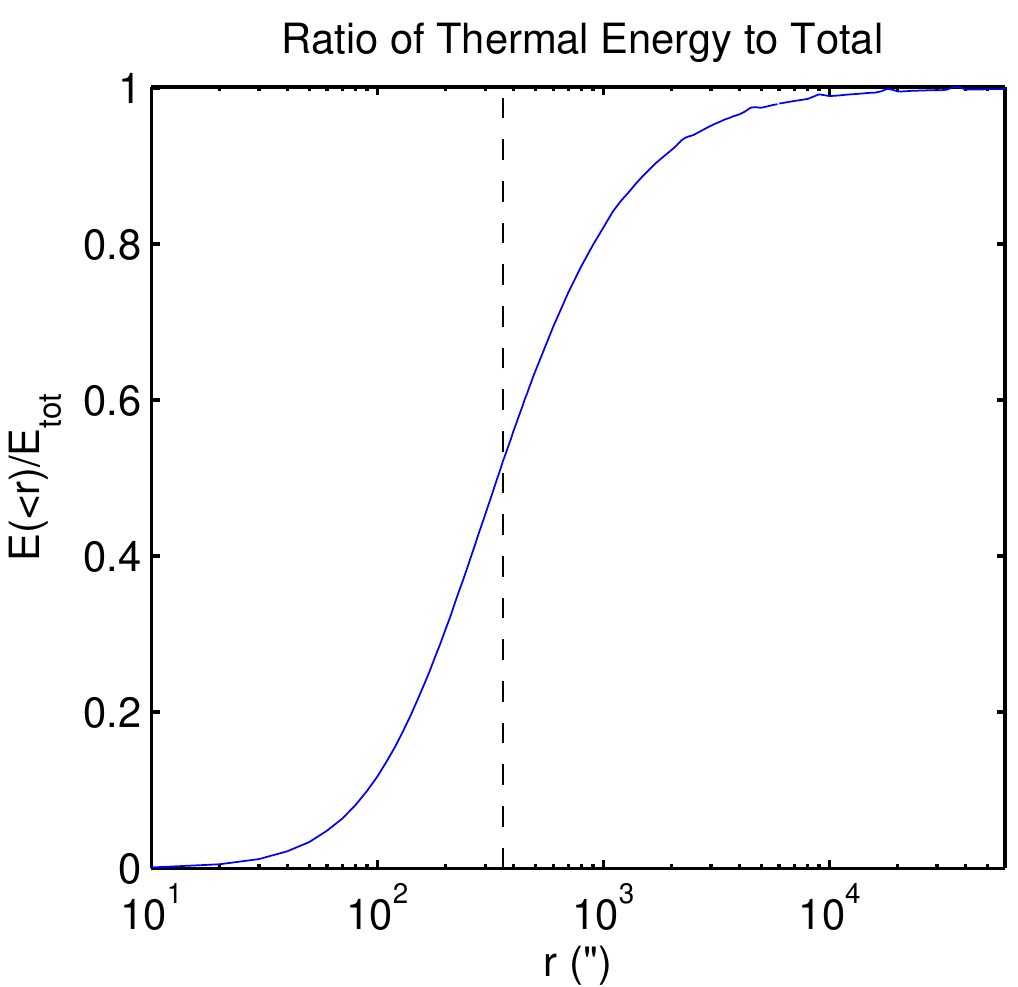}}
\caption{The ratio of the thermal energy content within a given radius to the total as $r\rightarrow\infty$, plotted
for the fit to A1835 of the N07 pressure profile.  The vertical line shows $r_{500} \simeq 360\arcsec
\simeq 1.4~{\rm Mpc}$.}
\label{fig:A1835_thermalE}
\end{figure}

The fact that the integral for the ICM thermal energy (Eq.~\ref{eq:thermalEnergy}) converges implies there is a single,
physical quantity that can be constrained by SZE data alone, without having to know $r_\Delta$
(which would otherwise have to be determined using independent, complementary observations).  
The total thermal energy is simply $\Yvol(r \rightarrow \infty)$ 
expressed in different units (c.f. Eqs.~\ref{eq:Yvol} and \ref{eq:thermalEnergy}), and is not -- by any means 
-- a new concept.  
However, one could envision measuring how $E_{tot}$ scales with $\Mtot(r_{500})$ for a small number 
of clusters that have high quality SZE and X-ray or lensing observations.  Then, as a larger SZE survey 
(such as those planned for ACT and SPT) discovers new clusters through the SZE, this scaling 
relation could be directly applied to obtain an $\Mtot(r_{500})$ estimate for each cluster in 
that sample, which is a useful step in constraining cosmology using clusters (see \S \ref{clustercosmology}).

\subsubsection{Comparison to the $\beta$-model for $E_\text{tot}$\label{Etotbeta}}

In contrast to $E_\text{tot,genNFW}$, we note that the integral in Eq.\ \ref{eq:thermalEnergy} when applied to the $\beta$-model
diverges for $\beta \leq 1$.  For $\beta>1$, the total thermal energy of the cluster gas is
\begin{equation}
E_{\rm tot,\beta} = \frac{3}{2} \pi^{3/2} r_c^3 \, P_{e,0} \, \left(\frac{\mu_e}{\mu}\right)
\frac{\Gamma(\frac{3\beta-3}{2})}{\Gamma(\frac{3\beta}{2})}.
\label{eq:thermalEnergy_beta}
\end{equation}
Since $\beta\approx0.7$ for X-ray observations, the isothermal $\beta$-model generally predicts 
an infinite thermal energy content in clusters; this is a product of assuming isothermality 
where the assumption is no longer valid.
As mentioned in \S \ref{genNFW_cases}, the generalized NFW profile can be simplified to the 
$\beta$-model by choosing $(a,b,c)=(2,3\beta,0)$.  
The condition that $\beta\!>\!1$ is required to obtain a finite total thermal energy within a cluster is consistent
with the condition that the genNFW parameter $b\!>\!3$ (i.e. $\beta\!<\!1 \Rightarrow b\!<\!3$).

\subsubsection{Comparison of Total Thermal Energy to an X-ray Proxy\label{Y_X}}

Recent work by \citet{kravtsov2006} has demonstrated an X-ray quantity that, like the 
SZE quantity \Yint, provides a robust proxy for \Mtot.  
This quantity is defined $Y_X \equiv \Mgas(r_{500}) \, T_X$.\footnote{Note that \cite{maughan2007}
demonstrated this quantity on a sample of clusters using an approximate formula for $r_{500}$,
so the X-ray data did not have to be of sufficient quality to perform a detailed HSE mass
estimate.
An overestimate of $r_{500}$ yields an overestimate of $\Mgas(r_{500})$, but also lowers
the estimate of $T_X$.  Similarly, an underestimate of $r_{500}$ lowers the estimate of
$\Mgas(r_{500})$, but raises $T_X$.  In this way, $Y_X$ is surprisingly robust to errors,
both systematic and statistical in nature. See \citet{kravtsov2006} for a more detailed
analysis.}
Here $T_X$ is measured within an annulus in radial range $[0.15, 1]\, r_{500}$.  
This range was chosen because cluster temperature profiles are most self-similar over this 
range \citep{vikhlinin2005a,nagai2007}. This large core cut makes the
measured $T_X$ insensitive to effects within cluster cores.

Using the ideal gas law (Eq.~\ref{eq:idealgaslaw}) and noting the similarity between the integral
to obtain the gas mass \Mgas\ (Eq.~\ref{eq:gasmass}) and that to obtain the thermal energy $E$ 
(Eq.~\ref{eq:thermalEnergy}), it is clear that $Y_X$ scales as the thermal energy content within
$r_{500}$, assuming isothermality. 
The implication of this is that the SZE-determined thermal energy content could provide the basis 
for a powerful, SZE-only observable that could readily be applied to upcoming SZE cluster surveys, 
without the need for corroborating observations other than those required for redshift determination.

\chapter{Applications of the Models}\label{model_application}

In this chapter, I demonstrate the application of the N07+SVM (Eqs.~\ref{eq:N07} \& \ref{eq:svm})
to observations of real clusters.
I initially tested the \emph{markov} fitting code and new models on mock SZA observations of 
several of the \cite{kravtsov2005} simulated clusters.  These mock SZA observations were also
jointly fit with mock \emph{Chandra} observations.  The tests provided sufficient confidence
that the modeling code was correctly implemented, but the mock SZA observations suffered systematic
biases due to the unrealistically cuspy cores of these simulated clusters (see Fig.~\ref{fig:n07_temperature} and 
discussion in \S \ref{genNFW_motivation}).
Any attempt to excise these cluster cores, which often dominated the SZE signal,
from the input simulation data would involve several assumptions about the cluster core and physics missing from
the simulations.
Instead I chose to use the mock observations to verify the functionality of the
modeling code, and then focused on tests involving real observations of real clusters.

I selected three massive clusters, well studied at X-ray wavelengths, 
spanning a wide range of redshifts ($z$~=~0.17--0.89) and cluster morphologies. 
These clusters are used test the joint analysis of {\em Chandra} and SZA observations, using the models and
methods described in Chapter \ref{modeling}.  Specifically, I test the N07 pressure 
profile (Eq.~\ref{eq:N07}) in conjunction with the SVM density profile (Eq.~\ref{eq:svm}).
I compare the results of these fits with results from a detailed, X-ray-only analysis, as well as with
a joint SZE+X-ray analysis using the traditional isothermal $\beta$-model.
I assume a \LCDM\ cosmology throughout the analysis presented in this chapter, with 
$\Omega_M=0.3$, $\Omega_\Lambda=0.7$, and $\Omega_k=0$.
The angular diameter distances to each of these clusters, as well as the redshifts and 
spectroscopic temperatures of the clusters, are listed in Table \ref{cluster_params}.


\begin{table}[t]
\centerline{
\begin{tabular}{|l|cc|c|}
\hline
{Cluster} & $z$\tablenotemark{a} & $d_A$ & $T_X$\tablenotemark{b}\\ 
 & & (Mpc) & (keV) \\
\hline 
& & & \\[-0.95pc]
A1835  & 0.25 & ~806.5 & $9.95^{+0.36}_{-0.37}$ \\[.2pc]
CL1226 & 0.89 & 1601.8 & $9.62^{+1.69}_{-1.25}$ \\[.2pc]
A1914  & 0.17 & ~600.6 & $8.63^{+0.47}_{-0.41}$ \\[.1pc]
\hline
\end{tabular}
}
\caption{Clusters chosen for testing the models.  Angular diameter distances
were computed assuming $\Omega_M=0.3$, $\Omega_\Lambda=0.7$, \& $\Omega_k=0$.}
\tablenotetext{a}{Redshifts for A1914 and A1835 are from \cite{struble1999}. 
Redshift for CL1226 is from \cite{ebeling2001}. 
All are in agreement with XSPEC fits to iron emission lines presented in \cite{laroque2006}.}
\tablenotetext{b}{Global X-ray spectroscopic temperatures were determined in the range $r \in [0.15,1.0] \, r_{500}$.
These temperatures were used in the isothermal $\beta$-model analysis.}
\label{cluster_params}
\end{table}


\begin{figure}
\begin{center}
\includegraphics[height=3.5in]{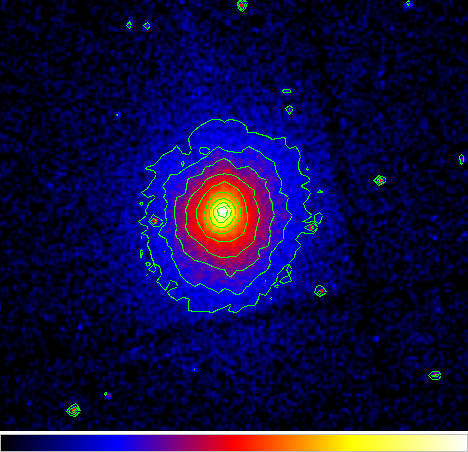}
\end{center}
\caption{X-ray image of A1835, showing it to be relaxed.  The X-ray analysis of A1835 relies on a single {\em Chandra} ACIS-I exposure,
with 85.7~ks of good time (unflagged exposure time).  The pixels of the ACIS-I detector are binned to be 1.968\arcsec on a side.
The X-ray image shown here is smoothed with a Gaussian that is 2 pixels in width (for display purposes only).
See Table \ref{xrayTable} for more details on the X-ray observation.  The inner 100~kpc (core) and all detected 
X-ray point sources were excluded from the X-ray surface brightness and spectroscopic analyses.}
\label{fig:xray_image_a1835}
\end{figure}

\begin{figure}
\begin{center}
\includegraphics[height=3.5in]{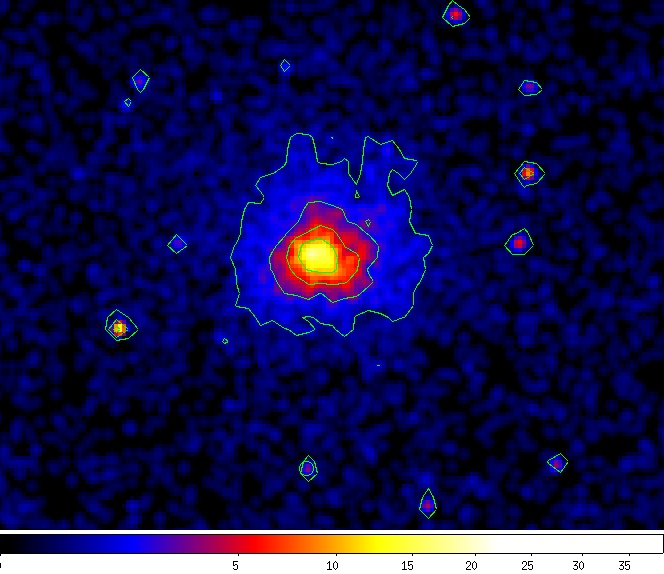}
\end{center}
\caption{X-ray image of CL1226, showing this high redshift cluster to be approximately circular on the sky,
and thus apparently relaxed.  This image combines two \emph{Chandra} exposures. The inner 100~kpc (core) 
and all detected X-ray point sources were excluded from the X-ray analysis.}
\label{fig:xray_image_cl1226}
\end{figure}

\begin{figure}
\begin{center}
\includegraphics[width=3.5in]{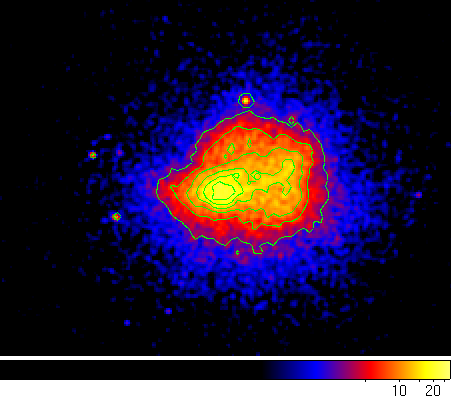}
\end{center}
\caption{X-ray image of A1914, showing it to be disturbed. 
This image combines two \emph{Chandra} exposures. 
A large subclump, measured to be hot by M08, is located to the east (left) of the cluster center.
This subclump, the inner 100~kpc (core), and all detected X-ray point sources were excluded from 
the X-ray analysis.}
\label{fig:xray_image_a1914}
\end{figure}


\section{Cluster Sample}\label{clusterselection}

Abell 1835 (A1835) is an intermediate-redshift ($z=0.25$)
cluster (see Fig.~\ref{fig:xray_image_a1835}, located at the mean redshift of a sample of clusters being used by the 
SZA collaboration to determine the SZE flux scaling relations (versus \Mgas, \Mtot, $T_X$, 
etc.).  This cluster sample is X-ray flux-limited, and is located in the redshift range $z \sim$ 0.2--0.3.
The sample has both X-ray observations and optical lensing mass estimates available, 
and will appear in the thesis of Ryan Hennessey as well as forthcoming SZA papers.  
A1835 is relaxed, as evidenced by both its circular morphology in the X-ray images
and its cool core (see Fig.~\ref{fig:xray_image_a1835}, and \cite{peterson2001}
for more details).  
To demonstrate the applicability of the joint SZE+X-ray analysis to high redshift 
clusters, I analyzed CL~J1226.9+3332 (CL1226), an apparently relaxed cluster at $z=0.89$
(see \cite{maughan2007b} -- hereafter referred to as M07 --and Fig.~\ref{fig:xray_image_cl1226}). Note 
that M07 argues, despite appearing spherically symmetric, that the lack of a cool core in this 
cluster is likely due to a recent merger.  
Finally, to assess how applicable this pressure profile is to disturbed clusters, I also analyzed Abell 1914 (A1914), an
intermediate redshift ($z=0.171$) cluster with a hot subclump near the core (see Fig.~\ref{fig:xray_image_a1914});
when the subclump is not excluded from the X-ray analysis, a large centroid shift is present in the fit to the
surface brightness (see \cite{maughan2008} -- hereafter referred to as M08 -- who use the size of the error in the fit X-ray 
centroid as a measure of cluster dynamical state).  I exclude this subclump from the X-ray analysis of A1914
by defining a $\sim 100$~kpc region centered on the peak brightness of the clump and excluding this region
from the X-ray likelihood calculation (see Eq.~\ref{eq:xray_L} and discussion of the X-ray likelihood in 
\S \ref{markov}).

Figures \ref{fig:sz_image_a1835}--\ref{fig:sz_image_a1914} show the \emph{clean}-ed\footnote{
\emph{clean}-ing interferometric data involves iteratively fitting point sources within a specified 
region and removing their flux.  See e.g., \citet{thompson2001} for more details.} interferometric maps, produced
from the SZA observations analyzed here, of these three clusters. 
Details of the SZA and \emph{Chandra} observations, including the X-ray fitting regions, the unflagged, on-source 
integration times, and the pointing centers used for the SZE observations, are 
presented in Tables \ref{obsTable} and \ref{xrayTable}.
Note that 90-GHz SZA data were included in the fits to CL1226, which is only partially resolved
at 30~GHz due to its small angular extent (see \S \ref{sze_visfits}).


\begin{table}[t]
\centerline{
{\scriptsize
\begin{tabular}{|l|cc|cc|cc|c|}
\hline
{Cluster} &
\multicolumn{2}{c}{\underline{SZA Pointing Center (J2000)}} &
\multicolumn{2}{c}{\underline{Short Baselines (0.34--1.5~$\rm k\lambda$)}} &
\multicolumn{2}{c}{\underline{Long Baselines (3--7.5~$\rm k\lambda$)}} &
{$\rm{t_{int,red}}$\tablenotemark{a}}
\\
{} & 
{$\alpha$} & {$\delta$} & 
{beam($\arcsec\times\arcsec$)\tablenotemark{b}} & {$\sigma$(mJy)\tablenotemark{c}} &
{beam($\arcsec\times\arcsec$)\tablenotemark{b}} & {$\sigma$(mJy)\tablenotemark{c}} &
{(hrs)} \\
\hline
A1914 & $14^h 26^m 00^s\!.8$ & $+37^{\circ}49^{\prime}35\arcsec\!.7$ & 117.5$\times$129.9 & 0.30 & 23.5$\times$17.4 & 0.35 & 11.5\\
A1835 & $14^h 01^m 02^s\!.0$ & $+02^{\circ}52^{\prime}41\arcsec\!.7$ & 116.6$\times$152.1 & 0.25 & 17.5$\times$23.5 & 0.33 & 18.6\\
CL1226 (30~GHz)& $12^h 26^m 58^s\!.0$ & $+33^{\circ}32^{\prime}45\arcsec\!.0$ & 117.4$\times$125.4 & 0.20 & 16.0$\times$21.2 & 0.20 & 22.0 \\
CL1226 (90~GHz)& `` '' & `` '' & 42.3$\times$39.1\tablenotemark{d}&  0.42 & 9.74$\times$7.46\tablenotemark{d}& 0.32 & 29.2\\
\hline
\end{tabular}
}
}
\caption{SZA Cluster Observations}
\tablenotetext{a}{Unflagged data after reduction.}
\tablenotetext{b}{Synthesized beam FWHM and position angle measured from North through East}
\tablenotetext{c}{Achieved {\em rms} noise in corresponding maps}
\tablenotetext{d}{The short and long baselines of the 90-GHz observation probe 1--4.5 and 8--22 $\rm k\lambda$.}
\label{obsTable}
\end{table}


\begin{table}[t]
 \centerline{
  \begin{tabular}{|l|cc|cc|c|}
   \hline
   Cluster & \multicolumn{2}{c}{\underline{Joint SZE+X-ray Analysis}} & \multicolumn{2}{c}{\underline{Maughan X-ray Analysis}} & ObsID\\
	& (ks)\tablenotemark{a} & \arcsec\tablenotemark{b} & (ks)\tablenotemark{a} &  \arcsec\tablenotemark{b} & \\
   \hline
   A1914	& 26.0  & 34.4--423.1 & 23.3  & 0.0--462.5 & 542+3593  \\
   A1835	& 85.7  & 25.6--344.4 & 85.7  & 0.0--519.6 & 6880      \\
   CL1226	& 64.4  & 12.9--125.0 & 50    & 0.0--125.0 & 3180+5014 \\
   CL1226\tablenotemark{c} & ~~N/A & N/A         & 75+68 & 17.1--115 & 0200340101 \\
    \hline
  \end{tabular}
}
\caption{Details of X-ray Observations.  The X-ray analysis presented here, as part of the joint SZE+X-ray modeling,
was performed independently from that performed by Maughan.}
\label{xrayTable}
\tablenotetext{a}{Good (unflagged, cleaned) times for X-ray observations after respective calibration pipelines.}
\tablenotetext{b}{X-ray image fitting region.}
\tablenotetext{c}{{\em XMM-Newton} observation of CL1226, presented in M07, was used only in the spectroscopic analysis.
The exposure times are, respectively, those of the MOS and PN camera.}
\end{table}


In the following sections, I discuss the results of the joint analysis of the SZE and X-ray data. 
I show how the models fit the X-ray surface brightness and radio interferometric SZE data,
and use these fits to place constraints on cluster astrophysical properties.
Finally, I compare these results with an independent, detailed, X-ray-only analysis performed by 
Ben Maughan, following the techniques presented in M07 and M08 and described in \S \ref{maughan_analysis}.
The high quality of the X-ray data on these three clusters allows this X-ray-only analysis, which is used
in assessing the new models.


\begin{figure}
\begin{center}
\includegraphics[width=5.25in]{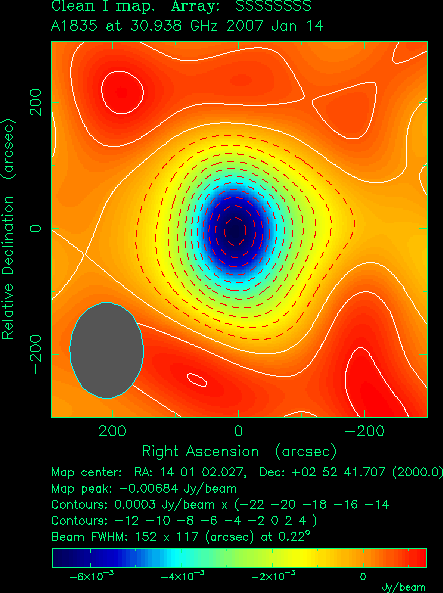}
\end{center}
\caption{Cleaned SZA 30-GHz map of A1835, showing the detection to have a high significance of $\sim$ 22-$\sigma$ 
at the peak.  This image, made with \emph{Difmap} using only the short baseline data ($\sim$ 0.35--1.5~k$\lambda$), 
is for presentation purposes only; we fit our data directly in \emph{u,v}-space, and do not use any interferometric
mapping to determine cluster properties.  
Note that for a given \emph{u,v}-space coverage, any unresolved structure in an interferometric SZE 
map of a cluster essentially ``looks'' like the synthesized beam; this determines the effective resultion
of the observation.
The synthesized beam is depicted in gray in the lower left corner.
Contours are overlaid at 2-$\sigma$ intervals. 
}
\label{fig:sz_image_a1835}
\end{figure}

\begin{figure}
\begin{center}
\includegraphics[width=5.25in]{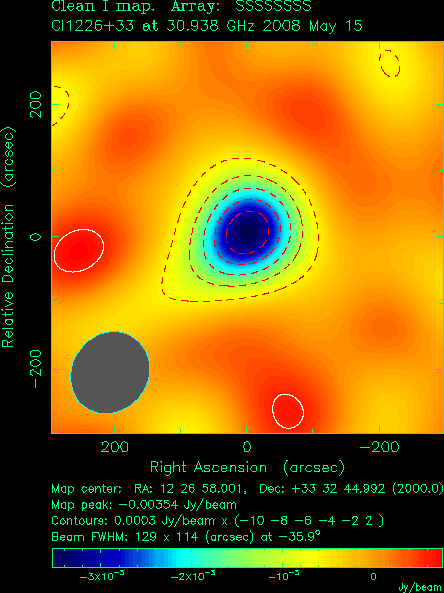}
\end{center}
\caption{Cleaned SZA 30-GHz map of CL1226, showing the detection to have a significance of $\sim$ 10-$\sigma$ at the peak.
See caption of Fig.~\ref{fig:sz_image_a1835} for further details.
}
\label{fig:sz_image_cl1226_30}
\end{figure}

\begin{figure}
\begin{center}
\includegraphics[width=5.25in]{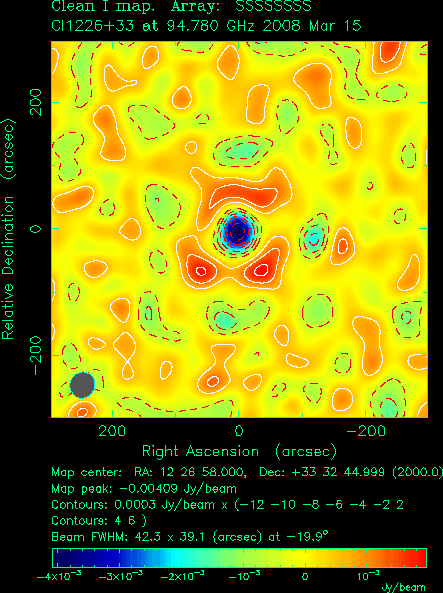}
\end{center}
\caption{Cleaned SZA 90-GHz map of CL1226, showing the detection with this higher resolution instrument to have a 
significance of $\sim$ 12-$\sigma$ at the peak.  This observation was included since 30-GHz data alone could not
constrain the radial profile of this high-redshift, small angular extent cluster (see Figures \ref{fig:uv_coverage_cl1226_30_90}
and \ref{fig:uv_space_hist_CL1226_30_90}).
See Fig.~\ref{fig:sz_image_a1835} for further details, noting the short baselines of the 90-GHz instrument are
$\sim$ 1--4.5~k$\lambda$
}
\label{fig:sz_image_cl1226_90}
\end{figure}

\begin{figure}
\begin{center}
\includegraphics[width=5.25in]{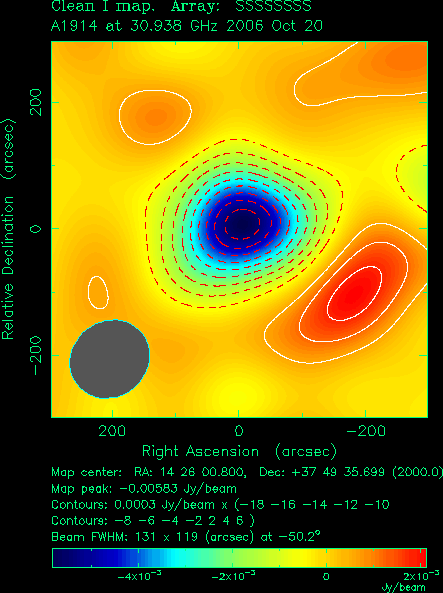}
\end{center}
\caption{Cleaned SZA 30-GHz map of A1914, showing the detection to have a high significance of $\sim$ 18-$\sigma$ at the peak.
See Fig.~\ref{fig:sz_image_a1835} for detail.
}
\label{fig:sz_image_a1914}
\end{figure}


\section{Unresolved Radio Sources}\label{point_sources}

As discussed in \S \ref{phasecal}, an unresolved, compact radio source (``point source'') can be represented
mathematically by a 2-dimensional Dirac delta function in image space.  In Fourier space, such unresolved
sources -- which can contaminate an SZE cluster observation -- have an analytic transform: a constant,
equal flux on all scales, as measured by all baselines. 
The flux from these sources is modeled analytically using a power law to describe
the frequency-dependence of the source flux, with flux $f_0$ normalized at $\nu_0=30.938~\rm GHz$, 
the center of the SZA 30-GHz band.
I also account for attenuation by the primary beam, $A_\nu(x,y)$, at each band's frequency. 
Therefore, the measured flux $f(\nu)$ from a point source at map location $(x,y)$ is fit using
\begin{equation}
f(\nu) = A_\nu(x,y) f_0 (\nu/\nu_0)^\alpha
\label{ptsrc_flux}
\end{equation}
where $\alpha$ is the spectral index.

When modeling a compact source detected in an SZE observation, I fix its
spectral index $\alpha$ so that it simultaneously fits the mean flux measured by the SZA at 30.938~GHz 
and that measured in the NVSS \citep{condon1998} or FIRST \citep{white1997} surveys,
which were performed at 1.4~GHz.\footnote{This is a sufficient approximation since the signal 
to noise in each individual band 
is too low to leave flux a free parameter to be fit by SZA data alone, and even if $\alpha$ were 
under or overestimated by $\sim 1$, the difference at each end of the band would only be 
biased by $\sim 12\%$. Even for the strongest point sources in these fields ($\sim 4$~mJy, 
as discussed below) this would result in an error smaller than any 1-$\sigma$ fluctuation 
in a particular band.
Furthermore, this method of point source modeling was tested by adding artificial point sources
to both real and simulated observations, and was found to have a negligible impact on cluster
parameters extracted from these observations.}
The mean point source flux and the source's approximate coordinates are first identified
from the SZA observation(s) using the interferometric imaging package \emph{Difmap}.
This position serves as the input location in \emph{markov}, which is then refined in 
trial \emph{markov} runs.  Such MCMC trials would sufficiently constrain the location of any 
newly-discovered point source, at least for the purposes of constraining point source
contamination in an SZE observation.
However, for all detected sources in the observations presented here, I found their trial 
positions agreed to within $\sim 2 \arcsec$ of point sources already found by NVSS or FIRST. 
Since the VLA has superior resolution to the SZA, which constrains sources
to within $\sim 7 \arcsec$ positional uncertainty, I simply fixed each source's location 
to the catalogued NVSS or FIRST position.
I leave the source flux a free parameter in the Markov chains used to constrain cluster
properties, so that the cluster SZE flux and any compact sources are simultaneously fit.
This marginalizes over uncertainties in the SZE and point source fluxes.

In the SZA cluster observations presented here, the A1835 field
contains two detectable compact sources at 30~GHz: a $2.8\pm0.3$~mJy
central source, and a $1.1\pm0.3$~mJy source $\sim 1\arcmin$ from
the cluster center (using the fixed position and spectral index for each source,
as discussed above).  
The central source was detected in both the NVSS and the FIRST survey,
while the weaker was only detected by FIRST.   
The SZA observation of CL1226 contains one detectable compact source,
identified in both FIRST and NVSS, with a flux at 30~GHz of $4.0\pm0.8$~mJy, located
$\sim 6\arcmin$ from the cluster center \citep[see also][]{muchovej2007}.   
Flux from three compact sources, with positions constrained by NVSS and FIRST, was detected 
at 30~GHz in the A1914 field.  The fluxes of the sources are $2.2\pm0.3$~mJy, $1.3\pm0.3$~mJy,
and $0.6\pm0.2$~mJy. The strongest was detected in both the NVSS and the FIRST survey,
while the second strongest was only detected in the FIRST survey, and the weakest was only 
detected in NVSS.  Figures \ref{fig:sz_image_a1835}--\ref{fig:sz_image_a1914} depict the
SZA data with these point sources removed.


\section{Independent X-ray Analysis}\label{maughan_analysis}

For each of these clusters, I compare the results of the joint SZE+X-ray 
analysis to the results of independent X-ray analyses, performed by Ben Maughan and 
described in detail in M07 and M08, which respectively presented the observations of CL1226 and A1914. 
The ACIS-I observation of A1835 used in this work became public more
recently than M08 was published, and is therefore presented here -- and in \citet{mroczkowski2009} -- for the first time.  
The observation of A1835 was calibrated and
analyzed using the same methods and routines described in M08.
Several key differences exist between both the M07/M08 data reduction and fitting methods, 
and those used in the joint SZE+X-ray analysis here (described in \cite{bonamente2004,bonamente2006},
as well as in Appendix~\ref{app2}). 
I discuss a few salient differences here.  

In the X-ray-only analysis, Maughan used blank-sky fields to 
estimate the background for both the imaging and spectral analysis. 
The X-ray images were binned into 1.97$\arcsec$ pixels (as are the images used in the
joint SZE+X-ray analysis; see Appendix \ref{app2}).
The imaging analysis (primarily used to obtain the gas emissivity profile)
is performed in the 0.7--2~keV energy band in order to maximize cluster signal to noise.
Both the \emph{Chandra}'s efficiency and the (redshifted) cluster emission are highest 
in this range (see Fig.~\ref{fig:effarea}); since spectral information is not preserved
in the X-ray images used in the surface brightness fit, binning noisier photons from outside 
the 0.7--2~keV energy band can degrade the cluster signal, as the binning is unweighted
(Appendix \ref{app2} discusses this further).

In Ben Maughan's spectral analysis (used to derive both the global temperature $T_X$ and the
temperature profiles), spectra extracted from each region of interest were
fit in the 0.6--9~keV band\footnote{
A larger range of photon energies can be used for the temperature determination
since photon energy information is preserved when performing the spectral fit.}
with an absorbed, redshifted APEC \citep{smith2001} model.
This absorption was fixed at the Galactic value.
The global $T_X$, used in the isothermal $\beta$-model fits, is determined within an 
annulus with radius $r \in [0.15,1.0] \, r_{500}$.

An important consideration when using a blank-sky background method is
that the count rate at soft energies can be significantly different in
the blank-sky fields than the target field, due to differences between
the level of the soft Galactic foreground emission in the target field
and that in the blank-sky field.
Ben Maughan accounted for this in the imaging analysis by normalizing the
background image to the count rate in the target image in regions
far from the cluster center. In his spectral analysis, this was modeled by an 
additional thermal component that was fit
to a soft residual spectrum (the difference between spectra extracted in
source free regions of the target and background datasets; see \citet{vikhlinin2005a}). 
The exception to this was the \emph{XMM-Newton} data
used in addition to the {\em Chandra} data for CL1226. 
A local background was found by M07 to be more reliable
for the CL1226 spectral analysis, thus required no correction for the soft Galactic foreground.

The M07/M08 X-ray analysis methods exploit the full V06 density and 
temperature models (Eqs.~\ref{eq:Vikhlinin} \& \ref{eq:V06_tprof}) 
to fit the emissivity and temperature profiles of each cluster. 
The results of these fits are used to derive the total hydrostatic mass profiles of each system. 
Uncertainties for the independent, X-ray-only analysis are derived by Ben Maughan using a 
Monte Carlo randomization process.   These fits involved typically
$\sim$1000 realizations of the temperature and surface brightness profiles, fit
to data randomized according to the measured noise.
For further details about this fitting procedure, see M07 and M08.


\section{SZE Cluster Visibility Fits}\label{sze_visfits}

As mentioned in both \S \ref{interf} and \S \ref{final_data_product}, interferometric SZE
 data are in the form of visibilities, $V(u,v)$. 
Equation~\ref{eq:visibility2}, which is an integral over all space ($\int \!\! \int \! dx \, dy$),
relates the visibilities (in flux) to the spatial intensity pattern (in units of flux per 
solid angle on the sky, or flux density). 
Each visibility is thereby a measure of the flux within the Fourier mode 
probed by a given baseline. 

Using relations provided in, e.g., \citet{carlstrom2002}, we can rescale the visibilities to a new 
quantity $Y(u,v)$, which is the integral of line-of-sight Compton $y$ within the Fourier modes probed 
by an interferometer.\footnote{Note that the Fourier transform of the line-of-sight Compton $y$, $Y(u,v)$, 
scales like \Yint\ but is not to be confused with the actual \Yint\ -- in the form of either \Ylos\ or 
\Yvol (see Eqs.~\ref{eq:Yint} \& \ref{eq:Yvol})  -- the recovery of which is discussed in \S \ref{derived_quants}.}  
In the image plane, we can convert intensity to line-of-sight Compton $y$ using the
CMB intensity $I_0$, the derivative of the blackbody function, and the spectrum of the SZE distortion $f(x)$ 
as a function of dimensionless frequency $x$ (Eq.~\ref{eq:fx}).  
The derivative of the blackbody function multiplied by $f(x)$ yields the 
spectral dependence of the intensity shift in primary CMB due to the SZE, which is
\begin{equation}
g(x) = \frac{x^4 e^x}{(e^x-1)^2} \left(x \frac{e^x + 1}{e^x - 1} - 4\right).
\end{equation}
The shift in primary CMB intensity due to the SZE, $\Delta I_{SZE}$, is
\begin{equation}
\label{eq:delta_Isze}
\Delta I_{SZE} = g(x) \, I_0 \, y,
\end{equation}
where the primary CMB intensity $I_0$ is
\begin{equation}
\label{eq:I0_cmb}
I_0 = \frac{2 (k_B \Tcmb)^3}{(h c)^2}.
\end{equation}
Here $k_B$ is Boltzman's constant, \Tcmb\ is the primary CMB temperature, $h$ is Planck's constant, a
nd $c$ is the speed of light.

\begin{figure}
\centerline{
\includegraphics[width=6in]{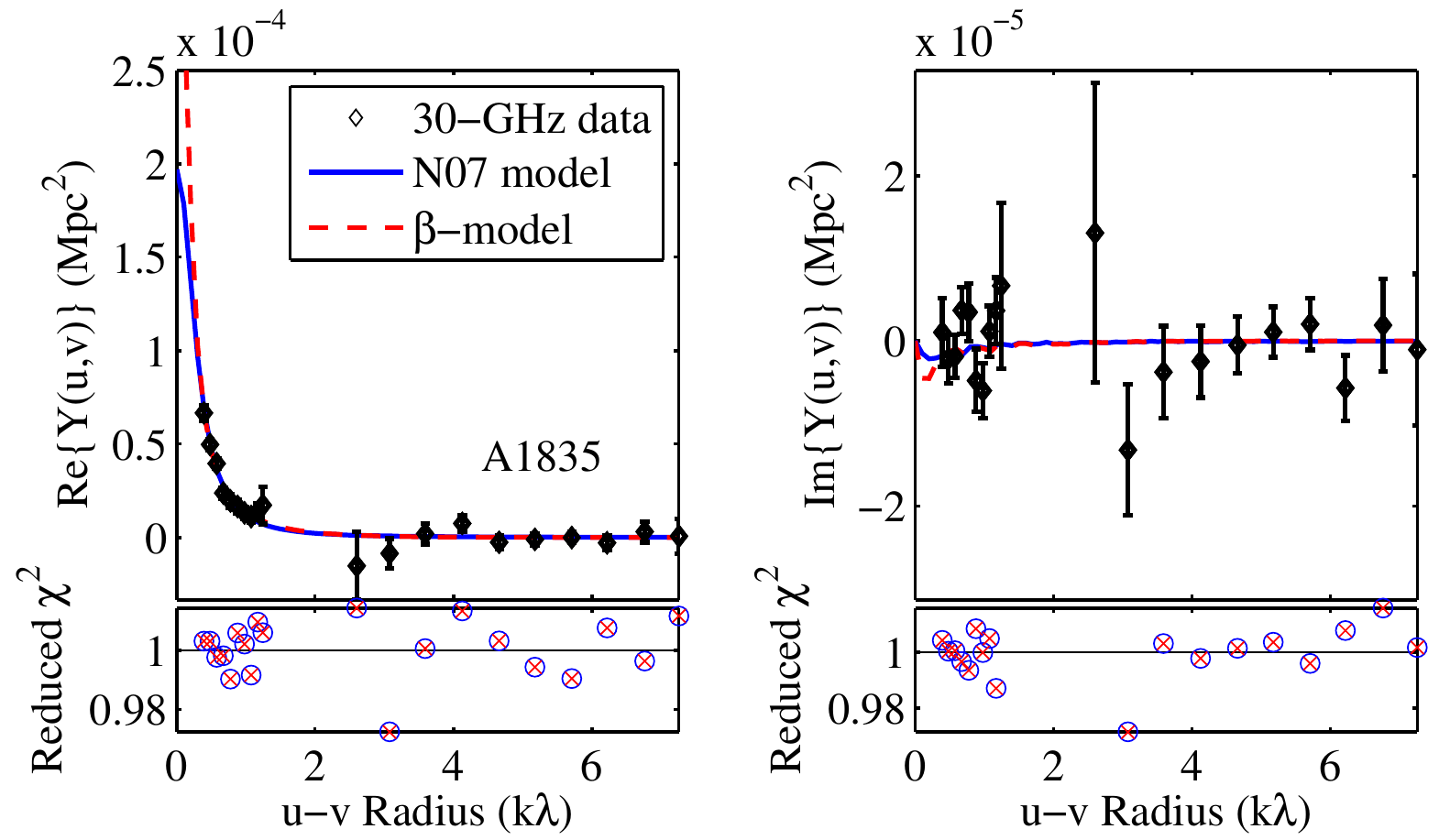}
}
\caption{Radially-averaged SZE model fits to A1835 in \emph{u,v}-space, from the jointly-fit N07+SVM and 
isothermal $\beta$-model.  
The upper panels show the real (left) and imaginary (right) components of the visibilities, radially-averaged
to be a function of $(u,v)$ radius, and rescaled to units of intrinsic, line-of-sight integrated Compton $y$;
I plot here the frequency-independent quantity $Y(u,v) \, d_A^2 = V(u,v)  \, d_A^2/g(x) I_0$, where each
band is scaled appropriately before binning, and the angular diameter distance is computed using the assumed \LCDM\ cosmology. 
The lower panels show the reduced $\chi^2$ of the fits to the data for the chosen binning.
The black points with error bars (1-$\sigma$) are the binned $Y(u,v) \, d_A^2$ data, 
with the point source models first subtracted from the cluster visibilities.  
The blue, solid line is a high likelihood N07 model fit, while the red, dashed line is a similarly-chosen
fit of the $\beta$-model.
For the available data points, both SZE models fit equally well (see lower panels, which shows the $\chi^2$
for each model is indistinguishable).  However, note that as the 
\emph{u,v}-radius approaches zero k$\lambda$ -- where there are no data to constrain the models
 -- the $\beta$-model predicts a much higher integrated Compton $y$ than the N07 model.
For cluster data centered on the phase center of the observation (with both cluster and primary beam
sharing this center), the mean imaginary component would equal zero.  Since these model fits include
the primary beam, which is not necessarily centered on the cluster, the small but non-zero imaginary 
component in the upper right panel is expected (note the smaller units for the $y$-axis of the right hand plot)
\citep[See, e.g.][for comparison]{reese2002}.
}
\label{fig:uvplot_a1835}
\end{figure}

\begin{figure}
\centerline{
\includegraphics[width=6in]{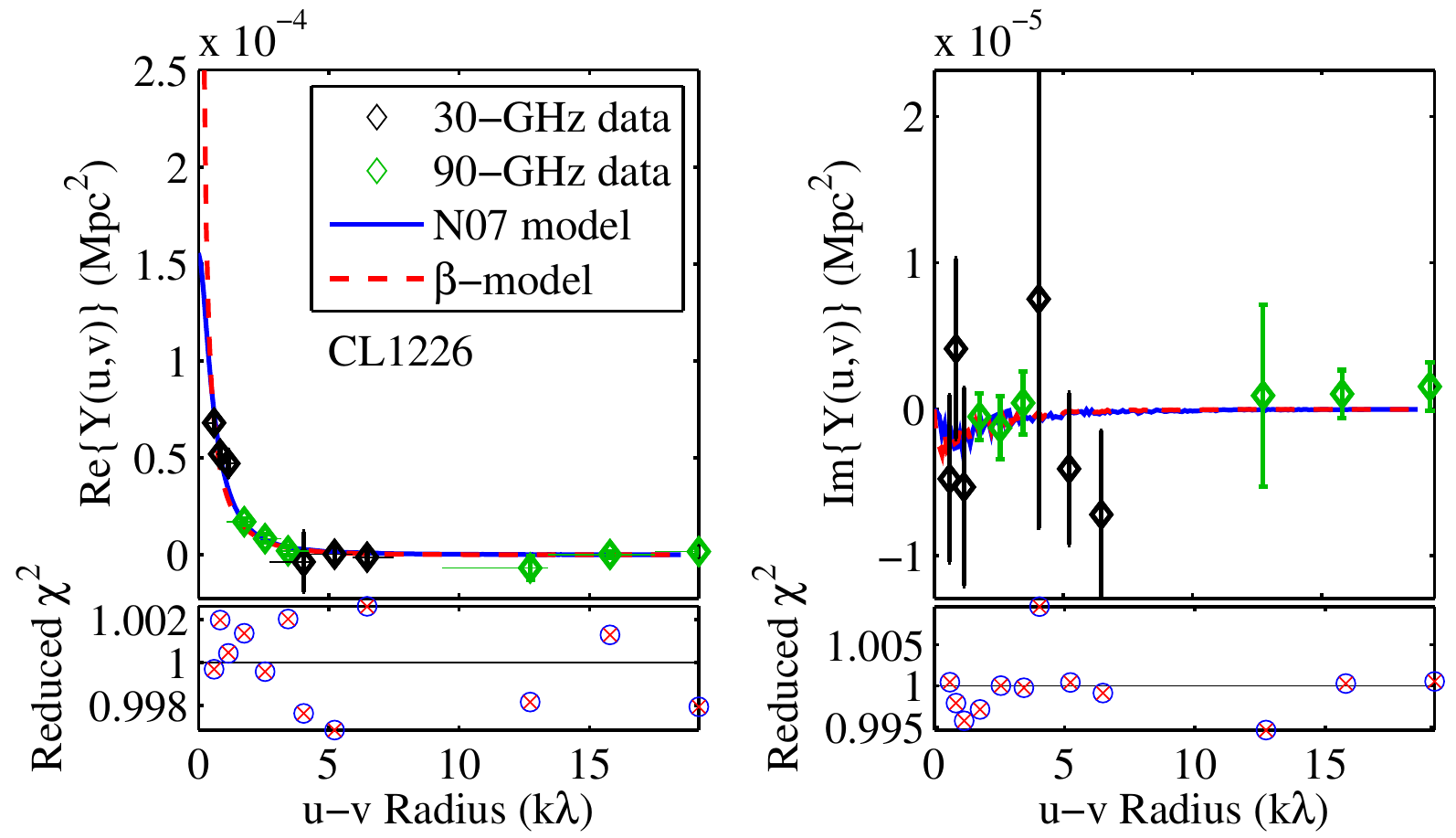}
}
\caption{Radially-averaged SZE model fits for CL1226 in \emph{u,v}-space.  See Figure \ref{fig:uvplot_a1835}
for additional caption details.  Note that the 30-GHz \emph{u,v}-space coverage alone does not probe a sufficient
range of cluster scales (see Figures \ref{fig:uv_coverage_cl1226} and \ref{fig:uv_space_hist_CL1226_30_90}) to determine
where the SZE signal falls to zero. This results in poor constraints on the cluster's radial profile when
using 30-GHz data alone, and is a result of this high-redshift source being relatively compact on the sky 
(compared to A1835 and A1914; note that $r_{2500}$ is on the order of 1 arcminute, 
as shown in Table \ref{table:derivedQuants_r2500}, \S \ref{derived_quants}).  
I therefore included 90-GHz SZA data (green) in the joint SZE+X-ray fit, since
the 90-GHz instrument was designed to complement the \emph{u,v}-coverage provided at 30-GHz
(see Figures \ref{fig:uv_coverage_cl1226_30_90} \& \ref{fig:uv_space_hist_CL1226_30_90}).
Again, the small but non-zero imaginary component is expected, but note the smaller units for the 
plot of the imaginary component of the radially-binned $Y(u,v)$.}
\label{fig:uvplot_cl1226}
\end{figure}

\begin{figure}
\centerline{
\includegraphics[width=6in]{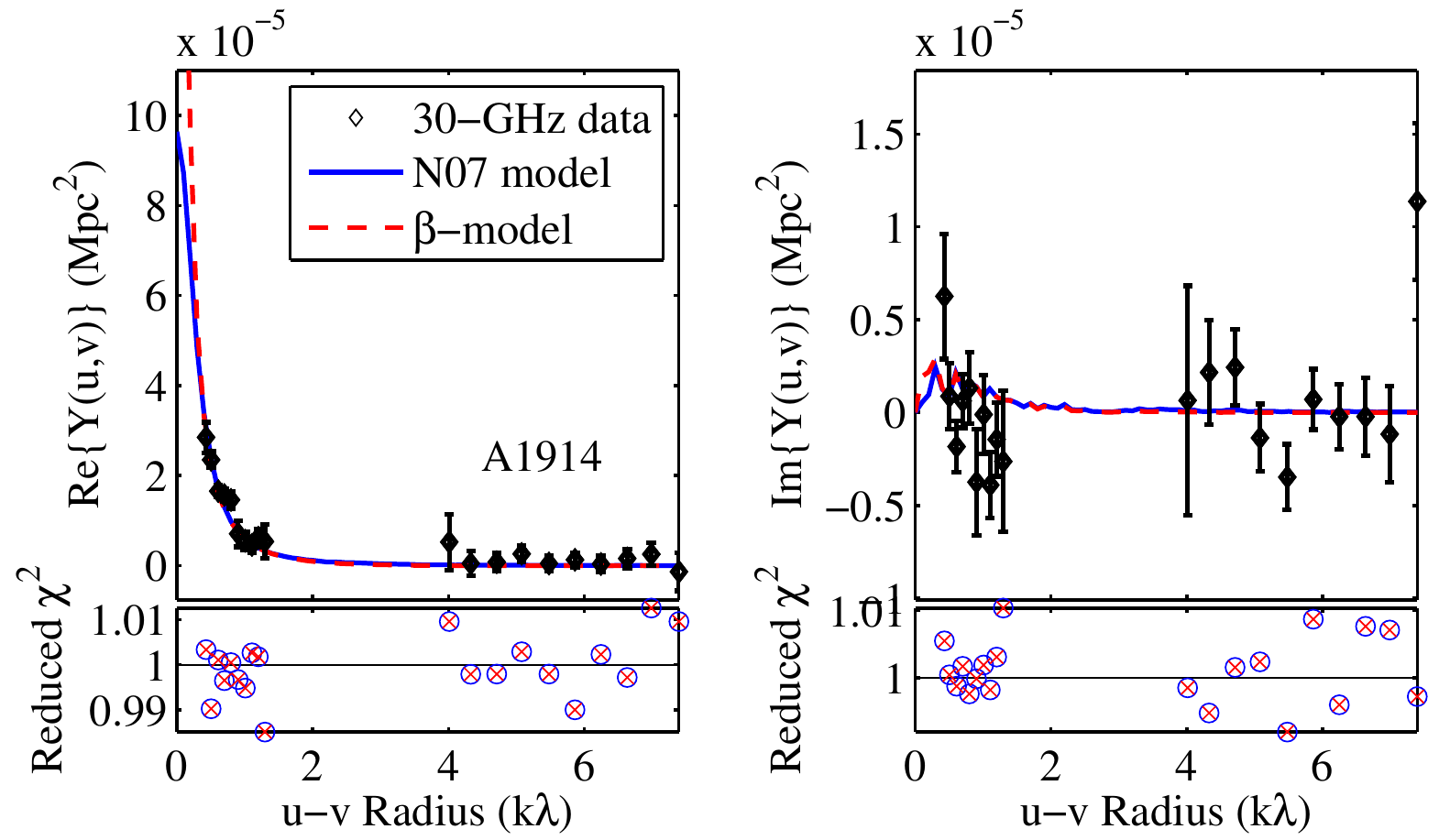}
}
\caption{Radially-averaged SZE model fits for A1914 in \emph{u,v}-space.  See Figure \ref{fig:uvplot_a1835}
for details.  Note that the slightly larger imaginary component in the upper right panel could be 
due to cluster asymmetry, since A1914 is disturbed and elliptical (see Fig.~\ref{fig:sz_image_a1914}).
Note that the reduced $\chi^2$ of this fit is no larger than those for the other two clusters (Figures \ref{fig:uvplot_a1835} \& \ref{fig:uvplot_cl1226}).
}
\label{fig:uvplot_a1914}
\end{figure}

Figures \ref{fig:uvplot_a1835}--\ref{fig:uvplot_a1914} show representative, high-significance fits of the 
N07 and isothermal $\beta$-model to each cluster's radially-averaged SZE visibility data (i.e. binned
according to \emph{u,v} distance).  This quantity is then rescaled by $d_A^2 /[g(x) I_0]$ to obtain the 
radially-averaged $Y(u,v)$ intrinsic to the cluster.\footnote{These fits were performed jointly on the 
X-ray imaging and SZE interferometric data.  However, the inclusion of X-ray data does not significantly 
affect the high-likelihood fits of the N07 profile; rather, it excludes some of the 
low likelihood N07 pressure fits (see discussion in \S \ref{derived_quants}).} 
This rescaling removes both the redshift and frequency dependence from the
cluster visibility data, after having first subtracted all point source models 
(as modeled simultaneously with cluster in the MCMC fitting code)
from the visibility data in \emph{u,v}-space.  

Note that for Figure \ref{fig:uvplot_cl1226}, I combine both 30 and 90-GHz data to increase
both the dynamic range and \emph{u,v}-coverage on the high-redshift cluster CL1226.
The \emph{u,v}-coverage provided by the 30-GHz observation alone was insuffient to constrain 
the cluster's profile.  Figure \ref{fig:uvplot_cl1226} shows the 30-GHz data points
at low \emph{u,v}-radii agree to within their 1-$\sigma$ error bars, meaning the SZE signal
from the cluster was not resolved by those baselines, nor does the 30-GHz data for CL1226
accurately determine where the signal falls to zero.  The inclusion of 90-GHz data provides 
the necessary \emph{u,v}-coverage and dynamic range to resolve the cluster profile.
As detailed in \S \ref{uvspace}, the 90-GHz instrument was designed to complement the 
\emph{u,v}-coverage provided at 30-GHz
(see Figures \ref{fig:uv_coverage_cl1226_30_90} \& \ref{fig:uv_space_hist_CL1226_30_90}).

In Fig.~\ref{fig:uvplot_a1914}, it can be seen that the average imaginary component of the 
radially-averaged $Y(u,v)$ of A1914 is slightly larger than that seen for the other two clusters
(Figures \ref{fig:uvplot_a1835} \& \ref{fig:uvplot_cl1226}).  
As mentioned in the caption, I attribute this to the disturbed, 
asymmetric nature of A1914, which is also apparent in the SZE image (see Fig.~\ref{fig:sz_image_a1914}).


\section{X-ray Surface Brightness Fits}\label{xray_surffits}

The X-ray data reduction used in the joint SZE+X-ray analysis follows that from \citet{bonamente2004,bonamente2006}, 
and is discussed in detail in Appendix \ref{app2}.  Each image was fit, excluding the inner 100~kpc from the X-ray imaging 
likelihood calculation (see \S \ref{markov}), out to $\sim r_{500}$, 
using the Markov Chain Monte Carlo technique, described in \S \ref{markov}, which jointly fits
the X-ray and SZE data.\footnote{Recall that the SZE data are used here to obtain temperature 
information in the joint N07+SVM fitting procedure.}  
Table \ref{xrayTable} lists the precise fitting regions for each X-ray observation.

\begin{figure}
\begin{center}
\includegraphics[width=1.66in,angle=270]{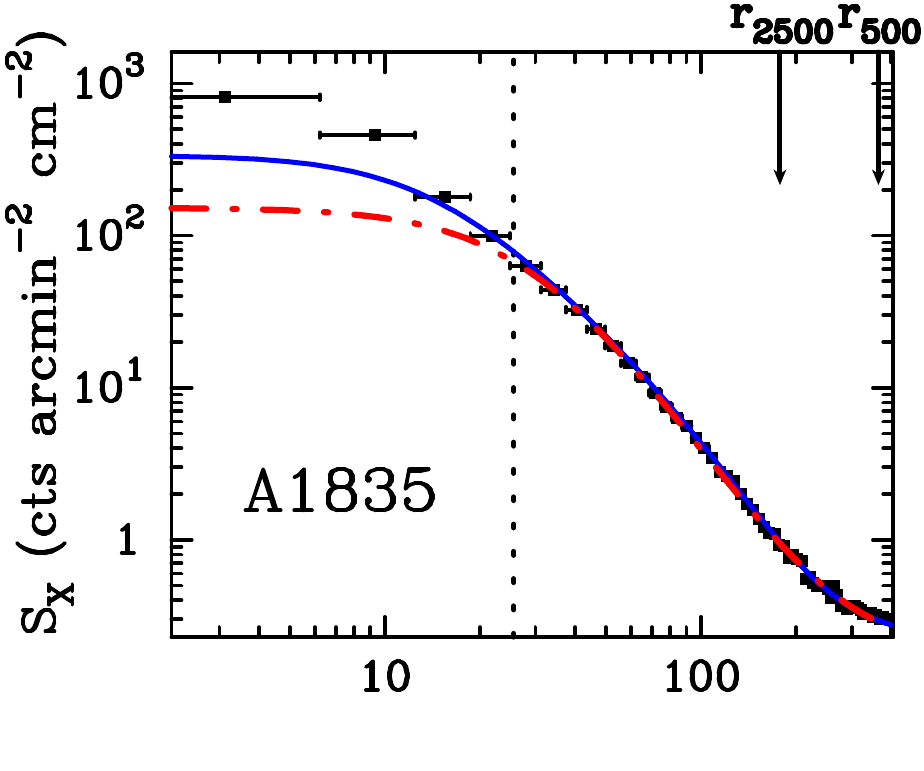}
\includegraphics[width=1.66in,angle=270]{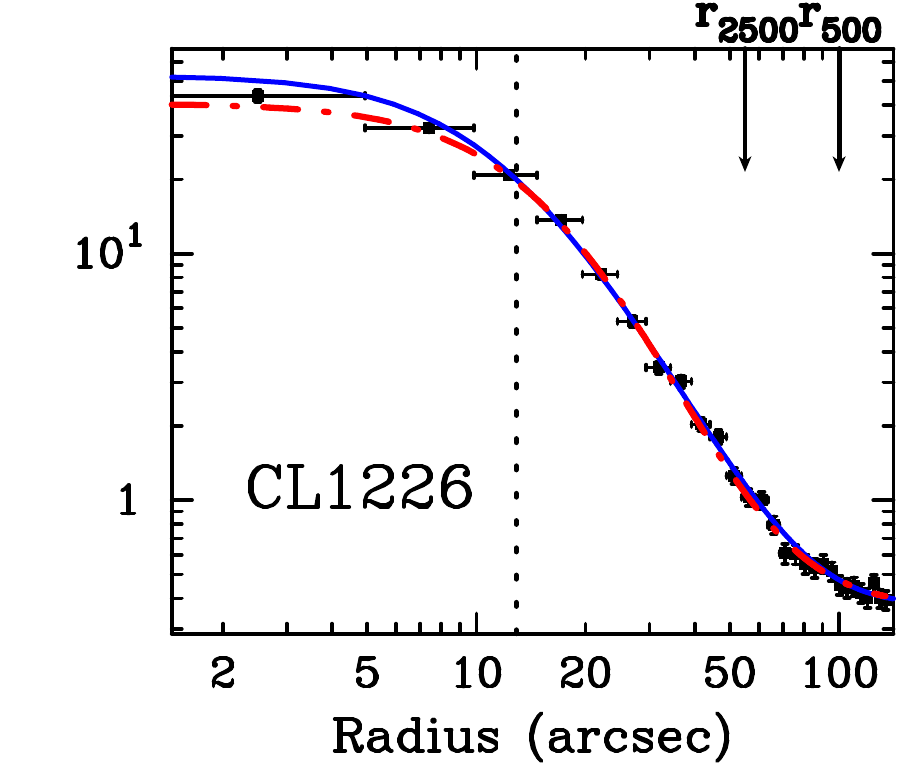}
\includegraphics[width=1.66in,angle=270]{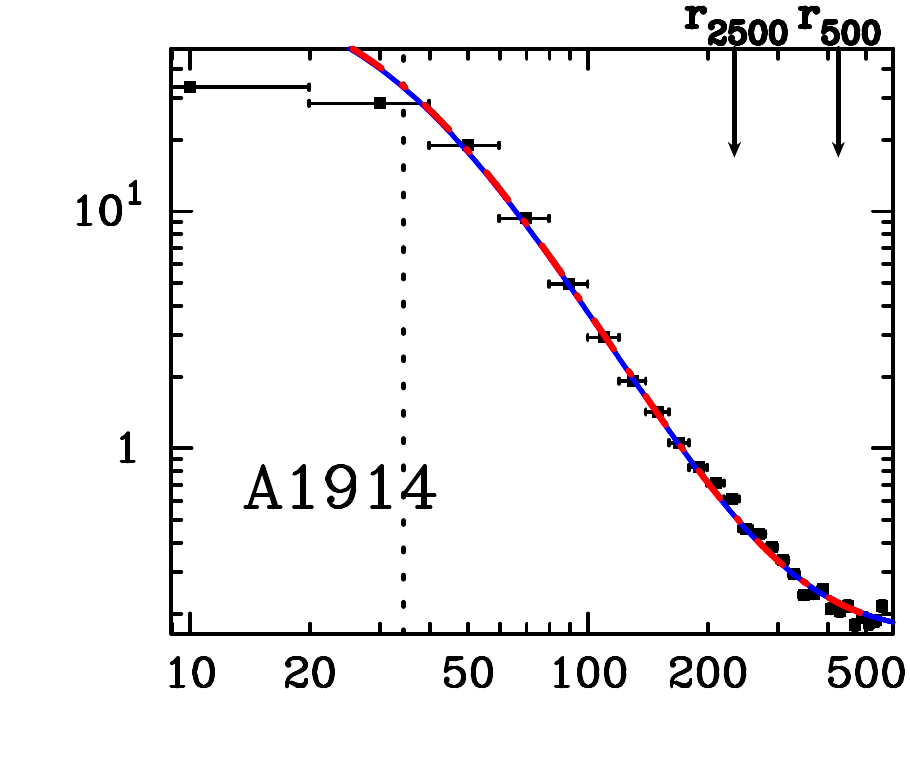}
\end{center}
\caption{X-ray surface brightness profile fits to the clusters.
The vertical dashed line denotes the 100~kpc core cut.
The blue, solid line is the surface brightness computed using a high-likelihood fit of the 
N07+SVM profiles (analogous to the SZE fits plotted in Figures \ref{fig:uvplot_a1835}--\ref{fig:uvplot_a1914}), 
while red, dot-dashed line is the surface brightness fit of a $\beta$-model.  Both model lines include 
the X-ray background that was fit simultaneously with the cluster model (i.e. the plotted lines are the superpositions
of each set of the background and cluster models, which were fit simultaneously to the X-ray imaging data).
The black squares are the annularly-binned X-ray data, where the widths of the bins are denoted
by horizontal bars.  The vertical error bars are the 1-$\sigma$ errors on the binned measurements.
Arrows indicate $r_{2500}$ and $r_{500}$ derived from the N07+SVM profiles (see \S \ref{derived_quants}).}
\label{fig:Sx_profiles}
\end{figure}

The surface brightness (Eq.~\ref{eq:xray_sb}) was modeled separately with both the isothermal $\beta$-model, using the 
spectroscopically-determined, global $T_X$ (measured within $r \in [0.15,1.0] \, r_{500}$), and the SVM, 
with temperature derived using the ideal gas law from the N07 pressure profile fit to the SZE data 
(as discussed in \S \ref{n07+svm}). Since the SVM was developed explicitly to be jointly fit with some 
form of the generalized NFW pressure profile, requiring both SZE and X-ray data,
the X-ray surface brightness fit for the SVM contains information determined by the N07 
pressure profile, which is primarily determined by SZE data. However, since 
X-ray surface brightness is a much stronger function of density than of temperature (Eq.~\ref{eq:xray_sb}), the SVM 
primarily determines the shape of the surface brightness fit.

Fig.~\ref{fig:Sx_profiles} shows representative, high-likelihood (low $\chi^2$) fits to the surface
brightness of each cluster, for both the SVM and isothermal $\beta$-model.  For plotting purposes, the X-ray data 
are radially-averaged around the cluster centroid, which is determined by fitting the two-dimensional X-ray imaging data 
with the spherically-symmetric SVM and isothermal $\beta$-model profiles.  


\section{Fit Cluster Gas Profiles}\label{fit_params}

The profiles parameterized in the models, which I apply in fitting the SZE visibilities and 
X-ray surface brightness data, are $P_e(r)$,  $n_e(r)$, and $T_e(r)$.  
In this section, I discuss the recovered model fits of the parameters, and compare them
with those fit in the detailed, independent X-ray analysis (\S \ref{maughan_analysis}).
For the N07+SVM fits presented here, metallicity is fixed to 
the best-fit value determined from X-ray spectroscopic data.  

\begin{figure}
\begin{center}
\includegraphics[height=2.22in]{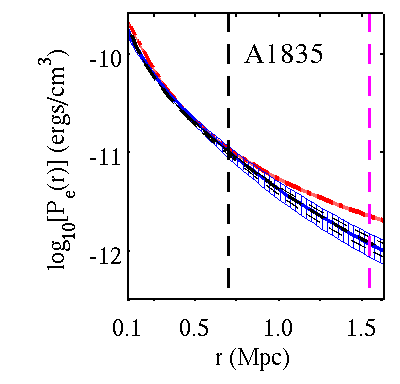}
\includegraphics[height=2.22in]{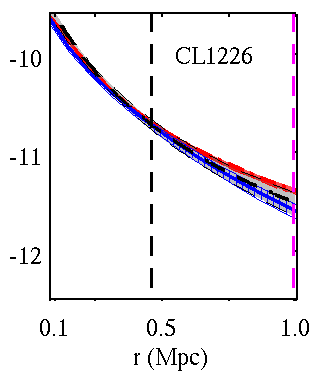}
\includegraphics[height=2.22in]{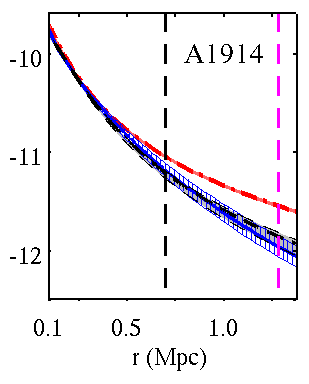}
\end{center}
\caption{$P_e(r)$ for each set of models fit to each cluster. The pressure from the jointly-fit 
N07+SVM is plotted in blue with vertical hatching.  Pressure constrained by the SZE fit of the 
isothermal $\beta$-model is plotted using red, dot-dashed lines; note that the isothermal $\beta$-model's
shape is constrained by X-ray imaging data, and the only unique parameter to the SZE data
in this fit is the central decrement (SZE normalization, see \S \ref{beta_model}).
Pressure derived from the density and temperature fits of the V06 profiles in the independent X-ray 
analysis is shown in black with grey shaded regions.  See text in \S \ref{fit_params} for details.
The vertical, black dashed line shows $r_{2500}$ derived from the N07+SVM fits, while the magenta 
dashed line is for $r_{500}$ (see Tables \ref{table:derivedQuants_r2500} and \ref{table:derivedQuants_r500}).}
\label{fig:pressure}
\end{figure}

Figures \ref{fig:pressure}--\ref{fig:temp} show the three-dimensional ICM radial profiles fit
in the joint analysis of the SZA visibility + \emph{Chandra} imaging data for A1835, CL1226, and
A1914 (from left to right).
In each figure, I compare the results, shown with their respective 68\% confidence intervals, of 
each of the models and fitting procedures tested.
The resulting constraints on $P_e(r)$,  $n_e(r)$, and $T_e(r)$, using the MCMC routine
to determine the probability density of each of the fits, are as follows:

\begin{itemize}

\item {\bf Pressure:} Fig.~\ref{fig:pressure} shows the pressure profile fit to the SZE data (blue for the N07
pressure profile fit jointly with the SVM profile to describe X-ray imaging),
red for the isothermal $\beta$-model), and compares it with pressure derived from the V06 density and temperature
profiles, which is constrained by the independent X-ray analysis (black), which utilizes X-ray spectroscopic
information not used in the N07+SVM fitting.\footnote{The ideal gas law is
assumed when deriving the electron pressure from the V06 density and temperature fits from the independent, 
X-ray-only analysis.}  
For all three clusters, the N07 pressure profile fits agree within their 68\% confidence intervals with 
those predicted by the independent X-ray analysis.

The pressure predicted by the isothermal $\beta$-model is in good agreement with the other 
model fits at $r\lesssim r_{2500}$, but is systematically  higher in the cluster outskirts ($\gtrsim r_{2500}$).  
This is likely to be due to the fact that cluster temperature declines with radius (see Fig.~\ref{fig:temp}),
since the density estimates agree within $r\lesssim r_{500}$ (discussed in next subsection; see Fig.~\ref{fig:density}).
Since the isothermal temperature $T_X$ is emission-weighted, the temperature is biased to favor the inner 
regions of a cluster (even when the core is excluded), where the density (and therefore the emission) is higher.  
The isothermal $\beta$-model predicts a large SZE signal in Fourier modes that are longer (lower \emph{u,v}--radii)
than the SZA interferometer can probe (see Figures \ref{fig:uvplot_a1835}--\ref{fig:uvplot_a1914}).
In contrast, the extra degree of freedom in the N07 model allows it to capture the true shape of the
cluster's pressure profile, since it fits pressure directly without relying on X-ray imaging to constrain 
the pressure profile's shape. The X-ray imaging data are of much higher significance
than the SZE visibility data, so they determine the shape of the isothermal $\beta$-model. 
In contrast, the parameters of the N07 model are only linked to the X-ray fit through the 
temperature derived from the N07+SVM profiles (see \S \ref{n07+svm}).

\begin{figure}
\begin{center}
\includegraphics[height=2.22in]{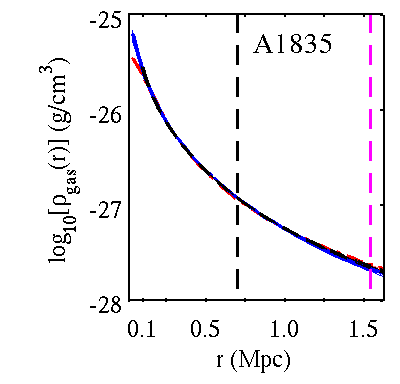}
\includegraphics[height=2.22in]{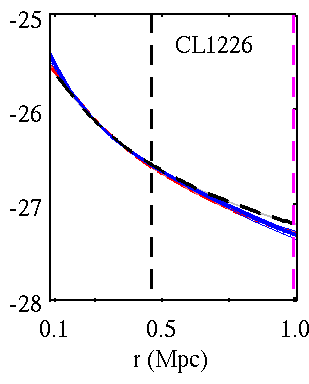}
\includegraphics[height=2.22in]{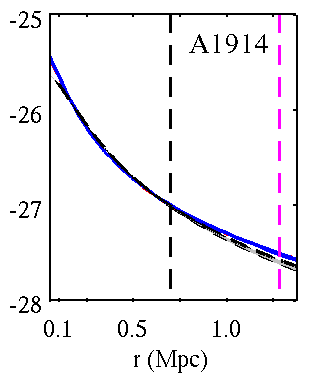}
\end{center}
\caption{$\rho_\text{gas}(r)$ for each set of models fit to each cluster.  
Colors and line styles are the same as in Fig~\ref{fig:pressure}.
See text in \S \ref{fit_params} for details.
Note that the vertical, black and magenta dashed lines show $r_{2500}$ and $r_{500}$, respectively, 
derived from the N07+SVM fits.}
\label{fig:density}
\end{figure}

\item {\bf Density:}  Fig.~\ref{fig:density} shows the density fits to the X-ray imaging data from
each cluster (where the SVM and isothermal $\beta$-model were jointly fit with the SZE data).  
Since the X-ray imaging data provide high significance measurements of $S_X$ 
(see Fig.~\ref{fig:Sx_profiles}), and since they are most sensitive to density 
(see Eq.~\ref{eq:xray_sb}), the X-ray data dominate the fit density.  
Since all three clusters have excellent X-ray data, density is tightly constrained for all 
three types of model fits.  The additional flexibility in the SVM and V06 density profiles, 
which were built to fit departures from the $\beta$-model density profile
in the cluster outskirts, were not necessary for any of these clusters.

For all three clusters, the fits of the SVM density generally agree to within 1-2\% of the full V06 density profile over
the radial range $r \in [100~{\rm kpc}, r_{2500}]$.
Differences in the fit densities arise at larger radii, due to the differing X-ray background fitting 
procedures.\footnote{This could explain Maughan's higher fit V06 density at $r>r_{2500}$ 
(see Fig.~\ref{fig:density}), which would arise if the X-ray background was underestimated; 
as one approaches a signal-to-noise ratio of 1, any systematic errors in the X-ray background 
(noise level) are fit by the cluster density profile. Although it appears negligible 
in the logarithmic plot, any discrepancy integrates to produce non-negligible contribution to, e.g.,
the Compton-$y$ computed for this cluster from the X-ray only analysis (see Fig.~\ref{fig:compy}).}

The strong agreement between the densities derived from the SVM and the isothermal $\beta$-model 
at large radii is due to the fact that they use the same X-ray data and X-ray background levels,
and none of the clusters had density profiles that diverged significantly from a $\beta$-model
for the fitting regions considered here ($r \in [100~{\rm kpc}, r_{500}]$).  
Note that the density fit from the isothermal $\beta$-model is overlaid in the plots by the SVM density
fit (due to the strong agreement).
This supports the argument the SZE data do not influence the shape parameters 
of the isothermal $\beta$-model ($r_c$ and $\beta$, Eq.~\ref{eq:beta}).

\begin{figure}
\begin{center}
\includegraphics[height=2.22in]{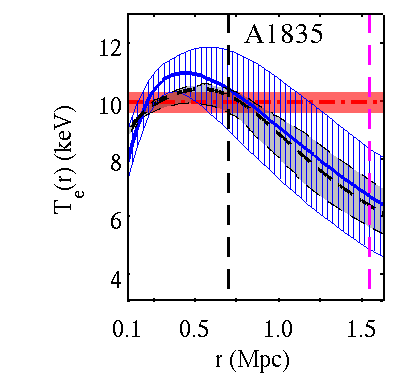}
\includegraphics[height=2.22in]{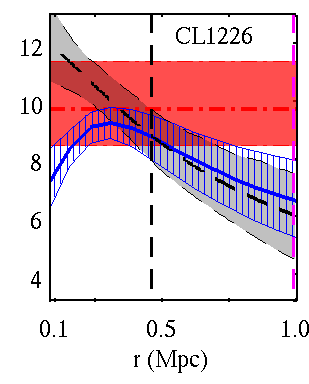}
\includegraphics[height=2.22in]{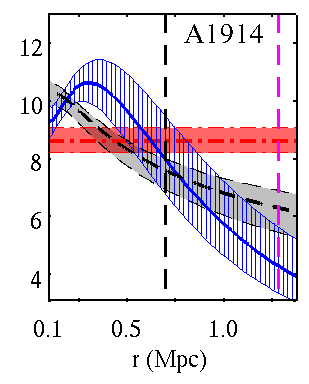}
\end{center}
\caption{$T_e(r)$ for each set of models fit to each cluster.
Colors and line styles are the same as in Fig~\ref{fig:pressure}.
Note that the isothermal $\beta$-model's constant $T_e(r)=T_X$, constrained by the X-ray data, is 
plotted using red, dot-dashed lines and dark red shading.
See text in \S \ref{fit_params} for details.
Note that the vertical, black and magenta dashed lines show $r_{2500}$ and $r_{500}$, respectively, 
derived from the N07+SVM fits.}
\label{fig:temp}
\end{figure}

\item {\bf Temperature:} By not relying on X-ray spectroscopic temperature information when fitting
the N07+SVM profiles to the SZE+X-ray data, $T_e(r)$ can be derived. This derived temperature profile 
provides a means by which we can diagnose how well the X-ray+SZE derived and X-ray spectroscopic 
temperatures independently agree.
As shown in Fig.~\ref{fig:temp}, the N07+SVM derived temperature profiles broadly agree with the 
V06 temperature profile fit to the X-ray spectroscopic data.  For the relaxed
cluster A1835, the median derived N07+SVM temperature profile is well within 1-$\sigma$ of the fit V06 
$T_e(r)$ for nearly all radii outside the 100~kpc core.  For CL1226, which is likely spherically-symmetric
but has not had enough time to form a cool core (as argued in M07), the temperature agrees within 1-$\sigma$
at $r\in[r_{2500},r_{500}]$, but does not agree in the core.  

For the disturbed cluster A1914, the N07+SVM derived $T_e(r)$ agrees with 
Maughan's fit $T_e(r)$ over a larger range than the isothermal $T_X$ does, but fails to capture the overall
slope of $T_e(r)$.  Since the N07 pressure profile was derived using relaxed clusters, it may not be surprising
that the derived $T_e(r)$ could be biased; however, note that $P_e(r)$ does agree for the N07 and Maughan derived
pressure (see Fig.~\ref{fig:pressure}), so the bias in the derived $T_e(r)$ could arise from the assumption
of spherical symmetry being applied to the density profile (which in the X-ray image is clearly clumpy; 
see Fig.~\ref{fig:xray_image_a1914}).

The assumption of isothermality shows good agreement (typically within 1-$\sigma$) within
$r < r_{2500}$ with Maughan's fit of the V06 temperature profile.
As expected, the isothermal $T_X$ does not agree with the more sophisticated analysis
 in the cluster outskirts ($r > r_{2500}$).  
This implies that isothermality is a relatively poor assumption at large radii, as
the temperature universally declines at large radii in fits that have more degrees of freedom.

The radial electron temperature profiles $T_e(r)$ derived from  
the N07+SVM joint-fit model (\S \ref{n07+svm}) reproduce the spectroscopically-determined 
$T_e(r)$ from deep \emph{Chandra} observations of clusters quite
well.  This approach provides a unique and potentially powerful probe
of $T_e(r)$ for high-redshift clusters, for which X-ray spectroscopic 
temperatures are difficult and often expensive to obtain.

\end{itemize}

\section{Derived Cluster Properties}\label{derived_quants}

In Tables \ref{table:derivedQuants_r2500} and \ref{table:derivedQuants_r500}, 
I report the global properties of individual clusters derived from the N07+SVM model fits to the
SZE+X-ray data.  I calculate all quantities (following the 
methods outlined in \S \ref{mass_deriv} \& \ref{sz_quants}) at overdensity radii
$r_{2500}$ and $r_{500}$, and compare them to results from both the
jointfit isothermal $\beta$-model analysis and to the Maughan X-ray-only analysis.

At both $r_{2500}$ and $r_{500}$, the measurements of \Ylos\ derived from the joint
N07+SVM and the X-ray-only analysis are consistent at the 1-$\sigma$
level, for all three clusters.  The isothermal $\beta$-model analysis,
however, overestimates \Ylos\ by $\sim$~20\%--40\% at $r_{2500}$, and
by $\sim$~30\%--115\% at $r_{500}$.  This is due to the large
contribution to \Ylos\ (at every projected radius) from the cluster
outskirts, where the $\beta$-model significantly overestimates the
pressure.  

In contrast, the determinations of \Yvol\ at $r_{2500}$ are generally
consistent among the three analyses, except in the case of CL1226, due to the lower
fit pressure.  The excellent agreement at this radius between \Yvol\
derived using fits of either the isothermal $\beta$-model or the N07 profile 
arises directly from the constraints on SZE flux provided by $Y(u,v)$ (see \S \ref{sze_visfits}).  
At $r_{500}$, however, the median \Yvol\ values from the isothermal $\beta$-model are
$\sim$20\%--60\% higher than either the N07+SVM or M08 results, due to the fact that 
isothermality is a poor description of the cluster outskirts.  With more data points 
probing cluster scales $\sim r_{2500}$, the good agreement between the isothermal $\beta$-model's 
determination of \Yvol\ and that from the N07+SVM fit is expected at $r_{2500}$.
However, with the X-ray data determining the shape parameters of the $\beta$-model (in
particular, the value of $\beta$ that can describe density is too shallow to 
describe pressure in the cluster outskirts), 
the over-constrained isothermal $\beta$-model fails to capture accurately $\Yvol(r_{500})$.

\begin{table}[t]
\centerline{
{\scriptsize
\begin{tabular}{|l|cc|cc|ccc|}
\hline
Cluster & {$\theta_{2500}$} &  {$r_{2500}$} & {$Y_{\rm los}$} & {$Y_{\rm vol}$} &{$M_{\rm gas}$} &{$M_{\rm tot}$} &{$f_{\rm gas}$} \\
~Model fit  & (\arcsec) & (Mpc) & {($10^{-5} {\rm Mpc}^2$)}  & {($10^{-5} {\rm Mpc}^2$)} &{($10^{13} {\rm M_\odot}$)} &{($10^{14} {\rm M_\odot}$)} & \\
\hline
\underline{\bf Abell 1835} & & & & & & & \\[.25pc]
~N07+SVM              & 178$^{+6.0}_{-5.7}$   & 0.70$^{+0.02}_{-0.02}$ & 12.40$^{+1.58}_{-1.28}$ & 8.85$^{+0.89}_{-0.79}$ & 5.98$^{+0.27}_{-0.26}$ & 6.18$^{+0.65}_{-0.57}$ & 0.097$^{+0.005}_{-0.005}$ \\[.25pc]
~Maughan (this work)  & 169$^{+5.5}_{-8.0}$   & 0.66$^{+0.02}_{-0.03}$ & 11.58$^{+0.61}_{-0.67}$ & 7.88$^{+0.49}_{-0.72}$ & 5.77$^{+0.25}_{-0.35}$ & 5.30$^{+0.53}_{-0.72}$ & 0.109$^{+0.009}_{-0.006}$ \\[.25pc]
~isothermal $\beta$-model    & 159$^{+3.0}_{-2.9}$   & 0.62$^{+0.01}_{-0.01}$ & 13.85$^{+0.72}_{-0.67}$ & 7.94$^{+0.43}_{-0.40}$ & 4.96$^{+0.13}_{-0.12}$ & 4.38$^{+0.25}_{-0.24}$ & 0.113$^{+0.004}_{-0.004}$ \\[.5pc]
\underline{\bf CL~J1226+3332.9} & & & & & & & \\[.25pc]
~N07+SVM              & 53.0$^{+1.9}_{-2.0}$  & 0.41$^{+0.01}_{-0.02}$ & ~5.45$^{+0.53}_{-0.50}$ & 3.55$^{+0.37}_{-0.37}$ & 2.92$^{+0.14}_{-0.15}$ & 2.71$^{+0.30}_{-0.30}$ & 0.108$^{+0.008}_{-0.007}$ \\[.25pc]
~\citet{maughan2007b} & 57.3$^{+1.6}_{-1.5}$  & 0.45$^{+0.01}_{-0.01}$ & ~7.57$^{+0.33}_{-0.34}$ & 5.04$^{+0.31}_{-0.28}$ & 3.25$^{+0.14}_{-0.13}$ & 3.41$^{+0.30}_{-0.26}$ & 0.095$^{+0.004}_{-0.004}$ \\[.25pc]
~isothermal $\beta$-model    & 54.7$^{+4.7}_{-4.4}$  & 0.43$^{+0.04}_{-0.03}$ & ~7.37$^{+1.15}_{-1.02}$ & 4.25$^{+0.73}_{-0.64}$ & 2.97$^{+0.34}_{-0.30}$ & 2.98$^{+0.83}_{-0.66}$ & 0.100$^{+0.016}_{-0.013}$ \\[.5pc]
\underline{\bf Abell 1914} & & & & & & & \\[.25pc]
~N07+SVM              & 233$^{+12.8}_{-10.4}$ & 0.68$^{+0.04}_{-0.03}$ & ~8.71$^{+1.52}_{-1.09}$ & 6.66$^{+1.03}_{-0.76}$ & 4.83$^{+0.33}_{-0.27}$ & 5.28$^{+0.91}_{-0.67}$ & 0.092$^{+0.008}_{-0.008}$ \\[.25pc]
~\citet{maughan2008}  & 218$^{+7.1~}_{-5.7~}$ & 0.63$^{+0.02}_{-0.02}$ & ~7.87$^{+0.56}_{-0.55}$ & 5.69$^{+0.37}_{-0.38}$ & 4.64$^{+0.17}_{-0.16}$ & 4.31$^{+0.43}_{-0.33}$ & 0.107$^{+0.005}_{-0.006}$ \\[.25pc]
~isothermal $\beta$-model    & 204$^{+5.7~}_{-5.1~}$ & 0.59$^{+0.02}_{-0.01}$ & 11.28$^{+0.59}_{-0.56}$ & 6.24$^{+0.34}_{-0.32}$ & 4.04$^{+0.15}_{-0.14}$ & 3.52$^{+0.30}_{-0.26}$ & 0.115$^{+0.005}_{-0.005}$ \\[.25pc]
\hline
\end{tabular}
}
}
\caption{$Y_{\rm los}$, $Y_{\rm vol}$, $M_{\rm gas}$, $M_{\rm tot}$, and $f_{\rm gas}$ for each model, computed within each model's 
estimate of $r_{2500}$.}
\label{table:derivedQuants_r2500}
\end{table}

\begin{table}[t]
\centerline{
{\scriptsize
\begin{tabular}{|l|cc|cc|ccc|}
\hline
Cluster & {$\theta_{500}$} &  {$r_{500}$} & {$Y_{\rm los}$} & {$Y_{\rm vol}$} &{$M_{\rm gas}$} &{$M_{\rm tot}$} &{$f_{\rm gas}$} \\
~Model fit  & (\arcsec) & (Mpc) & {($10^{-5} {\rm Mpc}^2$)}  & {($10^{-5} {\rm Mpc}^2$)} &{($10^{13} {\rm M_\odot}$)} &{($10^{14} {\rm M_\odot}$)} & \\
\hline
\underline{\bf Abell 1835} & & & & & & & \\[.25pc]
~N07+SVM              & 365$^{+32}_{-22}$ & 1.42$^{+0.13}_{-0.09}$ & 20.87$^{+4.66}_{-3.00}$ & 17.79$^{+3.72}_{-2.45}$ & 13.66$^{+1.29}_{-0.90}$ & 10.68$^{+3.11}_{-1.85}$ & 0.128$^{+0.017}_{-0.020}$ \\[.25pc]
~Maughan (this work)  & 363$^{+17}_{-12}$ & 1.42$^{+0.07}_{-0.05}$ & 21.37$^{+2.45}_{-1.58}$ & 17.41$^{+1.61}_{-0.99}$ & 13.94$^{+0.64}_{-0.52}$ & 10.68$^{+1.54}_{-1.01}$ & 0.133$^{+0.009}_{-0.012}$\\[.25pc]
~isothermal $\beta$-model    & 361$^{+7~}_{-6~}$ & 1.41$^{+0.03}_{-0.03}$ & 34.53$^{+1.78}_{-1.68}$ & 21.29$^{+1.09}_{-1.02}$ & 13.29$^{+0.27}_{-0.27}$ & 10.30$^{+0.58}_{-0.54}$ & 0.129$^{+0.005}_{-0.005}$ \\[.5pc]
\underline{\bf CL~J1226+3332.9} & & & & & & & \\[.25pc]
~N07+SVM              & 125$^{+12}_{-9}$ & 0.97$^{+0.09}_{-0.07}$ & 11.6$^{+1.9}_{-1.6}$ & ~9.53$^{+1.50}_{-1.25}$ & 8.44$^{+0.64}_{-0.57}$ & ~7.17$^{+2.22}_{-1.48}$ & 0.118$^{+0.021}_{-0.022}$\\[.25pc]
~\citet{maughan2007b} & 115$^{+3~}_{-3~}$ & 0.89$^{+0.02}_{-0.02}$ & 13.9$^{+1.3}_{-1.1}$ & 10.59$^{+0.69}_{-0.68}$ & 8.30$^{+0.32}_{-0.36}$ & ~5.49$^{+0.46}_{-0.47}$ & 0.151$^{+0.008}_{-0.008}$\\[.25pc]
~isothermal $\beta$-model    & 127$^{+10}_{-10}$ & 0.99$^{+0.08}_{-0.08}$ & 18.3$^{+2.7}_{-2.3}$ & 11.91$^{+1.71}_{-1.51}$ & 8.32$^{+0.68}_{-0.60}$ & ~7.49$^{+1.93}_{-1.60}$ & 0.111$^{+0.020}_{-0.016}$\\[.5pc]
\underline{\bf Abell 1914} & & & & & & & \\[.25pc]
~N07+SVM              & 425$^{+37}_{-27}$ & 1.24$^{+0.11}_{-0.08}$ & 12.69$^{+2.76}_{-1.87}$ & 11.01$^{+2.32}_{-1.58}$ & 10.11$^{+0.94}_{-0.67}$ & ~6.38$^{+1.83}_{-1.13}$ & 0.159$^{+0.021}_{-0.024}$\\[.25pc]
~\citet{maughan2008}  & 448$^{+24}_{-21}$ & 1.29$^{+0.07}_{-0.06}$ & 13.47$^{+1.68}_{-1.77}$ & 10.78$^{+1.03}_{-1.09}$ & 10.24$^{+0.45}_{-0.57}$ & ~7.49$^{+1.29}_{-1.00}$ & 0.138$^{+0.015}_{-0.018}$\\[.25pc]
~isothermal $\beta$-model    & 461$^{+13}_{-11}$ & 1.34$^{+0.04}_{-0.03}$ & 29.08$^{+1.52}_{-1.44}$ & 17.07$^{+0.87}_{-0.84}$ & 11.05$^{+0.36}_{-0.33}$ & ~8.14$^{+0.69}_{-0.59}$ & 0.136$^{+0.007}_{-0.007}$\\[.25pc]
\hline
\end{tabular}
}
}
\caption{$Y_{\rm los}$, $Y_{\rm vol}$, $M_{\rm gas}$, $M_{\rm tot}$, and $f_{\rm gas}$ for each model, computed within each model's
estimate of $r_{500}$.}
\label{table:derivedQuants_r500}
\end{table}

\begin{figure}
\centerline{\includegraphics[width=4in]{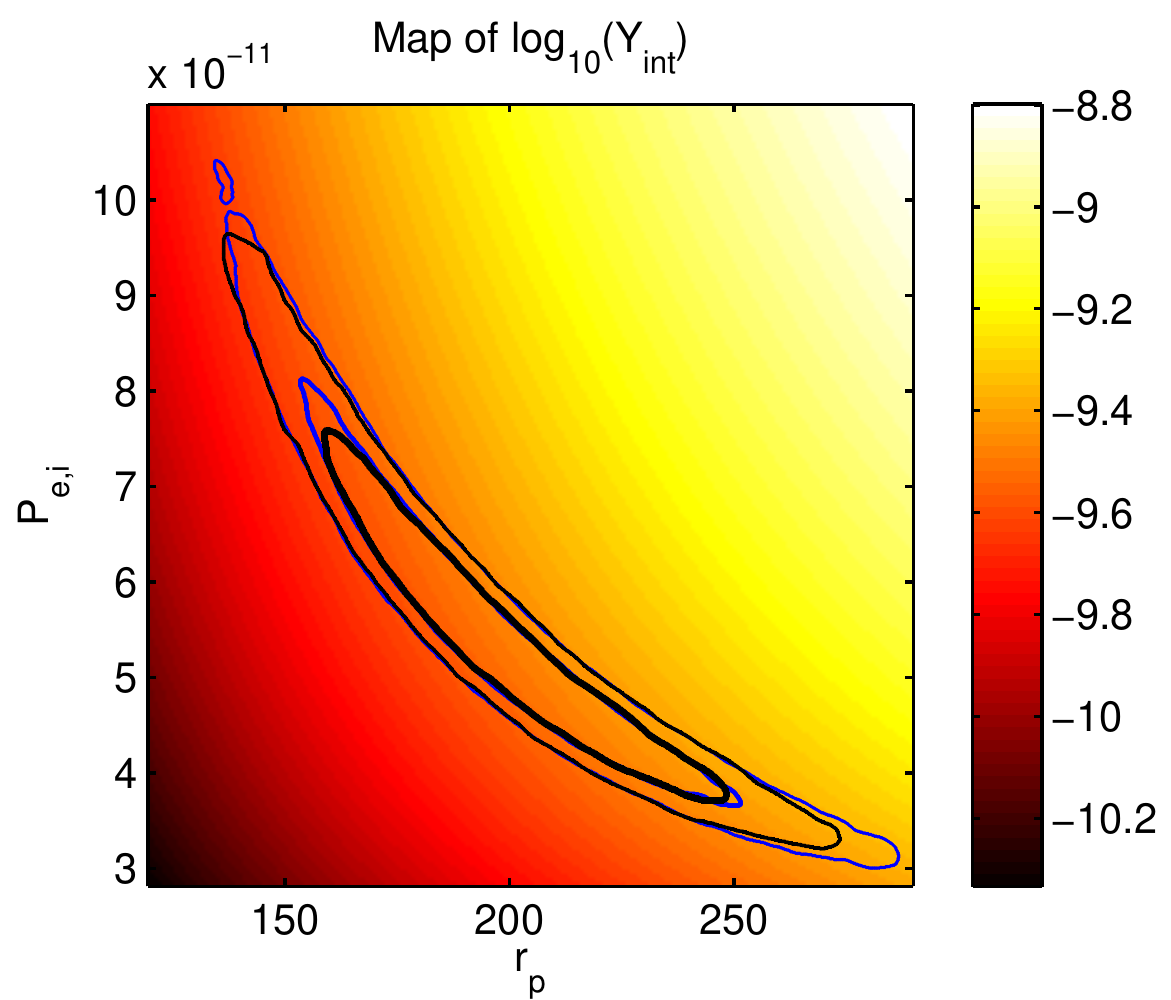}}
\caption{\Ylos\ computed within $6^\prime$ ($\sim 1.4~\rm Mpc$, which is $\approx r_{500}$
for this cluster) for fits to SZA observation of A1835. The bold, black contours contain 68\% and
95\% of the accepted iterations to the jointly-fit {\em Chandra} + SZA data, while the
thinner, blue contours are those for fits to SZA data alone.
Note that the vertical, black and magenta dashed lines show $r_{2500}$ and $r_{500}$, respectively, 
derived from the N07+SVM fits.}
\label{fig:a1835_yint_joint_SZ_degen}
\end{figure}

The N07 pressure profile (Eq.~\ref{eq:N07}) -- fit to SZE data alone
or fit jointly to SZE+X-ray data -- has just two free parameters to 
describe the SZE data: $P_{e,i}$ and $r_p$.  
In the joint fit N07+SVM, these parameters are only linked 
to the X-ray imaging data through the derived temperature (see discussion 
in \S \ref{n07+svm}). 
Figure~\ref{fig:a1835_yint_joint_SZ_degen} shows the
degeneracy between $P_{e,i}$ and $r_p$ when fitting the SZA observations of A1835,
comparing the SZE-only and the joint SZE+X-ray fits of the N07 profile. 
This degeneracy is similar to that between $r_c$ and $\beta$, when fitting the $\beta$-model
to SZE data alone \citep[see][for example]{grego2001}. These two quantities are not
individually constrained by our SZE observations, but they are tightly
correlated.  The preferred region in the $P_{e,i}-r_p$ plane encloses
approximately constant \Ylos. As a result, the 68\%
confidence region allows less than $\pm$25\% variation in \Ylos, 
despite the much larger variation individually in $P_{e,i}$ and $r_p$.  Since \Ylos\
at large radii scales with the cluster SZE flux probed by an interferometer 
(e.g.\ $Y(u,v)$ discussed in \S \ref{sze_visfits}), this parameter is most directly
constrained when fitting models to the SZE data.

The gas mass estimates -- computed by integrating the gas density fit with either 
the jointly-fit SVM or the isothermal $\beta$-model (as discussed in \S \ref{gas_mass}) 
-- agree with the gas mass estimates derived from the Maughan X-ray fits 
(Tables \ref{table:derivedQuants_r2500} and \ref{table:derivedQuants_r500}).  This agreement is not surprising,
given that the gas mass is determined in all cases from density fits
to the X-ray surface brightness data.  It demonstrates, however, that the 100~kpc core
makes a negligible contribution to the total gas mass even at
$r_{2500}$, and that excluding the core from the joint analysis does
not therefore introduce any significant bias in our estimate of
\Mgas. Incidentally, it also shows that the additional components
in the SVM and the full V06 density models were not necessary to fit
these clusters.

Tables \ref{table:derivedQuants_r2500} and \ref{table:derivedQuants_r500} also present estimates of the total
masses, computed using each model's estimate of the overdensity radius
($r_\Delta$, Eq.~\ref{eq:r_delta}) for each Monte Carlo realization of the fit parameters.
For two of the clusters, I find that the error bars are significantly
larger for \Mtot\ determined from the N07+SVM fits than for the
isothermal $\beta$-model or M08 fits.  This is a consequence of the fact
that the $\beta$-model analysis, with fewer free parameters and the
assumption of isothermality, typically places strong but poorly-motivated
priors on the total mass, while the M08 fits make use of the spatially
resolved temperature profile afforded by the deep X-ray observations.
CL1226 is the exception, with both the N07+SVM and M07 constraints on \Mtot\
being tighter than those provided by the isothermal $\beta$-model; the N07+SVM
fits rely on high-significance SZE constraints provided by combining 30+90~GHz SZA data, 
while the M07 constraints are narrowed by using both \emph{XMM-Newton} and \emph{Chandra}
spectroscopy. 
I find that the N07+SVM and M08 total mass estimates broadly agree
at both $r_{2500}$ and $r_{500}$, leading to good overall agreement
between gas fractions computed using the N07+SVM profiles and those
from the Maughan X-ray fits.
As discussed in \S \ref{fgas}, I compute the gas mass fraction \fgas\ in 
a way that takes advantage of the fact that MCMC explores the probability 
density distribution of model fits to the data.
Both the Maughan V06 and N07+SVM estimates of \fgas\ within $r_{2500}$ are 
all consistent with the constant fraction found in \citet{allen2004, allen2007}.
The isothermal $\beta$-model, on the other hand, is too constrained to
agree with the non-isothermal fits at both $r_{2500}$ and $r_{500}$; 
its estimate of \Mtot\ is moreover sensitive to the annulus within
which $T_X$ is determined.  This trend can also be seen in Figure \ref{fig:mtot}, 
which shows \Mtot$(r)$ for each cluster.

\begin{figure}
\begin{center}
\includegraphics[height=2.22in]{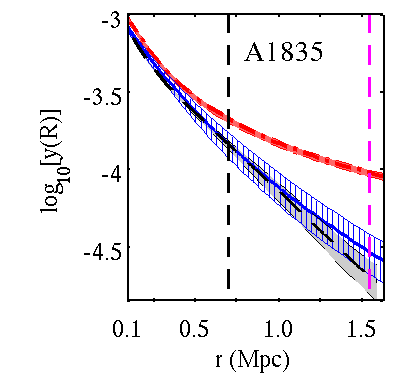}
\includegraphics[height=2.22in]{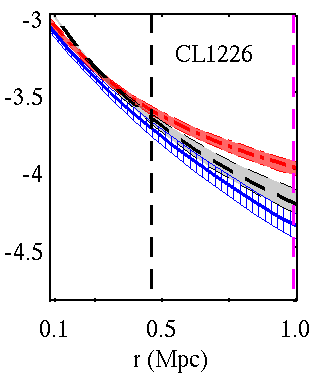}
\includegraphics[height=2.22in]{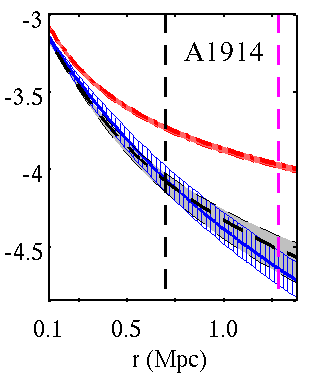}
\end{center}
\caption{$y(R)$ --  Compton $y$ (integrated along the line of sight, Eq.~\ref{eq:compy}) as a function 
of sky radius $R$.  Colors and line styles are the same as in 
Fig~\ref{fig:pressure}.  See text in \S \ref{fit_params} for details.
Note that the vertical, black and magenta dashed lines show $r_{2500}$ and $r_{500}$, respectively, 
derived from the N07+SVM fits.}
\label{fig:compy}
\end{figure}

In addition to the data presented in Tables \ref{table:derivedQuants_r2500} and \ref{table:derivedQuants_r500},
which only show parameters computed at each model's estimate of the overdensity radii $r_{2500}$ and $r_{500}$,
I plot the derived cluster profiles and discuss them here:

\begin{itemize}

\item {\bf Compton $y$:} Figure \ref{fig:compy} shows $y(R)$ (integrated along the line of sight through
the cluster, Eq.~\ref{eq:compy}), for sky radius $R$, from the isothermal $\beta$-model
and the N07 model, and compares it to $y(R)$ derived from Maughan's fit density and temperature profiles.  This quantity
is simply the line-of-sight integral of electron pressure (see Eq.~\ref{eq:compy}).  It is worth noting that the excess
pressure predicted by the $\beta$-model can translate to an over-prediction of $y(R)$ even at the cluster core.  Because
this excess is nearly constant over sky radius, it is not constrained by an interferometer, which is not sensitive to
scales larger than a baseline probes (see discussion in \S \ref{interf}).
$y(R)$ from the N07 fits to A1835 and A1914 agrees well with that derived from the independent X-ray analysis,
but disagrees to the extent that the N07 pressure drops off more rapidly for CL1226 than the X-ray-derived pressure.  
Simply put, the V06 density could be biased by assumptions about the X-ray background, leading to an overestimate
of the density at large radii, which would contribute to the X-ray estimate of $y(R)$
(see discussion of the density fits in \S \ref{fit_params}).  While this may be disconcerting, the reader should
note this is one of the drawbacks when using quantities integrated along an infinite sight line through the cluster;
if the profile does not decline rapidly enough, the overestimate at large radii yields a non-negligible contribution.

\begin{figure}
\begin{center}
\includegraphics[height=2.22in]{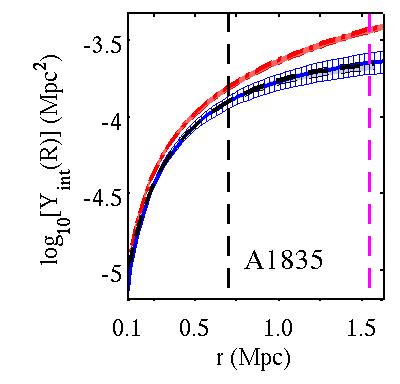}
\includegraphics[height=2.22in]{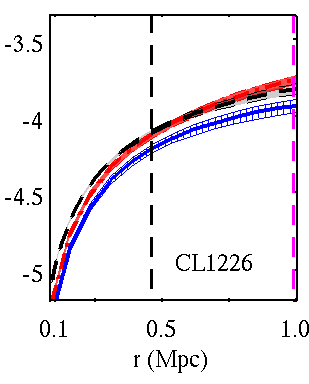}
\includegraphics[height=2.22in]{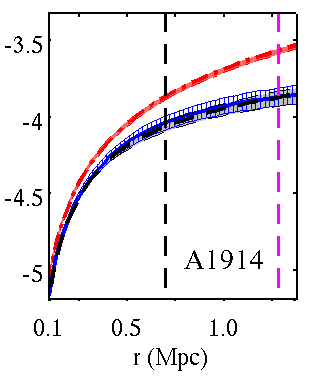}
\end{center}
\caption{\Ylos\ -- Compton $y$ integrated over sky radius $R$.  Colors and line styles are the same as in 
Fig~\ref{fig:pressure}.  See text in \S \ref{fit_params} for details.
Note that the vertical, black and magenta dashed lines show $r_{2500}$ and $r_{500}$, respectively, 
derived from the N07+SVM fits.}
\label{fig:yint_los}
\end{figure}

\item {\bf \Ylos\ -- the line-of-sight Integrated Compton $y$, integrated over
an area of the sky:} 
Figure \ref{fig:yint_los} shows the SZE scaling quantity \Ylos\, integrated along 
the line of sight through the cluster and within a region of the sky (see \S \ref{yint_deriv}, Eq.~\ref{eq:Yint}).  
Given the isothermal $\beta$-model's $y(R)$, and the fact that the integral for \Ylos\
diverges for $\beta<1$, it is unsurprising that there is a large systematic difference, 
at large radii, between the models' estimates.  
For A1835 and A1914, the estimates of $\Ylos(R)$ from the SVM+N07 fits agree 
with those calculated from Maughan's fits, which is expected given the agreement in $y(R)$.


\begin{figure}
\begin{center}
\includegraphics[height=2.22in]{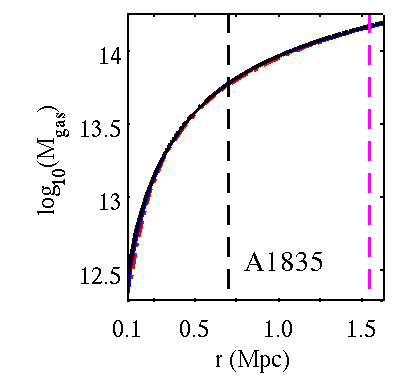}
\includegraphics[height=2.22in]{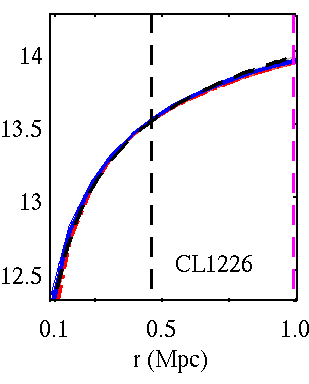}
\includegraphics[height=2.22in]{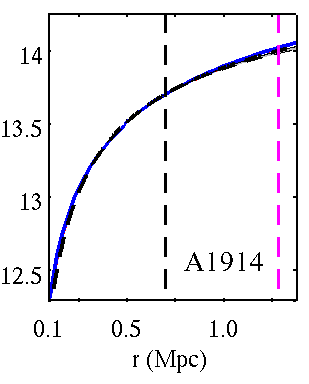}
\end{center}
\caption{\Mgas\ -- the gas mass integrated within a spherical volume defined by cluster radius $r$. 
Colors and line styles are the same as in Fig~\ref{fig:pressure}.  See text in \S \ref{fit_params} for details.}
\label{fig:mgas}
\end{figure}

\item {\bf \Mgas -- Gas Mass:} Figure \ref{fig:mgas} shows the gas mass obtained by integrating the gas density 
within a spherical volume (using Eq.~\ref{eq:gasmass}).  Since the densities obtained using each method of fitting 
were similar (see Fig.~\ref{fig:density}), \Mgas(r) obtained from each method is also similar.
Note, however, that the independent X-ray analysis also modeled the cluster core, which contributes negligibly 
to the overall gas mass at large radii (such as $r_{2500}$ and $r_{500}$).

\begin{figure}
\begin{center}
\includegraphics[height=2.22in]{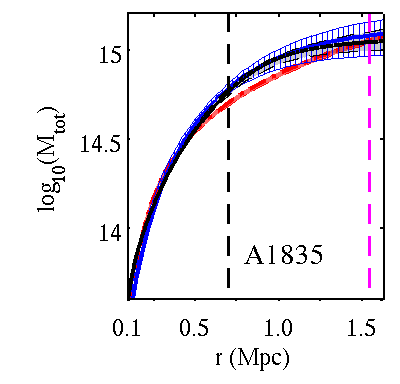}
\includegraphics[height=2.22in]{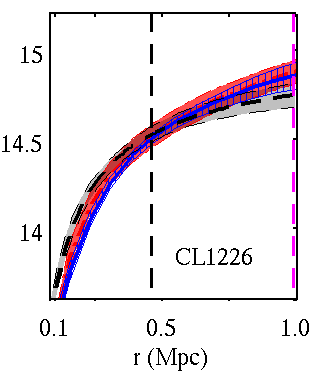}
\includegraphics[height=2.22in]{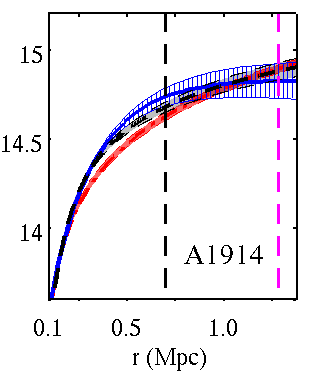}
\end{center}
\caption{\Mtot\ -- the total mass estimated assuming hydrostatic equilibrium at cluster radius $r$. 
Colors and line styles are the same as in Fig~\ref{fig:pressure}.  See text in \S \ref{fit_params} for details.}
\label{fig:mtot}
\end{figure}

\begin{figure}
\centerline{\includegraphics{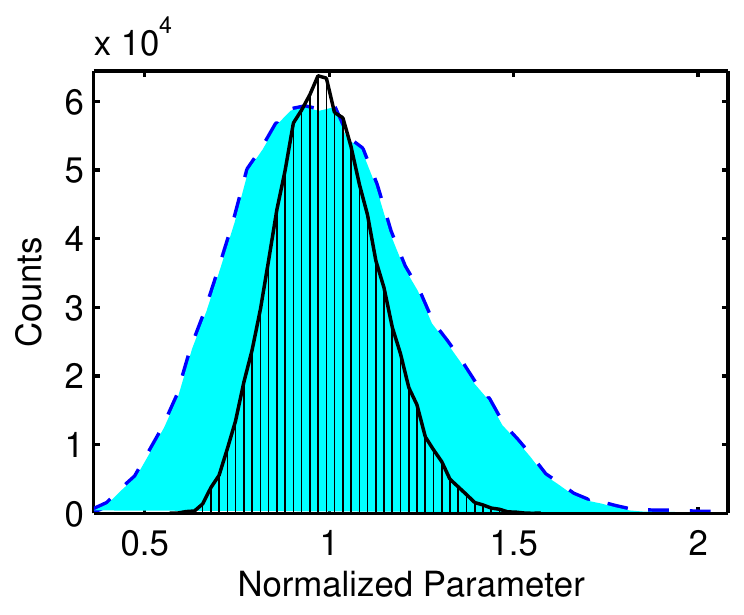}}
\caption{1-D histograms of the N07+SVM jointfit estimates, for A1835, of \Mtot\ and \Ylos\, 
normalized by their respective median values,
\Mtot is the cyan region with a dashed outline, and \Ylos\ is the vertically 
hatched region with a solid black outline.
Both \Mtot\ and \Ylos\ are computed within a fixed radius of $\theta=360\arcsec$.  
The derived \Ylos, which scales with integrated SZE flux, has a more tightly constrained 
and centrally peaked distribution than that of \Mtot, as \Mtot\ is sensitive to the change 
in slope of the pressure profile (see Eqs.~\ref{eq:N07_dPdr} \& \ref{eq:hse}).
Since the N07 model was developed primarily to recover \Yint\ from SZE observations, it
is useful to note how well this model performs in this capacity.} 
\label{fig:1dhists}
\end{figure}

\item {\bf \Mtot -- Total Mass:} Figure \ref{fig:mtot} shows the total mass profile 
estimated by applying hydrostatic equilibrium to each model (Eq.~\ref{eq:hse}).
The total mass from the N07+SVM agrees with that from the independent X-ray analysis
for most cluster radii, for all three clusters.  In contrast, the isothermal $\beta$-model
disagrees for A1835 and A1914 over a large range.
Due to the steepness of the derived N07+SVM temperature profile
of A1914 (see Fig.~\ref{fig:temp}), its estimate of \Mtot\ flattens before $r_{500}$.

Since the density fits are all well-constrained (see Fig.~\ref{fig:density}), 
the error bars from the isothermal $\beta$-model and the V06 fits performed by Ben Maughan 
arise almost entirely from the temperature constraints (including constraints on the derivative 
of temperature), while the error bars on \Mtot\ from the N07+SVM fits arise from 
those on the pressure  profile and its derivative.
Fig.~\ref{fig:1dhists} shows a comparison of the ability of the N07+SVM, when fit to X-ray imaging and
SZE data, to constrain \Mtot\ and \Ylos.  The sparsely sampled \emph{u,v}-data provide poorer
constraints on $dP/dr$ than they do on \Ylos, which scales with SZE flux.  The relatively poor constraints
on $dP/dr$ translate directly to error bars on \Mtot.

Unlike the N07+SVM, the isothermal $\beta$-model lacks the flexibility (by definition) to model a gas temperature
profile that varies over the cluster's radius.   In Fig.~\ref{fig:mtot}, one can see that the total mass estimate 
from the isothermal $\beta$-mode never agrees with the independent, detailed X-ray-only analysis at both 
$r_{2500}$ and $r_{500}$.
Where the isothermal $\beta$-model does agree depends mainly on the region within which $T_X$ was measured.  
Since the HSE estimate of \Mtot\ depends on the slope of the pressure profile (as discussed in \S \ref{eq:hse}),\footnote{
Assuming the ideal gas law, as we do when performing the HSE estimate of \Mtot, $dP/dr = n k_B (dT/dr) + k_B T (dn/dr)$.}  
the fact that the isothermal $\beta$-model assumes a constant temperature 
($dT/dr = 0$), but recovers density well (e.g. Fig.~\ref{fig:density}), indicates the isothermal $\beta$-model 
will provide a systematically-biased estimate of \Mtot\ at some radii.

\begin{figure}
\begin{center}
\includegraphics[height=2.22in]{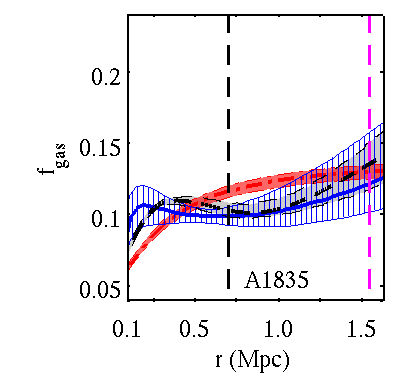}
\includegraphics[height=2.22in]{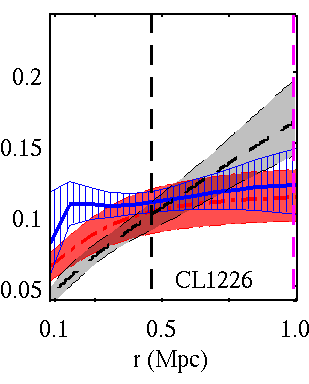}
\includegraphics[height=2.22in]{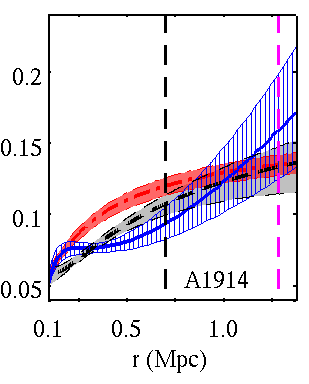}
\end{center}
\caption{\fgas\ -- the gas mass fraction, computed as $\Mgas/\Mtot$ at cluster radius $r$ for each accepted
iteration in the Markov chain. 
Colors and line styles are the same as in Fig~\ref{fig:pressure}.  See text in \S \ref{fit_params} for details.}
\label{fig:fgas}
\end{figure}

\item {\bf \fgas\ -- Hot Gas Fraction:} Figure \ref{fig:fgas} shows the gas fraction, computed using each accepted 
MCMC iteration's prediction for \Mgas\ and \Mtot (as discussed in \S \ref{fgas}). 
The N07+SVM estimates of \fgas\ at intermediate radii ($r \sim r_{2500}$) are consistent with 
previous, X-ray-only results \citep[such as][]{allen2004, allen2007}.
The flatness of the \Mtot\ profile derived from N07+SVM fits to A1914 leads 
to the prediction that \fgas$_{,500} > \Omega_b/\Omega_M \approx 0.165$ \citep[][assuming $h=0.7$]{hinshaw2008}, 
which seems unlikely.  This suggests that the N07+SVM estimate of \Mtot\ for A1914 is too low at $r\sim r_{500}$,
due to the steeply declining temperature profile discussed above.

\end{itemize}

\section{Conclusions}\label{conclusions}

I have applied a new model for the ICM pressure profile -- motivated both 
by theory and detailed cluster observations -- to fit the SZE signal from three
galaxy clusters.  
I have also developed and tested a complementary density model -- a simple extension 
to the $\beta$-model -- that can be used jointly with this new SZE model to fit 
X-ray observations. I show the new pressure profile accurately captures the bulk 
properties of relaxed clusters outside the core, and out to $r_{500}$. 

I also argue that this new model should supplant the 
isothermal $\beta$-model when attempting to use SZE data -- interferometric or otherwise -- to determine cluster 
SZE scaling relations.  The additional degree of freedom, versus the isothermal $\beta$-model, 
in the N07 pressure profile allows it to more accurately describe a cluster 
that has been only sparsely sampled in \emph{u,v}-space.

Finally, the derived temperature from the N07+SVM could prove to be a useful tool, either
in independently confirming X-ray spectroscopic temperature measurements, or in measuring
the temperatures of high-redshift clusters for which sufficiently deep X-ray exposures are 
unavailable, and are difficult to obtain.  I present in Chapter~\ref{Tsl_extension}
a way in which X-ray spectroscopic data can be used to provide additional constrains in the 
context of the N07+SVM profile.

\chapter{Extensions to the models\label{Tsl_extension}}

Throughout the cluster analysis presented in Chapter~\ref{model_application}, I tested the new pressure 
and density models by discarding all X-ray spectroscopic temperature information from the joint SZE+X-ray 
fit, and using this information as an independent test.
I used the SZE constraints on pressure to derive temperature, assuming the ideal gas law and 
a fixed \LCDM\ cosmology with $\Omega_{\rm M}=0.3$, $\Omega_\Lambda=0.7$, $\Omega_{\rm k}=0$.  
In this section, I consider what additional tests can be performed by including spectroscopic 
temperature information. 

For these proposed tests, we will choose a sample of relaxed clusters comprising many of the 
$\delta>-10^\circ$ clusters with publicly-available \emph{Chandra} observations.
By selecting apparently relaxed clusters, the impact of assuming spherical symmetry can be 
reduced.  

\section{Using X-ray Spectroscopic Data}

Medium exposure observations of high redshift clusters typically provide too few photons to constrain
their temperature profiles in detail.  The most robustly-determined spectroscopic temperature
for any cluster is a single, global $T_X$.  Measure $T_X$ within a core cut annulus.
This choice attempts simply to probe the more self-similar portions of a cluster, avoiding the systematic 
discrepancies between ``cool-core'' and ``non-cool-core'' clusters (see, for example, \cite{kravtsov2006,maughan2007}, 
where a core cut was used to make $Y_X$ and $L_X$ more robust proxies for \Mtot).

The observable, emission-weighted spectroscopic temperature $T_X$ can be predicted if a
cluster's density and temperature distributions are somehow known (e.g. for a simulated cluster,
or when using the N07+SVM profile without including spectroscopic information). 
\citet{mazzotta2004} did precisely this, and verified this method is reliable for temperatures $T\gtrsim3.5~\rm keV$.
\citet{vikhlinin2006a} provides a simple, integral form for the ``spectroscopic-like'' temperature $T_{\rm sl}$,
which for spherically-symmetric profiles reduces to
\begin{equation}
T_{\rm sl} = 
\frac{\int_{r_{\rm min}}^{r_{\rm max}} n^2(r) \, T(r)^{1-\alpha} \, r^2 \, dr}
{\int_{r_{\rm min}}^{r_{\rm max}} n^2(r) \, T(r)^{ -\alpha} \, r^2 \, dr}, 
\label{eq:Tsl}
\end{equation}
where the temperature weighting factor $\alpha=0.75$.

Using the ideal gas law, and assuming the electron temperature is equal to the overall ICM temperature
($T_e(r)=T(r)$), we can derive $T_{\rm sl}$ from fits of the N07+SVM profiles to X-ray and SZE imaging
alone.  Eq.~\ref{eq:Tsl} becomes
\begin{equation}
T_{\rm sl} = k_B^{-1}
\frac{\int_{r_{\rm min}}^{r_{\rm max}} n^{1+\alpha}(r) \, P(r)^{1-\alpha} \, r^2 \, dr}
{\int_{r_{\rm min}}^{r_{\rm max}} n^{2+\alpha}(r) \, P(r)^{ -\alpha} \, r^2 \, dr}.
\label{eq:Tsl_sz}
\end{equation}
By ensuring $T_{\rm sl}$ is computed over the same volume within which $T_X$ was measured, 
one can include the likelihood that $T_X = T_{\rm sl}$ in the MCMC fitting process 
(described in \S \ref{markov}).  This allows for the tests proposed in the following
sections, which rely on the combination of SZE data with X-ray imaging and spectroscopic data.\footnote{
Note that using $T_X = T_{\rm sl}$ does not force the temperature profile $T(r)$ to be isothermal;
it is simply stating that, with a known $T(r)$ and $n(r)$, one can compute the emission-weighted 
temperature $T_X$ that would be measured within a given region of an X-ray observation.}

\section{Refining Constraints on \Mtot\ and \fgas}

While the ICM gas density is well-constrained by X-ray imaging, large uncertainties in 
cluster pressure and temperature remain when using either X-ray spectroscopically-measured 
or X-ray+SZE-derived temperature alone (see \S \ref{fit_params}). 
These uncertainties dominate the uncertainty in \Mtot\ estimates (assuming systematics in the HSE
mass determination can be accounted for).  However, one can use the combined data to constrain 
more tightly the observable properties of a cluster, using the direct SZE pressure measurements to complement the
X-ray-derived pressure.  The combination of high-significance X-ray and SZE data on relaxed, 
spherically-symmetric clusters will allow for detailed astrophysical measurements of these systems.

\begin{figure}
\begin{center}
\includegraphics{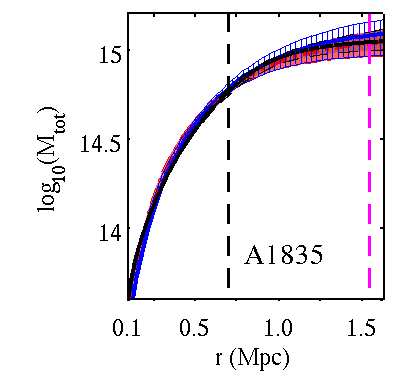}
\includegraphics{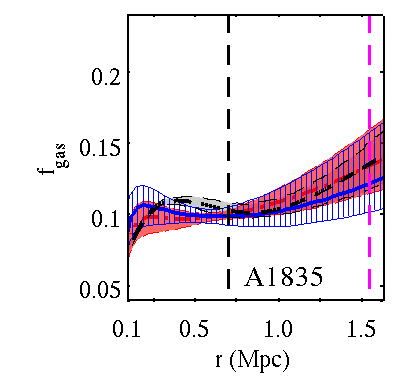}
\end{center}
\caption{Constraints on $\Mtot(r)$ and $\fgas(r)$ using the N07+SVM combined with
the spectroscopically-measured temperature, using $T_X=T_{\rm sl}$, to fit A1835.
The spectroscopically-measured $T_X$ used here is the same
as that used in the isothermal $\beta$-model analysis presented in Chapter~\ref{model_application}.  
The N07+SVM results that include X-ray spectroscopy are plotted using red, 
dot-dashed lines and red shading.
The N07+SVM results without spectroscopic constraints (i.e. the resulted detailed in 
Chapter~\ref{model_application}) are plotted in blue with vertical blue hatching.  
Results derived from the density and temperature fits of the V06 profiles, from the independent 
X-ray analysis, are shown in black with grey, shaded regions.
Both panels show that including spectroscopic information in fits of N07+SVM profile tightens
constraints and improves the already remarkable agreement between it and the independent X-ray analysis.}
\label{fig:Tsl}
\end{figure}

For these studies of cluster astrophysics, we continue to assume a fixed $d_A$ for each cluster.
A test of how well this performs is presented in Figure \ref{fig:Tsl}.


\section{Sensitivity to Angular Diameter Distance ($d_A$)}\label{Tsl_dA}

Rather than constraining \Mtot\ more precisely, we can relax the assumption of cosmology
by including spectroscopic information in the joint fit.  We consider here the cosmological
sensitivity of our fits to the data.

The Compton $y$ parameter (Eq.~\ref{eq:compy_Pe}), which scales as electron pressure integrated along sight line $\ell$, 
has an inverse linear dependence on $d_A$, since $d\ell = d_A \, d\theta$:
\begin{equation}
y = \frac{\sigT}{m_e c^2} \int \!\! P_e \,d\ell
~~~
\Rightarrow 
~~~
P_e \propto d_A^{-1}.
\label{eq:compy_dA}
\end{equation}
Because X-ray imaging is sensitive to the surface brightness $S_X$ (Eq.~\ref{eq:xray_sb}), X-ray-constrained density is
dependent on $d_A^{-1/2}$:
\begin{equation}
S_X = \frac{1}{4\pi (1+z)^4} \! \int \!\! n_e^2 \Lamee(T_e,Z) \,d\ell 
~~~
\Rightarrow 
~~~
n_e \propto {d_A}^{-1/2}.
\label{eq:xray_sb_dA}
\end{equation}
Combining these dependencies, we see
\begin{equation}
\label{eq:Tsl_dA}
T_{\rm sl} \propto P_e/n_e \propto d_A^{-1/2}.
\end{equation}
The spectroscopically-measured temperature $T_X$ does not have any 
dependence on $d_A$, so setting $T_X = T_{\rm sl}$ results in a ${d_A}^{-1/2}$ sensitivity 
to cosmology.  
This is the same cosmological sensitivity exploited in previous, joint X-ray+SZE $d_A$ 
determinations such as \citet{bonamente2004}, but relies on a more sophicated pair of models
for the cluster gas.

\begin{table}[t]
\centerline{
\begin{tabular}{|l|cc|ccc|}
\hline
{Cluster} & $z$ & $d_A$\tablenotemark{a} & $T_X$\tablenotemark{b} & $d_A$\tablenotemark{c} & $T_{\rm sl} $\\ 
 & & (Mpc) & (keV) & (Mpc) & (keV) \\
\hline 
& & & \\[-0.95pc]
A1835  & 0.25 & ~806.5 & $9.95^{+0.36}_{-0.37}$ & $~867^{+140}_{-152}$ & $9.99^{+0.57}_{-0.54}$\\[.2pc]
CL1226 & 0.89 & 1601.8 & $9.30^{+1.33}_{-1.25}$ & $1407^{+444}_{-321}$ & $9.44^{+1.36}_{-1.08}$\\[.2pc]
\hline
\end{tabular}
}
\caption{Clusters chosen for testing the ability of the upgraded N07+SVM profile to constrain
cosmology.}
\tablenotetext{a}{Computed angular diameter distance, assuming $\Omega_M=0.3$, $\Omega_\Lambda=0.7$, \& $\Omega_k=0$.}
\tablenotetext{b}{Global X-ray spectroscopic temperatures were determined in the range 
$r \in [0.15,1.0] \, r_{500}$.}
\tablenotetext{c}{Angular diameter distance fit to the SZE + X-ray imaging and spectroscopic data.}
\label{table:Tsl_dA}
\end{table}

The inverse square root cosmological sensitivity of a joint SZE+X-ray fit to a cluster implies that, 
for example, a single cluster observation with $\pm 10\%$ uncertainty in the measurements of 
$T_{\rm sl}$ and $T_X$ (the measurements of which should be independent) can constrain $d_A$ 
to within $\pm 37.6\%$. 
Here, I have added the errors on $T_X$ and $T_{\rm sl}$ in quadrature, 
since constraints provided by X-ray spectroscopic and the temperature derived from 
X-ray+SZE imaging should be independent.
A determination of the angular diameter distance benefits from a large sample of clusters 
\citep[e.g.][]{molnar2002, bonamente2006}, and we 
can expect a sample of $\sim 40$ relaxed clusters to yield $\sim 5\%$ statistical uncertainty on 
$d_A$ (where I have ignored systematics due to cluster asphericity and systematic effects in
X-ray temperature measurements; see for example \cite{ameglio2006}).

I tested this method on the relaxed clusters A1835 and CL1226, using the global $T_X$, measured within
$r \in [0.15,1.0] \, r_{500}$, from Chapter~\ref{model_application}.  The results of this test
are consistent with \LCDM and are shown in Table~\ref{table:Tsl_dA}.

It has been suggested (Alexey Vikhlinin, private communication) that a comparison between the X-ray
proxy $Y_X \equiv \Mgas T_X$ and the SZE quantity \Yvol\ could be used to constrain $d_A$ more robustly 
than previous X-ray+SZE attempts, since both $Y_X$ and \Yvol\ are integrated quantities that can be
robustly determined.  
However, since the N07+SVM profiles share a density profile, simple inspection of Eq.~\ref{eq:Tsl} 
reveals that setting $T_X = T_{\rm sl}$ is in fact the correct implementation of comparison between
$Y_X$ and \Yvol. The X-ray spectroscopic $T_X$ and the derived $T_{\rm sl}$ contain the same weighting 
by density, while $Y_X$ and \Yvol are not in general equivalent.  
Since \Yvol\ is the integral of pressure, it scales as $\Mgas T_{\rm mw}$, 
where $T_{\rm mw}$ is the mass-weighted temperature.  In general, the emission-weighted temperature $T_X$ is not
equivalent to the mass-weighted temperature, as $T_X$ favors denser regions more
heavily than $T_{\rm mw}$ does, and clusters are not isothermal.

Recently, \citet{allen2007} used X-ray observations of clusters to constrain $d_A$ by assuming
the gas fraction $\fgas(r_{2500})$ is constant with redshift.  They find $\fgas = \Mgas/\Mtot \propto
d_A^{3/2}$, which is far more sensitive to $d_A$ than the line-of-sight SZE+X-ray joint fit 
when no assumptions are made about the evolution of \fgas.  

For SZE+X-ray joint fits without spectroscopy, \Mgas\ has same dependence as it does when constrained by
X-ray data alone, which is
\begin{equation}
M_{\rm gas} \propto n_e V \propto d_A^{5/2}.
\end{equation}
Volume is proportional to the cube of angular diameter distance ($V \propto d_A^{3}$), 
and density is proportional to the inverse square-root of $d_A$ 
($n_e \propto {d_A}^{-1/2}$, as implied by Eq.~\ref{eq:xray_sb_dA}).  
For the \Mtot\ cosmological dependence, we have a slightly different cosmological dependence than 
that from X-ray alone:
\begin{equation}
M_{\rm tot}(r) = - \frac{r^2}{G \rho_{\rm gas}(r)} \left(\frac{dP_{\rm gas}}{dr}\right) 
\propto \frac{d_A^2}{d_A^{-1/2}} d_A^{-1} = d_A^{3/2},
\end{equation}
using $r = \theta \, d_A$ and Eqs.~\ref{eq:compy_dA} \& \ref{eq:xray_sb_dA}.  Since
the SZE-determined pressure exhibits a different dependence on cosmology than the 
X-ray-determined pressure, we arrive at
\begin{equation}
f_{\rm gas} \propto d_A.
\end{equation}

We conclude that an SZE+X-ray determination of $d_A$ using an assumed evolution of \fgas\ would 
provide precisely the same sensitivity as \citet{allen2007} exploit, since we recover an additional
$d_A^{1/2}$ dependence through Eq.~\ref{eq:Tsl_dA}.  However, the additional, independent constraints
afforded by the inclusion of SZE data could provide a more precise determination than possible with
X-ray data alone.





\section{Final Words}

In this thesis, I have presented an exciting new tool for observing the Sunyaev-Zel'dovich
effect from galaxy clusters.  I have shown how data from this -- or any interferometric SZE instrument --
can be combined with X-ray imaging to place reasonable constraints on galaxy clusters, including 
at high redshift where the SZE can contribute most significantly to the understanding of large scale
structure.\footnote{These methods are, of course, more broadly applicable, and could be applied
to non-interferometric SZE instruments.}  Finally, I have shown how the new models can be combined 
with X-ray spectroscopy to improve
constraints and perform key measurements of the expansion of the Universe.

\bibliography{tonythesis}

\begin{thebibliography}{91}
\expandafter\ifx\csname natexlab\endcsname\relax\def\natexlab#1{#1}\fi

\bibitem[{{Afshordi} {et~al.}(2007){Afshordi}, {Lin}, {Nagai}, \&
  {Sanderson}}]{afshordi2007}
{Afshordi}, N., {Lin}, Y.-T., {Nagai}, D., \& {Sanderson}, A.~J.~R. 2007,
  \mnras, 378, 293

\bibitem[{{Albrecht} {et~al.}(2006){Albrecht}, {Bernstein}, {Cahn}, {Freedman},
  {Hewitt}, {Hu}, {Huth}, {Kamionkowski}, {Kolb}, {Knox}, {Mather}, {Staggs},
  \& {Suntzeff}}]{albrecht2006}
{Albrecht}, A., {Bernstein}, G., {Cahn}, R., {Freedman}, W.~L., {Hewitt}, J.,
  {Hu}, W., {Huth}, J., {Kamionkowski}, M., {Kolb}, E.~W., {Knox}, L.,
  {Mather}, J.~C., {Staggs}, S., \& {Suntzeff}, N.~B. 2006, ArXiv Astrophysics
  e-prints

\bibitem[{{Allen} {et~al.}(2007){Allen}, {Rapetti}, {Schmidt}, {Ebeling},
  {Morris}, \& {Fabian}}]{allen2007}
{Allen}, S.~W., {Rapetti}, D.~A., {Schmidt}, R.~W., {Ebeling}, H., {Morris},
  G., \& {Fabian}, A.~C. 2007, ArXiv e-prints, 706

\bibitem[{{Allen} {et~al.}(2004){Allen}, {Schmidt}, {Ebeling}, {Fabian}, \&
  {van Speybroeck}}]{allen2004}
{Allen}, S.~W., {Schmidt}, R.~W., {Ebeling}, H., {Fabian}, A.~C., \& {van
  Speybroeck}, L. 2004, \mnras, 353, 457

\bibitem[{{Ameglio} {et~al.}(2006){Ameglio}, {Borgani}, {Diaferio}, \&
  {Dolag}}]{ameglio2006}
{Ameglio}, S., {Borgani}, S., {Diaferio}, A., \& {Dolag}, K. 2006, \mnras, 369,
  1459

\bibitem[{{Anders} \& {Grevesse}(1989)}]{anders1989}
{Anders}, E. \& {Grevesse}, N. 1989, \gca, 53, 197

\bibitem[{{Arnaud}(1996)}]{arnaud1996}
{Arnaud}, K.~A. 1996, in Astronomical Data Analysis Software and Systems V, ed.
  G.~H. {Jacoby} \& J.~{Barnes}, 17

\bibitem[{{Best} {et~al.}(1995){Best}, {Cowles}, \& {Vines}}]{best1995}
{Best}, N.~G., {Cowles}, M.~K., \& {Vines}, S.~K. 1995, CODA Manual Version
  0.30 (MRC Biostatic Unit, Cambridge, UK)

\bibitem[{{Bonamente} {et~al.}(2008){Bonamente}, {Joy}, {LaRoque}, {Carlstrom},
  {Nagai}, \& {Marrone}}]{bonamente2008}
{Bonamente}, M., {Joy}, M., {LaRoque}, S.~J., {Carlstrom}, J.~E., {Nagai}, D.,
  \& {Marrone}, D.~P. 2008, \apj, 675, 106

\bibitem[{{Bonamente} {et~al.}(2004){Bonamente}, {Joy}, {Carlstrom}, {Reese},
  \& {LaRoque}}]{bonamente2004}
{Bonamente}, M., {Joy}, M.~K., {Carlstrom}, J.~E., {Reese}, E.~D., \&
  {LaRoque}, S.~J. 2004, \apj, 614, 56

\bibitem[{{Bonamente} {et~al.}(2006){Bonamente}, {Joy}, {LaRoque}, {Carlstrom},
  {Reese}, \& {Dawson}}]{bonamente2006}
{Bonamente}, M., {Joy}, M.~K., {LaRoque}, S.~J., {Carlstrom}, J.~E., {Reese},
  E.~D., \& {Dawson}, K.~S. 2006, ApJ, 647, 25

\bibitem[{{Bracewell}(2000)}]{bracewell2000}
{Bracewell}, R.~N. 2000, The Fourier Transform and its Applications (The
  Fourier Transform and its Applications / Ronald N.\ Bracewell.\ Boston :
  McGraw Hill, c2000.\ (McGraw-Hill series in electrical and computer
  engineering.\ Circuits and systems))

\bibitem[{Carlstrom {et~al.}(1998)Carlstrom, Grego, Holzapfel, \&
  Joy}]{carlstrom1998}
Carlstrom, J.~E., Grego, L., Holzapfel, W.~L., \& Joy, M. 1998, {Eighteenth
  Texas Symposium on Relativistic Astrophysics and Cosmology}, ed A. Olinto, J.
  Frieman, and D. Schramm, World Scientific, 261

\bibitem[{{Carlstrom} {et~al.}(2002){Carlstrom}, {Holder}, \&
  {Reese}}]{carlstrom2002}
{Carlstrom}, J.~E., {Holder}, G.~P., \& {Reese}, E.~D. 2002, \araa, 40, 643

\bibitem[{{Carlstrom} {et~al.}(2000){Carlstrom}, {Joy}, {Grego}, {Holder},
  {Holzapfel}, {Mohr}, {Patel}, \& {Reese}}]{carlstrom2000}
{Carlstrom}, J.~E., {Joy}, M.~K., {Grego}, L., {Holder}, G.~P., {Holzapfel},
  W.~L., {Mohr}, J.~J., {Patel}, S., \& {Reese}, E.~D. 2000, Physica Scripta
  Volume T, 85, 148

\bibitem[{{Cavaliere} \& {Fusco-Femiano}(1976)}]{cavaliere1976}
{Cavaliere}, A. \& {Fusco-Femiano}, R. 1976, \aap, 49, 137

\bibitem[{{Cavaliere} \& {Fusco-Femiano}(1978)}]{cavaliere1978}
---. 1978, \aap, 70, 677

\bibitem[{{Condon} {et~al.}(1998){Condon}, {Cotton}, {Greisen}, {Yin},
  {Perley}, {Taylor}, \& {Broderick}}]{condon1998}
{Condon}, J.~J., {Cotton}, W.~D., {Greisen}, E.~W., {Yin}, Q.~F., {Perley},
  R.~A., {Taylor}, G.~B., \& {Broderick}, J.~J. 1998, \aj, 115, 1693

\bibitem[{{da Silva} {et~al.}(2004){da Silva}, {Kay}, {Liddle}, \&
  {Thomas}}]{dasilva2004}
{da Silva}, A.~C., {Kay}, S.~T., {Liddle}, A.~R., \& {Thomas}, P.~A. 2004,
  \mnras, 348, 1401

\bibitem[{{David} {et~al.}(1995){David}, {Jones}, \& {Forman}}]{david1995}
{David}, L.~P., {Jones}, C., \& {Forman}, W. 1995, \apj, 445, 578

\bibitem[{{Ebeling} {et~al.}(2001){Ebeling}, {Edge}, \& {Henry}}]{ebeling2001}
{Ebeling}, H., {Edge}, A.~C., \& {Henry}, J.~P. 2001, \apj, 553, 668

\bibitem[{{Ettori}(2001)}]{ettori2001}
{Ettori}, S. 2001, in Astronomical Society of the Pacific Conference Series,
  Vol. 245, Astrophysical Ages and Times Scales, ed. T.~{von Hippel},
  C.~{Simpson}, \& N.~{Manset}, 500--+

\bibitem[{{Gilks} {et~al.}(1996){Gilks}, {Richardson}, \&
  {Spiegelhalter}}]{gilks1996}
{Gilks}, W.~R., {Richardson}, S., \& {Spiegelhalter}, D.~J. 1996, Markov Chain
  Monte Carlo in Practice (Chapman and Hall)

\bibitem[{Grego(1999)}]{grego1999a}
Grego, L. 1999, PhD thesis, California Institute of Technology

\bibitem[{{Grego} {et~al.}(2000){Grego}, {Carlstrom}, {Joy}, {Reese}, {Holder},
  {Patel}, {Cooray}, \& {Holzapfel}}]{grego2000}
{Grego}, L., {Carlstrom}, J.~E., {Joy}, M.~K., {Reese}, E.~D., {Holder}, G.~P.,
  {Patel}, S., {Cooray}, A.~R., \& {Holzapfel}, W.~L. 2000, \apj, 539, 39

\bibitem[{{Grego} {et~al.}(2001){Grego}, {Carlstrom}, {Reese}, {Holder},
  {Holzapfel}, {Joy}, {Mohr}, \& {Patel}}]{grego2001}
{Grego}, L., {Carlstrom}, J.~E., {Reese}, E.~D., {Holder}, G.~P., {Holzapfel},
  W.~L., {Joy}, M.~K., {Mohr}, J.~J., \& {Patel}, S. 2001, \apj, 552, 2

\bibitem[{{Haiman} {et~al.}(2001){Haiman}, {Mohr}, \& {Holder}}]{haiman2001}
{Haiman}, Z., {Mohr}, J.~J., \& {Holder}, G.~P. 2001, \apj, 553, 545

\bibitem[{{Hawkins} {et~al.}(2004){Hawkins}, {Woody}, {Wiitala}, {Fredsti}, \&
  {Rauch}}]{hawkins2004}
{Hawkins}, D.~W., {Woody}, D.~P., {Wiitala}, B., {Fredsti}, J., \& {Rauch},
  K.~P. 2004, in Presented at the Society of Photo-Optical Instrumentation
  Engineers (SPIE) Conference, Vol. 5498, Millimeter and Submillimeter
  Detectors for Astronomy II. Edited by Jonas Zmuidzinas, Wayne S. Holland and
  Stafford Withington Proceedings of the SPIE, Volume 5498, pp. 567-578
  (2004)., ed. C.~M. {Bradford}, P.~A.~R. {Ade}, J.~E. {Aguirre}, J.~J. {Bock},
  M.~{Dragovan}, L.~{Duband}, L.~{Earle}, J.~{Glenn}, H.~{Matsuhara}, B.~J.
  {Naylor}, H.~T. {Nguyen}, M.~{Yun}, \& J.~{Zmuidzinas}, 567--578

\bibitem[{{Hinshaw} {et~al.}(2008){Hinshaw}, {Weiland}, {Hill}, {Odegard},
  {Larson}, {Bennett}, {Dunkley}, {Gold}, {Greason}, {Jarosik}, {Komatsu},
  {Nolta}, {Page}, {Spergel}, {Wollack}, {Halpern}, {Kogut}, {Limon}, {Meyer},
  {Tucker}, \& {Wright}}]{hinshaw2008}
{Hinshaw}, G., {Weiland}, J.~L., {Hill}, R.~S., {Odegard}, N., {Larson}, D.,
  {Bennett}, C.~L., {Dunkley}, J., {Gold}, B., {Greason}, M.~R., {Jarosik}, N.,
  {Komatsu}, E., {Nolta}, M.~R., {Page}, L., {Spergel}, D.~N., {Wollack}, E.,
  {Halpern}, M., {Kogut}, A., {Limon}, M., {Meyer}, S.~S., {Tucker}, G.~S., \&
  {Wright}, E.~L. 2008, ArXiv e-prints, 803

\bibitem[{{Holder} {et~al.}(2000){Holder}, {Mohr}, {Carlstrom}, {Evrard}, \&
  {Leitch}}]{holder2000}
{Holder}, G.~P., {Mohr}, J.~J., {Carlstrom}, J.~E., {Evrard}, A.~E., \&
  {Leitch}, E.~M. 2000, \apj, 544, 629

\bibitem[{{Itoh} {et~al.}(1998){Itoh}, {Kohyama}, \& {Nozawa}}]{itoh1998}
{Itoh}, N., {Kohyama}, Y., \& {Nozawa}, S. 1998, \apj, 502, 7

\bibitem[{{Jing} \& {Suto}(2000)}]{jing2000}
{Jing}, Y.~P. \& {Suto}, Y. 2000, \apjl, 529, L69

\bibitem[{{Joy} {et~al.}(2001){Joy}, {LaRoque}, {Grego}, {Carlstrom}, {Dawson},
  {Ebeling}, {Holzapfel}, {Nagai}, \& {Reese}}]{joy2001}
{Joy}, M., {LaRoque}, S., {Grego}, L., {Carlstrom}, J.~E., {Dawson}, K.,
  {Ebeling}, H., {Holzapfel}, W.~L., {Nagai}, D., \& {Reese}, E.~D. 2001,
  \apjl, 551, L1

\bibitem[{{Kaastra} \& {Mewe}(1993)}]{kaastra1993}
{Kaastra}, J.~S. \& {Mewe}, R. 1993, \aaps, 97, 443

\bibitem[{{Kravtsov} {et~al.}(2005){Kravtsov}, {Nagai}, \&
  {Vikhlinin}}]{kravtsov2005}
{Kravtsov}, A.~V., {Nagai}, D., \& {Vikhlinin}, A.~A. 2005, \apj, 625, 588

\bibitem[{{Kravtsov} {et~al.}(2006){Kravtsov}, {Vikhlinin}, \&
  {Nagai}}]{kravtsov2006}
{Kravtsov}, A.~V., {Vikhlinin}, A., \& {Nagai}, D. 2006, \apj, 650, 128

\bibitem[{{LaRoque} {et~al.}(2006){LaRoque}, {Bonamente}, {Carlstrom}, {Joy},
  {Nagai}, {Reese}, \& {Dawson}}]{laroque2006}
{LaRoque}, S.~J., {Bonamente}, M., {Carlstrom}, J.~E., {Joy}, M.~K., {Nagai},
  D., {Reese}, E.~D., \& {Dawson}, K.~S. 2006, \apj, 652, 917

\bibitem[{{LaRoque} {et~al.}(2003){LaRoque}, {Joy}, {Carlstrom}, {Ebeling},
  {Bonamente}, {Dawson}, {Edge}, {Holzapfel}, {Miller}, {Nagai}, {Patel}, \&
  {Reese}}]{laroque2003}
{LaRoque}, S.~J., {Joy}, M., {Carlstrom}, J.~E., {Ebeling}, H., {Bonamente},
  M., {Dawson}, K.~S., {Edge}, A., {Holzapfel}, W.~L., {Miller}, A.~D.,
  {Nagai}, D., {Patel}, S.~K., \& {Reese}, E.~D. 2003, \apj, 583, 559

\bibitem[{{Leitch} {et~al.}(2005){Leitch}, {Carlstrom}, {Davidson}, {Dragovan},
  {Halverson}, {Holzapfel}, {Laroque}, {Kovac}, {Pryke}, {Schartman}, \&
  {Yamasaki}}]{leitch2005}
{Leitch}, E.~M., {Carlstrom}, J.~E., {Davidson}, G., {Dragovan}, M.,
  {Halverson}, N.~W., {Holzapfel}, W.~L., {Laroque}, S., {Kovac}, J., {Pryke},
  C., {Schartman}, E., \& {Yamasaki}, M.~J. 2005, in IAU Symposium, Vol. 201,
  New Cosmological Data and the Values of the Fundamental Parameters, ed. A.~N.
  {Lasenby} \& A.~{Wilkinson}, 33--+

\bibitem[{{Liedahl} {et~al.}(1995){Liedahl}, {Osterheld}, \&
  {Goldstein}}]{liedahl1995}
{Liedahl}, D.~A., {Osterheld}, A.~L., \& {Goldstein}, W.~H. 1995, \apjl, 438,
  L115

\bibitem[{{Loken} {et~al.}(2002){Loken}, {Norman}, {Nelson}, {Burns}, {Bryan},
  \& {Motl}}]{loken2002}
{Loken}, C., {Norman}, M.~L., {Nelson}, E., {Burns}, J., {Bryan}, G.~L., \&
  {Motl}, P. 2002, \apj, 579, 571

\bibitem[{{Longair}(1998)}]{longair1998}
{Longair}, M.~S., ed. 1998, {Galaxy formation}

\bibitem[{{Markevitch} {et~al.}(1998){Markevitch}, {Forman}, {Sarazin}, \&
  {Vikhlinin}}]{markevitch1998}
{Markevitch}, M., {Forman}, W.~R., {Sarazin}, C.~L., \& {Vikhlinin}, A. 1998,
  \apj, 503, 77

\bibitem[{{Maughan}(2007)}]{maughan2007}
{Maughan}, B.~J. 2007, ArXiv Astrophysics e-prints

\bibitem[{{Maughan} {et~al.}(2008){Maughan}, {Jones}, {Forman}, \& {Van
  Speybroeck}}]{maughan2008}
{Maughan}, B.~J., {Jones}, C., {Forman}, W., \& {Van Speybroeck}, L. 2008,
  \apjs, 174, 117

\bibitem[{{Maughan} {et~al.}(2007){Maughan}, {Jones}, {Jones}, \& {Van
  Speybroeck}}]{maughan2007b}
{Maughan}, B.~J., {Jones}, C., {Jones}, L.~R., \& {Van Speybroeck}, L. 2007,
  \apj, 659, 1125

\bibitem[{{Mazzotta} {et~al.}(2004){Mazzotta}, {Rasia}, {Moscardini}, \&
  {Tormen}}]{mazzotta2004}
{Mazzotta}, P., {Rasia}, E., {Moscardini}, L., \& {Tormen}, G. 2004, \mnras,
  354, 10

\bibitem[{{Mewe} {et~al.}(1985){Mewe}, {Gronenschild}, \& {van den
  Oord}}]{mewe1985}
{Mewe}, R., {Gronenschild}, E.~H.~B.~M., \& {van den Oord}, G.~H.~J. 1985,
  \aaps, 62, 197

\bibitem[{{Molnar} {et~al.}(2002){Molnar}, {Birkinshaw}, \&
  {Mushotzky}}]{molnar2002}
{Molnar}, S.~M., {Birkinshaw}, M., \& {Mushotzky}, R.~F. 2002, \apj, 570, 1

\bibitem[{{Moore} {et~al.}(1999){Moore}, {Quinn}, {Governato}, {Stadel}, \&
  {Lake}}]{moore1999}
{Moore}, B., {Quinn}, T., {Governato}, F., {Stadel}, J., \& {Lake}, G. 1999,
  \mnras, 310, 1147

\bibitem[{{Mroczkowski} {et~al.}(2008){Mroczkowski}, {Bonamente}, {Carlstrom},
  {Culverhouse}, {Greer}, {Hawkins}, {Hennessy}, {Joy}, {Lamb}, {Leitch},
  {Loh}, {Maughan}, {Marrone}, {Miller}, {Nagai}, {Muchovej}, {Pryke}, {Sharp},
  \& {Woody}}]{mroczkowski2009}
{Mroczkowski}, T., {Bonamente}, M., {Carlstrom}, J.~E., {Culverhouse}, T.~L.,
  {Greer}, C., {Hawkins}, D., {Hennessy}, R., {Joy}, M., {Lamb}, J.~W.,
  {Leitch}, E.~M., {Loh}, M., {Maughan}, B., {Marrone}, D.~P., {Miller}, A.,
  {Nagai}, D., {Muchovej}, S., {Pryke}, C., {Sharp}, M., \& {Woody}, D. 2008,
  ArXiv e-prints

\bibitem[{{Muchovej}(2008)}]{muchovej2008}
{Muchovej}, S. 2008, PhD thesis, Columbia University

\bibitem[{{Muchovej} {et~al.}(2007){Muchovej}, {Mroczkowski}, {Carlstrom},
  {Cartwright}, {Greer}, {Hennessy}, {Loh}, {Pryke}, {Reddall}, {Runyan},
  {Sharp}, {Hawkins}, {Lamb}, {Woody}, {Joy}, {Leitch}, \&
  {Miller}}]{muchovej2007}
{Muchovej}, S., {Mroczkowski}, T., {Carlstrom}, J.~E., {Cartwright}, J.,
  {Greer}, C., {Hennessy}, R., {Loh}, M., {Pryke}, C., {Reddall}, B., {Runyan},
  M., {Sharp}, M., {Hawkins}, D., {Lamb}, J.~W., {Woody}, D., {Joy}, M.,
  {Leitch}, E.~M., \& {Miller}, A.~D. 2007, \apj, 663, 708

\bibitem[{{Mushotzky} \& {Scharf}(1997)}]{mushotzky1997}
{Mushotzky}, R.~F. \& {Scharf}, C.~A. 1997, \apjl, 482, L13+

\bibitem[{{Nagai}(2006)}]{nagai2006}
{Nagai}, D. 2006, \apj, 650, 538

\bibitem[{{Nagai} {et~al.}(2007{\natexlab{a}}){Nagai}, {Kravtsov}, \&
  {Vikhlinin}}]{nagai2007b}
{Nagai}, D., {Kravtsov}, A.~V., \& {Vikhlinin}, A. 2007{\natexlab{a}}, \apj,
  668, 1

\bibitem[{{Nagai} {et~al.}(2007{\natexlab{b}}){Nagai}, {Vikhlinin}, \&
  {Kravtsov}}]{nagai2007}
{Nagai}, D., {Vikhlinin}, A., \& {Kravtsov}, A.~V. 2007{\natexlab{b}}, \apj,
  655, 98

\bibitem[{{Navarro} {et~al.}(1997){Navarro}, {Frenk}, \& {White}}]{navarro1997}
{Navarro}, J.~F., {Frenk}, C.~S., \& {White}, S. D.~M. 1997, \apj, 490, 493

\bibitem[{{Neumann}(2006)}]{neumann2006}
{Neumann}, D.~M. 2006, in EAS Publications Series, Vol.~20, EAS Publications
  Series, ed. G.~A. {Mamon}, F.~{Combes}, C.~{Deffayet}, \& B.~{Fort}, 179--182

\bibitem[{NIST(2008)}]{nistweb}
NIST. 2008, Cryogenic Technologies Group -- Cryogenics Material Properties,
  \url{http://www.cryogenics.nist.gov/MPropsMAY/material%20properties.htm}

\bibitem[{Oppenheim {et~al.}(1999)Oppenheim, Schafer, \& Buck}]{oppenheim1999}
Oppenheim, A.~V., Schafer, R.~W., \& Buck, J.~R. 1999, Discrete-time signal
  processing (2nd ed.) (Upper Saddle River, NJ, USA: Prentice-Hall, Inc.)

\bibitem[{{Patel} {et~al.}(2000){Patel}, {Joy}, {Carlstrom}, {Holder}, {Reese},
  {Gomez}, {Hughes}, {Grego}, \& {Holzapfel}}]{patel2000}
{Patel}, S.~K., {Joy}, M., {Carlstrom}, J.~E., {Holder}, G.~P., {Reese}, E.~D.,
  {Gomez}, P.~L., {Hughes}, J.~P., {Grego}, L., \& {Holzapfel}, W.~L. 2000,
  \apj, 541, 37

\bibitem[{{Peterson} {et~al.}(2001){Peterson}, {Paerels}, {Kaastra}, {Arnaud},
  {Reiprich}, {Fabian}, {Mushotzky}, {Jernigan}, \& {Sakelliou}}]{peterson2001}
{Peterson}, J.~R., {Paerels}, F.~B.~S., {Kaastra}, J.~S., {Arnaud}, M.,
  {Reiprich}, T.~H., {Fabian}, A.~C., {Mushotzky}, R.~F., {Jernigan}, J.~G., \&
  {Sakelliou}, I. 2001, \aap, 365, L104

\bibitem[{{Pfrommer} {et~al.}(2007){Pfrommer}, {En{\ss}lin}, {Springel},
  {Jubelgas}, \& {Dolag}}]{pfrommer2007}
{Pfrommer}, C., {En{\ss}lin}, T.~A., {Springel}, V., {Jubelgas}, M., \&
  {Dolag}, K. 2007, \mnras, 378, 385

\bibitem[{{Piffaretti} {et~al.}(2005){Piffaretti}, {Jetzer}, {Kaastra}, \&
  {Tamura}}]{piffaretti2005}
{Piffaretti}, R., {Jetzer}, P., {Kaastra}, J.~S., \& {Tamura}, T. 2005, \aap,
  433, 101

\bibitem[{{Plummer} {et~al.}(2006){Plummer}, {Best}, {Cowles}, \&
  {Vines}}]{plummer2006}
{Plummer}, M., {Best}, N.~G., {Cowles}, M.~K., \& {Vines}, S.~K. 2006, R News,
  6, 7

\bibitem[{{Pratt} \& {Arnaud}(2002)}]{pratt2002}
{Pratt}, G.~W. \& {Arnaud}, M. 2002, \aap, 394, 375

\bibitem[{{Pratt} {et~al.}(2007){Pratt}, {B{\"o}hringer}, {Croston}, {Arnaud},
  {Borgani}, {Finoguenov}, \& {Temple}}]{pratt2007}
{Pratt}, G.~W., {B{\"o}hringer}, H., {Croston}, J.~H., {Arnaud}, M., {Borgani},
  S., {Finoguenov}, A., \& {Temple}, R.~F. 2007, \aap, 461, 71

\bibitem[{{Raftery} \& {Lewis}(1992)}]{raftery1992}
{Raftery}, A.~L. \& {Lewis}, S. 1992, in Bayesian Statistics IV, ed. J.~M.
  {Bernardo} \& M.~H. {DeGroot} (Oxford University Press), 763

\bibitem[{{Rapetti} \& {Allen}(2007)}]{rapetti2007}
{Rapetti}, D. \& {Allen}, S.~W. 2007, ArXiv e-prints, 710

\bibitem[{{Raymond} \& {Smith}(1977)}]{raymond1977}
{Raymond}, J.~C. \& {Smith}, B.~W. 1977, \apjs, 35, 419

\bibitem[{{Reese} {et~al.}(2002){Reese}, {Carlstrom}, {Joy}, {Mohr}, {Grego},
  \& {Holzapfel}}]{reese2002}
{Reese}, E.~D., {Carlstrom}, J.~E., {Joy}, M., {Mohr}, J.~J., {Grego}, L., \&
  {Holzapfel}, W.~L. 2002, \apj, 581, 53

\bibitem[{{Reese} {et~al.}(2000){Reese}, {Mohr}, {Carlstrom}, {Joy}, {Grego},
  {Holder}, {Holzapfel}, {Hughes}, {Patel}, \& {Donahue}}]{reese2000}
{Reese}, E.~D., {Mohr}, J.~J., {Carlstrom}, J.~E., {Joy}, M., {Grego}, L.,
  {Holder}, G.~P., {Holzapfel}, W.~L., {Hughes}, J.~P., {Patel}, S.~K., \&
  {Donahue}, M. 2000, \apj, 533, 38

\bibitem[{{Riess} {et~al.}(1998){Riess}, {Filippenko}, {Challis},
  {Clocchiatti}, {Diercks}, {Garnavich}, {Gilliland}, {Hogan}, {Jha},
  {Kirshner}, {Leibundgut}, {Phillips}, {Reiss}, {Schmidt}, {Schommer},
  {Smith}, {Spyromilio}, {Stubbs}, {Suntzeff}, \& {Tonry}}]{riess1998}
{Riess}, A.~G., {Filippenko}, A.~V., {Challis}, P., {Clocchiatti}, A.,
  {Diercks}, A., {Garnavich}, P.~M., {Gilliland}, R.~L., {Hogan}, C.~J., {Jha},
  S., {Kirshner}, R.~P., {Leibundgut}, B., {Phillips}, M.~M., {Reiss}, D.,
  {Schmidt}, B.~P., {Schommer}, R.~A., {Smith}, R.~C., {Spyromilio}, J.,
  {Stubbs}, C., {Suntzeff}, N.~B., \& {Tonry}, J. 1998, \aj, 116, 1009

\bibitem[{{Rohlfs} \& {Wilson}(1996)}]{rohlfsw96}
{Rohlfs}, K. \& {Wilson}, T.~L. 1996, {Tools of Radio Astronomy}, 2nd edn.
  (Berlin: Springer)

\bibitem[{{Sarazin}(1988)}]{sarazin1988}
{Sarazin}, C.~L. 1988, X-ray Emission From Clusters of Galaxies (Cambridge
  University Press)

\bibitem[{{Scott} \& {Pound}(2006)}]{scott2006}
{Scott}, S.~L. \& {Pound}, M.~W. 2006, in Astronomical Society of the Pacific
  Conference Series, Vol. 351, Astronomical Data Analysis Software and Systems
  XV, ed. C.~{Gabriel}, C.~{Arviset}, D.~{Ponz}, \& S.~{Enrique}, 670--+

\bibitem[{{Shepherd}(1997)}]{shepherd1997}
{Shepherd}, M.~C. 1997, in Astronomical Society of the Pacific Conference
  Series, Vol. 125, Astronomical Data Analysis Software and Systems VI, ed.
  G.~{Hunt} \& H.~{Payne}, 77--+

\bibitem[{{Smith} {et~al.}(2001){Smith}, {Brickhouse}, {Liedahl}, \&
  {Raymond}}]{smith2001}
{Smith}, R.~K., {Brickhouse}, N.~S., {Liedahl}, D.~A., \& {Raymond}, J.~C.
  2001, \apjl, 556, L91

\bibitem[{{Spergel} {et~al.}(2003){Spergel}, {Verde}, {Peiris}, {Komatsu},
  {Nolta}, {Bennett}, {Halpern}, {Hinshaw}, {Jarosik}, {Kogut}, {Limon},
  {Meyer}, {Page}, {Tucker}, {Weiland}, {Wollack}, \& {Wright}}]{spergel2003}
{Spergel}, D.~N., {Verde}, L., {Peiris}, H.~V., {Komatsu}, E., {Nolta}, M.~R.,
  {Bennett}, C.~L., {Halpern}, M., {Hinshaw}, G., {Jarosik}, N., {Kogut}, A.,
  {Limon}, M., {Meyer}, S.~S., {Page}, L., {Tucker}, G.~S., {Weiland}, J.~L.,
  {Wollack}, E., \& {Wright}, E.~L. 2003, \apjs, 148, 175

\bibitem[{{Struble} \& {Rood}(1999)}]{struble1999}
{Struble}, M.~F. \& {Rood}, H.~J. 1999, \apjs, 125, 35

\bibitem[{{Thompson} {et~al.}(2001){Thompson}, {Moran}, \&
  {Swenson}}]{thompson2001}
{Thompson}, A.~R., {Moran}, J.~M., \& {Swenson}, G.~W. 2001, {Interferometry
  and Synthesis in Radio Astronomy} (Wiley-Interscience, 2nd ed.)

\bibitem[{{van Speybroeck}(1999)}]{vanspeybroeck1999}
{van Speybroeck}, L. 1999, American Astronomical Society Meeting, 31, 917

\bibitem[{{Vikhlinin}(2006)}]{vikhlinin2006a}
{Vikhlinin}, A. 2006, \apj, 640, 710

\bibitem[{{Vikhlinin} {et~al.}(2008){Vikhlinin}, {Burenin}, {Ebeling},
  {Forman}, {Hornstrup}, {Jones}, {Kravtsov}, {Murray}, {Nagai}, {Quintana}, \&
  {Voevodkin}}]{vikhlinin2008}
{Vikhlinin}, A., {Burenin}, R.~A., {Ebeling}, H., {Forman}, W.~R., {Hornstrup},
  A., {Jones}, C., {Kravtsov}, A.~V., {Murray}, S.~S., {Nagai}, D., {Quintana},
  H., \& {Voevodkin}, A. 2008, ArXiv e-prints, 805

\bibitem[{{Vikhlinin} {et~al.}(2006){Vikhlinin}, {Kravtsov}, {Forman}, {Jones},
  {Markevitch}, {Murray}, \& {Van Speybroeck}}]{vikhlinin2006}
{Vikhlinin}, A., {Kravtsov}, A., {Forman}, W., {Jones}, C., {Markevitch}, M.,
  {Murray}, S.~S., \& {Van Speybroeck}, L. 2006, \apj, 640, 691

\bibitem[{{Vikhlinin} {et~al.}(2005){Vikhlinin}, {Markevitch}, {Murray},
  {Jones}, {Forman}, \& {Van Speybroeck}}]{vikhlinin2005a}
{Vikhlinin}, A., {Markevitch}, M., {Murray}, S.~S., {Jones}, C., {Forman}, W.,
  \& {Van Speybroeck}, L. 2005, \apj, 628, 655

\bibitem[{{White} {et~al.}(1997){White}, {Becker}, {Helfand}, \&
  {Gregg}}]{white1997}
{White}, R.~L., {Becker}, R.~H., {Helfand}, D.~J., \& {Gregg}, M.~D. 1997,
  \apj, 475, 479

\bibitem[{{White} {et~al.}(1993){White}, {Navarro}, {Evrard}, \&
  {Frenk}}]{white1993a}
{White}, S. D.~M., {Navarro}, J.~F., {Evrard}, A.~E., \& {Frenk}, C.~S. 1993,
  \nat, 366, 429

\bibitem[{{Woodcraft}(2005)}]{woodcraft2005}
{Woodcraft}, A.~L. 2005, Cryogenics, 45, 626

\bibitem[{{Woody} {et~al.}(2004){Woody}, {Beasley}, {Bolatto}, {Carlstrom},
  {Harris}, {Hawkins}, {Lamb}, {Looney}, {Mundy}, {Plambeck}, {Scott}, \&
  {Wright}}]{woody2004}
{Woody}, D.~P., {Beasley}, A.~J., {Bolatto}, A.~D., {Carlstrom}, J.~E.,
  {Harris}, A., {Hawkins}, D.~W., {Lamb}, J., {Looney}, L., {Mundy}, L.~G.,
  {Plambeck}, R.~L., {Scott}, S., \& {Wright}, M. 2004, in Presented at the
  Society of Photo-Optical Instrumentation Engineers (SPIE) Conference, Vol.
  5498, Millimeter and Submillimeter Detectors for Astronomy II. Edited by
  Jonas Zmuidzinas, Wayne S. Holland and Stafford Withington Proceedings of the
  SPIE, Volume 5498, pp. 30-41 (2004)., ed. C.~M. {Bradford}, P.~A.~R. {Ade},
  J.~E. {Aguirre}, J.~J. {Bock}, M.~{Dragovan}, L.~{Duband}, L.~{Earle},
  J.~{Glenn}, H.~{Matsuhara}, B.~J. {Naylor}, H.~T. {Nguyen}, M.~{Yun}, \&
  J.~{Zmuidzinas}, 30--41

\end{thebibliography}

\part{Appendix \label{part-end}}
\appendix
\chapter{\emph{Chandra} X-ray Data Analysis}\label{app2}
 
The X-ray data used in this analysis were obtained with the {\em
Chandra} ACIS-I detector, which provides spatially resolved X-ray
spectroscopy and imaging with an angular resolution of
$0.492\arcsec$ and with energy resolution of $\sim$ 100-200
eV.  Data analysis was performed with the CIAO\footnote{{\em Chandra} Interactive Analysis of
Observations, \url{http://cxc.harvard.edu/ciao/}.} software (version 3.2)
and the CALDB calibration information (version 3.1) provided by the
{\em Chandra} calibration team.


Both images and spectra of the low-redshift clusters were limited to the 0.7--7.0~keV
energy band in order to exclude the low-energy and high-energy data that are more 
strongly affected by background and by calibration uncertainties.  
For the high-redshift cluster, CL1226 ($z=0.89$), the image was limited to the 0.7--2.0~keV
band, where \emph{Chandra}'s efficiency peaks (see Fig.~\ref{fig:effarea}).
This range of energies was chosen because, with a spectroscopic temperature of $\sim$9.8~keV, only
$\sim 50 \%$ of the redshifted cluster emission is measured at photon energies $\gtrsim 2$~keV; However, the
X-ray background in a 0.7--7.0~keV image is $\sim 4$ times higher than that in a 0.7--2.0~keV
image.
Combining this with the fact that the ACIS-I efficiency peaks below 2~keV (see Fig.~\ref{fig:effarea}), 
the highest $S/N$ is obtained in the detector energy range 0.7--2.0~keV.

The X-ray images were binned in 1.968$\arcsec$ pixels; this sets the limiting
angular resolution of our X-ray data, as the {\em Chandra} point
spread function in the center of the X-ray image is smaller than our
adopted pixel size.  The X-ray background was measured for each
cluster exposure, using peripheral regions of the adjacent detector
ACIS-I chips that are source free.  Additional details of the {\em
Chandra} X-ray data analysis are presented in
\citet{bonamente2004,bonamente2006}.


\end{document}